\documentclass[a4paper,11pt]{article}
\pdfoutput=1
\usepackage{jheppub}

\input{./utilities/packages.tex}
\newcommand{\clicsid}{\textsc{CLIC}\_\text{SiD}\xspace}
\newcommand{\clicild}{\textsc{CLIC}\_\text{ILD}\xspace}

\newcommand{\roostats}{RooStats\xspace}
\newcommand{\guineapig}{\textsc{GuineaPig}\xspace}
\newcommand{\mokka}{\textsc{Mokka}\xspace}
\newcommand{\marlin}{\textsc{Marlin}\xspace}
\newcommand{\geant}{\textsc{Geant4}\xspace}
\newcommand{\slic}{SLIC\xspace}

\newcommand{\lcsim}{\texttt{org.lcsim}\xspace}
\newcommand{\pythia}{\textsc{PYTHIA}\xspace}
\newcommand{\whizard}{\textsc{WHIZARD}\xspace}
\newcommand{\pandora}{\textsc{PandoraPFA}\xspace}
\newcommand{\fastjet}{\textsc{FastJet}\xspace}
\newcommand{\madgraph}{\textsc{MadGraph5\_aMC@NLO}\xspace}

\newcommand{\ilcdirac}{\textsc{iLCDirac}\xspace}
\newcommand{\lcfiplus}{\textsc{LcfiPlus}\xspace}

\newcommand{\tauola}{\textsc{Tauola}\xspace}
\newcommand{\physsim}{\textsc{PhysSim}\xspace}

\newcommand\gamgam {\ensuremath{\PGg\PGg}\xspace}
\newcommand\gghadrons {\ensuremath{\gamgam \to \mathrm{hadrons}}\xspace}

\newcommand{\roots}{\ensuremath{\sqrt{s}}\xspace}
\newcommand{\rootsprime}{\ensuremath{\sqrt{s^\prime}}\xspace}
\newcommand{\rootsprimereco}{\ensuremath{\sqrt{s_{R}^\prime}}\xspace}

\newcommand{\code}{\texttt}

\newcommand{\pT}{\ensuremath{p_\mathrm{T}}\xspace}
\newcommand{\kT}{\ensuremath{k_{\mathrm{t}}}\xspace}

\newcommand{\fb}{\ensuremath{\text{fb}}\xspace}
\newcommand{\ab}{\ensuremath{\text{ab}}\xspace}
\newcommand{\abinv}{\ensuremath{\text{ab}^{-1}}\xspace}
\newcommand{\fbinv}{\ensuremath{\text{fb}^{-1}}\xspace}

\newcommand{\micron}{\ensuremath{\upmu\mathrm{m}}\xspace}

\newcommand{\mrad}{\ensuremath{\text{mrad}}\xspace}

\newcommand{\Dd}{\overset{\raisebox{-3pt}[0pt][0pt]{\tiny$\small\leftrightarrow$}}{D}}

\newcommand{\epem}{\ensuremath{\Pep\Pem}\xspace}   %e+e-  
\newcommand{\ttbar}{\ensuremath{\PQt\PAQt}\xspace}

\newcommand{\qqbar}{\ensuremath{\PQq\PAQq}\xspace}

\newcommand{\ttH}{\ensuremath{\PQt\PAQt\PH}\xspace} 

\def\sqsq #1 {\ensuremath{\PSQ_{\text{#1}}\PSQ_{\stext{#1}}}\xspace} %stau %stau

\def\cW {\ensuremath{\mathrm{c}_{\PW}}\xspace}
\def\sW {\ensuremath{\mathrm{s}_{\PW}}\xspace}

\newcommand{\tev}{\ensuremath{\mathrm{\,Te\kern -0.1em V}}\xspace}
\newcommand{\gev}{\ensuremath{\mathrm{\,Ge\kern -0.1em V}}\xspace}
\newcommand{\mev}{\ensuremath{\mathrm{\,Me\kern -0.1em V}}\xspace}
\newcommand{\kev}{\ensuremath{\mathrm{\,ke\kern -0.1em V}}\xspace}
\newcommand{\ev}{\ensuremath{\mathrm{\,e\kern -0.1em V}}\xspace}
\newcommand{\gevc}{\ensuremath{{\mathrm{\,Ge\kern -0.1em V\!/}c}}\xspace}
\newcommand{\mevc}{\ensuremath{{\mathrm{\,Me\kern -0.1em V\!/}c}}\xspace}
\newcommand{\gevcc}{\ensuremath{{\mathrm{\,Ge\kern -0.1em V\!/}c^2}}\xspace}
\newcommand{\gevgevcccc}{\ensuremath{{\mathrm{\,Ge\kern -0.1em V^2\!/}c^4}}\xspace}
\newcommand{\mevcc}{\ensuremath{{\mathrm{\,Me\kern -0.1em V\!/}c^2}}\xspace}
\newcommand{\tevcc}{\ensuremath{{\mathrm{\,Te\kern -0.1em V\!/}c^2}}\xspace}

\def\barn{\ensuremath{\rm \,b}\xspace}

\def\pb {\ensuremath{\rm \,pb}\xspace}

\def\fb   {\ensuremath{\mbox{\,fb}}\xspace}

\def\invab   {\ensuremath{\mbox{\,ab}^{-1}}\xspace}

\def\mrad{\ensuremath{\rm \,mrad}\xspace}
\def\rad{\ensuremath{\rm \,rad}\xspace}

\newcommand{\GeV}{\ensuremath{\text{GeV}}\xspace}
\newcommand{\MeV}{\ensuremath{\text{MeV}}\xspace}
\newcommand{\TeV}{\ensuremath{\text{TeV}}\xspace}

\newcommand{\tabt}[1]{\multicolumn{1}{c}{#1}}

\newcommand{\LumiIntDiff}{\ensuremath{\int\frac{d \mathcal{L}}{d s'} ds'}}

\newcommand{\eett}{\ensuremath{\Pep\Pem\!\to\PQt\PAQt}\xspace} 
\newcommand{\afb}{\ensuremath{A_\mathrm{FB}}\xspace}
\newcommand{\csttbar}{\sigma_{\,\ttbar}\xspace}

\newcommand{\tch}{\ensuremath{\HepProcess{\PQt \to \PQc \PH}}\xspace}
\newcommand{\tqh}{\ensuremath{\HepProcess{\PQt \to \PQq \PH}}\xspace}
\newcommand{\tcg}{\ensuremath{\HepProcess{\PQt \to \PQc \PGg}}\xspace}
\newcommand{\tcx}{\ensuremath{\HepProcess{\PQt \to \PQc \HepParticle{E}\!\!\!\!\!\!\!\slash}}\xspace}
\newcommand{\hbb}{\ensuremath{\HepProcess{\PH \to \PQb \PAQb}}\xspace}
\newcommand{\CLs}{\ensuremath{\mathrm{CL}_\mathrm{s}}\xspace}

\title{\boldmath Top-Quark Physics at the CLIC Electron-Positron Linear Collider}

\collaborationImg{\includegraphics[width=0.12\columnwidth]{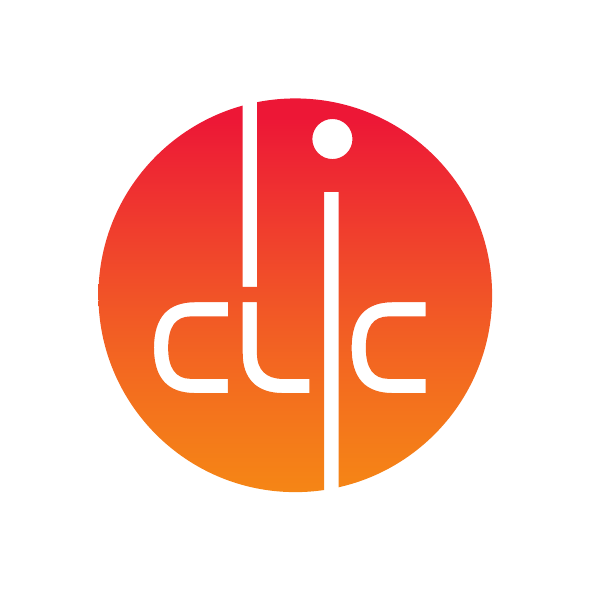}\\ \vspace{2mm} \boldmath{\Large{CLICdp collaboration}}}
\emailAdd{clicdp-top-paper-editors@cern.ch}

\def \TelAviv {1}
\def \CERN {2}
\def \Geneva {3}
\def \Annecy {4}
\def \Valencia {5}
\def \Birmingham {6}
\def \Belgrade {7}
\def \JINR {8}
\def \Siegen {9}
\def \IFJPANCracow {10}
\def \Liverpool {11}
\def \Oxford {12}
\def \Argonne {13}
\def \Sussex {14}
\def \AIBU {15}
\def \DESY {16}
\def \Bergen {17}
\def \Munich {18}
\def \Bristol {19}
\def \Cambridge {20}
\def \Humboldt {21}
\def \ViennaA {22}
\def \ViennaB {23}
\def \Warsaw {24}
\def \Minsk {25}
\def \Edinburgh {26}
\def \Salamanca {27}
\def \UAMCSIC {28}
\def \Glasgow {29}
\def \Santander {30}
\def \Michigan {31}
\def \DAMTP {32}
\def \IHEP {33}

\def \alsoWarsawUTech {a}
\def \alsoCERN        {b}
\def \nowBonn         {c}
\def \nowNikhef       {d}
\def \nowNikhefUtrecht{e}
\def \nowKIT          {f}
\def \alsoVienna      {g}
\def \alsoAachen      {h}
\def \nowPSI          {i}
\def \nowMainz        {j}
\def \alsoGlasgow     {k}

\author[\TelAviv]{H.~Abramowicz,}
\author[\CERN]{N.~Alipour~Tehrani,}
\author[\CERN, \alsoWarsawUTech]{D.~Arominski,}
\author[\TelAviv]{Y.~Benhammou,}
\author[\Geneva]{M.~Benoit,}
\author[\Annecy]{J.-J.~Blaising,}
\author[\Valencia]{M.~Boronat,}
\author[\TelAviv]{O.~Borysov,}
\author[\Birmingham]{R.R.~Bosley,}
\author[\Belgrade]{I.~Bo\v{z}ovi\'{c} Jelisav\v{c}i\'{c},}
% ...

\abstract{
The Compact Linear Collider (CLIC) is a proposed future high-luminosity linear electron-positron collider operating at three energy stages, with nominal centre-of-mass energies $\roots = 380\,\GeV$, $1.5\,\TeV$, and $3\,\TeV$. Its aim is to explore the energy frontier, providing sensitivity to physics beyond the Standard Model (BSM) and precision measurements of Standard Model processes with an emphasis on Higgs boson and top-quark physics. The opportunities for top-quark physics at CLIC are discussed in this paper. The initial stage of operation focuses on top-quark pair production measurements, as well as the search for rare flavour-changing neutral current (FCNC) top-quark decays. It also includes a top-quark pair production threshold scan around 350\,GeV which provides a precise measurement of the top-quark mass in a well-defined theoretical framework. At the higher-energy stages, studies are made of top-quark pairs produced in association with other particles. A study of $\ttH$ production including the extraction of the top Yukawa coupling is presented as well as a study of vector boson fusion (VBF) production, which gives direct access to high-energy electroweak interactions. Operation above 1\,TeV leads to more highly collimated jet environments where dedicated methods are used to analyse the jet constituents. These techniques enable studies of the top-quark pair production, and hence the sensitivity to BSM physics, to be extended to higher energies. This paper also includes phenomenological interpretations that may be performed using the results from the extensive top-quark physics programme at CLIC.}


\arxivnumber{1807.02441}

\begin{document}

%line numbers
%\setpagewiselinenumbers
%\linenumbers

%\SetWatermarkText{DRAFT}
%\SetWatermarkLightness{0.9}

\maketitle

\flushbottom

\section{Introduction}
\label{sec:intro}

As the heaviest known fundamental particle, the top quark provides a
unique probe of the Standard Model (SM) of particle physics and occupies
an important role in many theories of new physics beyond the SM (BSM).  
So far the top quark has been produced only in hadron collisions, at
the Tevatron and Large Hadron Collider (LHC);
however, top-quark production in electron-positron collisions would herald
a new frontier of complementary and improved precision measurements.
For example: a top-quark pair production threshold scan would provide a precise
measurement of the top-quark mass, which is a fundamental
SM parameter;  precise measurements of top-quark production observables 
could give unique sensitivity to new physics
effects, as could the search for rare top-quark decays; new particles
could be observed that couple preferentially to top quarks; and improved
measurements of the top Yukawa coupling could further illuminate the Higgs sector.

The Compact Linear Collider (CLIC) is a proposed multi-\TeV high-luminosity
linear \epem collider that is currently under development as a possible large-scale installation at CERN. 
It is based on a unique and innovative two-beam acceleration technique that can reach 
accelerating gradients of 100\,MV/m.
CLIC is proposed as a staged collider providing high-luminosity \epem
collisions at centre-of-mass energies, $\roots$, from a few hundred \GeV up to 3\,TeV \cite{staging_baseline_yellow_report}.
Top-quark pair production is accessible at the first energy stage,
and an energy scan over the $\PQt\PAQt$ production threshold is also proposed.  
The higher-energy stages will supplement the initial energy datasets with large
samples of top quarks, further enhancing the sensitivity to new physics.

The following sections describe the CLIC experimental conditions and give an overview of top-quark production at CLIC, the theoretical description of top-quark production and decay, and the simulation and event reconstruction used for the subsequent studies, including dedicated identification of boosted top quarks. 
Thereafter, sections are dedicated to measurements of the top-quark mass, top-quark pair production, the study of the associated production of top quarks and a Higgs boson, top-quark production through vector boson fusion, and searches for rare flavour-changing neutral current (FCNC) top-quark decays. 
Measurements are considered at all energy stages of the collider.
Most of these analyses are done using full event simulation and reconstruction, and are reported for the first time in this paper.
To demonstrate the wider implications of the CLIC top-quark physics programme, the final section is dedicated to phenomenological interpretations. These are based on the study of top pair-production in full simulation and consider a variety of different observables, including so-called ``statistically optimal observables''.
The work is carried out in the context of the CLIC Detector and Physics (CLICdp) collaboration.

\section{Experimental environment at CLIC}
\label{sec:clic}
The CLIC accelerator technology produces a unique beam structure that results in the need for specially-developed detector concepts to allow precise reconstruction of complex final states up to multi-TeV centre-of-mass energies. The accelerator, energy staging, and detector concepts are introduced in the following sections.

\subsection{Accelerator and beam conditions}
\label{ssec:accandbeams}
CLIC is based on room-temperature accelerating structures in a two-beam scheme. Power from a low-energy, high-current drive beam is extracted to generate radio-frequency power at 12\,GHz, which is used to accelerate the main particle beams.  Accelerating gradients exceeding 100\,MV/m have been demonstrated at the CLIC test facility, CTF3~\cite{Corsini:CTF3}, enabling a compact collider design.

\begin{figure}
\begin{center}
\includegraphics[width=0.65\columnwidth]{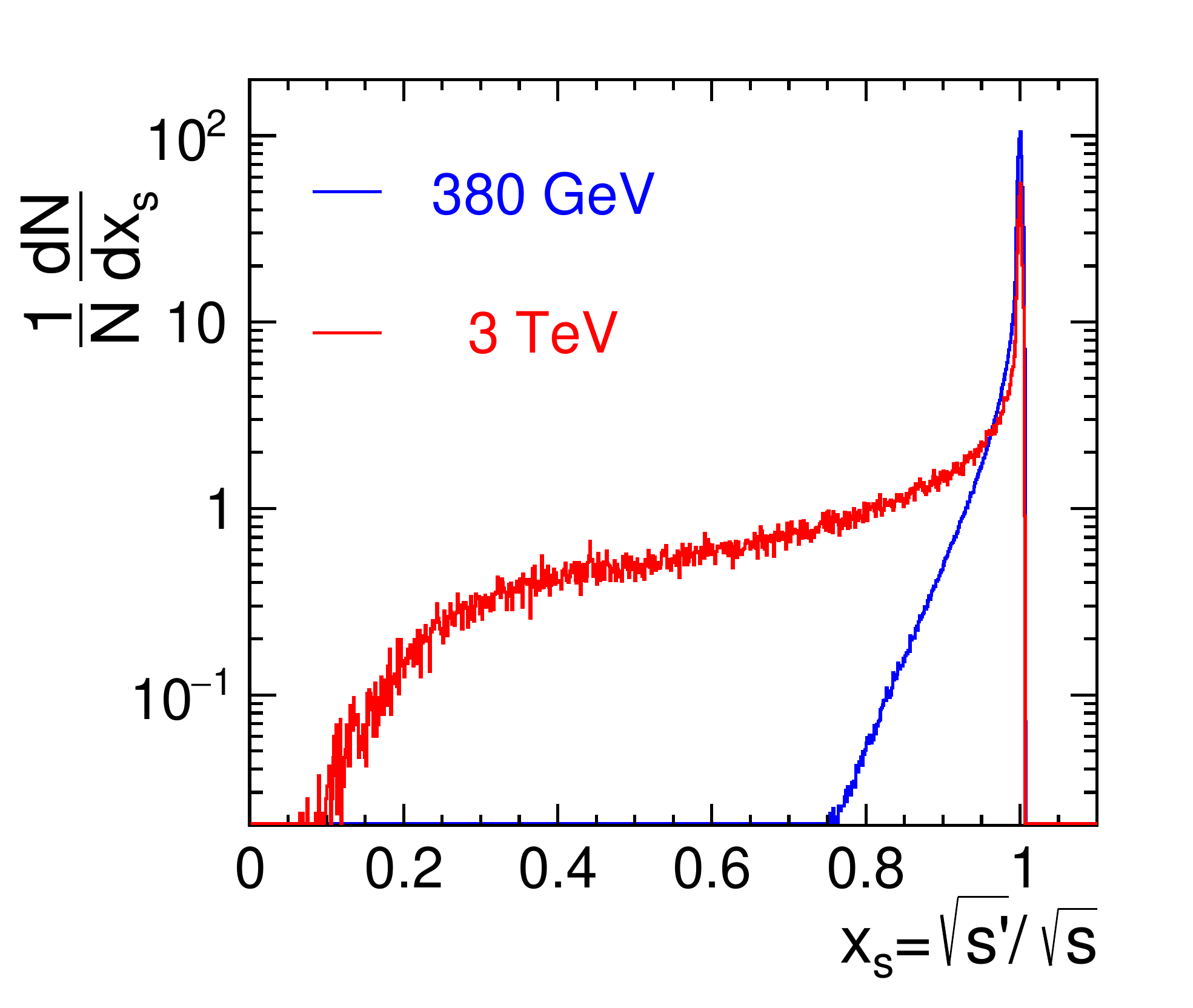}
\caption{The luminosity spectra for CLIC operating at $\roots=380\,\GeV$ and $3\,\TeV$, where $x_\textrm{s}$ denotes the ratio of the effective centre-of-mass energy after beamstrahlung, $\rootsprime$, to the nominal centre-of-mass energy $\roots$~\cite{CLIC_PhysDet_CDR}. \label{fig:luminosityspectrum}}
\end{center}
\end{figure}

Each bunch train consists of 312 bunches (352 bunches for the initial energy stage) with 0.5\,ns between bunch crossings at the interaction point, with a bunch train repetition rate of 50\,Hz.  The beam emittance is reduced in damping rings in the injector complex, and very small emittances are maintained through the accelerator chain, so that the resulting beams are highly-focused and intense in order to produce high instantaneous luminosities. This results in significant beamstrahlung\footnote{Radiation of photons from the colliding electrons or positrons in the electric field of the other beam.}~\cite{CLICCDR_vol1}, which means that although the average number of hard \epem interactions per single bunch train crossing is less than one, there are high rates of two-photon processes that deposit additional energy in the detector~\cite{CLIC_PhysDet_CDR}. Furthermore, the beamstrahlung results in a long lower-energy tail to the luminosity spectrum, as shown in \autoref{fig:luminosityspectrum} for operation at $\roots=380\,\GeV$ and $3\,\TeV$~\cite{CLIC_PhysDet_CDR}. The fractions of the total luminosity delivered above 99\% of the nominal $\roots$ are given in \autoref{tab:lumis}, and the effect is seen to be particularly significant at $\roots > 1$\,TeV.   The CLIC detector design and event reconstruction techniques are optimised to mitigate the influence of the beam-induced backgrounds, as discussed in \autoref{ssec:reco}.  The impact of initial-state radiation (ISR) on the effective centre-of-mass energy is similar to that of beamstrahlung.

\begin{table}[tb]\centering
 {\renewcommand{\arraystretch}{1.2}%
 \begin{tabular}{lrrr}
   \toprule 
                  $\roots=$                     & \tabt{380\,GeV} & \tabt{1.5\,TeV} & \tabt{3\,TeV} \\ \midrule
   Total instantaneous luminosity / $10^{34}$\,cm$^{-2}$s$^{-1}$  & 1.5 & 3.7 & 5.9 \\
   Total integrated luminosity / ab$^{-1}$                        & 1.0 & 2.5 & 5.0 \\
   Fraction of luminosity above 99\% of $\roots$  &  60\% & 38\% & 34\% \\
   \bottomrule
  \end{tabular}
  }%end of arraystretch
 \caption{Instantaneous and integrated luminosities for the baseline CLIC staging scenario, and fraction of the luminosity delivered above 99\% of $\roots$ \cite{Robson:2018zje,staging_baseline_yellow_report}. 
   \label{tab:lumis}}
\end{table}

\subsection{Staging scenario}
\label{ssec:staging}
To maximise the physics potential of CLIC, runs are foreseen at three energy stages \cite{staging_baseline_yellow_report}. Initial operation is at $\roots=380\,\GeV$, and will also incorporate an energy scan over the $\PQt\PAQt$ production threshold around $\roots = 350\,\GeV$. The second stage is at $\roots=1.5$\,TeV, which is the highest collision energy reachable with a single CLIC drive beam complex.  The second-stage energy of 1.5\,TeV has recently been adopted and will be used for future studies. In the work presented here, the previous baseline of 1.4\,TeV is used. The third stage of $\roots=3$\,TeV is the ultimate energy of CLIC, and requires two drive beam complexes. The expected instantaneous and total luminosities are given in \autoref{tab:lumis}. For the staging scenario assumed in this paper, each stage will consist of five to six years of operation at the nominal luminosity.

The baseline accelerator design foresees $\pm 80\%$ longitudinal electron spin polarisation by using GaAs-type cathodes \cite{CLICCDR_vol1}, and no positron polarisation. At the initial energy stage equal amounts of P($\Pem$) = -80\% and P($\Pem$) = +80\% running are foreseen as this improves the sensitivity to certain BSM effects~\cite{Robson:2018zje}. At the same time, the dominant Higgs production mechanism at the initial stage, Higgsstrahlung, is largely unaffected by the electron polarisation. At the higher-energy stages, the dominant single- and double-Higgs production mechanisms are through WW-fusion which is significantly enhanced (by around 80\%) for running with -80\% electron polarisation, owing to the underlying chiral structure of the electroweak interaction~\cite{Abramowicz:2016zbo}. However, some +80\% electron polarisation running is desired for improved BSM reach as illustrated in \autoref{sec:phenom_interp} of this paper. A baseline with shared running time for -80\% and +80\% electron polarisation in the ratio 80:20 is adopted for the two higher-energy stages~\cite{Robson:2018zje}.

\subsection{Detectors}
%CLIC_ILD, CLIC_SiD

\begin{figure*}
  \centering
  \includegraphics[scale=0.24,clip]{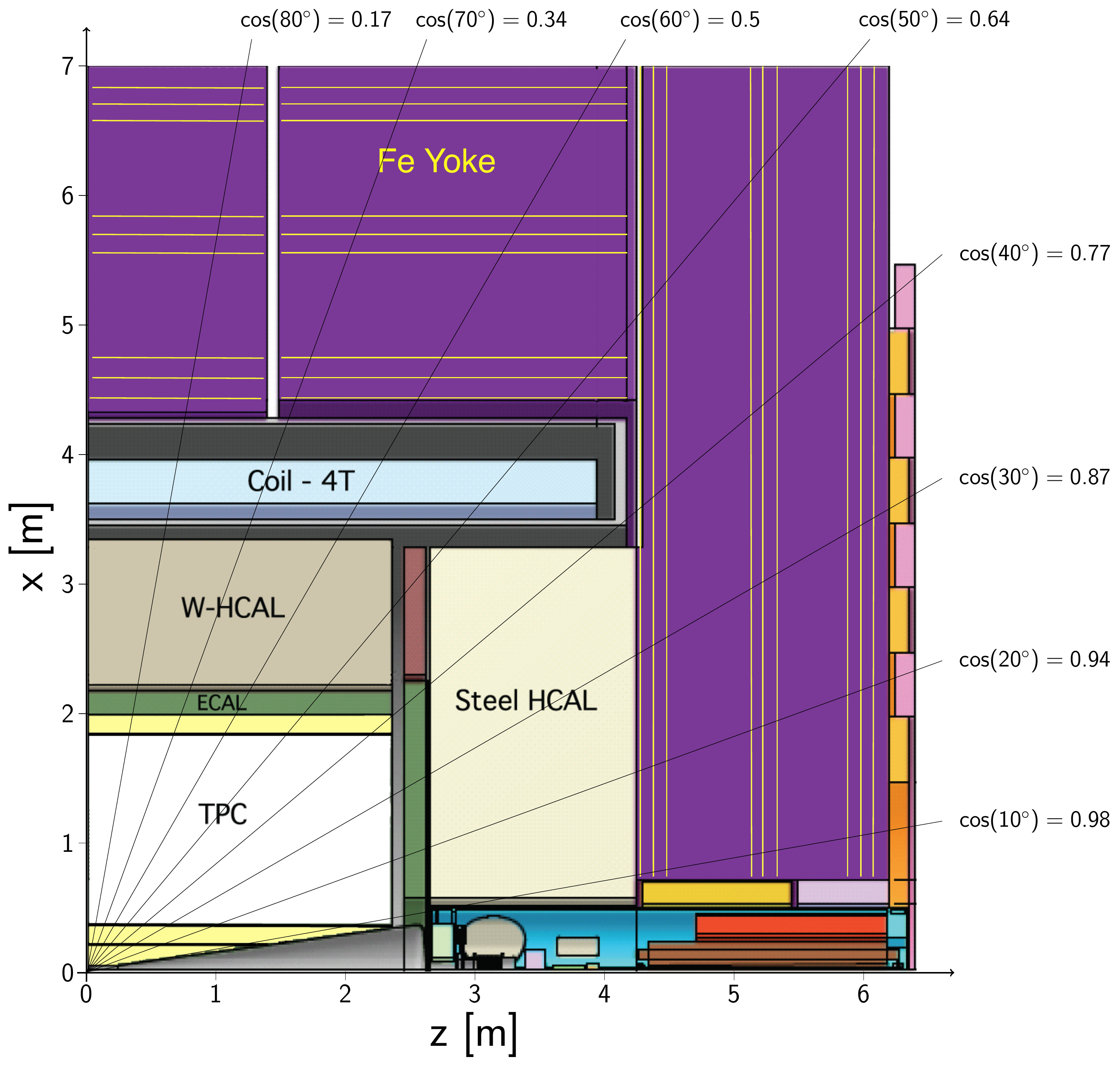}
  ~~~~
  \includegraphics[scale=0.24,clip]{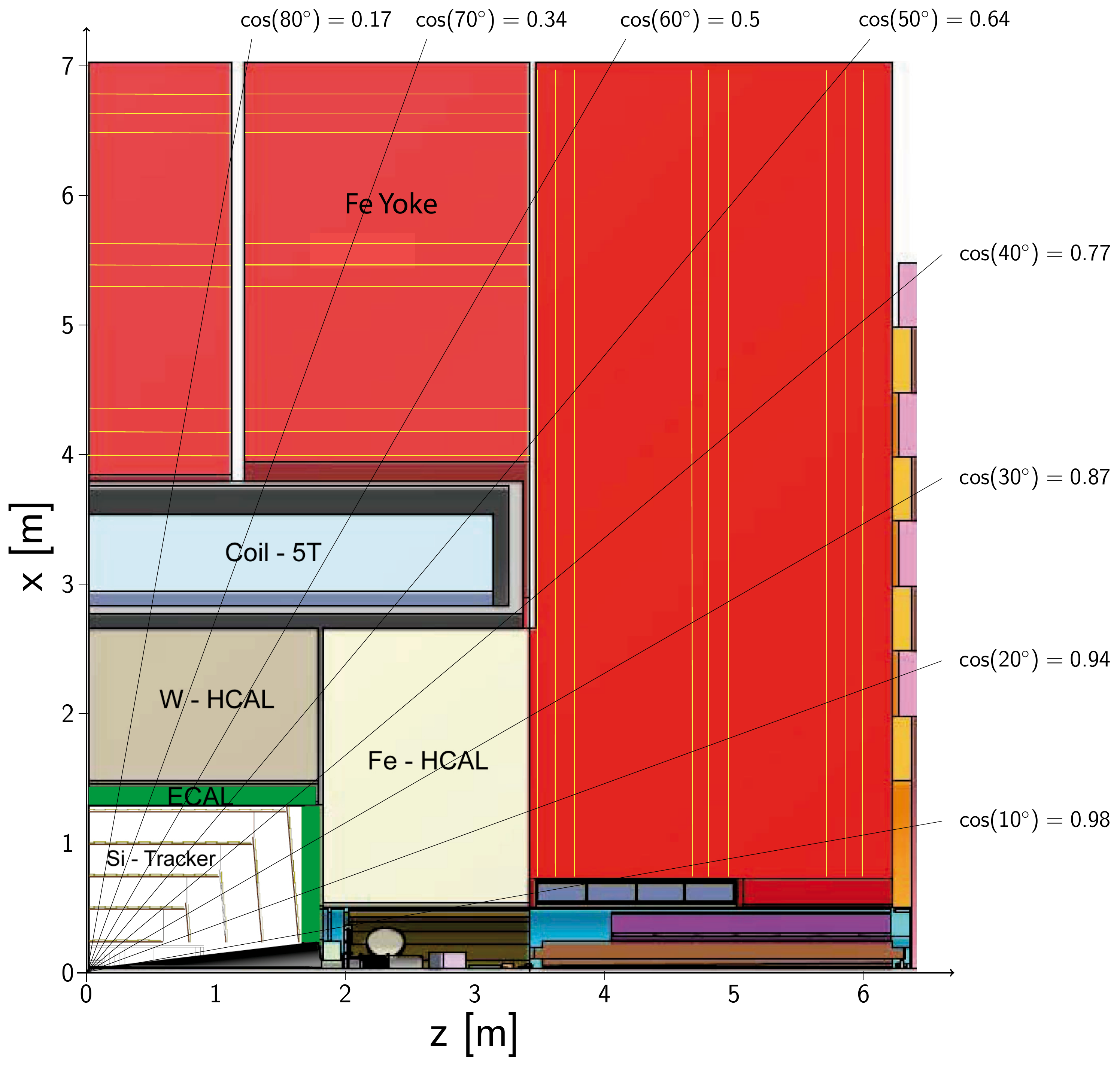}
  \caption{Longitudinal cross section of the top right quadrant of the
    \clicild (left) and \clicsid (right) detector concepts \cite{CLIC_PhysDet_CDR}.}
  \label{fig:detectors}
\end{figure*}

The detector concepts, \clicild and \clicsid, used for the CLIC physics studies described here and elsewhere~\cite{Abramowicz:2016zbo}, are adapted from the ILD~\cite{ildloi:2009,ilctdrvol4:2013} and SiD~\cite{Aihara:2009ad,ilctdrvol4:2013} detector concepts for the International Linear Collider (ILC). Design modifications are motivated by the smaller bunch spacing and different beam conditions as well as the higher-energy collisions at CLIC; both detectors are optimised for $3\,\TeV$. The two detector concepts, shown schematically in \autoref{fig:detectors}, are discussed in detail in~\cite{CLIC_PhysDet_CDR}. The detectors are described using a right-handed coordinate system with the $z$-axis along the electron beam direction, and $\theta$ denotes the polar angle w.r.t. the $z$-axis. \clicsid employs central silicon-strip tracking detectors, whereas \clicild includes a large central gaseous time projection chamber. In both concepts, the central tracking system is supplemented by silicon-pixel vertex detectors.

Vertex and tracking systems provide excellent track momentum resolution of $\sigma_{\pT}/\pT^2 \lesssim 2 \cdot 10^{-5}$ $\GeV^{-1}$ needed for the reconstruction of high-\pT charged leptons, as well as high impact parameter resolution, defined by $a \lesssim 5\,\micron$ and $b \lesssim 15\,\micron\,\GeV$ in $\sigma_{d_0}^2 = a^2 + b^2/(p^2\sin^3\theta)$. This allows accurate vertex reconstruction and enables flavour tagging with clean $\PQb$-, $\PQc$- and light-quark jet separation, crucial for top-quark identification and background rejection at the initial CLIC energy stage. In highly-boosted top-quark events, a significant fraction of the resulting $\PQb$-hadrons decay outside the vertex detector, and the jet environment is dense, motivating the development of alternative approaches to top-quark reconstruction that do not depend on flavour tagging.

The detector designs feature fine-grained electromagnetic and hadronic calorimeters (ECAL and HCAL) optimised for particle-flow reconstruction, which aims to reconstruct individual particles within a jet using the combined tracking and calorimeter measurements. The resulting jet-energy resolution, for isolated central light-quark jets with energy in the range $100\,\GeV$ to $1\,\TeV$, is $\sigma_E/E \lesssim 3.5\,\%$. The energy resolution for photons is approximately 16\%/$\sqrt{E/ \gev}$ with a constant term of 1\%.
Strong solenoidal magnets located outside the HCAL provide an axial magnetic field of 4\,T in \clicild and 5\,T in \clicsid. 
Two compact electromagnetic calorimeters in the forward region, LumiCal and BeamCal, allow electrons and photons to be measured down to around 10\,mrad in polar angle; this is particularly important for the determination of the luminosity spectrum via measurements of Bhabha scattering \cite{Poss:2013oea}.

The studies reported here assume that a single cell time resolution of 1\,ns will be reached in the calorimeters, and single strip or pixel time resolutions of 3\,ns in the silicon detectors. The integration times used for the formation of clusters are 10\,ns in the ECAL, the HCAL endcaps and in the silicon detectors, and 100 ns in the HCAL barrel. The latter is chosen to account for the more complex time structure of hadronic showers in the tungsten-based barrel HCAL. With these parameters,  sub-ns time resolution is achieved for reconstructed particle-flow objects consisting of tracks and calorimeter clusters. This allows energy deposits from hard physics events and those from beam-induced backgrounds in other bunch-crossings to be sufficiently distinguished.

\section{Overview of top-quark production at CLIC}
\label{sec:overview}

\begin{figure}
\begin{center}
\includegraphics[width=0.65\columnwidth]{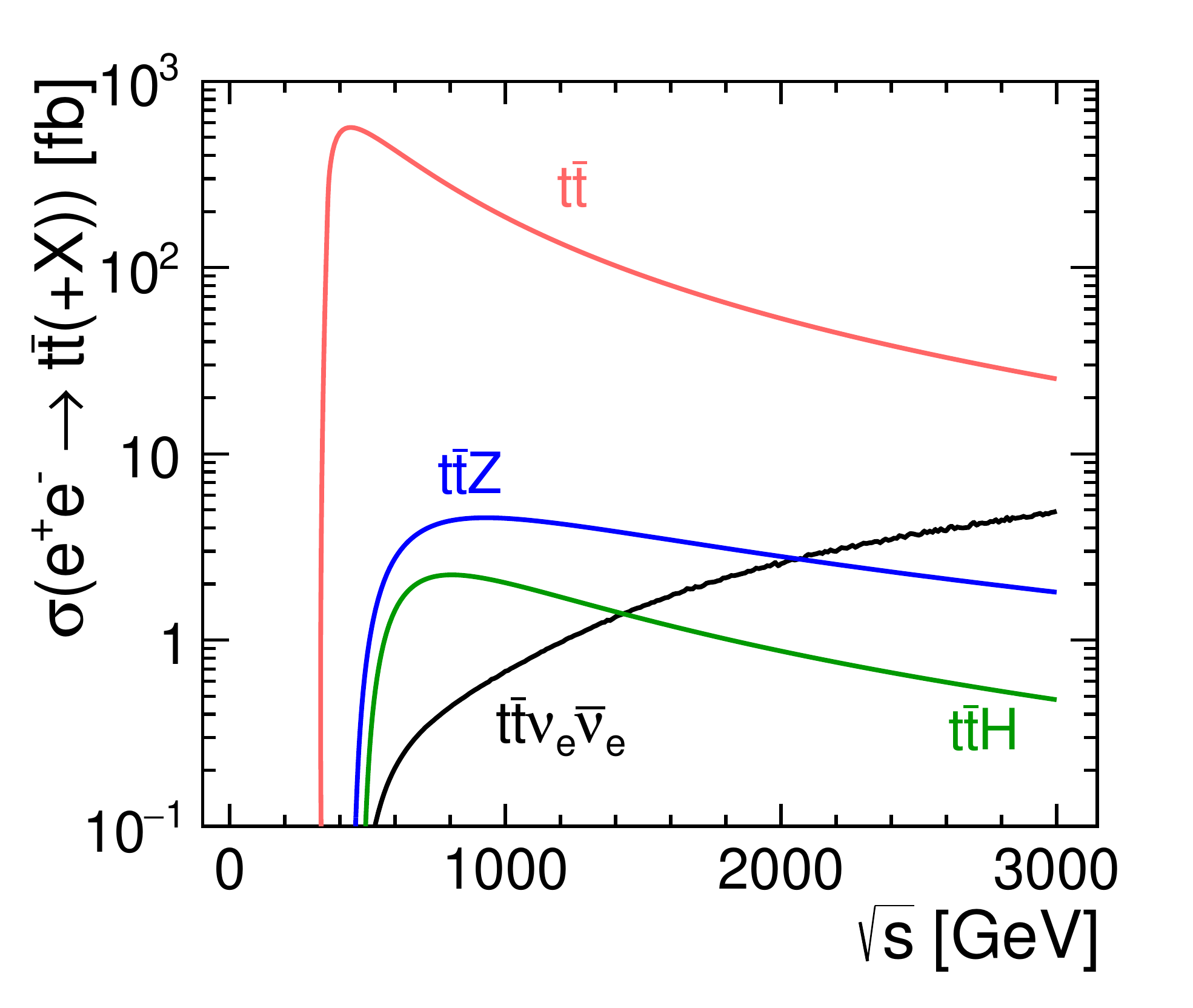}
\caption{Cross section as a function of centre-of-mass energy for the main top-quark pair production processes at an \epem collider for a top-quark mass of $m_{\PQt}$ = 174\,\GeV and a Higgs boson mass of $m_{\PH}$ = 125\,\GeV. The leading-order expectations for unpolarised beams with ISR are shown. The effect of beamstrahlung is not included.\label{fig:cross_sections_ttbar}}
\end{center}
\end{figure}

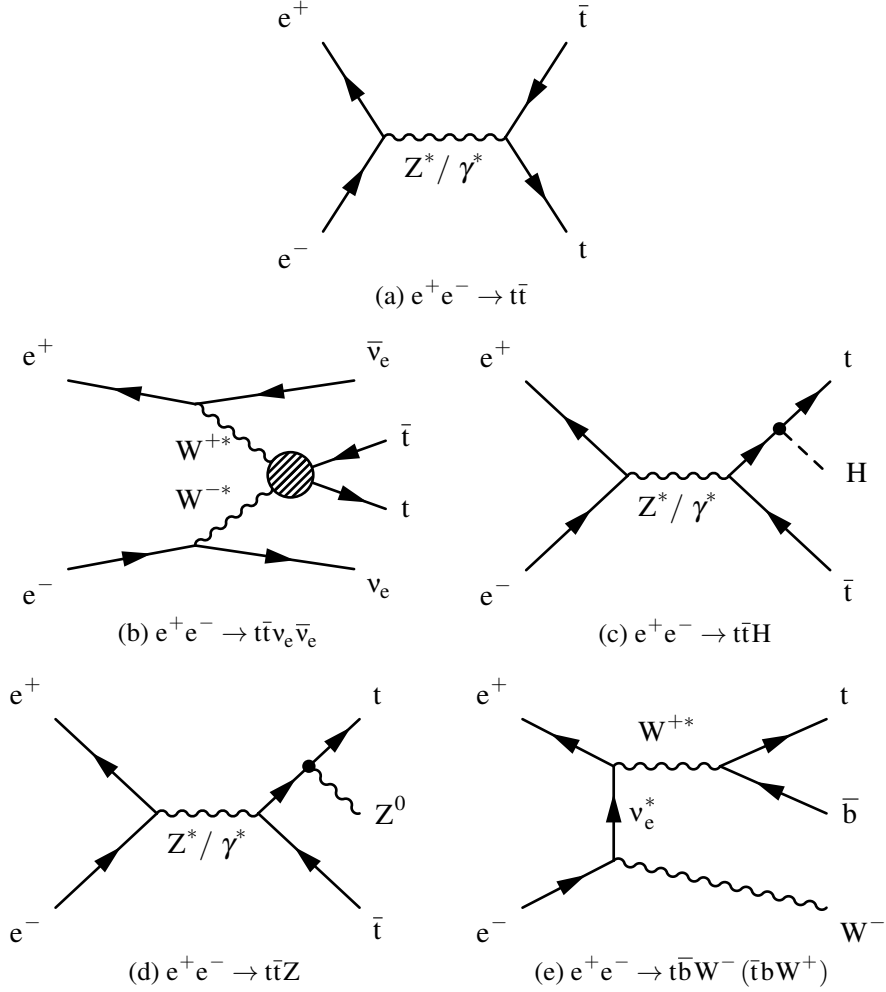
\begin{figure}
\unitlength = 1mm
\centering
\begin{subfigure}[b]{0.4\textwidth}
\centering
\begin{fmffile}{feyn_ttbar}
    \begin{fmfgraph*}(40,25)
    	\fmfset{wiggly_len}{3mm}
        \fmfleft{i1,i2}
        \fmfright{o1,o2}
        \fmflabel{$\Pem$}{i1}
        \fmflabel{$\Pep$}{i2}
        \fmflabel{$\PQt$}{o1}
        \fmflabel{$\PAQt$}{o2}
        \fmf{fermion}{i1,v1,i2}
        \fmf{fermion}{o2,v2,o1}
        \fmf{photon,label=$\PZ^{\ast}/~\gamma^{\ast}$}{v1,v2}
    \end{fmfgraph*}
\end{fmffile}
\bigskip
\caption{$\epem\to\ttbar$\label{fig:production_diagrams:ttbar}}
\end{subfigure}
\par\bigskip
\par\bigskip
\begin{subfigure}[b]{0.4\textwidth}
\centering
\begin{fmffile}{feyn_ttnunu}
    \begin{fmfgraph*}(47,25)
        \fmfset{wiggly_len}{3mm}
        \fmfleft{i1,i2}  
        \fmfright{o1,oh1,oh2,o2}
        \fmflabel{$\Pem$}{i1}
        \fmflabel{$\Pep$}{i2}  
        \fmflabel{$\PAGne$}{o2} 
        \fmflabel{$\PQt$}{oh1}            
        \fmflabel{$\PAQt$}{oh2}
        \fmflabel{$\PGne$}{o1}
        \fmf{fermion, tension=2.0}{i1,v1}
        \fmf{fermion, tension=1.0}{v1,o1}
        \fmf{fermion, tension=1.0}{o2,v2}
        \fmf{fermion, tension=2.0}{v2,i2}
         \fmf{photon, tension=1.0, label=$\PW^{-\ast}$,label.dist=0.5}{v1,vh}
        \fmf{photon, tension=1.0, label=$\PW^{+\ast}$, label.dist=0.5}{v2,vh}
        \fmf{fermion,  tension=1.0}{oh2,vh,oh1}
        	\fmfblob{17}{vh}
    \end{fmfgraph*}
\end{fmffile}
\bigskip
\caption{$\epem\to\ttbar\PGne\PAGne$\label{fig:production_diagram:vbf}}
\bigskip
\bigskip
\end{subfigure}
\begin{subfigure}[b]{0.4\textwidth}
\centering
\begin{fmffile}{feyn_tth} 
    \begin{fmfgraph*}(40,25)
        \fmfset{wiggly_len}{3mm}
        \fmfstraight
        \fmfleft{i1,i2}  
        \fmfright{o1,oh,o2}
        \fmflabel{$\Pem$}{i1}
        \fmflabel{$\Pep$}{i2}  
        \fmflabel{$\PQt$}{o2} 
        \fmflabel{$\PH$}{oh}
        \fmflabel{$\PAQt$}{o1}
        \fmf{photon,label=$\PZ^{\ast}/~\gamma^{\ast}$,tension=2.0}{v1,v2}
        \fmf{fermion,tension=1.0}{i1,v1,i2}
        \fmf{phantom,tension=1.0}{o1,v2,o2}
        \fmffreeze
        \fmf{fermion,tension=1.0}{o1,v2}
        \fmf{fermion,tension=1.0}{v2,vh,o2}
        \fmf{dashes,tension=0.0}{vh,oh}
        \fmfdot{vh}
    \end{fmfgraph*}
\end{fmffile}
\bigskip
\caption{$\epem\to\ttH$\label{fig:production_diagram:ttH}}
\bigskip
\bigskip
\end{subfigure}
\begin{subfigure}[b]{0.4\textwidth}
\centering
\begin{fmffile}{feyn_ttz} 
    \begin{fmfgraph*}(40,25)
        \fmfset{wiggly_len}{3mm}
        \fmfstraight
        \fmfleft{i1,i2}  
        \fmfright{o1,oh,o2}
        \fmflabel{$\Pem$}{i1}
        \fmflabel{$\Pep$}{i2}  
        \fmflabel{$\PQt$}{o2} 
        \fmflabel{$\PZ^{0}$}{oh}
        \fmflabel{$\PAQt$}{o1}
        \fmf{photon,label=$\PZ^{\ast}/~\gamma^{\ast}$,tension=2.0}{v1,v2}
        \fmf{fermion,tension=1.0}{i1,v1,i2}
        \fmf{phantom,tension=1.0}{o1,v2,o2}
        \fmffreeze
        \fmf{fermion,tension=1.0}{o1,v2}
        \fmf{fermion,tension=1.0}{v2,vh,o2}
        \fmf{photon,tension=0.0}{vh,oh}
        \fmfdot{vh}
    \end{fmfgraph*}
\end{fmffile}
\bigskip
\caption{$\epem\to\ttbar\PZ$\label{fig:production_diagram:ttZ}}
\end{subfigure}
\begin{subfigure}[b]{0.4\textwidth}
\centering
\begin{fmffile}{feyn_singletop}
  \begin{fmfgraph*}(40,25)
  \fmfset{wiggly_len}{3mm}
  \fmfstraight
  \fmfleft{i1,i2}
  \fmfright{o1,o2,o3}
  \fmflabel{$\Pem$}{i1}
  \fmflabel{$\Pep$}{i2}
  \fmf{fermion}{i1,v1}
  \fmf{fermion,label=$\PGn_\mathrm{e}^{\ast}$}{v1,v2}
  \fmf{fermion}{v2,i2}
  \fmf{photon,label=$\PW^{+\ast}$,label.dist=7}{v3,v2}
  \fmf{fermion}{v3,o3}
  \fmf{fermion}{o2,v3}
  \fmf{photon}{v1,o1}
  \fmflabel{$\PWm$}{o1}
  \fmflabel{$\PAQb$}{o2}
  \fmflabel{$\PQt$}{o3}
  \fmfforce{(0.3w,0.25h)}{v1}
  \fmfforce{(0.3w,0.75h)}{v2}
  \fmfforce{(0.65w,0.75h)}{v3}
  \end{fmfgraph*}
\end{fmffile}
\bigskip
\caption{$\epem\to\PQt\PAQb\PWm\,(\PAQt\PQb\PWp)$\label{fig:production_diagrams:singletop}}
\end{subfigure}
\vspace{3mm}
\caption{Representative diagrams for top-quark production processes relevant at CLIC; (a) $\ttbar$, (b) $\ttbar\PGne\PAGne$, (c) $\ttH$, (d) $\ttbar\PZ$, (e) single-top. The blob in \autoref{fig:production_diagram:vbf} represents the complete amplitude of the $\PWp\PWm\rightarrow\ttbar$ Feynman diagram, including potential new physics effects.}
\label{fig:production_diagrams}
\end{figure}

Operation at the initial CLIC energy stage, $\roots=380\,\GeV$, will allow top-quark pair production with close to maximal cross section as illustrated in \autoref{fig:cross_sections_ttbar}. The expected cross section, including higher-order quantum chromodynamics (QCD) effects and with ISR, is about $700\,\fb$ for unpolarised beams~\cite{Bach:2017ggt}.

Top-quark pair production is dominated by the $\PZ^{\ast}/\gamma^{\ast}$ exchange diagram shown in \autoref{fig:production_diagrams:ttbar}. The dominant top-quark decay mode in the SM is to a $\PQb$-quark and $\PW$ boson (about 99.8\%). The topology of the $\ttbar\to6\mathrm{\text{-}fermion}$ final state is defined by the decay channels of the two $\PW$ bosons. Most of the analyses described in this paper consider fully-hadronic events, where both $\PW$ bosons decay hadronically, or semi-leptonic events, where one of the $\PW$ bosons decays to a lepton and a neutrino and the other $\PW$ boson decays hadronically. Fully-leptonic events, which account for about 11\% of the events, have not been studied so far.

The contribution from $\text{non-}\ttbar$ processes, such as single-top production (see \autoref{fig:production_diagrams:singletop}) and triple gauge boson production, to the inclusive $\epem\to6\mathrm{\text{-}fermion}$ process cannot be fully separated due to interference. At $\roots =380$\,GeV its contribution to the final event sample is expected to be negligible. In contrast, at higher centre-of-mass energies where the fraction of $\text{non-}\ttbar$ events is significantly larger~\cite{Fuster2015}, such events make up the main part of the remaining background after all selections have been applied.

All three energy stages contribute to the global sensitivity to new physics from the precision measurement of $\ttbar$ production properties. These measurements make use of the electron beam polarisation available at CLIC: the cross section for $\epem\to\ttbar$ is enhanced (reduced) by $34\%$ at $380$\,\GeV for the -80\% (+80\%) polarisation configuration; and at the higher-energy stages, the cross section is 30\% larger (smaller) when operating with -80\% (+80\%) beam polarisation.

At higher energies, processes where the top-quark pair is produced in association with other particles are accessible, see for example \autoref{fig:production_diagram:ttH} and \autoref{fig:production_diagram:ttZ}. The $\ttH$ cross section has a maximum around $\roots =800$\,\GeV. This process enables direct measurements of the top Yukawa coupling and allows the study of CP properties of the Higgs boson in the $\PQt\PQt\PH$ coupling. As the luminosity of a linear collider increases with the centre-of-mass energy, the optimal energy in terms of yield at which to study this process is above the maximum of the cross section. The energy stage at 1.5\,\TeV (or the previous baseline of 1.4\,\TeV as used here) is ideally suited for studying this process as the production rate is close to its maximum.

The cross section for top-quark pair production in vector boson fusion (VBF), such as $\epem\to\PQt\PAQt\PGne\PAGne$ (see \autoref{fig:production_diagram:vbf}), has an approximately logarithmic increase with the centre-of-mass energy. Hence, studies of such processes benefit from the highest possible centre-of-mass energy available at CLIC.

The cross sections and expected numbers of events for some of the processes discussed above are summarised in \autoref{tab:top:events}.

\begin{table}[tb]\centering
 {\renewcommand{\arraystretch}{1.2}%
 \begin{tabular}{lrrr}
   \toprule 
                  $\roots=$                     & \tabt{380\,GeV} & \tabt{1.4\,TeV} & \tabt{3\,TeV} \\ \midrule
    $\sigma(\epem\to\ttbar)$          & 723\,\fb & 102\,\fb & 25.2\,\fb         \\
    $\sigma(\epem\to\ttH)$  & - & 1.42\,\fb & 0.478\,\fb       \\
    $\sigma(\epem\to\PQt\PAQt\PGne\PAGne)$     & - & 1.33\,\fb & 4.86\,\fb        \\
    \midrule
      \LumiIntDiff                             & 1.0\,\abinv     & 2.5\,\abinv    & 5.0\,\abinv  \\
    No.\@ $\ttbar$ events                   & 690,000 & 430,000 & 310,000       \\
    No.\@ $\ttH$ events           & - & 4,700 & 4,200      \\
    No.\@ $\PQt\PAQt\PGne\PAGne$ events              & - & 3,800 & 28,000       \\
    \bottomrule
  \end{tabular}
  }%end of arraystretch
    \caption{Unpolarised cross sections for 
       \ttbar, \ttH and $\PQt\PAQt\PGne\PAGne$
      production assuming $m_{\PQt}=174\,\GeV$ and $m_{\PH}=125\,\GeV$ at the three centre-of-mass
      energies studied in this paper. The numbers for 380\,\GeV include QCD corrections (see text) 
      while leading-order results are given for the higher energy stages. $\rootsprime$ is the 
      effective centre-of-mass energy of the $\epem$ collision.  
      The presented cross sections include the effects of ISR but not the effects of beamstrahlung. 
      Also given are numbers of expected events,
      including both effects. 
      The presented event numbers include the assumptions on electron beam polarisation
      described in \autoref{ssec:staging}.
    \label{tab:top:events}}
\end{table}

\section{Theoretical description of top-quark production and decay}
\label{sec:theory}

This section reports on the theoretical tools and concepts that we employ to describe top-quark physics within the SM and beyond. We start by summarising the status of SM calculations for top-quark production at the threshold and in the continuum regions. The choice of top-quark mass scheme plays a major role in the former. Next, we introduce the Effective Field Theory (EFT) framework that we use to parametrise new physics effects in the top-quark electroweak interactions. Its relation with the more canonical language of anomalous couplings is also discussed. Finally we discuss possible new physics effects inducing flavour changing neutral current top-quark decays. 

\subsection{Top-quark mass schemes}
\label{ssec:masschemes}

Observables with the highest sensitivity to the top-quark mass are related to production thresholds or resonances involving the top quark.  However, the fact that the top quark is unstable and coloured causes nontrivial and in general sizeable QCD and electroweak corrections, which currently can be systematically controlled only for a small number of observables (such as for the \ttbar threshold).  At the level of currently achievable experimental uncertainties for top-quark mass measurements these corrections, which significantly modify the simple leading-order picture of a particle with a definite mass that decays to an observable final state, cannot be neglected.  Most experimental studies of the top-quark mass therefore rely on multi-purpose MC event generators to measure a parameter of the generator associated with the top-quark mass. The interpretation of these top-quark mass measurements relies on the quality of the MC modelling of the observables used; it also suffers from the fact that the MC top-quark mass parameter is not fully understood at present from a quantum field theory perspective.

In theory calculations, different mass schemes are used, which are renormalisation-scale dependent. A common scheme is the ``pole mass'', defined as the pole of the quark propagator.  The top-quark pole mass is numerically close to the mass parameter of MC generators, but may not be identified with it; another scheme that has a close numerical relation to the generator mass parameter is the MSR mass, see for example \cite{Butenschoen:2016lpz}. In precision calculations at high energies, the $\overline{\rm MS}$ (modified minimal subtraction) mass scheme is frequently used. However, for the treatment of the threshold region (shown in \autoref{fig:Threshold:TopCrossSection}), neither the pole mass nor the  $\overline{\rm MS}$ mass is adequate, since they both show poor convergence and are subject to larger QCD corrections. At the threshold, two commonly used mass schemes are the 1S \cite{Hoang:1999zc} and the PS \cite{Beneke:1998rk} mass schemes, both of which result in stable behaviour of the calculated cross section in the threshold region and can also be related to the $\overline{\textrm{MS}}$ mass in a theoretically rigorous way with high precision \cite{Marquard:2015qpa}, for use in other perturbative calculations. Additional uncertainties from the precision of the strong coupling constant enter into this conversion. 

For the studies at the top-quark pair production threshold discussed in \autoref{ssec:threshold} and \ref{sec:Mass:Threshold}, the PS mass scheme is used, assuming a top-quark mass of  $m^{\mathrm{PS}}_{\mathrm{t}}$ = 171.5\,\GeV. With the assumed value of the strong coupling constant of 0.1185, this value corresponds to a top-quark pole mass of 173.3\,\GeV, which is consistent with measurements of the pole mass at the LHC~\cite{Aad:2015waa, Khachatryan:2016mqs}. Since the numerical value of the mass parameter in the event generator is close to the pole mass, the mass used in the threshold studies is also consistent with the top-quark mass used to generate event samples for the other analyses in this paper, as presented in \autoref{ssec:gen}.

\subsection{\texorpdfstring{\ttbar}{ttbar} production at threshold}
\label{ssec:threshold}

Top-quark pair production in the threshold region (340-355\,GeV) is characterised by a fast rise of the cross section induced by the formation of a quasi toponium bound state, and by additional higher-order effects from interactions of the quark pair, predominantly via the strong interaction~\cite{Fadin:1987wz, Fadin:1988fn, Strassler:1990nw}, but also via Higgs boson exchange.

\begin{figure}
  \centering
  \includegraphics[width=0.65\columnwidth]{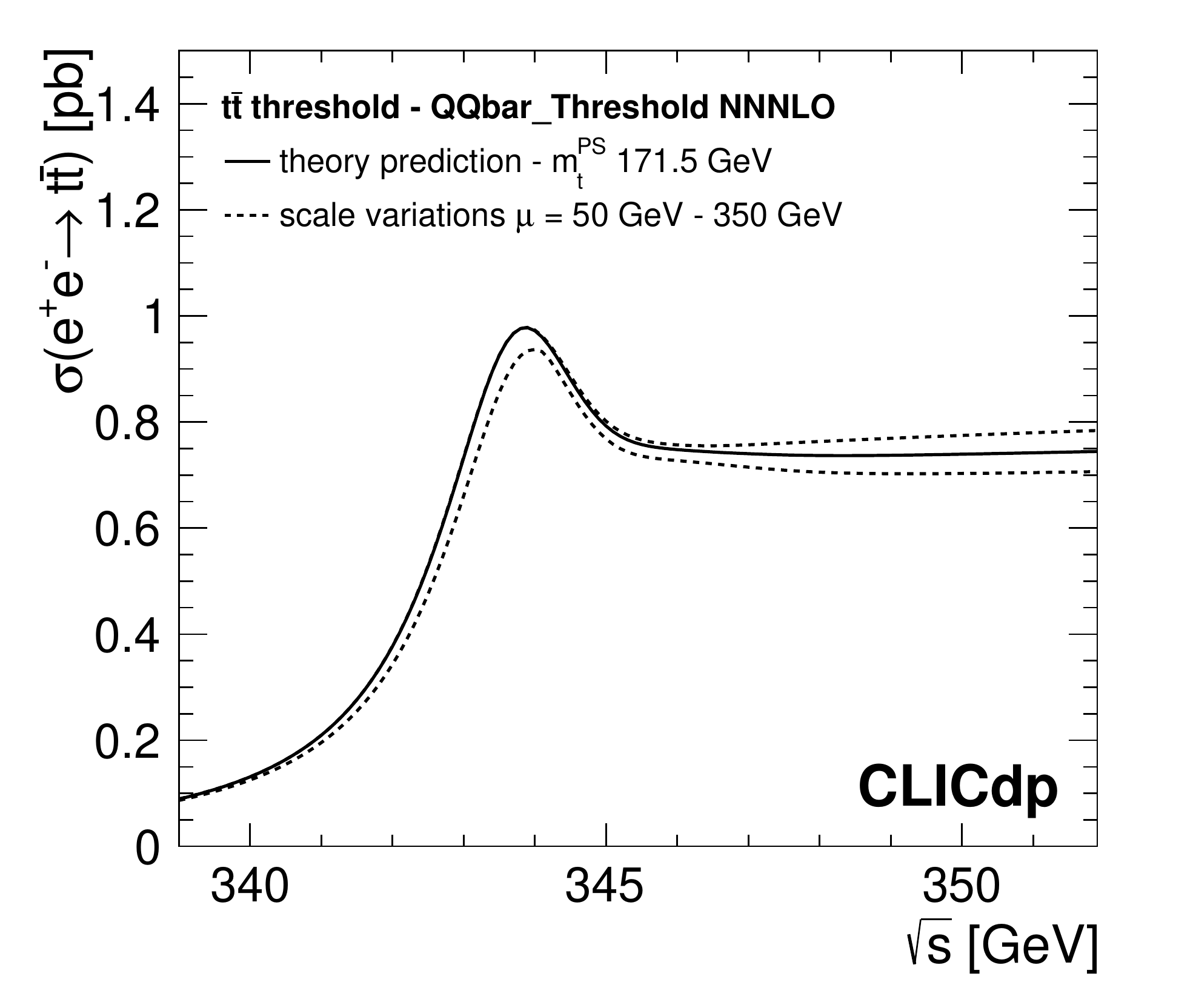}
  \caption{Cross section of top-quark pair production in the threshold region, showing the NNNLO QCD theory cross section obtained with \code{QQbar\_threshold} \cite{Beneke:2016kkb} and the associated renormalisation scale uncertainties. \label{fig:Threshold:TopCrossSection}}
\end{figure}

\autoref{fig:Threshold:TopCrossSection} shows the cross section of the process \eett as a function of centre-of-mass energy calculated at next-to-next-to-next-to-leading order (NNNLO) QCD \cite{Beneke:2016kkb}, taking next-to-leading order (NLO) Higgs effects and electroweak effects into account. Theoretical uncertainties obtained from variations of the renormalisation scale are also indicated. Consistent predictions with comparable uncertainties are provided also by NNLO + NNLL calculations containing logarithmic corrections to all orders not included in the NNNLO results \cite{Hoang:2013uda}. The observable cross section is obtained by including effects from ISR and the luminosity spectrum of the collider, as discussed in more detail in \autoref{sec:Mass:Threshold}.

The cross section, the position of the turn-on of the top-quark pair production, and the overall shape of the cross section as a function of collision energy are strongly dependent on the precise value of the top-quark mass as well as on the width, the Yukawa coupling, and the strength of the strong coupling \cite{Gusken:1985nf, Bigi:1986jk, Fadin:1987wz, Fadin:1988fn, Strassler:1990nw}.  A precise measurement of the top-quark pair threshold line shape can thus be used to extract the top-quark mass to excellent precision and with a rigorously defined mass scheme as introduced in \autoref{ssec:masschemes},  and can also be used to obtain other top-quark properties \cite{Martinez:2002st, Seidel:2013sqa,Horiguchi:2013wra}. 

\subsection{QCD and electroweak corrections to \texorpdfstring{\ttbar}{ttbar} and \texorpdfstring{\ttH}{ttH} in the continuum}
\label{ssec:continuum}

The fully differential cross section for top-quark pair production at lepton colliders was computed in \cite{Gao:2014eea,Gao:2014nva,Chen:2016zbz} at next-to-next-to-leading order (NNLO) in QCD. For collider energies in the continuum well above the top-quark pair production threshold, scale uncertainties on the relevant observables such as the total cross section, the top-quark forward-backward asymmetry (\afb), and the differential top-quark $\pT$ distribution are at the few per mille level \cite{Chen:2016zbz}. While top-quark decays can be directly included in these calculations by working in the narrow-width approximation, a full treatment of finite-width effects requires instead computing $\PWp\PWm\PQb\PAQb$ production, which is known only at NLO in QCD \cite{Lei:2008ii,Liebler:2015ipp,Nejad:2016bci}. Automated NLO computations of these processes are available in \whizard \cite{Weiss:2015npa} and \madgraph \cite{Alwall:2014hca}. The same tools also allow simulation of top-quark pair production in association with a Higgs or a $\PZ$ boson at NLO in QCD and including finite-width effects. Electroweak NLO corrections \cite{Beenakker:1991ca,Fleischer:2003kk,Hahn:2003ab} are known to be sizeable at high energy, reaching order $20\%$ on the total cross section and on \afb for a $1$\,TeV collider \cite{Khiem:2012bp}. They will thus play a role in the high-energy stages of CLIC. 
The resummation of log-enhanced QCD effects might also be important in the regime of boosted top quarks. Such calculations have been performed for the LHC~\cite{Pecjak:2016nee} and for lepton colliders~\cite{Fleming:2007qr, Fleming:2007tv}. It is expected that a complete treatment of these effects for all the relevant observables will be available for CLIC data analyses. A thorough study of the theoretical uncertainties associated with the different corrections outlined above has not been performed, but in general they are expected to be below the percent level, dominated by QCD scale uncertainties.

While the nominal centre-of-mass energy of the first CLIC stage of $\roots=380\,\gev$ is somewhat larger than the region where threshold effects are relevant (see \autoref{ssec:threshold}), the energy loss due to ISR and beamstrahlung reduces the effective centre-of-mass energy for a fraction of the top-quark pair production events to values close to the threshold. A combined approach to describe $\PWp\PWm\PQb\PAQb$ production matching NLO fixed-order continuum QCD calculations with NLL resummation of the threshold corrections is described in \cite{Bach:2017ggt}. While the scale uncertainties are well under control when including ISR, the addition of beamstrahlung requires further work.

\subsection{EFT in top-quark physics}
\label{ssec:interpretations_theory}

BSM effects induced by heavy new physics (above the direct reach of CLIC) are universally described by Effective Field Theory (EFT) operators of energy dimension ($d$) larger than $4$ that modify the low-energy dynamics with respect to SM predictions. Lower-dimensional operators normally~\cite{Grzadkowski:2010es} induce larger effects, and by assuming lepton (and baryon) number conservation the first EFT operators are those of dimension $d=6$. We thus restrict this study to $d=6$ operators and employ, whenever possible, the ``Warsaw basis'' notation of \cite{Grzadkowski:2010es}, introducing for the first time a complete non-redundant basis for these operators. The EFT Lagrangian is expressed as a sum over local operators $Q_i$ multiplied by coupling constants $C_i$, referred to as (dimensionful) Wilson coefficients:
\begin{equation*}
\mathcal{L}_{\,\mathrm{EFT}}=\mathcal{L}_{\,\mathrm{SM}}+\sum_i\,C_i\,Q_{\,i}^{\,d=6} \,.
\end{equation*}

The $d=6$ operators that contribute, at tree-level, to top-quark production at lepton colliders are conveniently classified as follows. ``Universal'' operators \cite{Barbieri:2004qk,Wells:2015uba,Wells:2015cre} emerge from the direct couplings of heavy BSM particles to the SM gauge and Higgs bosons. Given that such couplings are unavoidable in any BSM scenario that is connected with EW or EW symmetry-breaking physics, universal operators are very robust BSM probes. Universal operators do contribute to top-quark physics; however, they also produce correlated effects in a variety of other processes such as di-lepton, di-boson, associated Higgs boson production, and vector boson scattering processes. Since they are expected to be probed better in these other channels, we will not consider them here. Relevant operators are instead the ones, dubbed ``top-philic'', that emerge from the direct BSM coupling to the top-quark fields $\PQq=\{\PQt_\mathrm{L},\PQb_\mathrm{L}\}$ and $\PQt=\PQt_\mathrm{R}$.\footnote{Top-philic operators have also been adopted as one of the standards for top-quark measurements at the LHC~\cite{AguilarSaavedra:2018nen}.} There are valid reasons, supported by concrete BSM scenarios (see \autoref{sec:phenom_interp} for a discussion), to expect strong new physics couplings with the top quark, and consequently enhanced top-philic operator coefficients. Top-philic effects can thus be more effective indirect probes of new physics than the universal ones, where such an enhancement might not appear.

The top-philic operators are identified by first classifying all the $d=6$ gauge-invariant operators involving $\PQq$ and $\PQt$ fields, plus an arbitrary number of derivative and bosonic SM fields.\footnote{We ignore operators with gluon fields because they do not contribute at leading order to the final states considered in this paper.} Next, we apply Equations of Motion (EOM) and other identities to write each of them as a linear combination of Warsaw basis operators \cite{Grzadkowski:2010es} and we identify the independent combinations. This results in the nine top-philic operators, listed in \autoref{tab:operators}, which will be the focus of this paper. Note that because of the usage of the EOM for the gauge fields, some of the top-philic operators involve more than just $\PQq$, $\PQt$ and the bosonic fields. For instance $Q_{\Pl\PQt,\mathrm{B}\Xspace}$ is a four-fermion lepton-top-quark operator that emerges from
\begin{equation*}
\overline{\PQt}\PGg^{\PGm} t\, D^{\PGn} \mathrm{B}\Xspace_{\PGm\PGn}\overset{\textrm{\tiny{EOM}}}{=}-\frac{g^\prime}{2}Q_{\varphi{\PQt}}+g^\prime Q_{\Pl\PQt,\mathrm{B}\Xspace}+\ldots\,,
\end{equation*}
where $g^\prime$ is the hypercharge coupling, and the dots stand for four-fermion operators involving the top-quark, light quarks and leptons other than the electron. The latter ones can be safely ignored in the present analysis. Similarly one can construct ${{Q}}_{\Pl\PQq,\mathrm{B}}$ and $Q_{\Pl\PQt,\PW}$, for a total of 3 four-fermion operators that are specific linear combinations of the $4$ four-fermion operators that contribute to \eett, identified in \cite{AguilarSaavedra:2010zi}. Operators of this kind induce effects that grow quadratically with the centre-of-mass energy, hence they can be very efficiently probed by the high-energy stages of CLIC.

\begin{table}
\centering
\def\arraystretch{1.5}
\begin{tabular}{|c|}
\hline
%\phantom{\Large{H}}
$Q_{\varphi{\PQt}}\hspace{-2pt}=\hspace{-3pt}(\varphi^\dagger i\Dd_{\PGm}\varphi)({\overline{\PQt}}\PGg^{\PGm} \PQt)$ \\[.5ex]
\hline
%\phantom{\Large{H}}
$Q_{\PQt\varphi}\hspace{-2pt}=\hspace{-3pt}(\varphi^\dagger\varphi) (\overline{\PQq}\,\PQt\,\widetilde\varphi)$ \\[.5ex]
\hline
%\phantom{\Large{H}}
$Q_{\PQt\mathrm{B}}\hspace{-2pt}=\hspace{-3pt}(\overline{\PQq}\sigma^{\PGm\PGn}\PQt)\widetilde\varphi \mathrm{B}_{\PGm\PGn}$\\[.5ex]
\hline
%\phantom{\Large{H}}
$Q_{\varphi \PQq}^{(1)}\hspace{-2pt}=\hspace{-3pt}(\varphi^\dagger i\Dd_{\PGm}\varphi)({\overline{\PQq}}\PGg^{\PGm} \PQq)$\\[.5ex]
\hline
%\phantom{\Large{H}}
$Q_{\varphi \PQq}^{(3)}\hspace{-2pt}=\hspace{-3pt}(\varphi^\dagger i\Dd\vphantom{D}_{\PGm}^I\varphi)({\overline{\PQq}}\,\PGt^I\PGg^{\PGm} \PQq)$\\[.5ex]
\hline
%\phantom{\Large{H}}
$Q_{\PQt \PW}\hspace{-2pt}=\hspace{-3pt}({\overline{\PQq}}\sigma^{\PGm\PGn} \PQt)\PGt^I\widetilde\varphi\, \PW_{\PGm\PGn}^I$\\[.5ex]
\hline
\end{tabular}

\vspace{3mm}

\begin{tabular}{|c|}
\hline
%\vphantom{\Large{H}}
${{Q}}_{\Pl\PQt,\mathrm{B}}\hspace{-2pt}=\hspace{-3pt}({\overline{\PQt}}\PGg^{\PGm} \PQt)({\overline{\Pe}}\PGg_{\PGm} \Pe+\frac12 {\overline{\Pl}}\PGg_{\PGm} \Pl)\hspace{-3pt}\overset{\textrm{\tiny{EOM}}}{=}\hspace{-3pt}\frac12 Q_{\varphi{\PQt}}+\frac1{g^\prime}\overline{\PQt}\PGg^{\PGm} \PQt\, D^{\PGn} \mathrm{B}_{\PGm\PGn}+\ldots$ \\[.5ex]
\hline
%\vphantom{\Large{H}}
${{Q}}_{\Pl\PQq,\mathrm{B}}\hspace{-2pt}=\hspace{-3pt}({\overline{\PQq}}\PGg^{\PGm} \PQq)({\overline{\Pe}}\PGg_{\PGm} \Pe+\frac12 {\overline{\Pl}}\PGg_{\PGm} \Pl)\hspace{-3pt}\overset{\textrm{\tiny{EOM}}}{=}\hspace{-3pt}
\frac12 Q_{\varphi{\PQq}}^{(1)}+\frac1{g^\prime}\overline{\PQq}\PGg^{\PGm} \PQq\, D^{\PGn} \mathrm{B}_{\PGm\PGn}+\ldots$ \\[.5ex]
\hline
%\vphantom{\Large{H}}
${{Q}}_{\Pl\PQq,\PW}\hspace{-2pt}=\hspace{-3pt} ({\overline{\PQq}}\PGt^I\PGg^{\PGm} \PQq) ({\overline{\Pl}}\PGt^I\PGg_{\PGm} \Pl)
\hspace{-3pt}\overset{\textrm{\tiny{EOM}}}{=}\hspace{-3pt}
- Q_{\varphi{\PQq}}^{(3)}-\frac2{g}\overline{\PQq}\PGt^I \PGg^{\PGm} \PQq\, D^{\PGn} \PW_{\PGm\PGn}^I+\ldots$ \\[.5ex]
\hline
\end{tabular}
\caption{The nine top-philic $d=6$ operators considered in the present EFT analysis. All operators are those in \cite{Grzadkowski:2010es}, with the exception of $Q_{\Pl\PQt,\mathrm{B}\Xspace}$, $Q_{\Pl\PQq,\mathrm{B}\Xspace}$ and $Q_{\Pl\PQq,\PW}$, which are linear combinations of Warsaw basis four-fermion operators. Note that the Hermitian conjugate is added to the Lagrangian for the operators $Q_{\varphi{\PQt}\Xspace}$, $Q_{\varphi \PQq\Xspace}^{(1)}$, and $Q_{\varphi \PQq\Xspace}^{(3)}$, in spite of the fact that they are manifestly real. Hence, they effectively appear in the Lagrangian with an extra factor of 2. \label{tab:operators}}
\end{table}

Operators that belong neither to the universal nor to the top-philic categories are due to sizeable BSM couplings to the light fermions, a possibility that is generically disfavoured by flavour constraints for relatively light new physics, in the range of $10-100\,\tev$. Operators in this class can thus be generated only in BSM scenarios with exotic flavour structures, hence they would be more conveniently studied in the context of specific flavour models. For this reason we restrict the EFT analysis presented in this paper to top-philic BSM scenarios.\footnote{Note that when describing the CLIC capabilities to detect exotic top-quark decays we will implicitly be probing operators of the above mentioned type, however we will not phrase those results in the EFT language, but rather in terms of sensitivity to the branching ratios.}

\paragraph{Electroweak couplings and \texorpdfstring{\ttbar}{ttbar} production}
The operators listed in \autoref{tab:operators} produce correlated BSM effects in all the top-related processes at CLIC that are the subject of the present paper. BSM corrections arise from modifications of the SM Feynman vertices and from new interactions that are absent in the SM. For instance, the current-current operators $Q_{\varphi{\PQt}}$, $Q_{\varphi \PQq}^{(1)}$ and $Q_{\varphi \PQq}^{(3)}$ modify the $\ttbar\PZ$ SM vertex, but they also induce a new vertex, $\ttbar\PZ\PH$, that can be probed in $\ttH$ production. In contrast, the four-fermions operators ${{Q}}_{\Pl\PQt,\mathrm{B}}$, ${{Q}}_{\Pl\PQq,\mathrm{B}}$ and ${{Q}}_{\Pl\PQq,\PW}$ only produce new interactions and do not modify the SM vertices. This illustrates well that the formalism of anomalous couplings, that only includes corrections to the SM vertices, is inadequate to parametrise the effects induced by the EFT. Thus a direct comparison of the EFT prediction with data is needed, which is the approach we followed in this study.\footnote{Further note that the formalism of anomalous couplings, even when applicable, often hides relevant phenomenological aspects. For example, the sizeable and growing-with-energy contribution of the current-current operators to vector boson fusion top pair production is manifest in the EFT language thanks to the Equivalence Theorem, while in the anomalous couplings formalism it can be established only by direct computation.}

\begin{table}
\centering
\def\arraystretch{1.8}
\begin{tabular}{c|c|c|}
\multicolumn{1}{c}{}&
\multicolumn{1}{c}{$\PGg$} &
\multicolumn{1}{c}{$\PZ$} \\
\cline{2-3}
$F_{1V}$ & $\frac23$ & $\frac{\frac14-\frac23 \sW^2}{\sW\cW} - \frac{v^2}{2\sW\cW}\left[
C_{\varphi \PQt}+C_{\varphi \PQq}^{(1)}-C_{\varphi \PQq}^{(3)}
\right]$  \\
\cline{2-3}
$F_{1A}$ & $0$ & $\frac{-1}{4\sW\cW} - \frac{v^2}{2\sW\cW}\left[
C_{\varphi \PQt}-C_{\varphi \PQq}^{(1)}+C_{\varphi \PQq}^{(3)}
\right]$  \\
\cline{2-3}
$F_{2V}$ & $\frac{\sqrt{2}vm_{\PQt}}{e} \left[\cW C_{\PQt\mathrm{B}}+\sW C_{\PQt\PW} + {\rm{h.c.}}\right]$ & $\frac{\sqrt{2}vm_{\PQt}}{e} \left[\cW C_{\PQt\PW}-\sW C_{\PQt\mathrm{B}} + {\rm{h.c.}}\right]$ \\
\cline{2-3}
$F_{2A}$& $\frac{\sqrt{2}vm_{\PQt}}{e} \left[\cW C_{\PQt\mathrm{B}}+\sW C_{\PQt\PW} - {\rm{h.c.}}\right]$ & $\frac{\sqrt{2}vm_{\PQt}}{e} \left[\cW C_{\PQt\PW}-\sW C_{\PQt\mathrm{B}} - {\rm{h.c.}}\right]$  \\
\cline{2-3}
\end{tabular}
\caption{The contributions to the top-quark photon and $\PZ$ couplings induced by the operators in \autoref{tab:operators}. The sine and the cosine of the weak mixing angle are denoted \sW and \cW, respectively, while $v\simeq246\,\gev$ is the Higgs field vacuum expectation value.}\label{tab:ancoupt}
\end{table}

When focussing on specific processes and observables, it is in some cases possible to make partial contact between the EFT and the modified couplings approach. The $\epem\to\ttbar$ differential cross section, which will play an important role in \autoref{sec:pairprod}, is discussed below. Inspection of \autoref{tab:operators} reveals that two sources of new physics effects are present. One source is due to the modified $\PZ$ and photon top-quark vertices, which, in the parametrisation of \cite{Schmidt:1995mr}, read
\begin{equation}\label{eq:ancoup}
\displaystyle
ie\left[\PGg^{\PGm}(F_{1V}^{\PGg,\PZ}+\PGg^5 F_{1A}^{\PGg,\PZ})+\frac{i\sigma^{\PGm\PGn}q_{\PGn}}{2\,m_{\PQt}}(F_{2V}^{\PGg,\PZ}+\PGg^5 F_{2A}^{\PGg,\PZ})  \right]\,,
\end{equation}
where $e$ is the electric charge, $m_{\PQt}$ is the top mass and $q$ denotes the (incoming) vector boson momentum. The form-factor parameters $F_{1(2)V(A)}^{\gamma,Z}$ contain the SM vertices and the corrections proportional to the EFT Wilson coefficients as in \autoref{tab:ancoupt}. The second source of new physics effects are the four-fermions contact interactions with the generic structure
\begin{equation}\label{eq:ci}
\sum\limits_{i,j=\{L,R\}}C_{ij}(\overline{\Pe}_i\PGg^{\PGm} \Pe_i)(\overline{\PQt}_i\PGg_{\PGm} \PQt_j)\,,
\end{equation}
where 
\begin{align}
C_{LL} &= - C_{\Pl\PQq,\PW}+\frac12  C_{\Pl\PQq,\mathrm{B}}, \nonumber\\
C_{LR} & = \frac12  C_{\Pl\PQt,\mathrm{B}}, \nonumber\\
C_{RL} & = \frac{}{}C_{\Pl\PQq,\mathrm{B}}, \nonumber\\
C_{RR} &= \frac{}{}C_{\Pl\PQt,\mathrm{B}}. \nonumber
\end{align}
A proper description of the EFT thus requires the anomalous couplings in \autoref{eq:ancoup} to be supplemented with the contact interactions contributions in \autoref{eq:ci}.

The polarised $\epem\to\ttbar$ cross section, differential in the top-quark centre-of-mass scattering angle $\theta^*$ (defined with respect to the $\Pem$ beam), reads
\begin{equation}\label{eq:diffcsttbar}
\frac{d\sigma}{d(\cos(\theta^*))}(\Pe^-_{h_{\Pe^-}}\Pe^+_{h_{\Pe^+}}\rightarrow \PQt_{h_{\PQt}} {\overline{\PQt}}_{h_{\overline{\PQt}}})=\frac{\beta}{16\pi s}\left|
{\mathcal{\widehat{M}}}(h_{\Pe^-},h_{\Pe^+},h_{\PQt},h_{\overline{\PQt}})\right|^2\left(d^1_{h_{\Pe^-}-h_{\Pe^+},h_{\PQt}-h_{\overline{\PQt}}}\right)^2\,,
\end{equation}
where $h$ is the helicity in the centre-of-mass frame, $d^{j}_{m,m^\prime}(\theta^*)$ denotes the standard Wigner $d$-functions, $s$ is the centre-of-mass energy, and $\beta^2=1-4m_{\PQt}^2/s$ is the top-quark velocity. Properly normalised helicity amplitudes, with the dependence on $\theta^*$ factorised and encapsulated in the Wigner functions, are denoted as ${\mathcal{\widehat{M}}}$ and their explicit expressions are reported in \autoref{eq:amp} in \autoref{sec:hamp}. The contributions from the anomalous couplings in \autoref{eq:ancoup} (see also \cite{Schmidt:1995mr}) and from the contact interactions in \autoref{eq:ci} are clearly identifiable in these equations. It is worth emphasising that the latter contribution, unlike the former, produces terms that grow with the centre-of-mass energy, as $s$. This is the reason why the high-energy CLIC stages are so effective in probing the contact interaction operators, as we will see in \autoref{sec:phenom_interp}.

\subsection{Beyond Standard Model (BSM) top-quark decay}
\label{ssec:bsmtopdecay}

One of the possible ways to look for possible BSM physics effects in
top-quark physics at CLIC is the search for rare top-quark decays.
With the close to 1.4 million top quarks and anti-quarks expected at
the initial stage of $\roots=380\,\gev$, discoveries or limits down to branching fractions of about $10^{-5}$
are reachable.
FCNC top-quark decays,
$\PQt \rightarrow  \PQq \PX$ ($\PQq = \PQu, \; \PQc$; $\PX = \PGg, \; \Pg, \; \PZ, \; \PH$)\footnote{Charge conjugation is implied unless explicitly stated otherwise.}, are of
particular interest as they are very strongly suppressed in the SM. 
They are forbidden at tree level, 
and the loop level contributions are suppressed by the GIM-mechanism~\cite{Glashow:1970gm}. 
The suppression is not perfect because of the non-negligible $\PQb$-quark
mass; the  corresponding partial widths are proportional to the
square of the element $V_{\PQq\PQb}$ of the CKM-quark-mixing matrix~\cite{Cabibbo:1963yz,Kobayashi:1973fv}
and to the fourth power of the ratio of the $\Pb$ quark and $\PW$ boson masses.
These suppression factors\footnote{The GIM mechanism is not strictly
  applicable to the \tch channel as the Higgs coupling
  is proportional to the quark mass. Still, the expected FCNC 
branching ratio for this channel is the smallest in the SM.}
result in extremely small branching ratios. For decays involving a 
charm quark, SM expectations~\cite{Agashe:2013hma} are: 
\begin{eqnarray*}
\text{BR}(\PQt \to \PQc \Pg) & \sim & 5 \cdot 10^{-12}, \\
\text{BR}(\tcg) & \sim & 5 \cdot 10^{-14}, \\
\text{BR}(\PQt \to \PQc \PZ) & \sim & 1 \cdot 10^{-14}, \\
\text{BR}(\tch) & \sim & 3 \cdot 10^{-15}.
\end{eqnarray*}
The SM expectations for decays with an up quark in the final state decrease by another two orders of magnitude~\cite{Agashe:2013hma}.
Observation of decays involving either a charm or up quark would therefore constitute a direct signature
for BSM physics.

Many extensions of the SM predict significant enhancements
of the FCNC top-quark decays~\cite{Agashe:2013hma,deBlas:2018mhx}.
These enhancements can be due to FCNC couplings at tree level, but
in most models they result from contributions of new particles or from
modified particle couplings at the loop level. 
For most BSM scenarios, significant deviations in the
(light) Higgs boson couplings or contributions from additional
Higgs bosons to the loop diagrams result in the significant
enhancement of the \tch decay.
For the Two Higgs Doublet Model (2HDM), which is one of the simplest
extensions of the SM, loop contributions can be 
enhanced up to the level of $BR \sim 10^{-4}$~\cite{Bejar:2001sj}.
For the ``non-standard'' scenarios, 2HDM(III) or ``Top~2HDM'',
where one of the Higgs doublets only couples to the top quark, tree level
FCNC couplings are also allowed. Here an enhancement of up to
$10^{-2}$ is possible~\cite{DiazCruz:2006qy}.
BR(\tch) could be observable at CLIC also for the Randall-Sundrum
warped models or composite Higgs models with flavour violating Yukawa
couplings, provided the compositeness scale is sufficiently low (below
TeV scale)~\cite{deBlas:2018mhx}.
However, the possible observation of \tch should then be accompanied
by even more significant deviation of the measured Higgs boson
couplings to the vector bosons from the SM expectations.
Significant enhancement of FCNC top decays is also expected for SUSY
scenarios with $R$-parity violation. 
Enhancement up to the level of $10^{-5}$ is possible for both the \tch
\cite{Bardhan:2016txk} and the \tcg decay~\cite{Mele:1999zx}.
For an overview of top-quark FCNC predictions for different BSM scenarios
see \cite{Agashe:2013hma,deBlas:2018mhx}.

In the study presented here, the FCNC couplings involving the charm
quark are considered, as they are expected to be favoured in many BSM
scenarios. The three channels selected for
detailed study (see \autoref{sec:fcnc}) are: \tcg, \tch, and \tcx. In the latter,  
a top quark decays into a $\PQc$-jet and an invisible heavy scalar particle.
The existence of such particles, with masses in the 100\,GeV range, is
still allowed in many BSM scenarios, see for example \cite{Kalinowski:2018ylg}.

\section{Event generation, detector simulation, and reconstruction}
\label{sec:gendetreco}

The results reported here are based on detailed Monte Carlo (MC) simulation studies with 
\geant \cite{Agostinelli2003, Allison2006} 
based simulations of the CLIC detector concepts and a full event reconstruction,
unless indicated otherwise. 
All relevant background processes are included.  
Event simulation and reconstruction 
is performed using the \ilcdirac grid production tools \cite{Grefe:2014sca, Tsaregorodtsev:2008zz}.

\subsection{Event generation}
\label{ssec:gen}

The signal processes and main physics backgrounds, with up to six particles in the final state, are generated using the \whizard 1.95~\cite{Kilian:2007gr} program. ISR is described using the leading logarithmic approximation structure function~\cite{Skrzypek:1990qs} including hard collinear photons up to the third order. 
For many analyses only the backgrounds from $\epem$ collisions contribute. However, for some studies it is important also to include MC event samples from $\Pep\PGg$, $\PGg\Pem$, and $\PGg\PGg$ interactions, with photons originating from beamstrahlung.
In all cases the expected energy spectra for the CLIC beams, including the effects from beamstrahlung and the intrinsic energy spread, are used for the initial-state electrons, positrons and beamstrahlung photons. Low-$Q^2$ processes with quasi-real photons are described using the Weizs\"{a}cker-Williams approximation as implemented in \whizard.

The process of fragmentation and hadronisation is simulated using \pythia 6.4 \cite{Sjostrand2006} with a parameter set tuned to OPAL $\epem$ data recorded at LEP \cite{Alexander:1995bk} (see~\cite{CLIC_PhysDet_CDR} for details). The impact of other \pythia tunes in top-quark pair production events is illustrated in~\cite{Chekanov:2289960}. The decays of $\PGt$ leptons are simulated using \tauola~\cite{tauola}. MC samples with eight final-state fermions, for the study of the top Yukawa coupling measurement (see Section~\ref{sec:ttH}), are obtained using the \physsim~\cite{gen:physsim} package; again \pythia is used for fragmentation and hadronisation. The mass of the Higgs boson is taken to be $125.0\,\GeV$ and the decays of the Higgs boson are simulated using \pythia with the branching fractions listed in \cite{Dittmaier:2012vm}.
Apart from the special MC samples used for the threshold and radiative top-quark mass studies, the top-quark mass is set to  $m_{\PQt}=174.0$\,GeV.

\subsection{Detector simulation}
\label{ssec:detsim}

The \geant detector simulation toolkits \mokka~\cite{Mokka} and \slic~\cite{Graf:2006ei} are used to simulate the detector response to the generated events in the \clicild and \clicsid concepts,
respectively. The \texttt{QGSP\_BERT} physics list is used to model the hadronic interactions of particles in the detectors. The digitisation, i.e. the translation of the raw simulated energy
deposits into detector signals,
is performed using the \marlin~\cite{MarlinLCCD} and \lcsim \cite{Graf:2011zzc} software packages.

The most important beam-induced background are particles from the $\gghadrons$ process, a result of the high bunch charge density at high collision energy. These interactions are simulated separately using \pythia 6.4 \cite{Sjostrand2006} with the photon spectra from \guineapig \cite{guineapig}.  Events corresponding
to 60 bunch crossings are superimposed on the physics events before digitisation;
this is equivalent to 30\,ns and is much longer than the offline reconstruction
window, which is assumed to be 10\,ns around the hard physics event.
At $\sqrt s=380$\,GeV, the impact of this background is found to be small, but is larger 
at $\sqrt s=3$\,TeV, where approximately $1.2$\,TeV of energy is deposited in the calorimeters
during the 10\,ns time window \cite{CLIC_PhysDet_CDR}.
\subsection{Reconstruction}
\label{ssec:reco}

Track reconstruction is performed using the
\marlin and, for the \clicsid\ detector model, the \lcsim software packages.  
Calorimeter clustering and particle flow reconstruction is performed using
\pandora~\cite{thomson:pandora, Marshall2013153, Marshall:2015rfa},
creating a collection of so-called Particle-Flow Objects (PFOs).
Time-stamping information is used to suppress beam-related backgrounds.
To be used for further analysis, PFOs are required to have time stamps
of up to between 1 and 5\,ns around the reconstructed hard scattering interaction,
depending on the identified particle type, \pT, and detector region~\cite{CLIC_PhysDet_CDR}.
Three levels of timing selections are studied for each collision energy: \code{loose}, \code{default}, and \code{tight}, each applying a more stringent selection of the PFOs. In general, the more stringent selections are found to perform better for operation at higher centre-of-mass energy, where the beam backgrounds are more significant, and vice versa for operation at the initial CLIC stage.\footnote{Table B.1, B.2, and B.3 in \cite{CLIC_PhysDet_CDR} illustrate the timing selection cuts applied for the analyses at $\roots=1.4\,\tev$ and $\,3\,\tev$, while an adaption of the cuts presented in B.4 was applied for the analyses at the first CLIC stage.}

The classification of candidate top-quark events as fully-hadronic, semi-leptonic, or fully-leptonic requires efficient identification of high-energy, isolated charged leptons. Lepton finding is optimised to identify $\Pepm$ and $\PGmpm$ originating from the decay of $\PW$ bosons\footnote{$\PGtpm$ leptons are searched for using a dedicated TauFinder~\cite{LCDnote_TauFinder} algorithm implemented in \marlin. The algorithm studies the presence of highly energetic and low-multiplicity jets in the detector.}; these leptons are typically of much higher energy than those coming from hadronic decays inside quark jets, and are well-separated from other activity in the event. Isolated leptons candidates are identified by studying their energy depositions in the ECAL and HCAL, impact parameters, and isolation in a cone around each input track. The lepton charge is determined by the curvature of the helix from a standard Kalman-filter-based track reconstruction of the associated hits in the tracking system.

\begin{figure}[t!]
\centering
\includegraphics[width=0.48\columnwidth,clip]{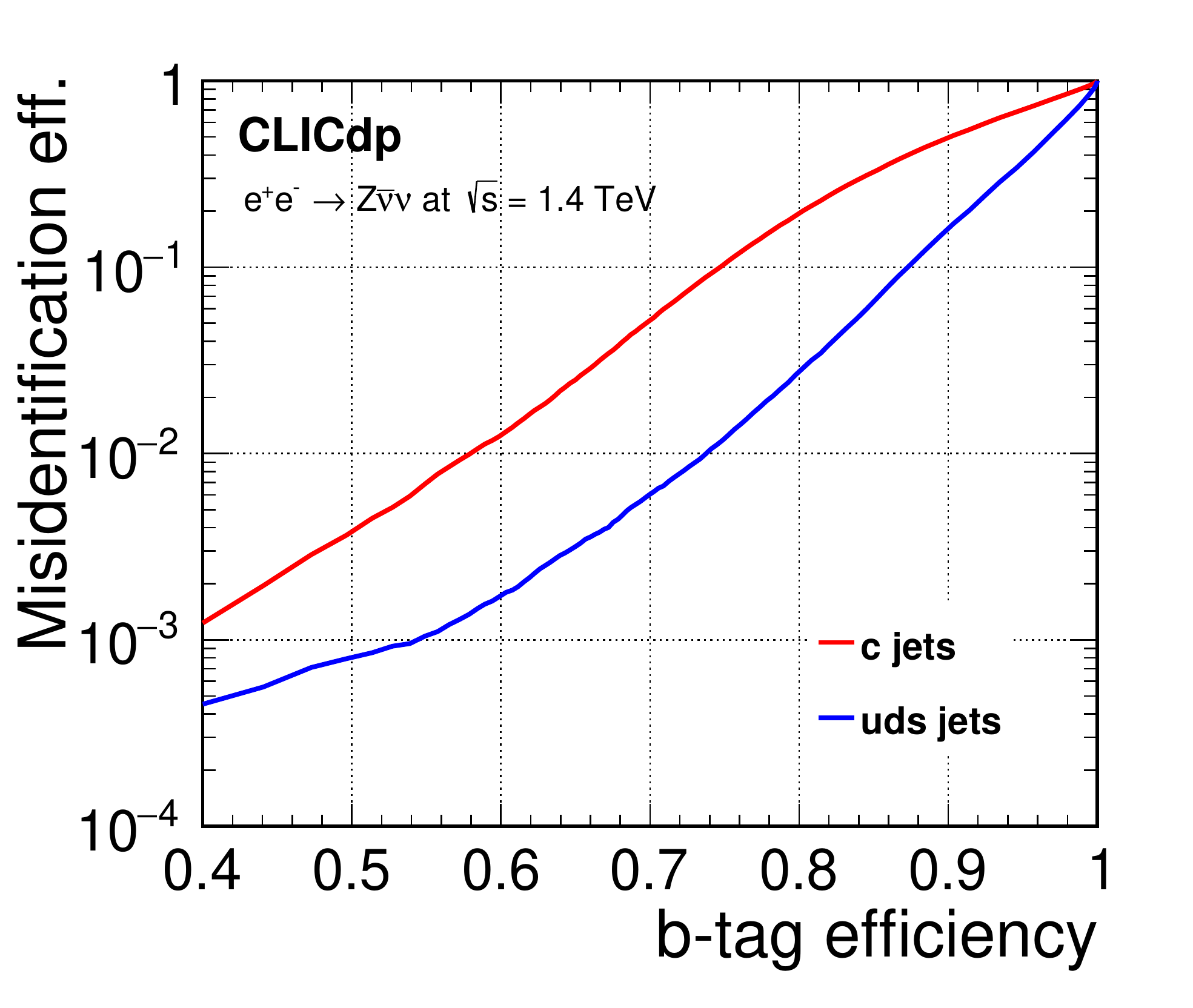}
~~
\includegraphics[width=0.48\columnwidth,clip]{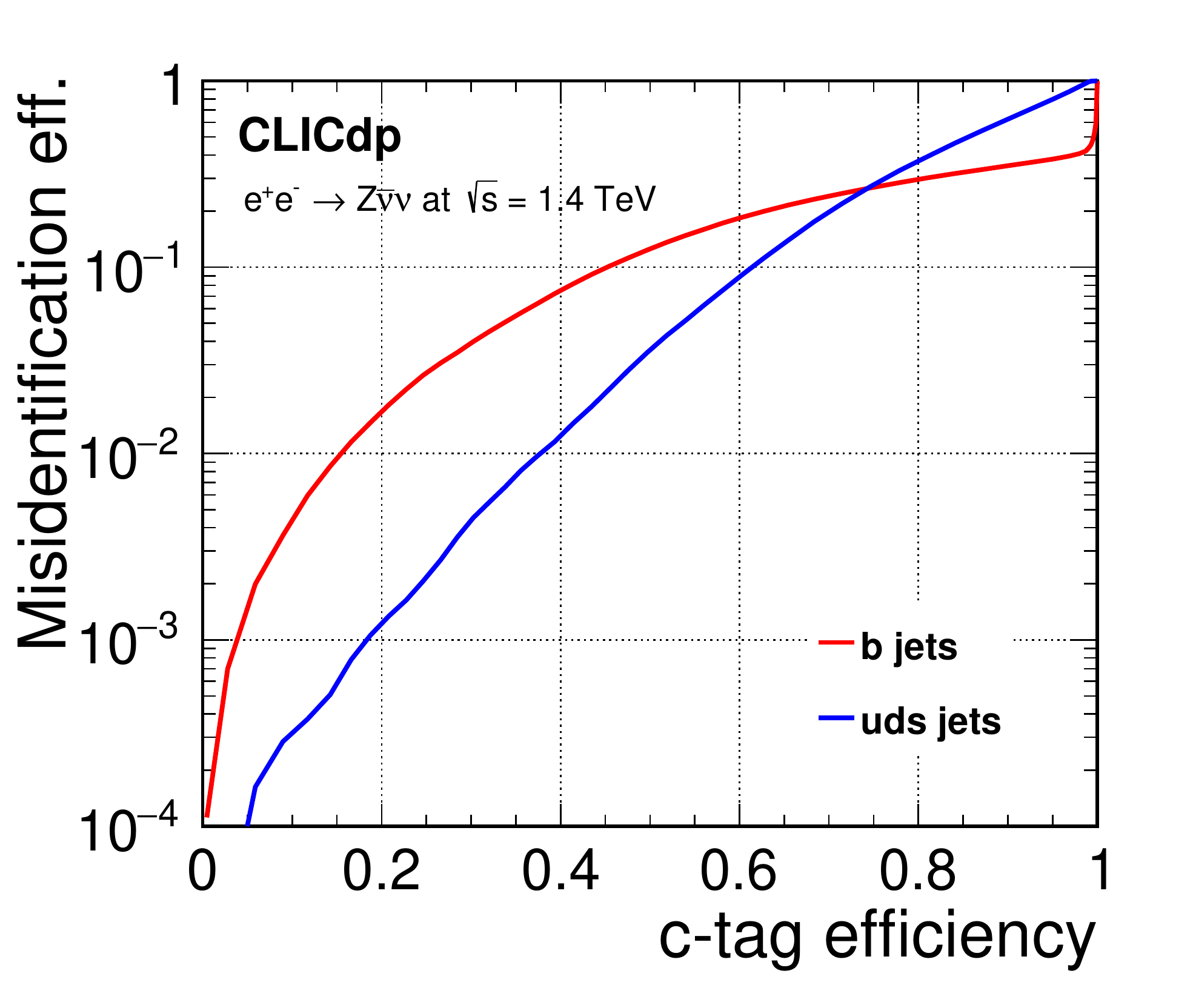}
\caption{Flavour tagging performance in $\epem\to\PZ\PAGn\PGn$ events at 1.4\,\tev reconstructed in the \clicild detector. The fake rates for jets of other flavours are shown functions of the b-tag (left) and c-tag (right) efficiencies.}
\label{fig:flavour_tagging_performance}
\end{figure}

In most cases, jet clustering is performed by the \fastjet package \cite{Fastjet}, in exclusive mode. Both the longitudinally-invariant \kT algorithm \cite{Catani:1993hr, Ellis:1993tq} and the VLC algorithm \cite{Boronat:2016tgd} are used;
these are sequential recombination algorithms that are found to give better robustness
against \gghadrons than traditional lepton collider
jet clustering algorithms \cite{Marshall2013153, CLIC_PhysDet_CDR, Simon2015, Boronat:2016tgd}.
The former uses the particle transverse momenta \pT and angular separation $\Delta R_{ij}^2 = (y_i - y_j)^2 + (\phi_i - \phi_j)^2$, where $y_i$ and $\phi_i$ are the rapidity and azimuth of particle $i$, to compute a clustering distance parameter $d_{ij}=\text{min}(p_{\mathrm{T}\,i}^2,p_{\mathrm{T}\,j}^2)\Delta R_{ij}^2/R^2$, where $R$ is the radius parameter that determines the maximum area of the jet. The VLC algorithm uses the particle energies $E$, and angular separation $\theta$, to compute a clustering distance parameter $d_{ij} = 2\min({E_i^{2\beta},E_j^{2\beta}})(1-\cos{\theta_{ij}})/R^2$.
Here, $\beta$ regulates the clustering order; the default choice is $\beta=1.0$ unless otherwise specified.

Both algorithms are effective for identifying particles that are likely to have originated
from beam-beam backgrounds; if particles are found to be closer to the beam axis than
to other particles then they are removed from the event, which mitigates the effect of \gghadrons pile-up.
For the \kT algorithm, the distance to the beam axis is measured by $d_{iB} = p_{\mathrm{T}\,i}^2$ 
and for the VLC algorithm by $d_{iB} = E_i^{2\beta}{(p_{\mathrm{T}\,i}/E_i)}^{2\gamma}$, where
the $\PGg$ parameter controls the rate of shrinking in jet size in the forward region\footnote{Here we apply the beam distance measure as implemented in the ValenciaPlugin of \fastjet `contrib' versions up to 1.039. Note that this differs slightly from the one quoted in \cite{Boronat:2016tgd}.};
the default choice is $\PGg=1.0$ unless otherwise specified.
The jet clustering algorithm is chosen and optimised for
each analysis to achieve the best balance between losing signal particles and
including extra background particles.  

Flavour tagging is essential for the identification and combinatoric assignment 
of top-quark events.  Vertex reconstruction and heavy-flavour tagging is
performed by the \lcfiplus package \cite{Suehara:2015ura}.
This contains a topological vertex finder that reconstructs the primary and 
secondary vertices. Several BDT classifiers provide
$\PQb$- and $\PQc$-jet probabilities for each jet reconstructed in the event. These are 
based on variables such as secondary vertex decay lengths, multiplicities and masses, as well as track impact parameters.
For analyses heavily dependent on flavour-tagging, \lcfiplus is also used for jet
clustering, using the same algorithms discussed above, but preventing tracks from a common 
secondary vertex to be split into different jets. This approach improves the flavour tagging performance in events 
with a large jet multiplicity.

As an example, the b- and c-tagging capabilities of the \clicild detector concept are 
shown in \autoref{fig:flavour_tagging_performance}. $\epem\to\PZ\PAGn\PGn$ events at \roots=1.4\,\tev were used 
for the training of the BDT's and for the performance evaluation. The jets in the considered process tend towards the 
beam direction where the flavour tagging is generally more difficult. The same training is used for the analysis 
of top-quark pair production at \roots=1.4\,\tev described in \autoref{ssec:pairprodboosted}.
\section{Boosted top-quark tagging}
\label{sec:boosted}
                                                                                                                                           
At the higher energy stages of CLIC, a large proportion of the top quarks in $\epem\to\ttbar$ events is produced with significant boosts leading to a more collimated jet environment where the separation between the individual top-quark decay products in general is small. In particular, the topology is very different from that of top quarks produced close to the production threshold. In this section we present a method exploiting the internal sub-structure of typically large-$R$ jets to tag top-quarks that decay hadronically.

The reconstruction of boosted top quarks was studied in full simulation using the \clicild detector model, including $\gghadrons$ background. The PFOs in each event are clustered in two subsequent steps following the approach described in~\cite{Nachman:2014kla}. In this study, a pre-clustering is done in an inclusive mode using the Generalised-\kT algorithm (with beam jets) for $\epem$ collisions (``gen-\kT algorithm'')~\cite{Fastjet} with a minimum $\pT$ threshold. The resulting PFOs are re-clustered into two exclusive jets using the VLC algorithm. The effect of this two-stage clustering is similar to that of grooming (and in particular trimming): the effective area of the jet is reduced and soft emission does not obscure the reconstruction of its substructure.

\begin{figure}
\centering
\includegraphics[width=0.48\columnwidth,clip]{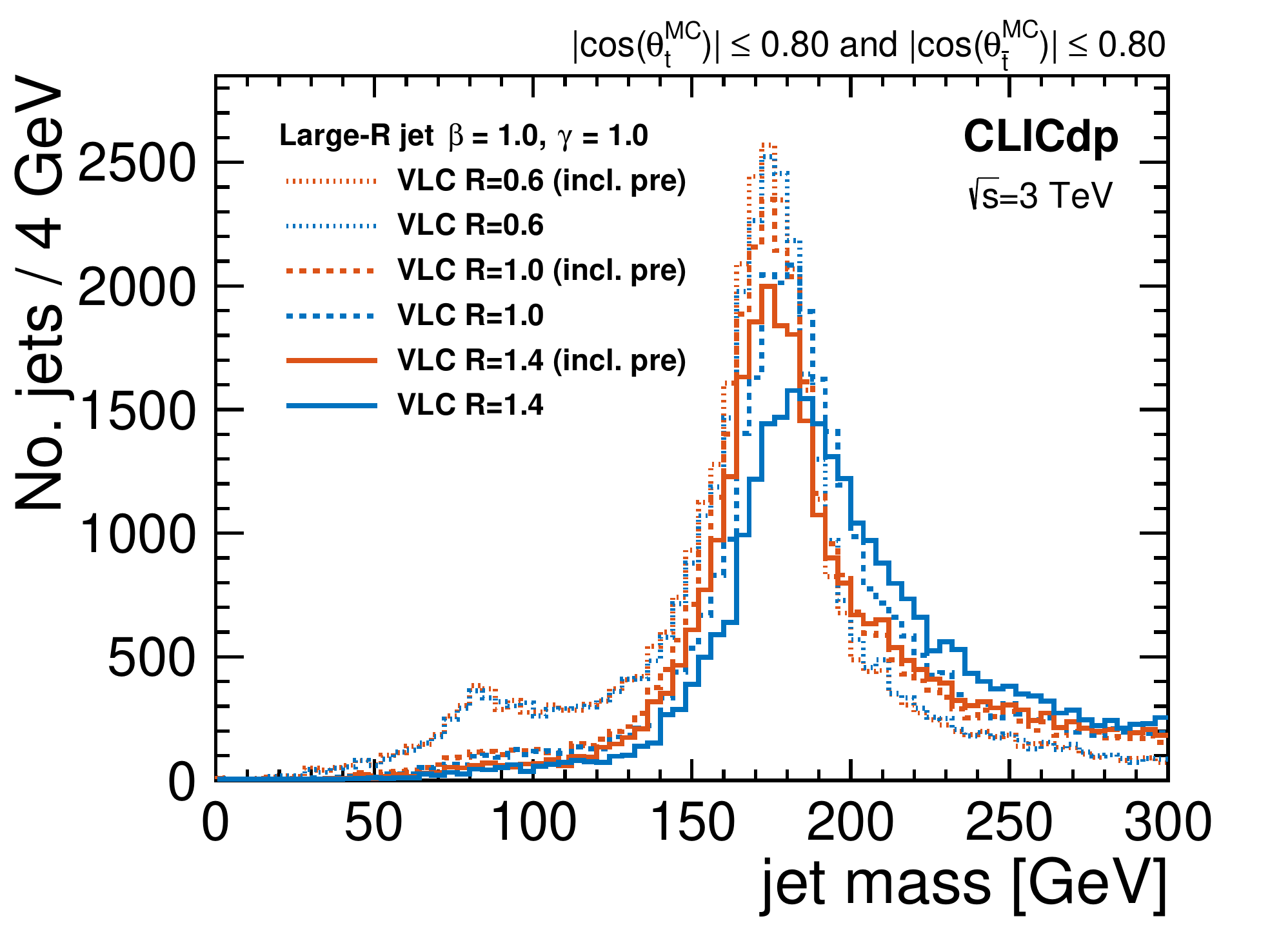}
~~
\includegraphics[width=0.48\columnwidth,clip]{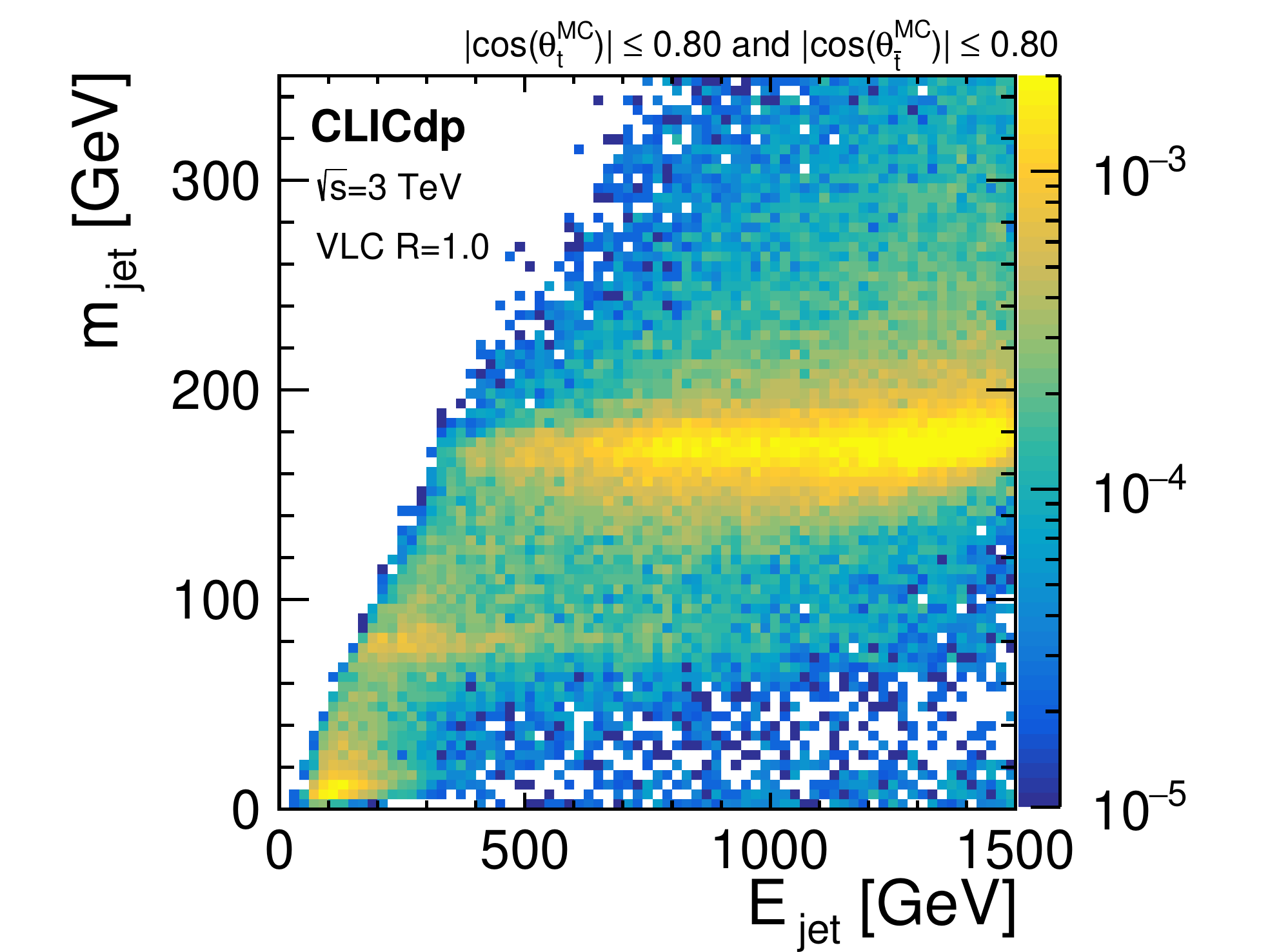}
\caption{Reconstructed large-$R$ jet mass for fully-hadronic $\ttbar$ events in CLIC at $\roots=3\,\tev$. Different choices of jet clustering radius $R$ including an illustration of the effect of applying a pre-clustering step (left). As a function of the corresponding large-$R$ jet energy for the optimal jet clustering parameters at $\roots=3\,\tev$ (right).}
\label{fig:massvspt}
\end{figure}

The left panel of \autoref{fig:massvspt} shows the reconstructed large-$R$ jet mass for different choices of jet clustering radius $R$ and also illustrate the effect of applying the pre-clustering step prior to the large-$R$ jet clustering, as described above. The figure is compiled using fully-hadronic $\ttbar$ events in CLIC at $\roots=3\,\tev$ with a reconstructed collision energy above 2.6\,TeV\footnote{Using the definition of reconstructed collision energy as outlined in \autoref{ssec:pairprodboosted}} and where both top-quarks, at parton-level, are located in the central region of the detector with a polar angle $\theta$ satisfying the condition\footnote{The detector coverage goes down to about $8^\circ$. Excluding a larger area in the forward direction for the optimisation reduces the effect of losing energy down the beam pipe and adds some margin for the finite size of the jets.} $37^\circ\leq\theta\leq143^\circ$. It is clear from the figure that too small a jet radius does not enclose the entire top-quark decay products, leading to a significant peak close to the mass of the $\PW$ boson. In contrast, larger jet radii include a growing contribution from background processes leading to a long tail in the distribution towards higher masses. The optimal jet clustering parameters, for both clustering stages, were selected as the best trade-off between achieving a narrow top-quark mass peak close to the generated parton-level top-quark mass, and minimising the contributions to the mass peak at $m_{\PW}$. In this context, we found that a jet radius of $R=0.4$ and a minimum $\pT$ threshold of $5\,\mathrm{GeV}$ were optimal in the pre-clustering step. Similarly we found that a large-$R$ jet radius of $R=1.4$ and $R=1.0$, each with $\beta=\gamma=1.0$, were optimal for operation at $\roots=\,1.4\,\TeV$ and $\roots=\,3\,\TeV$, respectively.

The right panel of \autoref{fig:massvspt} shows the reconstructed jet mass as a function of the reconstructed jet energy at $\roots=3\,\tev$, for the optimal clustering parameters in the two-step approach. Note that a cut on the reconstructed collision energy was not applied in this figure. The uppermost of the three visible yellow bands indicates top quarks that are fully captured within the large-$R$ jet, while the lower two bands represent partially captured top quarks close to the mass of $m_{\PW}$ and $m_{\PQb}$, respectively. As expected, the large-$R$ jet approach performs well for jets at higher energy, while the ability to capture the full top-quark jet is significantly reduced in the non-boosted regime, below $\sim500\,\gev$. The resulting large-$R$ jets serve as input for the top tagger algorithm described below.

\subsection{Top tagging algorithm and performance}

The tagging of boosted top quarks at CLIC is based on the Johns Hopkins top tagger~\cite{Kaplan:2008ie} as implemented in \fastjet~\cite{Fastjet, Fastjet:2006}. This tagger is explicitly designed for the identification of top quarks by recursively iterating through a jet cluster to search for up to three or four hard subjets and then imposing mass constraints on these subjets. This procedure provides strong discrimination power for hadronically decaying top quarks against QCD-induced light parton jets. Although the method was originally designed for fully-hadronic $\ttbar$ events in hadron colliders, in this paper it is applied to the hadronically decaying top quark in semi-leptonic $\ttbar$ events in CLIC, see \autoref{ssec:pairprodboosted}.

The tagging is based on an iterative de-clustering of the input jet and is carried out by reversing each step of the jet clustering. The algorithm is governed by two parameters: $\delta_r$, the subjet distance; and $\delta_p$, the fraction of subjet $\pT$ relative to the $\pT$ of the input jet. These parameters control whether to accept the objects, resulting from the split, as subjets for further de-clustering or whether, for example, the de-clustering should continue only on the harder of the two objects. An object is rejected if its $\pT$ fraction is lower than $\delta_p$ or if its distance to another object is smaller than $\delta_r$. The de-clustering loop is terminated when two successive splittings have been accepted resulting in two, three, or four subjets of the input jet. The case with two final subjets is rejected and the other cases are further analysed. The input jet is considered to be top-tagged if the total invariant mass of the subjets is within $\pm55$\,GeV of $m_{\PQt}$ and one subjet pair has an invariant mass within $\pm30$\,GeV of $m_{\PW}$.

\begin{figure}
\centering
\includegraphics[width=0.65\columnwidth,clip]{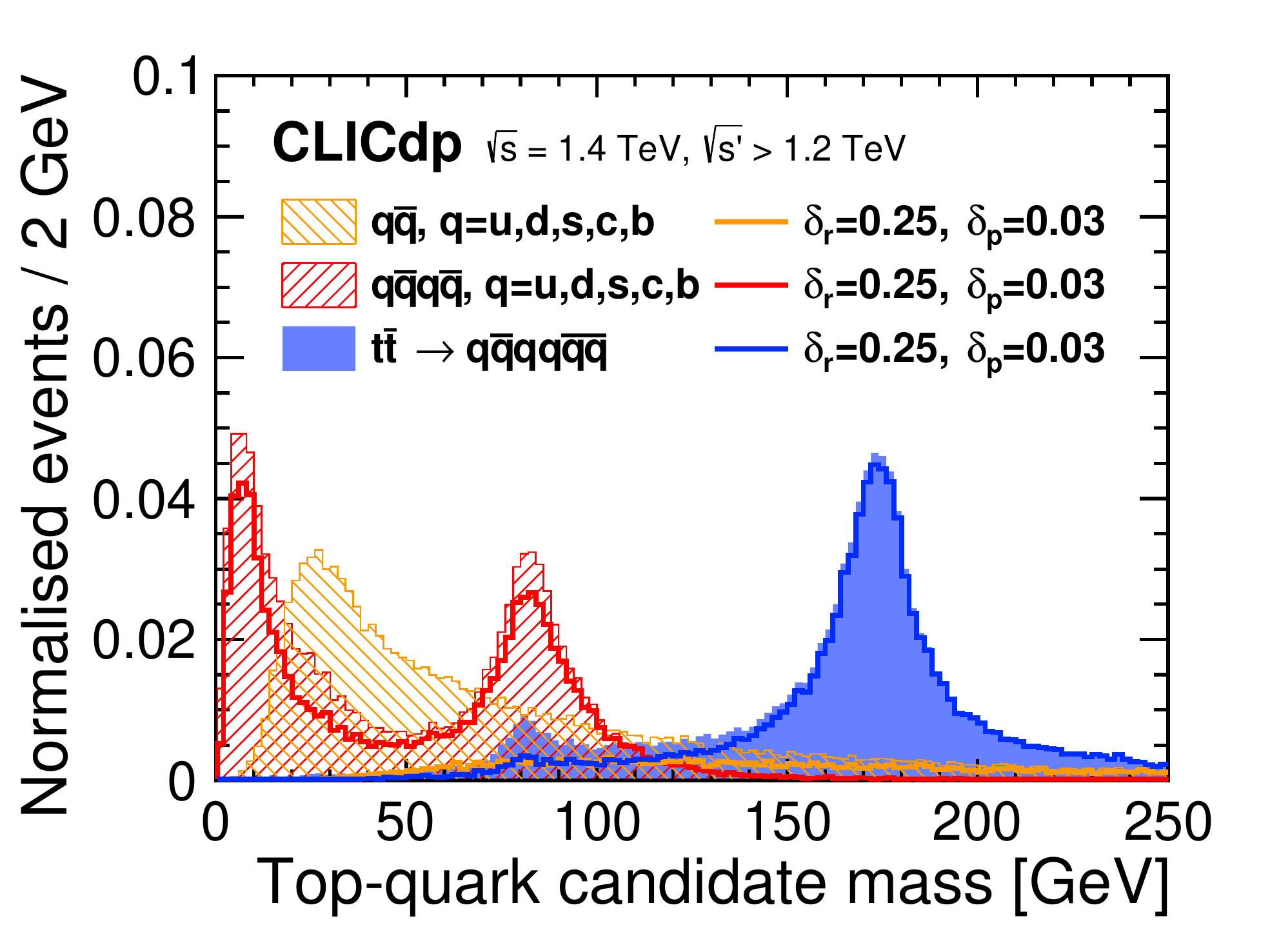}
\caption{Reconstructed top-quark candidate mass distributions at $\roots=1.4\,\tev$ for events with $\rootsprime\geq1.2\tev$. The filled distributions represent the top-quark candidate mass before application of the top tagger and are normalised to unity. The solid lines show the effect of applying the de-clustering procedure outlined in the text. Note that additional cuts on the invariant mass of both the top-quark and $\PW$ candidates are applied in a later step. Fully-hadronic $\ttbar$ events are shown in blue, four-jet events in red, and dijet events in orange. A cut, $|\cos\theta|\leq0.95$, is applied on the polar angle of the individual top and light quarks.}
\label{fig:toptagged:mass}
\end{figure}

\begin{figure}
\centering
\includegraphics[width=0.8\columnwidth,clip]{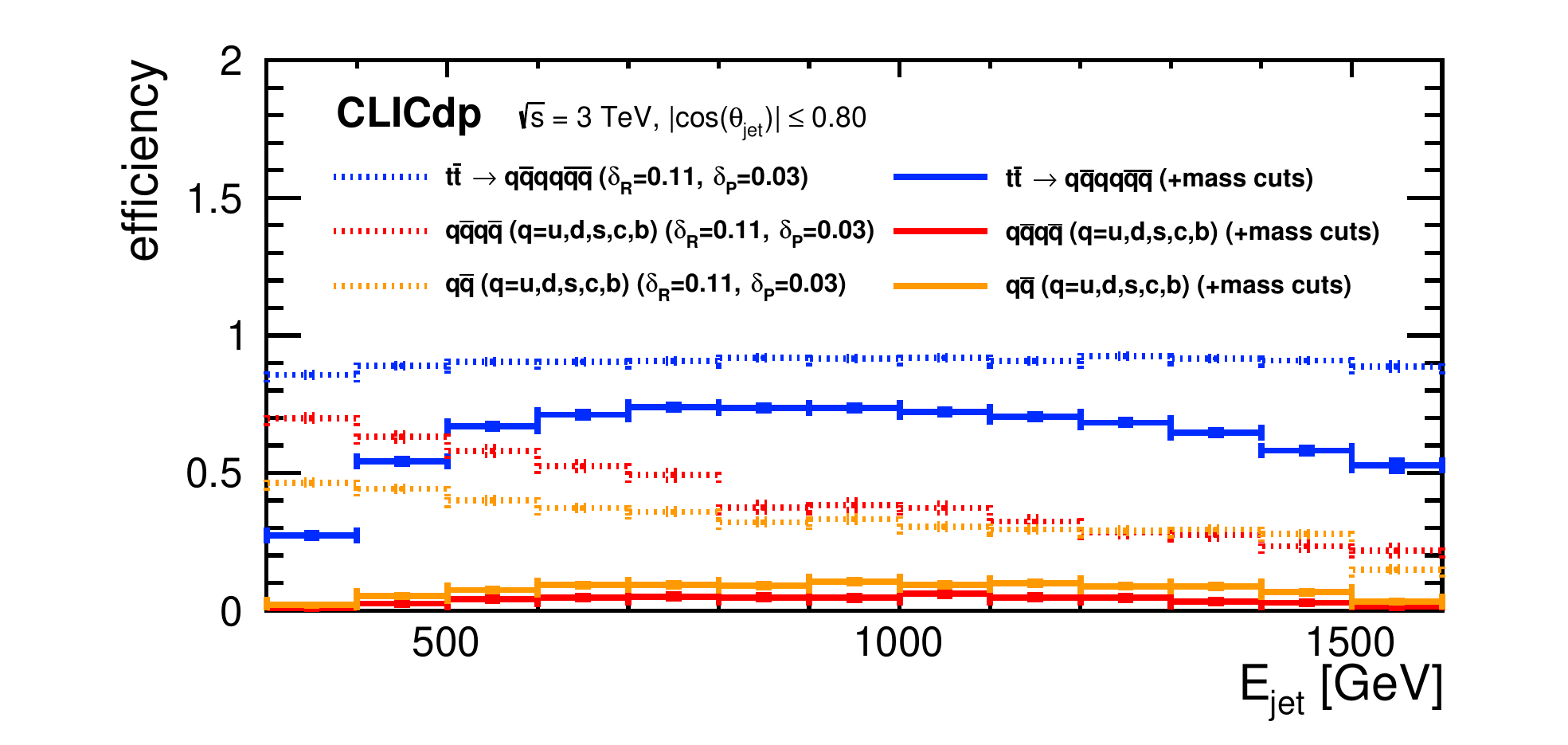}
\includegraphics[width=0.8\columnwidth,clip]{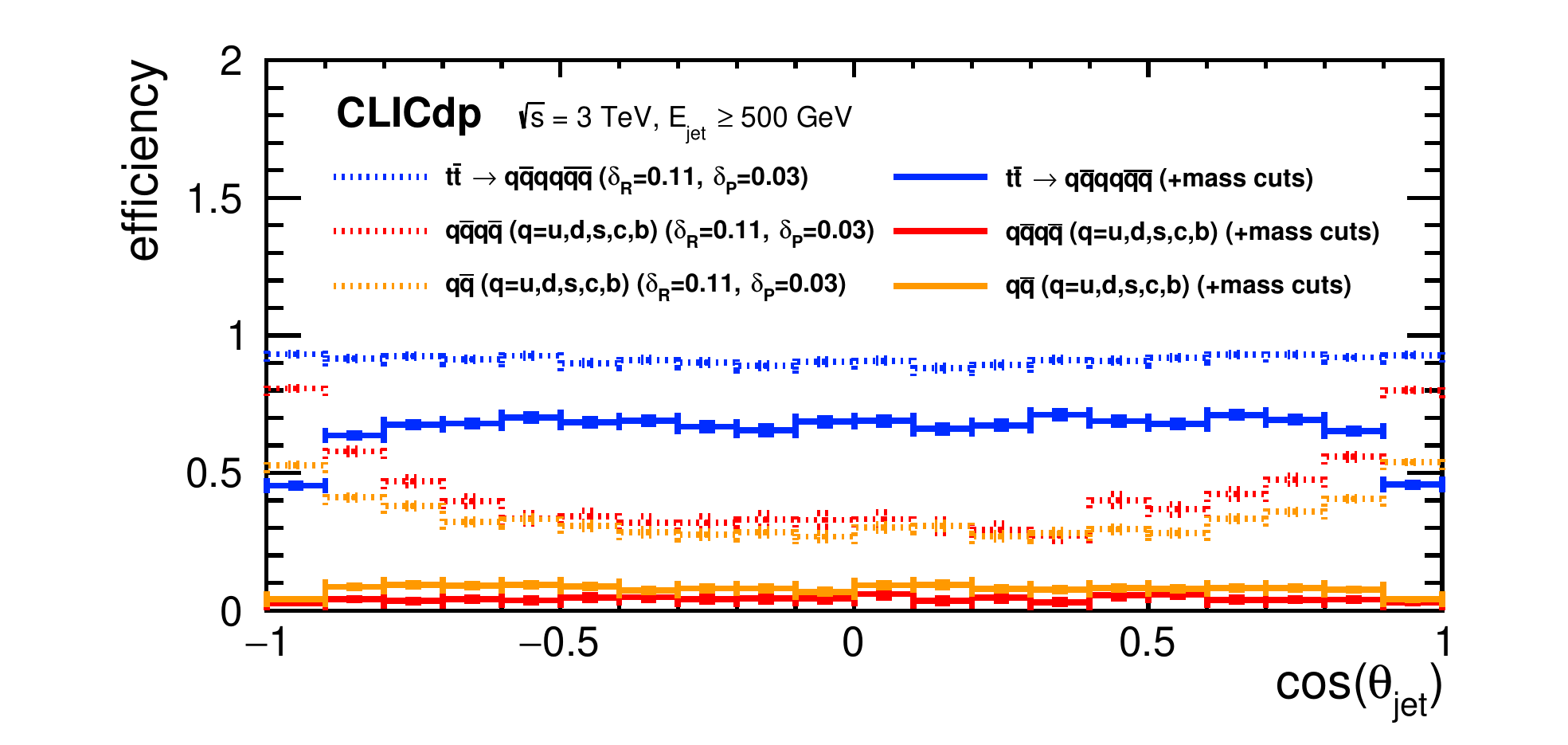}
\caption{Top tagger efficiency for fully-hadronic $\ttbar$ events (blue), four-jet events (red), and dijet events (orange) as function of jet energy (top) and jet polar angle $\theta$ (bottom). The dashed lines show the effect of applying the de-clustering procedure outlined in the text, while the solid lines show the efficiency including also the cuts on the reconstructed invariant mass of the top-quark and $\PW$ candidates.}
\label{fig:toptagged:eff}
\end{figure}

The optimisation and efficiency of the top tagging algorithm was studied using fully-hadronic $\ttbar$ events, four-jet events $\qqbar\qqbar\,(\mathrm{\PQu,\PQd,\PQs,\PQc,\PQb})$, and dijet events $\qqbar\,(\mathrm{\PQu,\PQd,\PQs,\PQc,\PQb})$. 
Since the background environment at a lepton collider is substantially lower than at a hadron collider, a somewhat higher rate of wrongly tagged light-quark $(\mathrm{\PQu,\PQd,\PQs,\PQc,\PQb})$ jets is acceptable and the optimisation of the algorithm is tuned to a high-efficiency operating point for the fully-hadronic $\ttbar$ sample; for the studies presented here we apply a benchmark efficiency of 70\%. The corresponding top tagger parameters, chosen by minimising the rate of wrongly tagged light-quark jets from the four-jet sample, are $\delta_r=0.25\,(0.11)$ and $\delta_p=0.03\,(0.03)$, for the samples at $\roots=1.4\,(3)\,\tev$, respectively. \autoref{fig:toptagged:mass} shows the reconstructed top-quark candidate mass before and after application of the top tagger declustering step for operation at $\roots=1.4\,\tev$. A small peak close to $m_{\PW}$ is clearly seen for the \ttbar distribution (blue) and is caused by top-quark events not fully captured by the large-$R$ jet.

The resulting tagging efficiency for top-quark jets from the $\roots=3\,\tev$ $\ttbar$ dataset is 69\% in the central region of the detector (defined as $|\cos\theta|\leq0.8$) and with an energy in the range from 500\,\gev to 1500\,\gev. The corresponding efficiency for wrongly tagged light-quark jets is substantially lower: 4.4\% and 8.8\% for the four-jet and di-jet background samples, respectively.\footnote{Alternatively, adopting a tighter operating point at $\roots=3\,\tev$ results in a top-quark jet efficiency of 54\% and an efficiency for wrongly tagged light-quark jets of 2.7\% (3.7\%)} The resulting efficiency for top-quark jets from the $\roots=1.4\,\tev$ dataset is 71\% in the central region of the detector (defined as $|\cos\theta|\leq0.8$) and with an energy in the range from 400\,\gev to 700\,\gev. The corresponding efficiency for wrongly tagged light-quark jets is 5.7\% (6.9\%) for jets from the four-jet (di-jet) background sample.
\autoref{fig:toptagged:eff} shows the top-quark tagging efficiency from the $\roots=3\,\tev$ dataset as a function of the large-$R$ jet energy (top) and polar angle $\theta$ (bottom). The dashed lines represent the distributions after the de-clustering step, while the solid lines include also the mass cuts. Note that the de-clustering step is particularly challenging in the forward region where hadrons from the larger beam-beam induced background, on top of the physics event, effectively mimic a prongy topology. As expected, the overall efficiency, including the mass cuts, drops at energies below $500\,\GeV$ where the jets are no longer sufficiently boosted to be contained within one large-$R$ jet. The slightly lower efficiency for large jet energies is also anticipated and is mainly due to a more challenging environment for the \pandora algorithm and the subjet de-clustering. Furthermore, the limited detector acceptance in the forward direction reduces the efficiency in the corresponding region significantly.

The top tagger algorithm outlined above increases the significance, estimated as $S/\sqrt{B}$ where $S$ represents the number of top-quark jets from the fully-hadronic $\ttbar$ sample and $B$ the number of wrongly tagged light-quark jets from either the four-jet or dijet sample, by between 18-26\% (depending on the background process and collision energy considered), compared to a simple cut on the reconstructed large-$R$ jet mass in the corresponding range (within $\pm55$\,GeV of $m_{\PQt}$). In addition, the declustering procedure provides additional handles on the jet substructure such as the mass and kinematic variables of the $\PW$ boson candidate and the reconstructed helicity angle $\theta_{\PW}$ that examine whether the subjets are consistent with a top decay.\footnote{The helicity angle is measured in the rest frame of the reconstructed W boson and is defined as the opening angle of the top quark to the softer of the two \PW boson decay subjets. Too shallow an angle would be an indication of a false splitting, where one of the pairs of subjets produces a small mass compatible with QCD-like emission.} As illustrated in \autoref{sec:pairprod}, these handles are useful to discriminate against the remaining background events. In conclusion, the use of dedicated techniques to reconstruct boosted topologies plays an important role in the physics programme of CLIC, extending the physics reach to higher energies.

\section{Top-quark mass measurements at the initial energy stage}
\label{sec:mass}

A precise measurement of the mass of the top quark is one of the key objectives of the top-physics programme at CLIC. Conceptually, there are two different approaches to this measurement.

The first is the determination of the top-quark mass from measurements of the top-quark pair production cross section. These measurements can either be carried out directly, in a dedicated energy scan of the top-quark pair production threshold (see \autoref{sec:Mass:Threshold}), or for radiative events at higher collision energies (see \autoref{sec:Mass:Radiative}). The advantage of this approach is that the top-quark mass is extracted in well-defined mass schemes, as introduced in \autoref{ssec:masschemes}.

The second approach is the measurement of the mass from kinematic observables reconstructed in continuum production, such as the measurement of the invariant mass of the decay products of top quarks (see \autoref{sec:Mass:cont}). Since the extracted mass value is obtained as a parameter of the event generators used in template fits, this technique suffers from ambiguities in the interpretation comparable to the issues encountered in most top-quark mass measurements at the LHC. On the other hand, the higher integrated luminosities collected well above the top-quark production threshold provide high statistics.

A combination of both classes of measurements may ultimately help to better constrain the systematics and to improve the theoretical understanding of the continuum reconstruction, also contributing to the interpretation of the top-quark measurements at hadron colliders.

\subsection{Threshold scan around 350\,GeV}
\label{sec:Mass:Threshold}

At \epem colliders, the top-quark mass is expected to be measured with high accuracy in a scan of the top-quark pair production threshold \cite{Bigi:1986jk, Fadin:1987wz, Fadin:1988fn, Strassler:1990nw}. Earlier studies have shown that a statistical precision of a few tens of MeV on the top-quark mass is achievable in such measurements when performed simultaneously with a fit to determine physical parameters such as the strong coupling constant or the top Yukawa coupling \cite{Martinez:2002st, Seidel:2013sqa,Horiguchi:2013wra}. 

This analysis is based on the study discussed in detail in \cite{Seidel:2013sqa}, which uses signal and background reconstruction efficiencies slightly above threshold, obtained from full detector simulations for the \clicild detector concept. The emphasis of the event selection is on maximising the signal significance and it considers both fully-hadronic as well as semi-leptonic events, the latter excluding $\PGt$ final states. The selection proceeds through the identification of isolated charged leptons, jet clustering into either six or four exclusive jets, flavour-tagging, and pairing of $\PW$ boson candidates and $\PQb$-jets into the two top-quark candidates via a kinematic fit. The constraints imposed by the kinematic fit already result in a substantial rejection of background. The kinematic fit is followed by an additional background rejection cut making use of a binned likelihood function combining flavour tagging information event shape and kinematic variables. After this selection, a highly pure sample of top-quark pair events is available for the measurement of the cross section. An overall signal selection efficiency of 70.2\%, including the relevant branching fractions, is achieved, whereas the dominant background channels are rejected at the 99.8\% level, resulting in an effective cross section of 73\,\fb for the remaining background.

The analysis is combined with higher order theory calculations of the signal process. Here, the latest NNNLO QCD calculations, available in the program {\code{QQbar\_threshold}} \cite{Beneke:2016kkb}, are used. The theory cross section is corrected for ISR and the luminosity spectrum of the collider using the techniques described in \cite{Seidel:2013sqa}. This corrected cross section is then used to generate pseudodata and the templates needed to fit the simulated data points to extract the top-quark mass. 

\begin{figure}[t!]
  \centering
  \includegraphics[width=0.48\columnwidth]{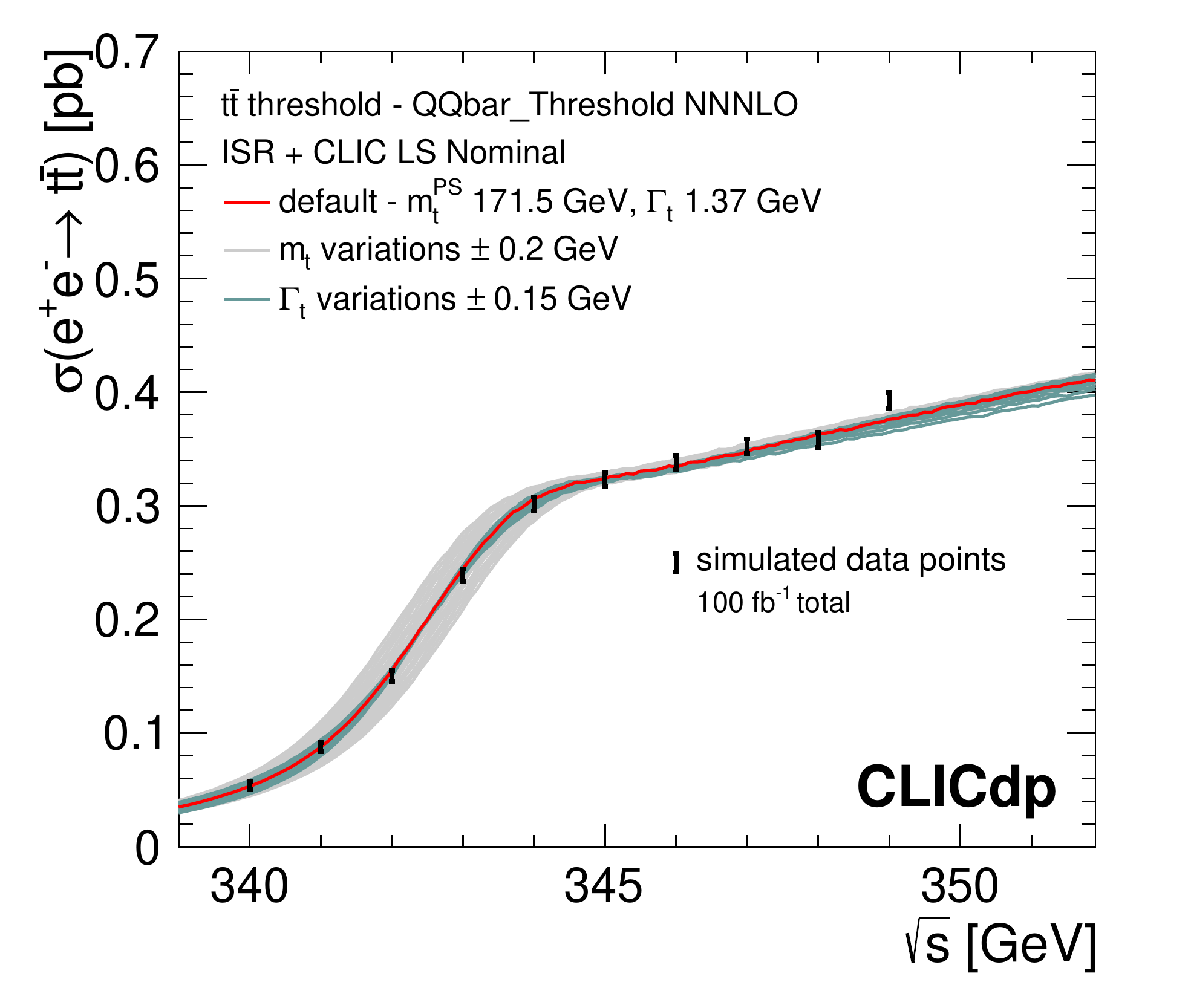}
  ~~~~
   \includegraphics[width=0.48\columnwidth]{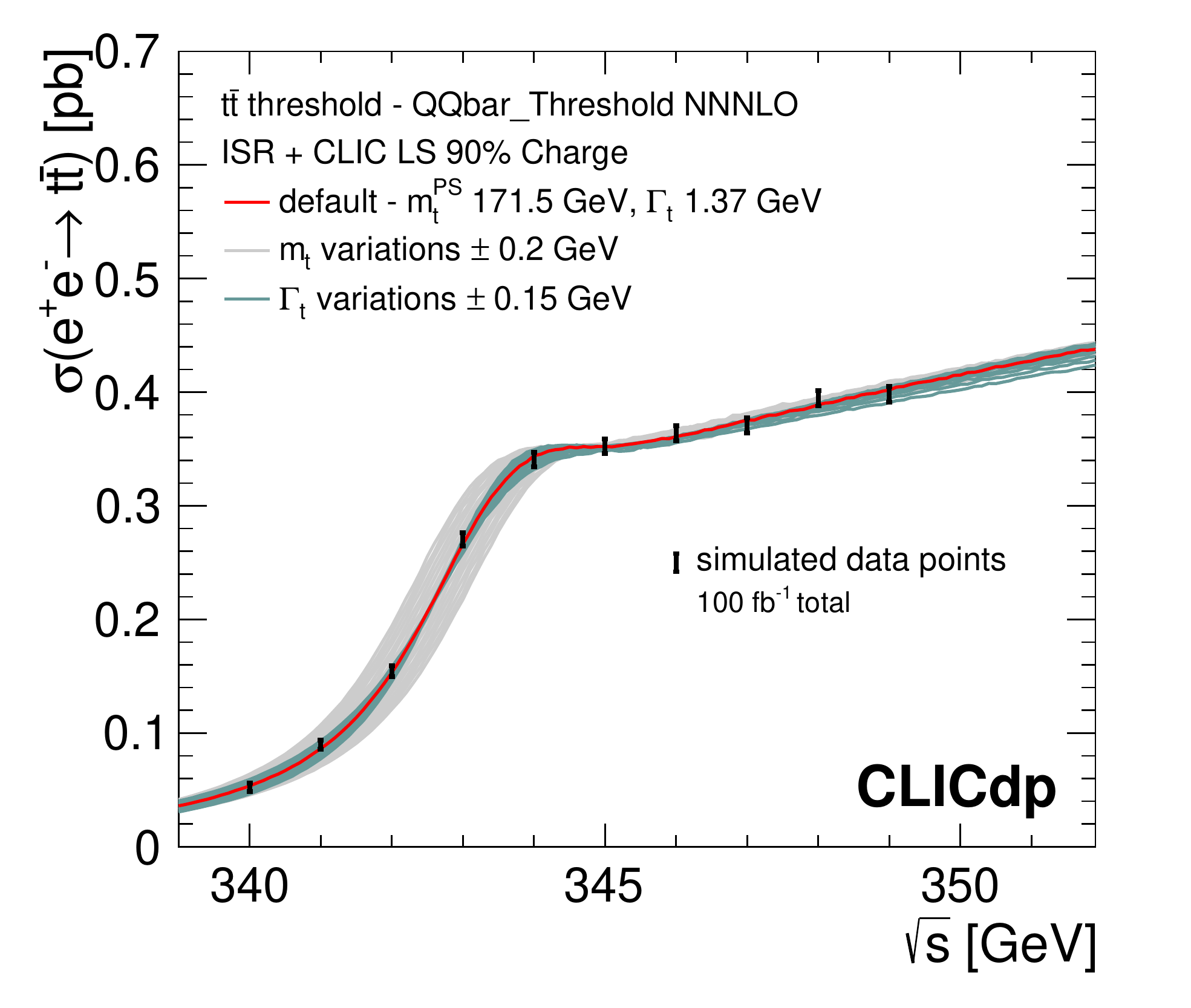}
  \caption{Illustration of a top-quark threshold scan at CLIC with a total integrated luminosity of 100\,fb$^{-1}$, for two scenarios for the luminosity spectrum, nominal (left) and `reduced charge' (right). The bands around the central cross section curve show the dependence of the cross section on the top-quark mass and width, illustrating the sensitivity of the threshold scan. The error bars on the simulated data points show the statistical uncertainties of the cross section measurement, taking into account signal efficiencies and background levels. \label{fig:TopThresholdScan}}
\end{figure}

In the context of the running scenario of CLIC discussed in \autoref{ssec:staging}, it is assumed that an integrated luminosity of 100\,fb$^{-1}$ of the first stage of CLIC would be devoted to a scan of the top pair production threshold. Here, a baseline scenario of ten equidistant points is assumed, with 10\,\fb per point and a point-to-point spacing of 1\,GeV, in the energy range from $2m^{\mathrm{PS}}_{\mathrm{t}} - 3\,\GeV$ to $2m^{\mathrm{PS}}_{\mathrm{t}} + 6\,\GeV$. Such a threshold scan is shown in \autoref{fig:TopThresholdScan}, for two luminosity spectrum scenarios discussed below. The bands illustrate the dependence of the cross section on the generated top-quark mass and width. The error bars on the data points are statistical, taking into account signal efficiencies and background levels. The top-quark mass is extracted using a template fit to the measured cross sections as a function of centre-of-mass energy. The cross section templates are simulated for different input mass values. The top-quark width is given by the SM expectation provided by \code{QQbar\_threshold}, which is around 1.37\,GeV for the range of masses considered here. For the calculation of the templates the width corresponding to the respective mass is used. The extraction of the mass is performed directly in the PS mass scheme.

\begin{figure}[b!]
  \centering
  \includegraphics[width=0.48\columnwidth]{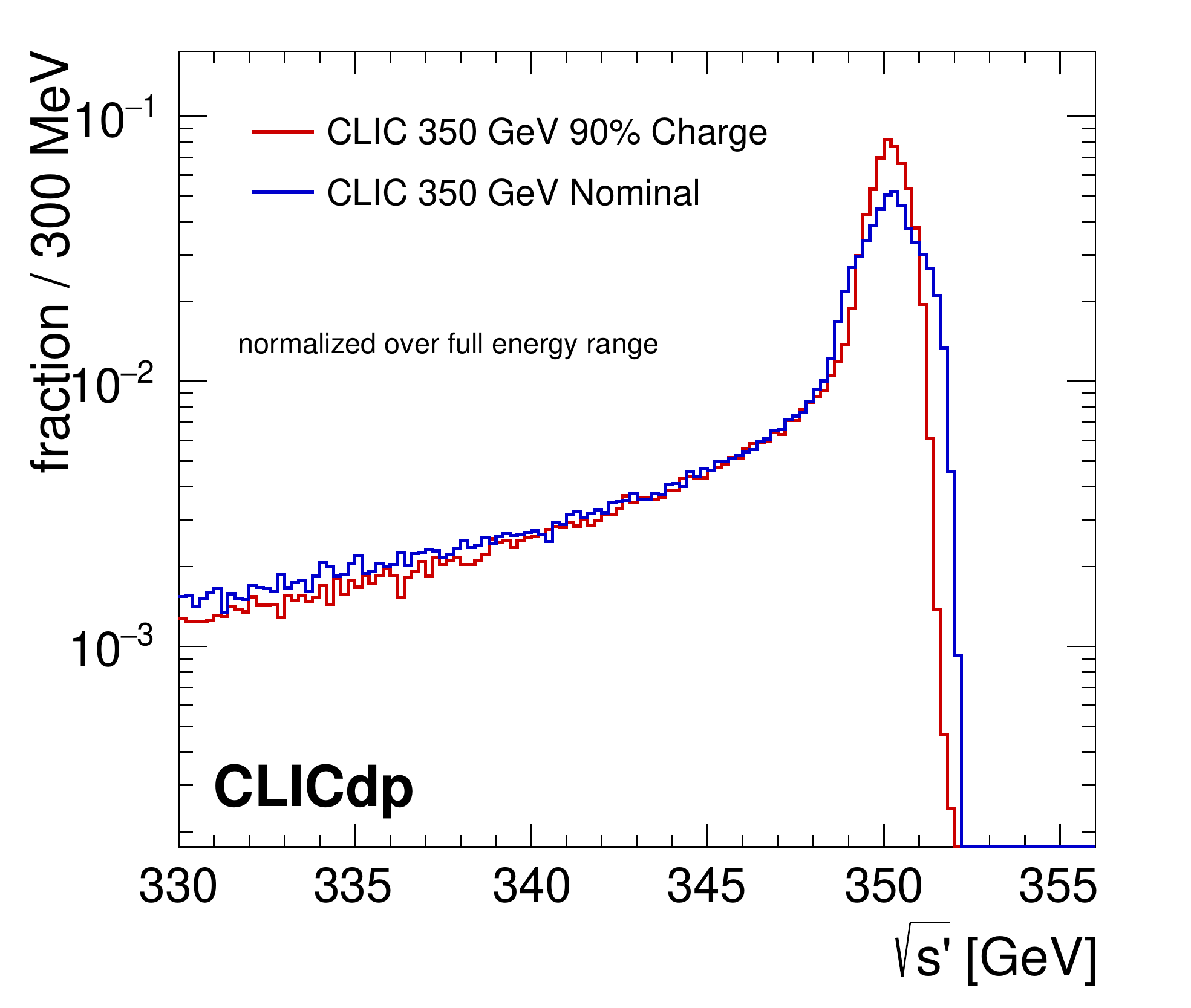}
  ~~~~
  \includegraphics[width=0.48\columnwidth]{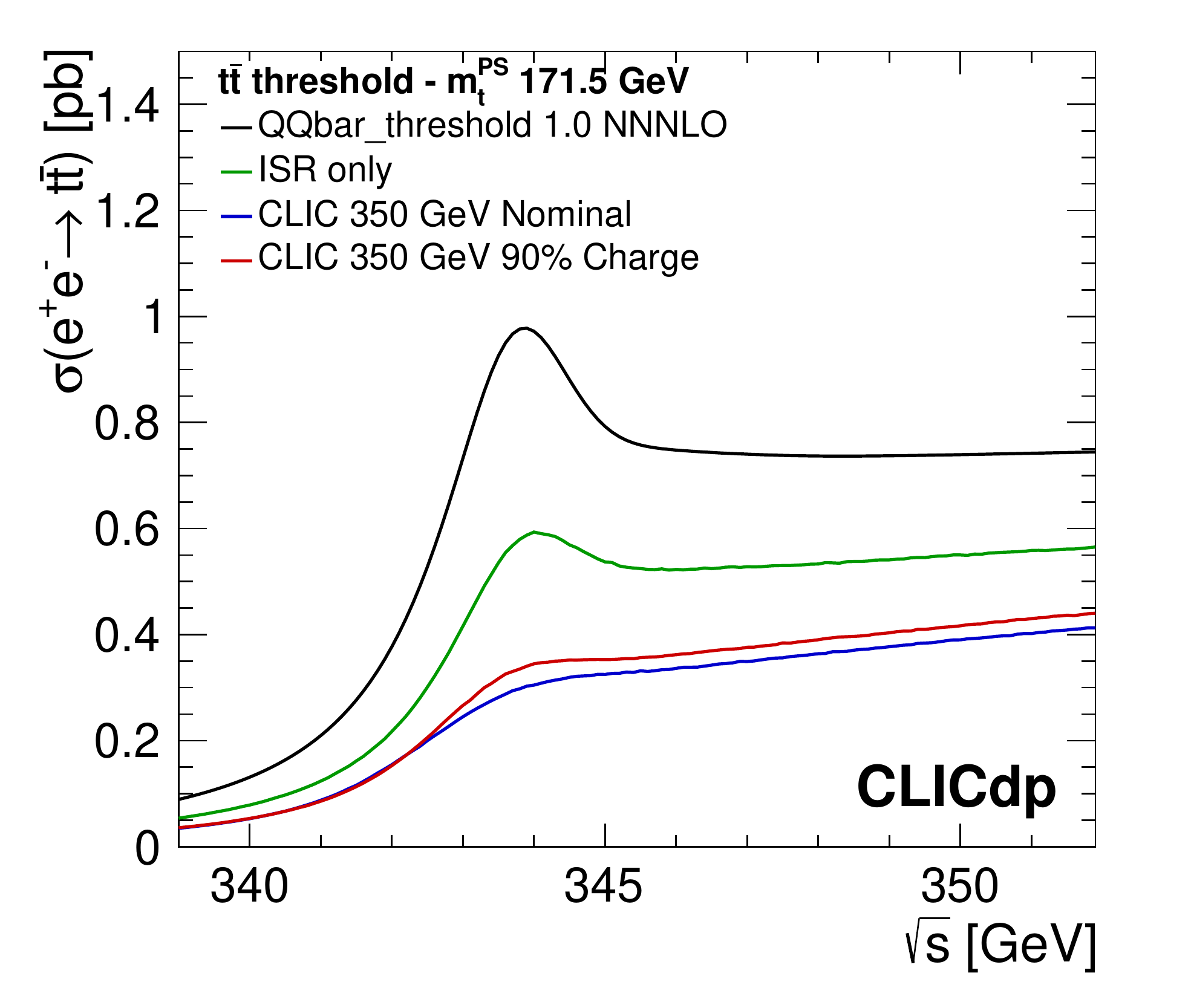}
  \caption{Two scenarios of the CLIC luminosity spectrum for a threshold scan (left); one based on the nominal accelerator parameters of the 380\,GeV initial stage of CLIC (optimised for instantaneous luminosity), and one optimised for reduced beamstrahlung (`reduced charge'). The impact of the luminosity spectra on the top-quark pair production cross section (right), where the blue and red curves show the observable cross section for the nominal and the `reduced charge' luminosity spectra, respectively. \label{fig:TopThreshold:LumiTuning}}
\end{figure}

The luminosity spectrum of CLIC has a strong impact on the shape of the cross section in the threshold region, which influences the extraction of top-quark properties. The smearing of the turn-on behaviour and the would-be 1S peak of the cross section depends on the level of beamstrahlung and the beam energy spread. A larger beam energy spread results in a more pronounced tail to lower energies while the level of beamstrahlung influences the behaviour in the resonance region and above, reducing the effective cross section. Both of these effects result in a broadening of the threshold curve. This in turn reduces the statistical sensitivity of a mass measurement for a given total integrated luminosity, and degrades the precision for the combined extraction of several top-quark properties, such as mass and width or mass and Yukawa coupling. The beam energy spread and the level of beamstrahlung can be tuned by modifying the bunch charge and the beam focusing, allowing optimisation of the spectrum specifically for a top-quark threshold scan. This illustrates well the flexibility of CLIC to optimise the luminosity spectrum without physically changing the accelerator. This aspect might also be useful for other physics applications such potential threshold scans for newly discovered particles. However, an improvement of the quality of the luminosity spectrum also results in a reduction of the instantaneous luminosity. 

\autoref{fig:TopThreshold:LumiTuning} shows the effects of ISR only, and of ISR and the luminosity spectrum combined on the top-quark pair production cross section. Here, two scenarios for the luminosity spectrum at the threshold are considered: one based on the nominal accelerator parameters optimised for luminosity (denoted ``nominal luminosity spectrum''), and one with a reduced beam energy spread and correspondingly a narrower and more pronounced main luminosity peak, using a bunch charge reduced to 90\% of the nominal charge (denoted `reduced charge' luminosity spectrum) \cite{CLIC_beam_web}. For the latter scenario, the instantaneous luminosity is reduced by 24\% compared to the nominal parameters, resulting in a 31\% increase of the required running time for a 100\,\fbinv threshold scan.

\begin{figure}
  \centering
  \includegraphics[width=0.48\columnwidth]{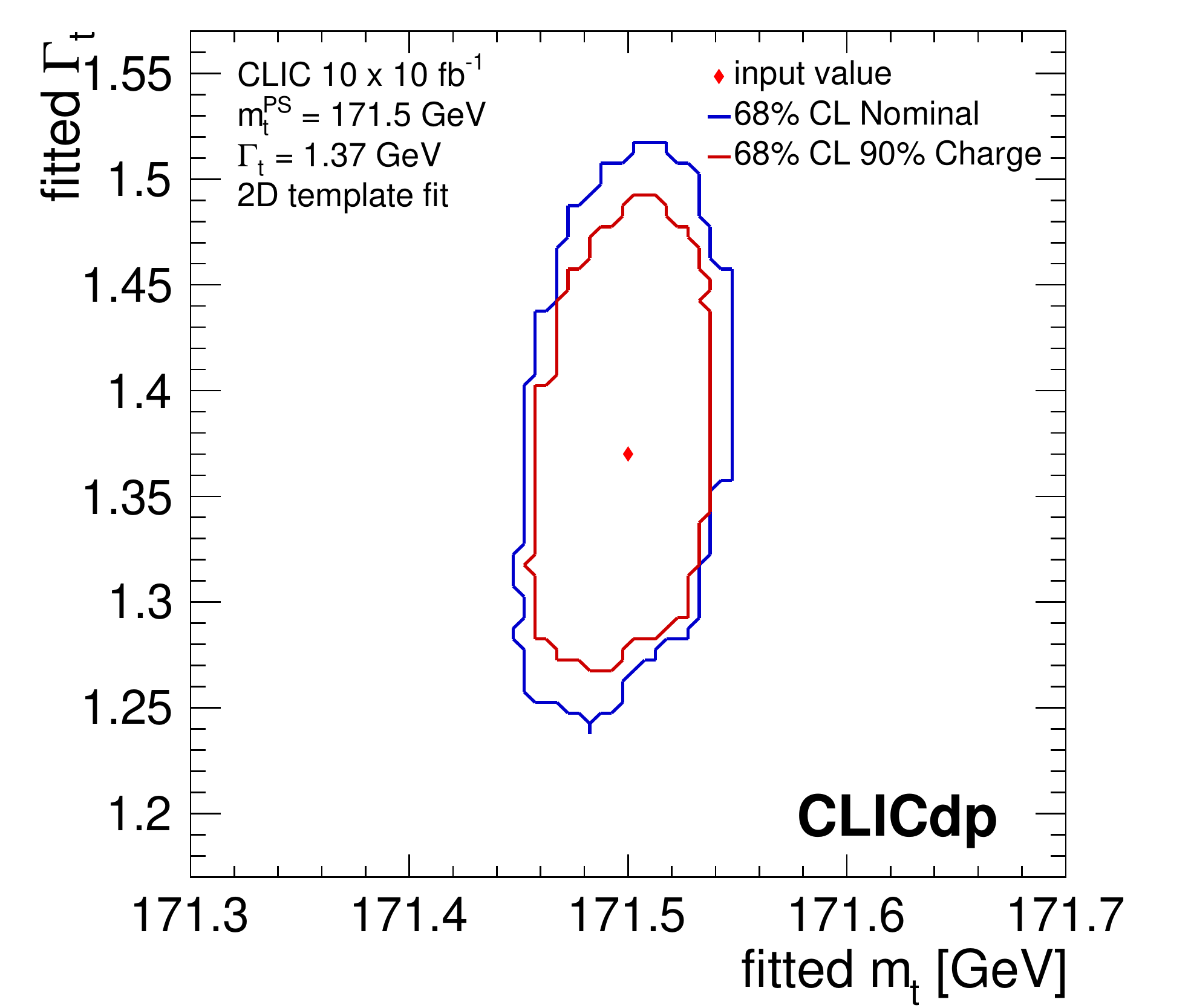}
  ~~~~
  \includegraphics[width=0.48\columnwidth]{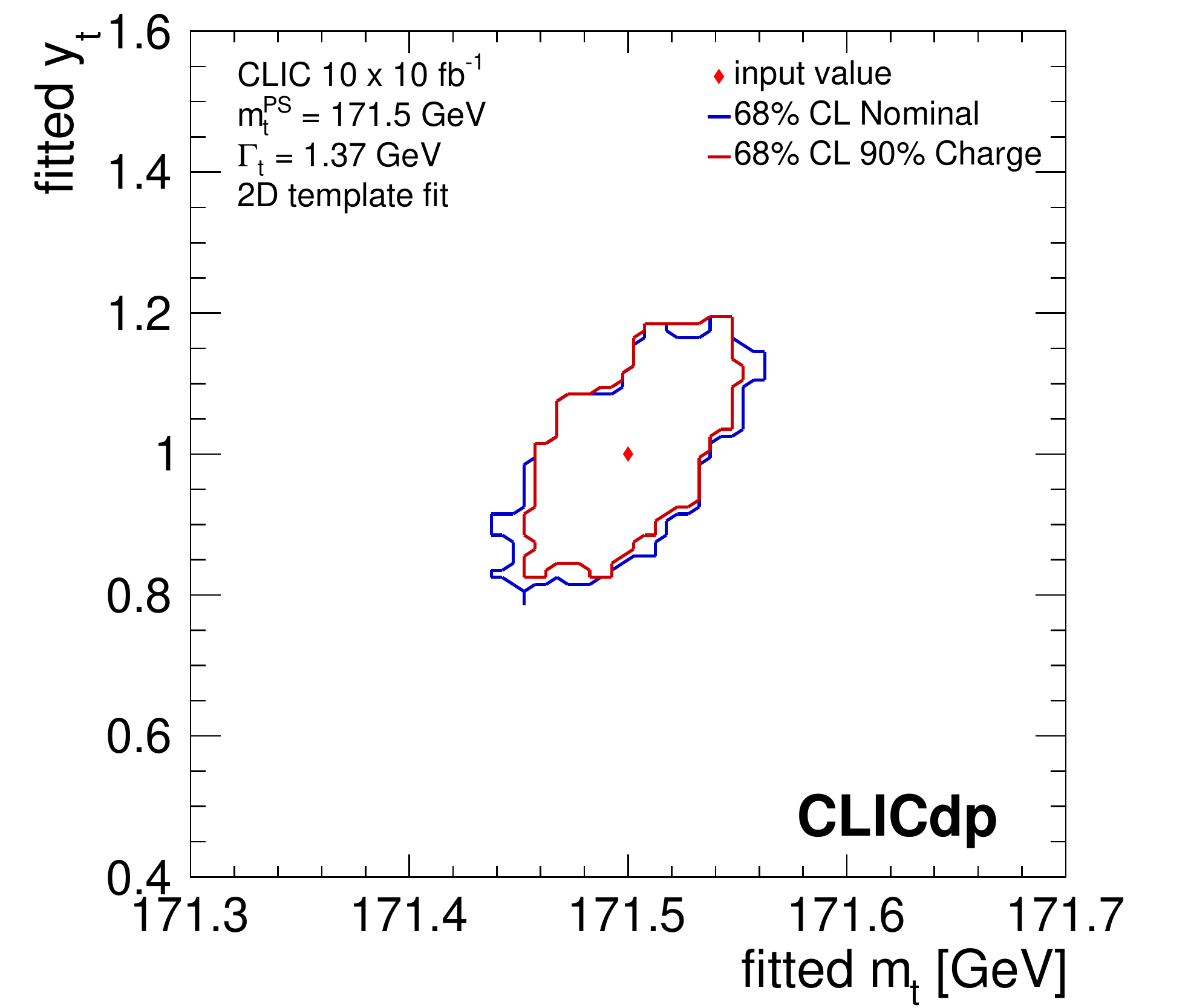}
  \caption{68\% CL statistical uncertainty contours of two-parameter fits to the top threshold region, combining the top-quark mass and width (left) and the top-quark mass and the top Yukawa coupling (right). The contours are shown for both the nominal luminosity spectrum and the `reduced charge' option, in both cases assuming an integrated luminosity of 100\,\fbinv. \label{fig:TopThreshold:2DFits}}
\end{figure}

The expected statistical uncertainty for the top-quark mass in the PS scheme, assuming equal integrated luminosity of 100\,\fbinv, is 22\,(20)\,\mev for the nominal\,(`reduced charge') luminosity spectrum. From running time considerations alone, the `reduced charge' luminosity spectrum does not offer advantages for the top-quark mass measurement. This conclusion changes when extending the analysis to other parameters such as the top-quark width or Yukawa coupling. As is apparent from the width of the green band representing the effect of changes in top-quark width in \autoref{fig:TopThresholdScan}, the sensitivity to the width is considerably lower using the nominal luminosity spectrum compared with the `reduced charge' scenario. \autoref{fig:TopThreshold:2DFits} shows the 68\% CL contours for a simultaneous fit of the top-quark mass and width (left) and top-quark mass and the Yukawa coupling (right). The marginalised 1\,$\sigma$ statistical uncertainties for the two dimensional mass and width fit are 24\,(21)\,\mev for $m_t$ and 57\,(51)\,\mev for $\Gamma_t$  for the nominal\,(`reduced charge') luminosity spectrum. For the two-dimensional mass and Yukawa coupling fit, the corresponding uncertainties are 28\,(24)\,\mev for $m_t$ and 7.5\,(8.4)\% for $y_t$.
In particular for the combined extraction of the mass and the width, the `reduced charge' option provides an improved resolution that largely compensates for the penalty of the reduced luminosity.

It should also be noted that the energy points for the threshold scan, and the integrated luminosities recorded at each point, can be optimised to maximise the precision for a given observable. Owing to the steeper turn-on behaviour of the cross section in the `reduced charge' option, the potential for this optimisation is expected to be bigger in this case, in particular for measurements of the mass and width. 

\subsubsection*{Systematic uncertainties in a threshold scan}

Given the high statistical precision of the top-quark mass measurement at threshold, systematic uncertainties are likely to limit the ultimate precision. Various sources of uncertainties have been investigated, including beam energy \cite{Seidel:2013sqa}, knowledge of the luminosity spectrum \cite{Simon:2014hna}, selection \mbox{efficiencies} and residual background levels \cite{Seidel:2013sqa}, non-resonant contributions \cite{Fuster2015, Hoang:2004tg, Hoang:2008ud, Hoang:2010gu, Beneke:2010mp, Beneke:2017rdn}, parametric uncertainties from the strong coupling \cite{Simon:2016htt}, and theoretical uncertainties estimated from factorisation and renormalisation scale variations \cite{Simon:2016htt, Simon:2016pwp}.

The combined theoretical and parametric uncertainties are expected to be in the range 30\,\MeV to 50\,\MeV, depending on assumptions on the expected improvement in the theoretical description and the knowledge of input parameters such as the strong coupling constant. They have been evaluated for CLIC in the context of the different scenarios for the luminosity spectrum. The results are summarised in \autoref{tab:Mass:ThresholdSystematics}. Similarly, the combined experimental systematic uncertainties are expected to be around 25\,\MeV to 50\,\MeV.  The beam energy is expected to be known with a relative uncertainty of approximately $10^{-4}$, both from machine parameter measurements and from detector measurements of the luminosity spectrum peak from Bhabha scattering, where the momentum scale can be calibrated using $\PZ$ boson decays with sufficient accuracy. The precision of the measurement of the total luminosity, which has a direct impact on the precision of the cross section measurement used to extract the top quark mass, is expected to be in the  few per mille range \cite{Lukic:2013fw, Bozovic-Jelisavcic:2013aca}. This results in an uncertainty on the top-quark mass of a few MeV, substantially smaller than other uncertainties considered here. As discussed above, the luminosity spectrum plays an important role in the analysis of a threshold scan, so the uncertainties of the knowledge of the spectrum are highly relevant. The studies discussed in \cite{Simon:2014hna} make use of a study scaled from 3\,TeV \cite{Poss:2013oea}. A dedicated study for the 380\,GeV case has recently been performed in the context of the analysis discussed in \autoref{sec:Mass:Radiative}, which will be used in the future to further refine the uncertainty estimate for a threshold scan.

\begin{table}
\centering
\caption{The impact of QCD scale uncertainties at NNNLO and of uncertainties of the strong coupling constant on the measured top-quark mass in a threshold scan. The  parametric uncertainty originating from the strong coupling corresponds to an uncertainty of 0.001 in $\alpha_{\mathrm{s}}$. The sign of the change in mass is opposite to the sign of the change in $\alpha_{\mathrm{s}}$. \label{tab:Mass:ThresholdSystematics}}
\begin{tabular}{lcc}
\toprule
& $\Delta\,m^{\mathrm{PS}}_{\mathrm{t}}$ nominal spectrum&  $\Delta\,m^{\mathrm{PS}}_{\mathrm{t}}$ `reduced charge' spectrum \\
\midrule
QCD scale uncertainties & $\pm$42\,MeV & $\pm$41\,MeV \\
parametric $\alpha_{\mathrm{s}}$  & $\mp$ 31\,MeV & $\mp$ 30\,MeV\\
\bottomrule
\end{tabular}
\end{table}

Systematic uncertainties also play an important role for the two-parameter studies shown in \autoref{fig:TopThreshold:2DFits}. Here, the symmetrised theory uncertainties given by scale variations are 60\,MeV (41\,MeV) for the top-quark width and 15\% (14\%)  for the top Yukawa coupling for the nominal (`reduced charge') spectrum. The Yukawa coupling is also sensitive to parametric uncertainties from the strong coupling constant, with an uncertainty of 0.001 in $\alpha_{\mathrm{s}}$ leading to an uncertainty of 6.8\% on the top Yukawa coupling, independent of the luminosity spectrum.

%%%%%%%%%%%%%%%%%%%%%%%%%%%%%%%%%%%%
%%%%%%%%%%%%%%%%%%%%%%%%%%%%%%%%%%%%

\subsection{Top-quark mass from radiative events at 380\,GeV}
\label{sec:Mass:Radiative}

\begin{figure}
\centering
\includegraphics[width=0.48\columnwidth]{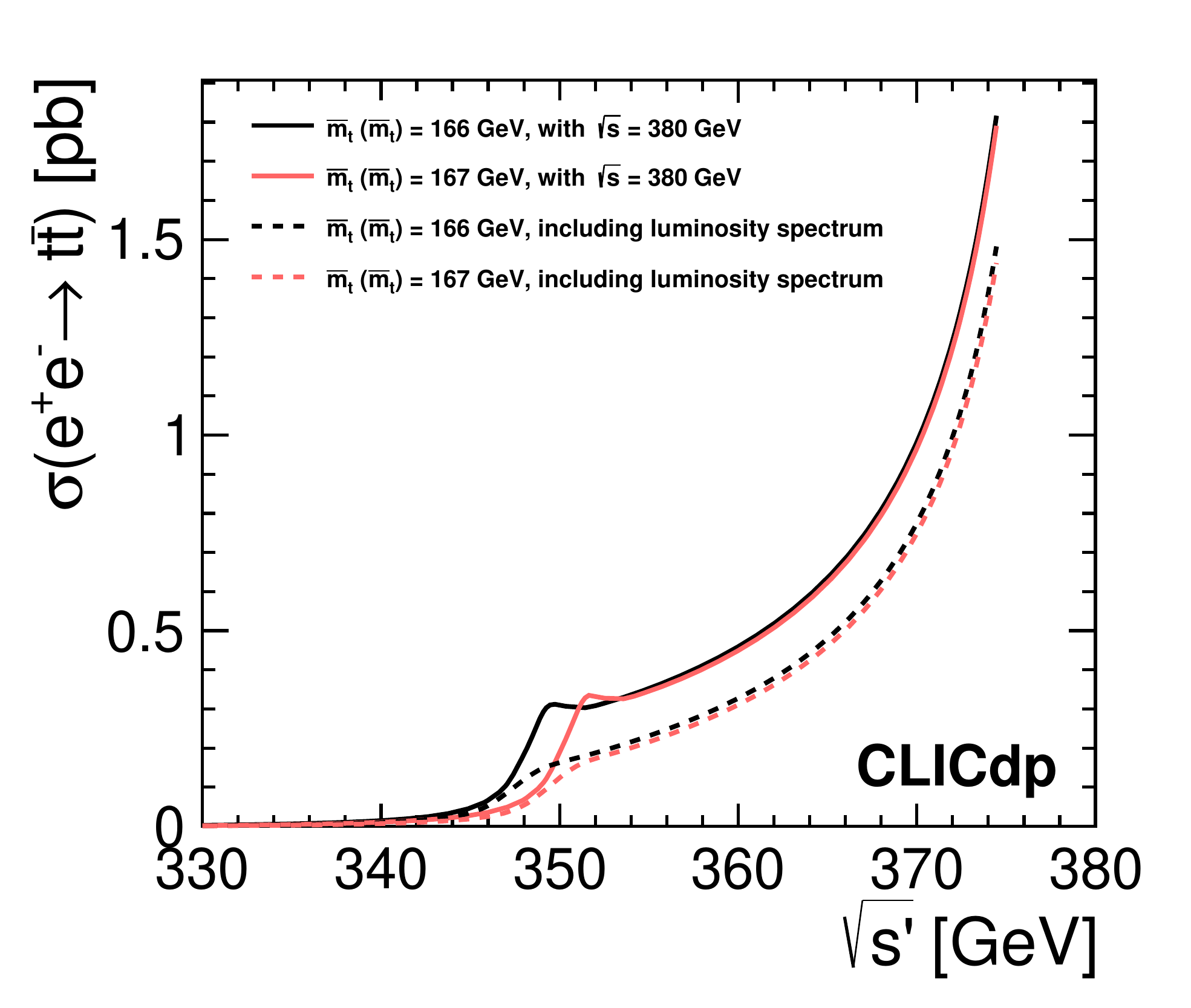}
~~~~
\includegraphics[width=0.48\columnwidth,trim=0 0 2cm 11.5cm,clip]{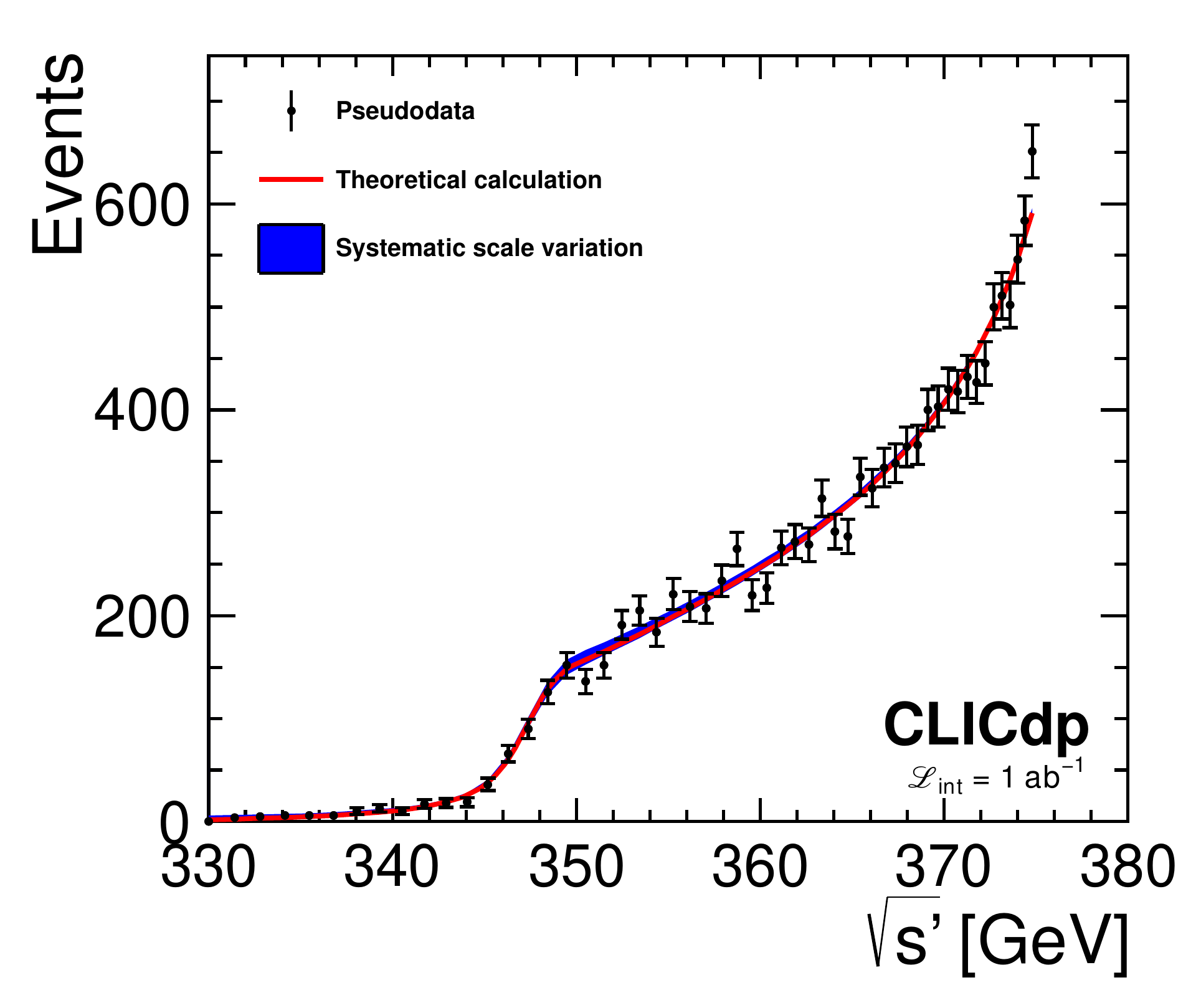}
\caption{Prediction of the observable (left) for $\overline{m}_{\PQt}(\overline{m}_{\PQt}) = 166,\,167$\,GeV (where $\overline{m}_{\PQt}(\overline{m}_{\PQt})$ denotes the top-quark mass in the $\overline{\textrm{MS}}$ scheme, evaluated at the top-quark mass in the $\overline{\textrm{MS}}$ scheme) with the matched NNLL threshold and NNLO continuum calculation for $\roots = 380$\,GeV (solid line) and folded with the CLIC luminosity spectrum (dashed line). Pseudodata (right) generated with the matched NNLL threshold and NNLO continuum calculation for $\roots = 380$\,GeV and folded with the CLIC luminosity spectrum. The markers give a statistical uncertainty estimated from the $\pm \,1\,\sigma$ envelope of $1500$ datasets of $1.0\,\abinv$. The shaded area gives the envelope of the scale variation presented in \autoref{tab:table1}. \label{fig:mass:radiative}}
\end{figure}

In the continuum, the top-quark mass can be extracted from the cross section of radiative events, $\epem\to\ttbar\PGg$, where a top-quark pair is produced in association with an energetic ISR photon radiated from the incoming electron or positron beam. This method is illustrated here using a parton-level study at $\roots = 380$\,GeV. As with the threshold scan, the top-quark mass is extracted directly in theoretically well-defined mass schemes, avoiding interpretation uncertainties. \autoref{fig:mass:radiative} illustrates the dependence of the cross section on the top-quark mass as a function of the effective $\ttbar$ centre-of-mass energy,
\begin{equation*} 
s^\prime = s \left( 1 - \frac{2 E_{\PGg}}{\sqrt{s}}\right),
\end{equation*}
where $E_{\PGg}$ is the energy of the ISR photon. 
The top-quark mass is extracted from a measurement of ${\mathrm{d}\sigma_{\ttbar\PGg}}/{\mathrm{d}\sqrt{s^{\prime}}}$, by fitting templates computed from:
\begin{equation*}
\frac{\mathrm{d}\sigma_{\ttbar\PGg}}{\mathrm{d}\cos\theta\,\mathrm{d}\sqrt{s^{\prime}}} = \frac{\alpha_{\mathrm{em}}}{\pi^2}\,g(x,\theta)\,
\sigma_{\ttbar}(s^\prime).
\end{equation*}
Here, $g(x,\theta)$ is a calculable function of the polar angle $\theta$ of the emitted photon, and the photon energy fraction $x=E_{\PGg}/\sqrt{s}$. The polar angle is integrated over a range in which the photon can be measured in the detector, which excludes the photon being collinear with the incoming electron or positron.  This method requires only identification, rather than complete kinematic reconstruction, of the top-quark candidates.

An accurate prediction of the \rootsprime distribution requires a matched calculation that includes the enhancement of the cross section at the $\ttbar$ production threshold from bound-state effects and remains valid at centre-of-mass energies well above threshold. The theoretical predictions used in this study are based on the NNLL renormalization group improved threshold cross section of \cite{Hoang:2013uda}, and ${\cal O}(\alpha_s^3)$ predictions for the continuum production \cite{Hoang:2008qy,Maier:2017ypu}, which have been smoothly matched together~\cite{Widl:2018}. 
The cross section for $\epem\to\ttbar+X+\PGg_{\mathrm{ISR}}$ factorises into the ISR photon emission from the incoming leptons and the $\epem\to\ttbar +X$ inclusive production.

The differential cross section of the $\epem\to\ttbar+X+\PGg_{\mathrm{ISR}}$ process is given as a function of \rootsprime (or, equivalently, $E_{\PGg}$) for specific values of $s$ and $m_{\PQt}$. The input mass for the cross section is expressed in the $\overline{\textrm{MS}}$ scheme, although for the calculation itself the 1S \cite{Hoang:1998ng,Hoang:1998hm,Hoang:1999ye} and the MSR~\cite{Hoang:2008yj,Hoang:2017suc,Hoang:2017btd} schemes are used. The polar angle $\theta$ of the emitted photon is limited to the interval $10^\circ \!<\! \theta \!<\! 170^\circ$, which agrees with the acceptance of the CLIC detector. The differential distribution in \rootsprime is shown on the left hand side of \autoref{fig:mass:radiative} for two different values of the top-quark mass. The maximum sensitivity of the observable is reached at the $\ttbar$ pair production threshold. % 

The CLIC luminosity spectrum has an important effect on the observable distribution. The two dashed curves on the left hand side of \autoref{fig:mass:radiative} represent the distribution weighted by the luminosity spectrum. The binning in $\rootsprime$ corresponds to the energy resolution of the CALICE silicon-tungsten electromagnetic calorimeter physics prototype: $16.53\,/\sqrt{E(\mathrm{GeV})}\,\oplus\,1.07\,(\%)$ \cite{Adloff:2008aa}. 
Compared with the ideal calculation shown in solid lines, the threshold peak is smeared out considerably. The loss of sensitivity leads to an increase of the statistical uncertainty on the top-quark mass of $\sim 60\%$ for an integrated luminosity of $1.0\,\abinv$.
An estimate of the statistical precision is obtained by fitting large numbers of pseudo-experiments, each corresponding to an integrated luminosity of $1.0\,\abinv$, to the theoretical prediction with the mass as a free parameter. Pseudodata corresponding to one mass point are shown on the right hand side of \autoref{fig:mass:radiative}. The distribution includes the effect of the CLIC luminosity spectrum.
Assuming a selection and reconstruction efficiency of $50\,$\% for $\ttbar\,X\,\PGg$ radiative events, consistent with the expected $\ttbar$ event selection and photon reconstruction efficiency, the resulting statistical precision on the top-quark mass is $100$\,MeV.  The propagation of the luminosity spectrum uncertainty adds an uncertainty less than 10\,MeV on the top-quark mass determination.

The uncertainty on the mass measurement from theoretical uncertainties is estimated by varying the renormalisation scales used in the non-relativistic QCD (NRQCD) calculation~\cite{Hoang:2012us}. Two parameters, $h$ and $f$, are used to vary the scales; factors of $h$, $hf$, and $hf^2$ are applied to the hard, soft, and ultra-soft scales, respectively.  These scales correspond to the top-quark mass, top-quark 3-momentum, and kinetic energy of the \ttbar system, respectively.  These parameters are varied in the intervals given in \autoref{tab:table1} and the corresponding cross-section distributions are generated, folded with the CLIC luminosity spectrum, and fitted using the nominal distribution with the $\overline{\textrm{MS}}$ mass $\overline{m}_{\mathrm{t}}(\overline{m}_{\mathrm{t}})$ as a free parameter. The results are shown in \autoref{tab:table1} and combined results in a theoretical uncertainty estimate of $\pm 100$\,MeV. The final precision on the top-quark mass is around $140$\,MeV for $1.0\,\abinv$.

\begin{table}
\begin{center}
\begin{tabular}{c|ccccccccc}
\toprule
$h$ & $1/2$ & $1/2$ & $1/2$ & $1$ & $1$ & $1$ & $2$ & $2$ & $2$ \rule{0pt}{3ex} \\
\midrule
$f$ & $1$ & $3/2$ & $2$ & $\sqrt{1/2}$ & $1$ & $\sqrt{2}$ & $1/2$ & $3/4$ & $1$ \rule{0pt}{3ex} \\
\midrule
$\Delta\overline{m}_{\PQt}(\overline{m}_{\PQt})$\,[MeV] & $-120$ & $-113$ & $-113$ & $+1$ & $0$ & $+1$ & $+77$ & $+63$ & $+63$ \rule{0pt}{3ex} \\
\bottomrule
\end{tabular}
\end{center}
\caption{Parameter variation and associated shifts in the extracted value of $\overline{m}_{\PQt}(\overline{m}_{\PQt})$ when fitting to the observable with
the default values $(h,f) = (1,1)$. \label{tab:table1}}
\end{table}

%%%%%%%%%%%%%%%%%%%%%%%%%%%%%%%%%%%%
%%%%%%%%%%%%%%%%%%%%%%%%%%%%%%%%%%%%

\subsection{Direct top-quark mass reconstruction in the continuum at 380\,GeV}
\label{sec:Mass:cont}

The top-quark invariant mass can be extracted from the 
large sample of top-quark pairs 
collected above the threshold, in the continuum at 380\,GeV.
For this study only hadronic and semi-leptonic final states are considered.
In these final states the top-quark mass can be directly reconstructed for
the hadronic top-quark decay(s), without applying kinematic constraints.
The VLC algorithm is applied using a radius of 1.6 ($\beta,~\gamma = 0.8$)
to cluster the final state hadrons into six or four exclusive jets,
for hadronic and semi-leptonic event reconstruction, respectively. 
For suppression of four-fermion production and quark-pair production processes,
which are the dominant background contributions, two jets are required to be flavour-tagged as b-jets by \lcfiplus.
This pre-selection removes about 80\% of the quark-pair and 92\%
of the four-fermion backgrounds, while removing only about 12\% of the
top-pair production events.

\begin{figure}
  \centering
  \includegraphics[width=0.48\columnwidth]{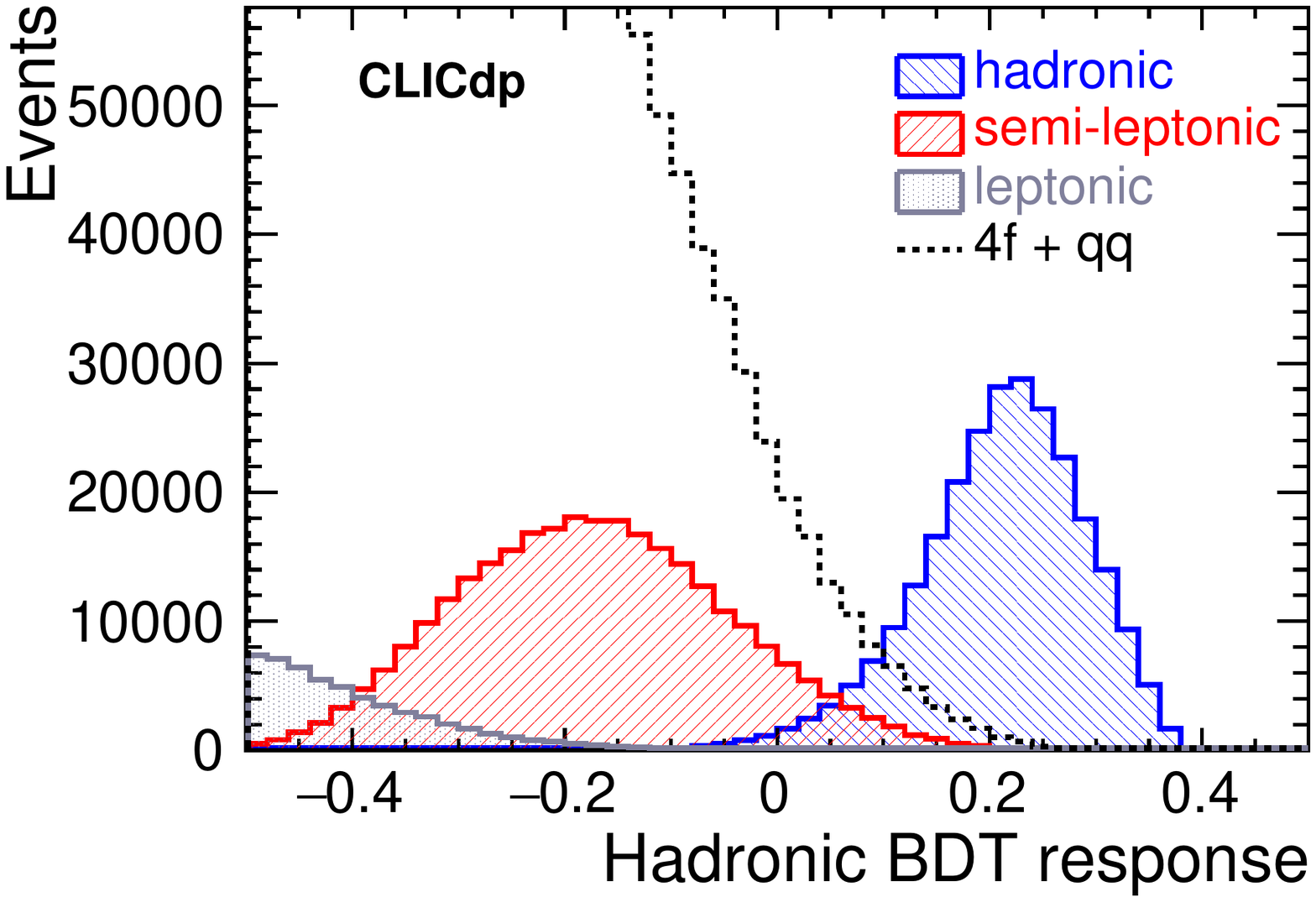}
  \includegraphics[width=0.48\columnwidth]{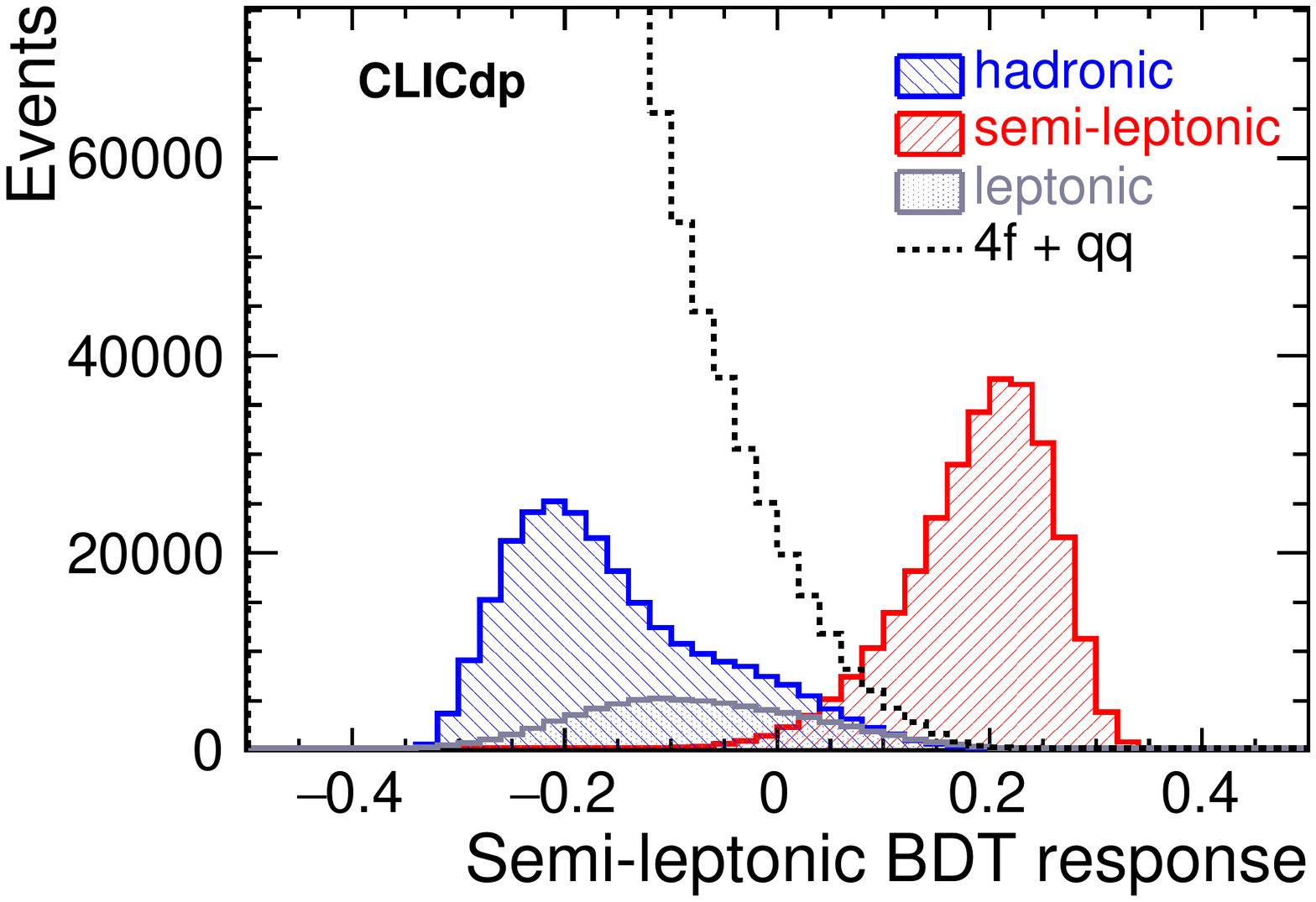}
  \caption{Response distributions for BDT classifiers trained to recognise
    hadronic top-quark pair events (left) and semi-leptonic top-quark pair events
    (right). Distributions for different samples of
    $\ttbar$ events and other SM backgrounds are compared for $1.0\,\abinv$ at 380\,GeV CLIC. 
\label{fig:mass_cont_hadlep}}
\end{figure}

Multivariate BDT (Boosted Decision Tree) classifiers are used
for additional suppression of the non-\ttbar background and
classification of the \ttbar candidate events as either hadronic
or semi-leptonic events. The algorithms are
trained separately for hadronic and semi-leptonic event selection.
The classification is based on the following variables:
  total energy of the event, 
  total transverse and 
        longitudinal momenta, 
  reconstructed missing mass, 
  sphericity and acoplanarity of the event,
  number of isolated leptons,
  energy of isolated lepton with highest transverse momentum, 
  minimum jet energy for the six-jet final state,
  minimum and maximum distance cuts
  for six-, four-, and two-jet reconstruction with the VLC
  algorithm.\footnote{For the four- and two-jet clustering, the identified isolated leptons
  are not included.} 
Response distributions of the BDT classifier trained for selection of
fully-hadronic and of semi-leptonic events are shown in
\autoref{fig:mass_cont_hadlep}. Events having at least one of the
classifier responses greater than zero are selected for mass extraction.
Events which are selected in both channels are assigned to the category 
corresponding to the higher BDT response.
The BDT classification efficiency for top-pair production events is about
90\%, while the four-fermion and quark-pair production backgrounds are
suppressed by a factor of about 20 and 100, respectively.

\begin{figure}
  \centering
  \includegraphics[width=0.48\columnwidth]{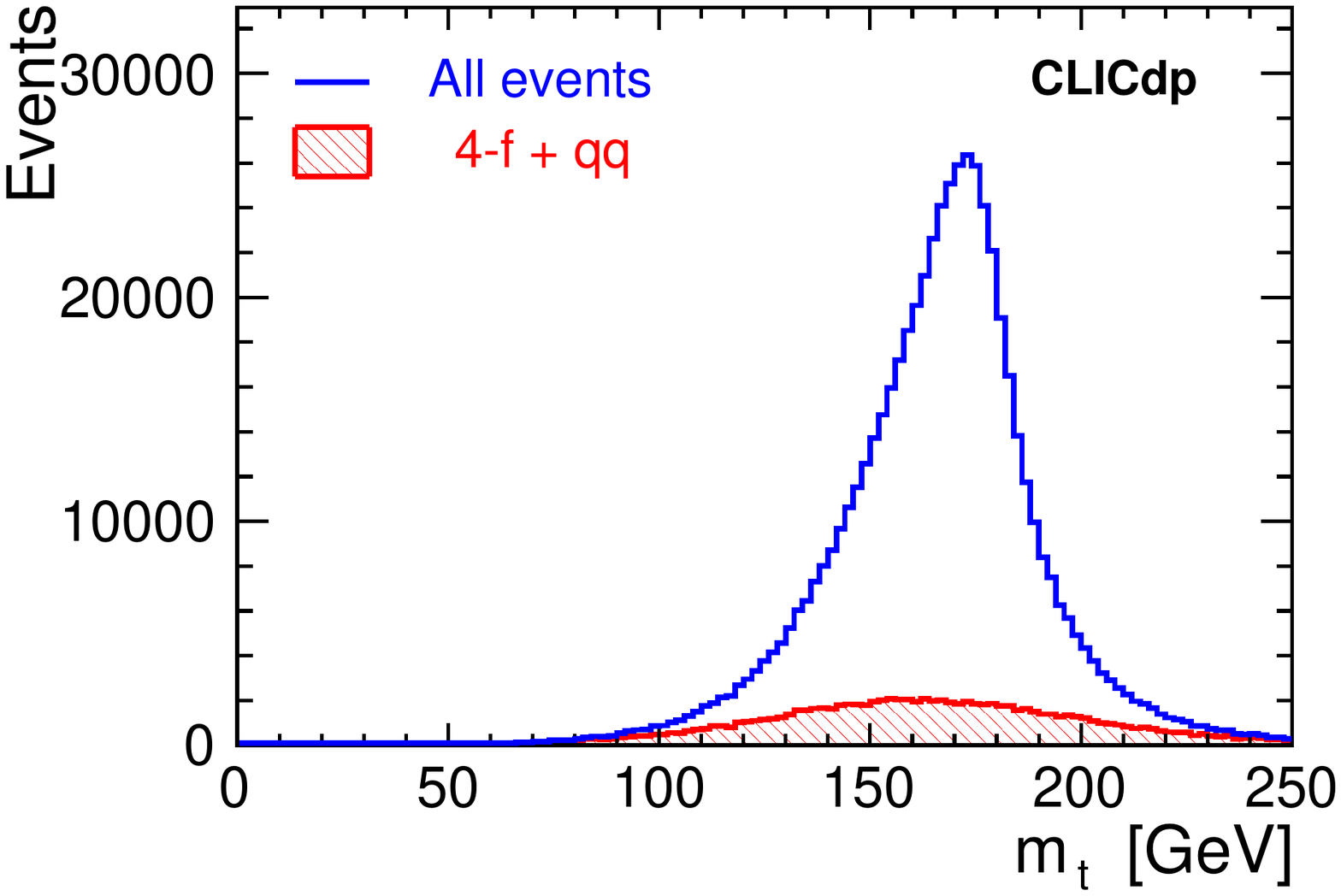}
  \includegraphics[width=0.48\columnwidth]{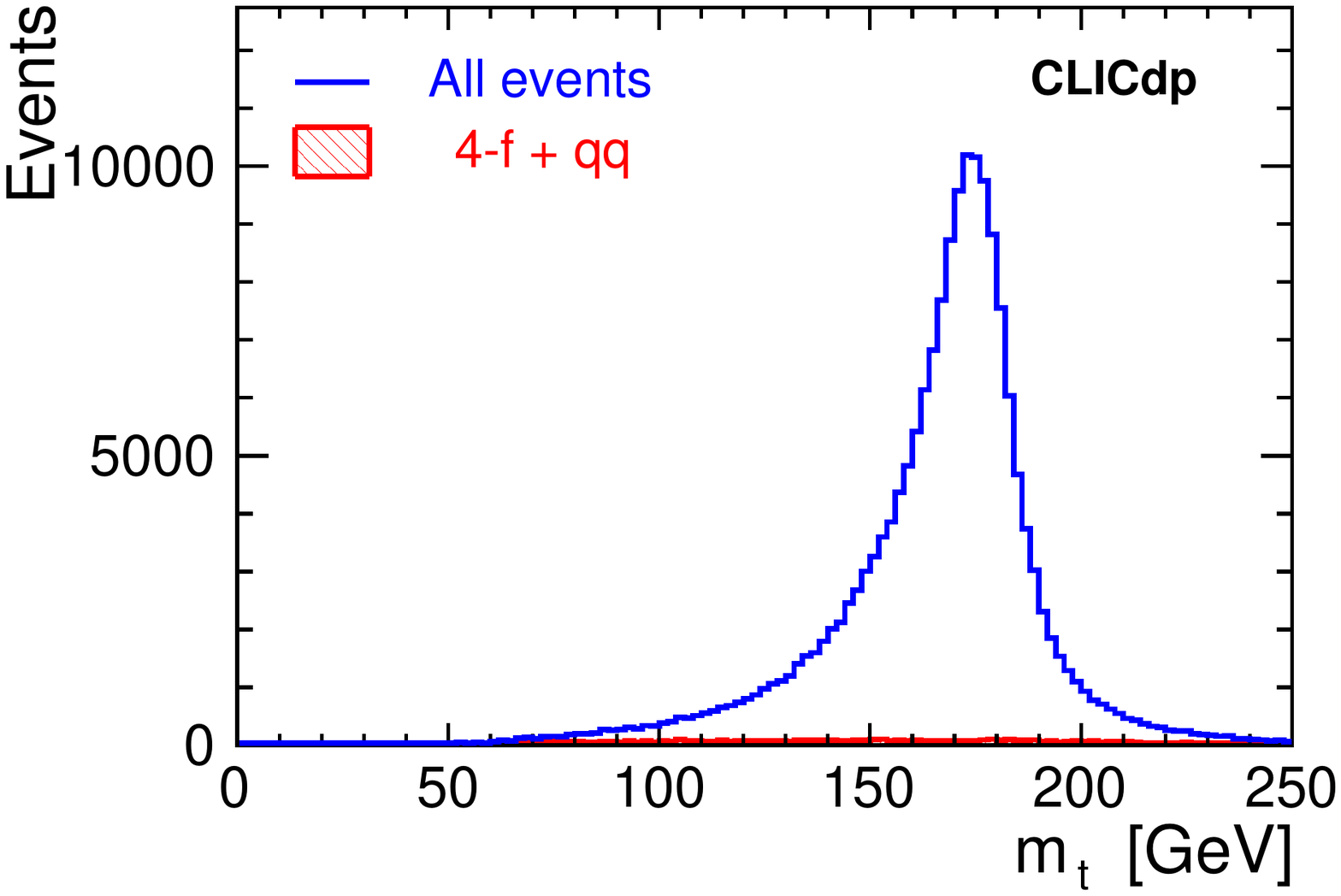}
  \caption{Distributions of the top-quark mass reconstructed from the hadronic top-quark decays
    for hadronic (left) and semi-leptonic (right) events, for $1.0\,\abinv$ at 380\,GeV CLIC. 
\label{fig:mass_cont_distr}}
\end{figure}

For the mass reconstruction, the jet combination that
minimises a $\chi^2$ value for the event is selected.  
The $\chi^2$ formula includes constraints on the invariant masses and
Lorentz boosts of the two reconstructed top-quark candidates, as well
as on the two ratios of the reconstructed $\PW$ boson and the parent
top-quark masses. 
The use of the mass ratio instead of the $\PW$ mass constraint is
motivated by the correlation between the reconstructed masses of the
$\PW$ boson and the parent top quark.
For semi-leptonic events exactly one isolated lepton (electron or muon)
with energy of at least 15\,GeV is required.  
Distributions of the reconstructed top-quark mass for hadronic and semi-leptonic top-quark pair production events
are shown in \autoref{fig:mass_cont_distr}.
Using a template fit method the position of the maximum in the invariant mass distribution can be extracted with a statistical uncertainty of 30\,MeV and 40\,MeV, for hadronic and semi-leptonic events respectively.
Varying the value of the top-quark mass assumed in the $\chi^2$ minimisation for the event reconstruction has little influence on the reconstructed peak position.
The expected statistical precision on the top-quark mass, taking into account both the hadronic
and the semi-leptonic channels and the dilution due to the use of the fixed mass in the $\chi^2$ formula,
is about 30\,MeV.

With high statistical precision of the measurement, systematic effects become the
dominant source of the uncertainty.
In particular, to match the expected level of statistical precision,
the absolute jet energy scale should be controlled at the level of 0.02\%.
Preliminary studies suggest that this level of precision could be achieved by including a short calibration run at the $\PZ$-pole at the start of each year.
A more detailed analysis is required to give a quantitative estimate of the expected
jet energy scale resolution.
An additional theoretical uncertainty of at least a few hundred MeV is also expected
when converting the extracted mass value to a particular renormalisation scheme.

Systematic effects resulting from the uncertainty of the jet energy scale
can be significantly reduced by relating the reconstructed top-quark mass to the mass of the $\PW$ boson. 
The statistical uncertainty on the extracted ratio of the top-quark and $\PW$ boson masses corresponds to a
top-quark mass uncertainty of about 30\,MeV.
The measurement is hardly sensitive to the absolute jet energy scale.
However, the energy scale of $\PQb$-jets, relative to light-quark jets,
should still be controlled to about 0.05\%, to match the statistical precision. 

\section{Kinematic properties of top-quark pair production}
\label{sec:pairprod}

Top-quark production is precisely predicted in the SM but may receive substantial modifications from new physics effects; for example, theories with extra dimensions \cite{Randall:1999ee} and compositeness \cite{Pomarol:2008bh} can modify the couplings significantly.
A deviation from the SM expectation of the forward-backward asymmetry for $\PQb$-quarks at the $\PZ$ pole was observed by the experiments operating at the electron-positron colliders SLC and LEP. This measurement is in tension with the SM prediction at the level of $2.8\,\sigma$~\cite{ALEPH:2005ab}, and it is the most significant discrepancy of the electroweak precision data fit. Since these measurements directly involve the third family of quarks, they reinforce the importance of further precision studies of the top quark counterpart.

Precision studies of observables such as the $\ttbar$ production cross section, $\csttbar$, and the top-quark forward-backward asymmetry, $\afb$, provide a simple way to probe the operators presented in \autoref{tab:operators} and thus constitute a powerful tool for discovery and a deeper understanding of the nature of the electro-weak symmetry breaking. The differential $\ttbar$ cross section, as a function of polar angle $\theta^*$ of the top quark in the $\ttbar$ centre-of-mass system (defined with respect to the electron beam), is here described by
\begin{equation}\label{eq:AFBFit}
   \frac{d\sigma}{d(\cos(\theta^*))} = \sigma_1(1+\cos(\theta^*))^2 + \sigma_2(1-\cos(\theta^*))^2 + \sigma_3(1-\cos^2(\theta^*)).
\end{equation}
At tree level the three terms can be related to the top-quark pair production cross sections for different helicity combinations in the final state, $\sigma_{1,2,3}$. The coefficients in front of the helicity amplitudes can be expressed using \autoref{eq:diffcsttbar} and \autoref{eq:amp} by taking into account the polarisation factors and summing over the different helicity states of the initial and final states. The forward and backward cross sections, $\sigma_{\mathrm{F}}$ and $\sigma_{\mathrm{B}}$, can be obtained by integrating the differential cross section over the top-quark polar angle ranges, $0<\theta^*<\pi/2$ and $\pi/2<\theta^*<\pi$, respectively. The total production cross section, $\csttbar$, can be expressed as 
\begin{equation} \label{eq:totcs}
\csttbar = \sigma_{\mathrm{F}} + \sigma_{\mathrm{B}} = (4/3)(2\,\sigma_1+2\,\sigma_2+\sigma_3),
\end{equation}
while the top-quark forward-backward asymmetry is defined as
\begin{equation} \label{eq:afb}
\afb \equiv \frac {\sigma_{\mathrm{F}} - \sigma_{\mathrm{B}}} {\sigma_{\mathrm{F}} + \sigma_{\mathrm{B}}} = \frac{1}{\csttbar}\,2\,(\sigma_1-\sigma_2).
\end{equation}
The latter is particularly important to probe and disentangle EFT operators that have a strong angular dependence. Measurements with different beam polarisation, enriching the event samples in either left-handed or right-handed top-quarks, allow the photon and $Z$-boson contributions~\cite{Amjad:2015mma} to be disentangled, while data from two (or more) different centre-of-mass energies effectively constrain BSM operators whose effects grow with energy~\cite{AguilarSaavedra:2012vh,PerelloVosZhang}. Extracting $\csttbar$ and $\afb$ for the full CLIC staging programme, thus allows all degrees of freedom in a global fit to be constrained, as will be seen in \autoref{ssec:topphilicinterp}.

In addition, the clean environment of lepton colliders is well suited for the accurate measurement of observables that characterise the differential distributions of the top-quark scattering and decay kinematics. The extra information contained in such observables can improve the sensitivity to certain EFT operators. For example, the corrections induced by anomalous dipole moment operators (i.e., $Q_{\PQt\mathrm{B}}$ and $Q_{\PQt \PW}$ in \autoref{tab:operators}) to the distribution of the azimuthal decay angle of the top quark grow with energy, while the corresponding effects on $\csttbar$ and $\afb$ are essentially energy-independent~\cite{PerelloVosZhang}.\footnote{Similar considerations hold for CP-violating EFT operators that could be efficiently probed by specifically designed CP-odd observables~\cite{Bernreuther:2017cyi}.} 
To best exploit these and other differential features of the signal a multivariate statistical framework is required. For the $\ttbar$ analyses presented in \autoref{ssec:pairprod380gev} and \ref{ssec:pairprodboosted}, we adopt an approach based on statistically optimal observables~\cite{Atwood:1991ka, Davier:1992nw, Diehl:1993br}. This method has been used in the context of top-quark pair production at lepton colliders in \cite{Atwood:1991ka, Grzadkowski:2000nx, Janot:2015yza, Khiem:2015ofa,PerelloVosZhang}; more details are reported in \autoref{ssec:topphilicinterp}.

The following sections describe the event selection and extraction of $\csttbar$ and $\afb$ in $\ttbar$ production at the different CLIC stages. The results are further used in a global EFT fit to constrain the top-philic operators in \autoref{tab:operators}, a study presented in \autoref{sec:phenom_interp}, where we also present the results for the corresponding study using a set of statistically optimal observables.

\subsection{General analysis strategy}
\label{ssec:ttbar:analysis:strategy}

The analyses presented in this section use the \clicild detector concept and focus on ``lepton+jets'' final states ($\ttbar\to\PQq\PQq\PQq\PQq\Pl\PGn$), where the reconstructed charged lepton is used to determine the charge of the hadronically decaying top quark. Isolated lepton identification hence constitutes an important part of the analyses. Events without any identified leptons are discarded along with events with more than one reconstructed lepton. Further, we do not consider semi-leptonic $\ttbar$ events with a tau lepton as signal since these are more difficult to reconstruct because of the additional missing energy. The investigation of such events is left for future study.

After having removed the identified isolated lepton, the VLC algorithm is used to cluster the remaining particles into either two or four exclusive jets. While the former configuration is suitable at the higher energy stages where a boosted topology is expected, the latter is used for the initial stage of CLIC at 380\,GeV.

For the analyses presented here, we adopt an operation time split between the two polarisation states, P($\Pem$) = -80\% and P($\Pem$) = +80\%, consistent with the updated CLIC luminosity staging baseline~\cite{Robson:2018zje}. At the initial energy stage of 380\,\GeV an equal amount of time is assumed for each of the two polarisations states considered, while at the higher-energy stages a larger fraction of data (80\%) is foreseen at P($\Pem$) = -80\%, as motivated by the significant enhancement of important Higgs physics production mechanisms discussed in \autoref{ssec:staging}. For the general top-philic interpretation discussed in \autoref{ssec:topphilicinterp} some data with P($\Pem$) = +80\% is needed, but a fraction lower than 50\% does not degrade the results significantly.

As discussed in \autoref{sec:overview}, final states with six fermions are generally dominated by the $\ttbar$ production process, but have a contribution from non-$\ttbar$ processes such as single-top production and triple gauge boson production. An irreducible number of $\text{non-}\ttbar$ events is expected in the final analyses at the high-energy stages of CLIC since these contributions cannot be fully separated due to interference. While no algorithm can separate them completely, attempts are made in the event selection to reduce the fraction of $\text{non-}\ttbar$ events using some of the characteristic features of the $\ttbar$ process.

Top-quark pair production is simulated as part of an inclusive six-fermion sample. For the analyses at $\roots=1.4\,\tev$ and $\roots=3\,\tev$, the $\ttbar$ events are extracted using a parton-level categorisation requiring two on-shell top-quark candidates. Each candidate consists of three of the six final state particles and should have a mass within $\sim$7.6\,GeV of the generated top-quark mass; this value corresponds to five times the generated top-quark width. To stay conservative we treat the $\text{non-}\ttbar$ contributions as background in the following. The analyses further consider a range of additional relevant background processes, including di-quark final states and final states resulting from $\PW\PW$- and $\PZ\PZ$-fusion events.

\begin{figure}
  \centering
      \includegraphics[width=0.65\columnwidth]{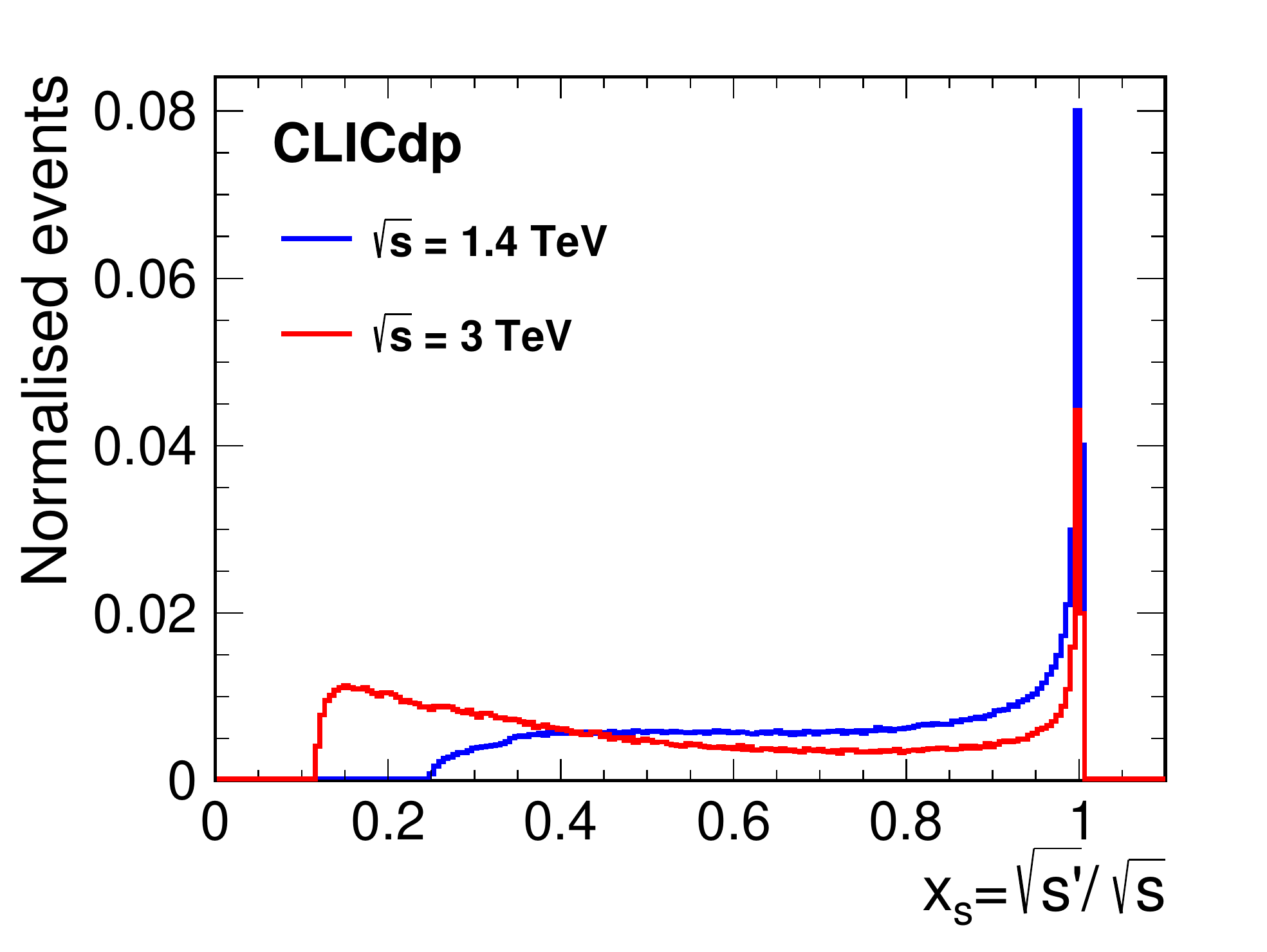}
     \caption{The $\rootsprime$ distribution for $\ttbar$ events at $\roots=1.4\,(3)\,\TeV$ in blue (red), including QED ISR, electroweak corrections, and the CLIC luminosity spectrum. \label{fig:ttbar:radiative:rootsprime}}
\end{figure}

The unique beam conditions at CLIC give rise to a luminosity spectrum with a peak at the nominal collision energy as shown in \autoref{fig:luminosityspectrum}. This results in a distribution of effective collision energies $\rootsprime$ as illustrated in \autoref{fig:ttbar:radiative:rootsprime}, where the $\ttbar$ cross section is shown including the effects of beamstrahlung and ISR. This enables an extension of the $\ttbar$ analyses to include radiative events, with a collision energy below the nominal collision energy $\roots$. Such events are studied for the 1.4\,TeV dataset. In addition, we study the $\ttbar$ production at the nominal collision energies of 380\,GeV, 1.4\,TeV, and 3\,TeV.

The top-quark forward-backward asymmetry and the total production cross section are extracted by fitting \autoref{eq:AFBFit} to the reconstructed polar-angle distribution of the hadronically decaying top quark (or anti-top quark) as calculated in the $\ttbar$ centre-of-mass system. Note that the sign of $\cos(\theta^{*})$ is inverted for events with hadronically decaying anti-top quarks. The fit is performed after background subtraction and correction for finite selection efficiencies. The measured cross sections represent a convolution of $\csttbar$ with the luminosity spectrum. In the analyses at $\roots=1.4\,\tev$ and $3\,\tev$ the extraction of $\csttbar$ is performed in a range of effective collision energies close to the nominal collision energy.

The analysis at 380\,GeV is presented in \autoref{ssec:pairprod380gev}. Radiative events are used to extract the top-quark production observables in three intervals of $\rootsprime$ for operation at $\roots=1.4\,\TeV$. Here the $\ttbar$ events are selected using a multivariate classifier including variables sensitive to the top-quark sub-structure; the analysis is presented in detail in \autoref{ssec:pairprodradiative}. Events that are either partially or substantially boosted are studied at 1.4 and 3\,TeV, where we apply a dedicated tagger for identification of boosted top quarks; see a description of the tagger in \autoref{sec:boosted} and of the analysis in \autoref{ssec:pairprodboosted}. The results of the three analyses are presented in \autoref{ssec:pairprodresults}, while \autoref{ssec:pairprodresults:sys} includes a discussion of the dominant systematic uncertainties.

%===================================================================================================
%===================================================================================================

\subsection{\texorpdfstring{\ttbar}{ttbar} production at 380\,GeV}
\label{ssec:pairprod380gev}

\begin{figure}
  \centering
  \includegraphics[width=0.48\columnwidth]{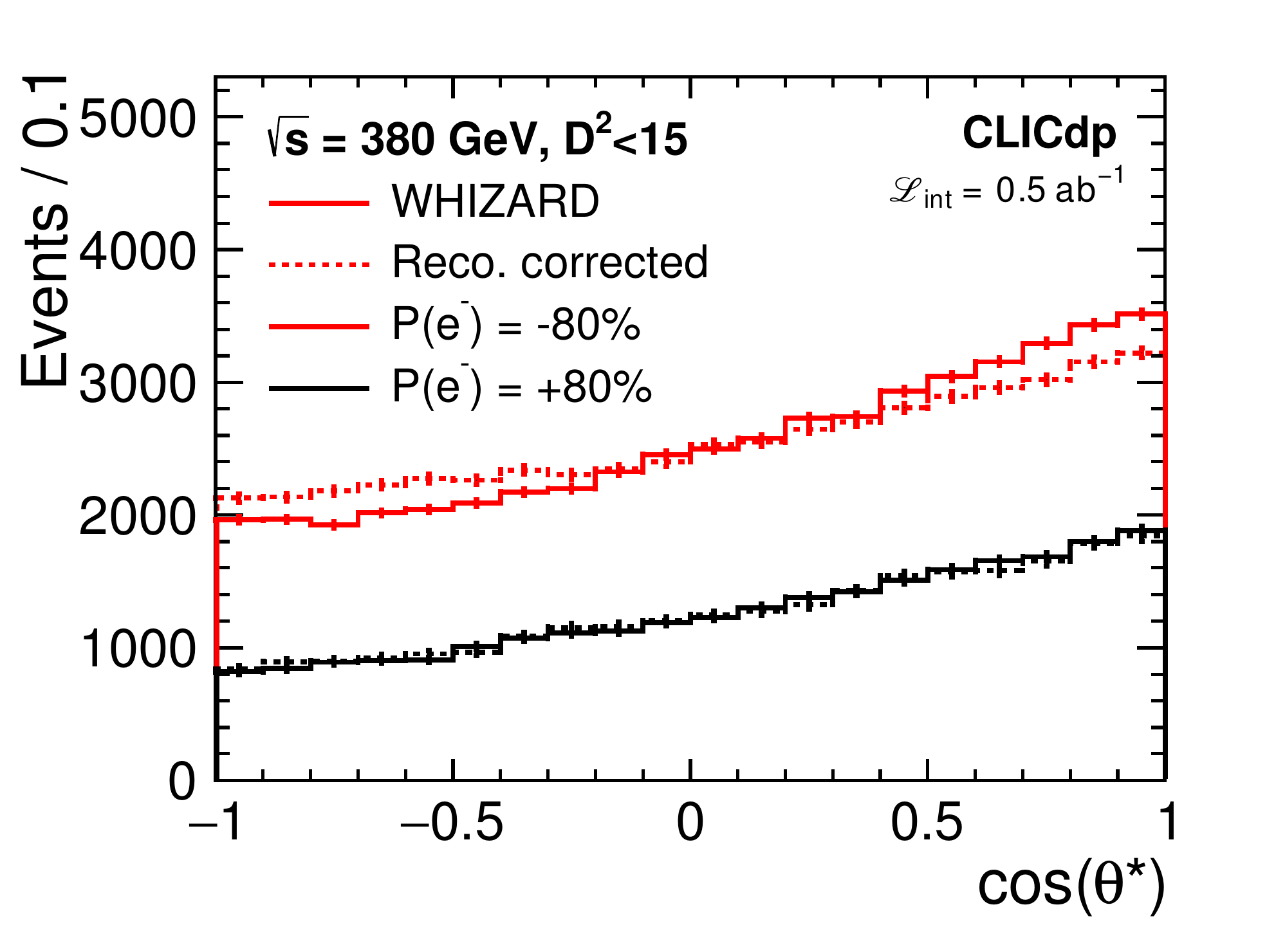}
  ~~~~
  \includegraphics[width=0.48\columnwidth]{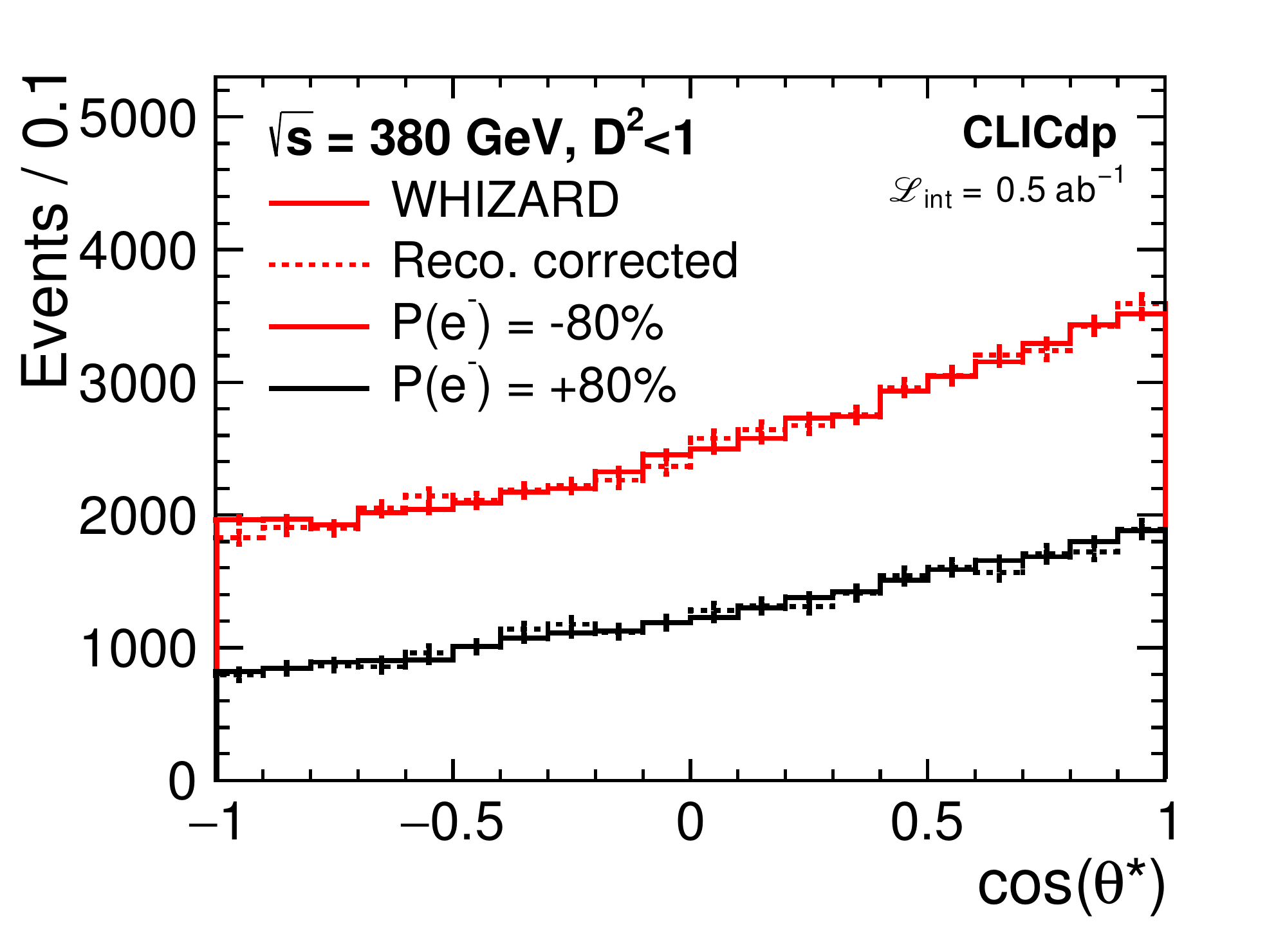}
  \caption{Top-quark polar angle distributions for operation at $\roots=380\,\GeV$ after the application of a quality cut based on the kinematic variable $D^2$. A cut of $D^2<15\,(1)$ was applied for the left (right) figure. The solid lines show the reconstructed distributions including the effects of detector modelling, event reconstruction and candidate selection, while the dashed lines show the \whizard parton-level distributions, for the two beam polarisation configurations considered. Note that efficiency corrections have been applied, corresponding to the parton-level expectation for $D^{2}<15$. \label{fig:ttbar:clic380:costheta}}
\end{figure} 

At the 380\,GeV stage of CLIC, top-quark candidates are formed by combining jets into larger objects. The input jets result from an initial clustering of PFOs with \code{loose} timing cuts as explained in \autoref{ssec:reco}. The PFOs are clustered into exactly four jets using the VLC algorithm with a radius of 1.6 ($\beta,~\gamma = 0.8$). The use of such a large jet radius is made possible by the low level of beam-induced background at 380\,GeV. Note that this analysis is based on the one developed in \cite{Amjad:2015mma}. A summary of the analysis is presented in this section; further details are available in \cite{garcia:2016}.

The general selection relies extensively on $\PQb$-tagging as well as the identification of one isolated lepton. Two jets must satisfy high and intermediate purity b-tagging selection criteria. In addition, the non-b-tagged jets are required to have an energy above 15\,\gev.
The $\Pb$-tagging criteria are applied using the standard flavour-tagging tools described in \autoref{ssec:reco} and alone suppress about 97\% of the dominant $\PWp\PWm$ background.
Lepton candidates are selected as outlined in \autoref{ssec:reco}, with a resulting efficiency of about 85\% for $\ttbar$ events with a leptonic decay ($\Pe,\,\PGm$). The lepton candidate is removed from the list of PFOs considered for jet clustering. Additionally, we require that the $\pT$ of the isolated lepton candidate fulfils $\pT\geq10\,\mathrm{GeV}$.

Top-quark candidates are formed by merging the two non-$\PQb$-tagged jets, which form the hadronically decaying $\PW$ boson, with each of the two $\PQb$-tagged jets. The ambiguity in this reconstruction is resolved by minimising the kinematic variable $d^2$ defined as
\begin{equation*}
d^2 = \biggl( \frac{m_{\PQt}-174\,\mathrm{GeV}}{\sigma_{m_{\PQt}}} \bigg)^2 + \biggl( \frac{E_{\PQt}-190\,\mathrm{GeV}}{\sigma_{E_{\PQt}}} \bigg)^2 + \biggl( \frac{E^{*}_{\PQb}-68\,\mathrm{GeV}}{\sigma_{E^{*}_{\PQb}}} \bigg)^2 + \biggl( \frac{\cos\theta_{\,\PQb\PW}-\langle \cos\theta_{\,\PQb\PW} \rangle}{\sigma_{\cos\theta_{\,\PQb\PW}}} \bigg)^2,
\end{equation*}
where $m_{\PQt}$ and $E_{\PQt}$ are the invariant mass and energy of the hadronically decaying top-quark candidate, $E^{*}_{\PQb}$ is the energy of the $\PQb$-quark in the centre-of-mass frame of the top quark, and $\cos\theta_{\,\PQb\PW}$ is the angle between the $\PQb$-tagged jet and the $\PW$ boson candidate in the lab frame. The reference values for the two first quantities correspond to the simulated values of the top-quark mass and energy, while the third quantity is the expectation value from the two-body decay kinematics of the top quark. The fourth quantity, $\langle \cos\theta_{\,\PQb\PW} \rangle = -0.67$, is the mean of the corresponding distribution from studies using full simulation. The denominator in each term represents the root mean square (RMS) of the observed distribution. Achieving a good pairing of the jets from the hadronically decaying $\PW$ boson with the associated $\PQb$-tagged jet from the top-quark decay is particularly important at 380\,GeV, where the event topology is isotropic and substantial mixing between the jets from the top- and anti-top quark occur.

The above jet pairing constitutes a source of mis-reconstruction that can lead to severe effects predominately for the P($\Pem$) = -80\% sample that is enriched with top quarks of left-handed helicity. For top quarks with left-handed helicity, the $\PW$ boson is emitted opposite to the flight-direction of the top quark and decays nearly at rest. The resulting final state has two hard jets from the $\PQb$-quarks and soft jets from the hadronically decaying $\PW$ boson; a configuration that leads to substantial migrations in the top-quark polar angle distribution when paired wrongly. Since the directional measurement depends very strongly on the correct association of top-quark decay particles, the final step of the analysis is carried out separately for the two polarisation states with stricter quality cuts applied for the extraction of $\afb$ for the P($\Pem$) = -80\% sample.

The selection criteria $40\,\gev< m_{\PW}<190\,\gev$ and $100\,\gev<m_{\PQt}<250\,\gev$ are applied to the reconstructed top candidates. Note that the loose upper cut on $m_{\PW}$ is mainly applied to reject mis-reconstructed events. The preselection requirements on the lepton, jets and top-quark candidate, as defined above, efficiently reduce the number of background events as seen in \autoref{tab:ttbar:clic380:selection:polneg} and \autoref{tab:ttbar:clic380:selection:polpos}.

\begin{table}
\centering
\begin{tabular}{lcccccc}
\toprule
\vspace{1.5mm}
Process & $\sigma\,[\fb]$ & $\epsilon_{\,\mathrm{Pre}}\,[\%]$ & $\epsilon_{D^2<15}\,[\%]$ & $\epsilon_{D^2<1}\,[\%]$ & $N_{D^2<15}$ & $N_{D^2<1}$ \\
\midrule
$\epem\to\PQq\PQq\PQq\PQq\Pl\PGn\,(\Pl=\Pe,\PGm)$ & 161 & 69 & 93 & 34 & 51,080 & 18,802 \\
\midrule
$\epem\to\PQq\PQq\PQq\PQq\Pl\PGn\,(\Pl=\PGt)$ & 80.5 & 16 & 90 & 29 & 5,946 & 1900 \\
$\epem\to\PQq\PQq\PQq\PQq\PQq\PQq$ & 215 & 0.61 & 67 & 6.6 & 440 & 44 \\
$\epem\to\PQq\PQq\Pl\PGn\Pl\PGn$ & 61.8 & 5.6 & 58 & 5.5 & 1006 & 96 \\
$\epem\to\PQq\PQq\PQq\PQq$ & 8,910 & 0.01 & 44 & 2.9 & 226 & 16 \\
$\epem\to\PQq\PQq\Pl\PGn$ & 9,800 & 0.03 & 20 & 1.6 & 302 & 24 \\
$\epem\to\PQq\PQq\Pl\Pl$ & 1,840 & 0.25 & 43 & 3.4 & 970 & 76 \\
$\epem\to\PQq\PQq$ & 26,100 & 0.01 & 32 & 0.57 & 360 & 6 \\
\bottomrule
\end{tabular}
\caption{Pre-selection and final event selection efficiencies and expected number of events for the $\ttbar$ analysis of CLIC at a nominal collision energy of 380\,GeV. The numbers are shown for $\mathrm{P}(\Pem) = \text{-}80\%$ assuming $0.5\,\abinv$. \label{tab:ttbar:clic380:selection:polneg}}
\end{table}

\begin{table}
\centering
\begin{tabular}{lcccccc}
\toprule
\vspace{1.5mm}
Process & $\sigma\,[\fb]$ & $\epsilon_{\,\mathrm{Pre}}\,[\%]$ & $\epsilon_{D^2<15}\,[\%]$ & $\epsilon_{D^2<1}\,[\%]$ & $N_{D^2<15}$ & $N_{D^2<1}$ \\
\midrule
$\epem\to\PQq\PQq\PQq\PQq\Pl\PGn\,(\Pl=\Pe,\PGm)$ & 76 & 72 & 93 & 34 & 25,320 & 9,398 \\
\midrule
$\epem\to\PQq\PQq\PQq\PQq\Pl\PGn\,(\Pl=\PGt)$ & 38.1 & 17 & 89 & 28 & 2,854 & 912 \\
$\epem\to\PQq\PQq\PQq\PQq\PQq\PQq$ & 102 & 0.65 & 70 & 9.1 & 234 & 30 \\
$\epem\to\PQq\PQq\Pl\PGn\Pl\PGn$ & 29.2 & 5.7 & 56 & 5.2 & 466 & 44 \\
$\epem\to\PQq\PQq\PQq\PQq$ & 1,240 & 0.03 & 39 & 1.4 & 68 & 2 \\
$\epem\to\PQq\PQq\Pl\PGn$ & 1,360 & 0.03 & 17 & 0.52 & 34 & 2 \\
$\epem\to\PQq\PQq\Pl\Pl$ & 1,690 & 0.21 & 42 & 2.0 & 738 & 36 \\
$\epem\to\PQq\PQq$ & 16,400 & 0.01 & 26 & 2.0 & 162 & 12 \\
\bottomrule
\end{tabular}
\caption{Pre-selection and final event selection efficiencies and expected number of events for the $\ttbar$ analysis of CLIC at a nominal collision energy of 380\,GeV. The numbers are shown for $\mathrm{P}(\Pem) = +80\%$ assuming $0.5\,\abinv$. \label{tab:ttbar:clic380:selection:polpos}}
\end{table}

The effect of migration in the top-quark polar angle distribution is reduced by applying a quality cut on $D^2$, defined as 
\begin{equation*} 
D^2 = \biggl( \frac{\gamma_{\PQt}-\langle \gamma_{\PQt} \rangle}{\sigma_{\gamma_{\PQt}}} \bigg)^2 + \biggl( \frac{E^*_{\PQb}-68\,\mathrm{GeV}}{\sigma_{E^*_{\PQb}}} \bigg)^2 + \biggl( \frac{\cos\theta_{\,\PQb\PW}-\langle \cos\theta_{\,\PQb\PW} \rangle}{\sigma_{\cos\theta_{\PQb\PW}}} \bigg)^2,
\end{equation*}
where $\langle \gamma_{\PQt} \rangle = \roots/2m_{\PQt} \approx 1.09$ and $\sigma_{\gamma_{\PQt}}$ is the mean and RMS, respectively, of the observed distribution of the top-quark Lorentz factor obtained from studies using full simulation.
\autoref{fig:ttbar:clic380:costheta} shows the top-quark polar angle distributions for the hadronically decaying top quarks in the signal sample after application of the quality cut $D^2<15$ ($D^2<1$). As expected, the effect of migration is most clearly seen for P($\Pem$) = -80\%, for which a strict cut has to be placed to retrieve the MC $\afb$ value.\footnote{Note that the normalisation of $D^2$ is different from a pure $\chi^2$. This is caused by tails in the distributions for which the $\sigma$'s in the denominators are derived.}

The signal selection efficiency for the dataset at P($\Pem$) = -80\%, relevant for the extraction of $\afb$ as introduced in \autoref{eq:afb}, is 69\% after the initial pre-selection and drops to 23\% when applying the quality cut $D^2<1$. The corresponding numbers for P($\Pem$) = +80\% are 72\% after the initial pre-selection and 67\% after the application of the quality cut $D^2<15$. 
As the extraction of $\csttbar$, introduced in \autoref{eq:totcs}, is less sensitive to mis-reconstructions, a cut at $D^2<15$ is applied for both polarisations since it already  suppresses background events efficiently.
The efficiencies for signal and dominant background processes are presented in \autoref{tab:ttbar:clic380:selection:polneg} and \autoref{tab:ttbar:clic380:selection:polpos}. Contributions from other backgrounds such as $\epem\to\PQq\PQq\PGn\PGn$ and additional six-fermion processes are found to be negligibly small.

A somewhat looser cut is also applied in the construction of the statistically optimal observables. Here we apply a modified quality cut for P($\Pem$) = -80\%, that puts constraints on both the leptonic and the hadronic side of the event; this reduces the efficiency by $\sim40\%$ compared to a cut at $D^2<15$, resulting in 31,032 events in the final event sample.

%===================================================================================================
%===================================================================================================

\subsection{Radiative events at 1.4\,TeV}
\label{ssec:pairprodradiative}

\begin{figure}
  \centering
  \includegraphics[width=0.51\columnwidth,clip]{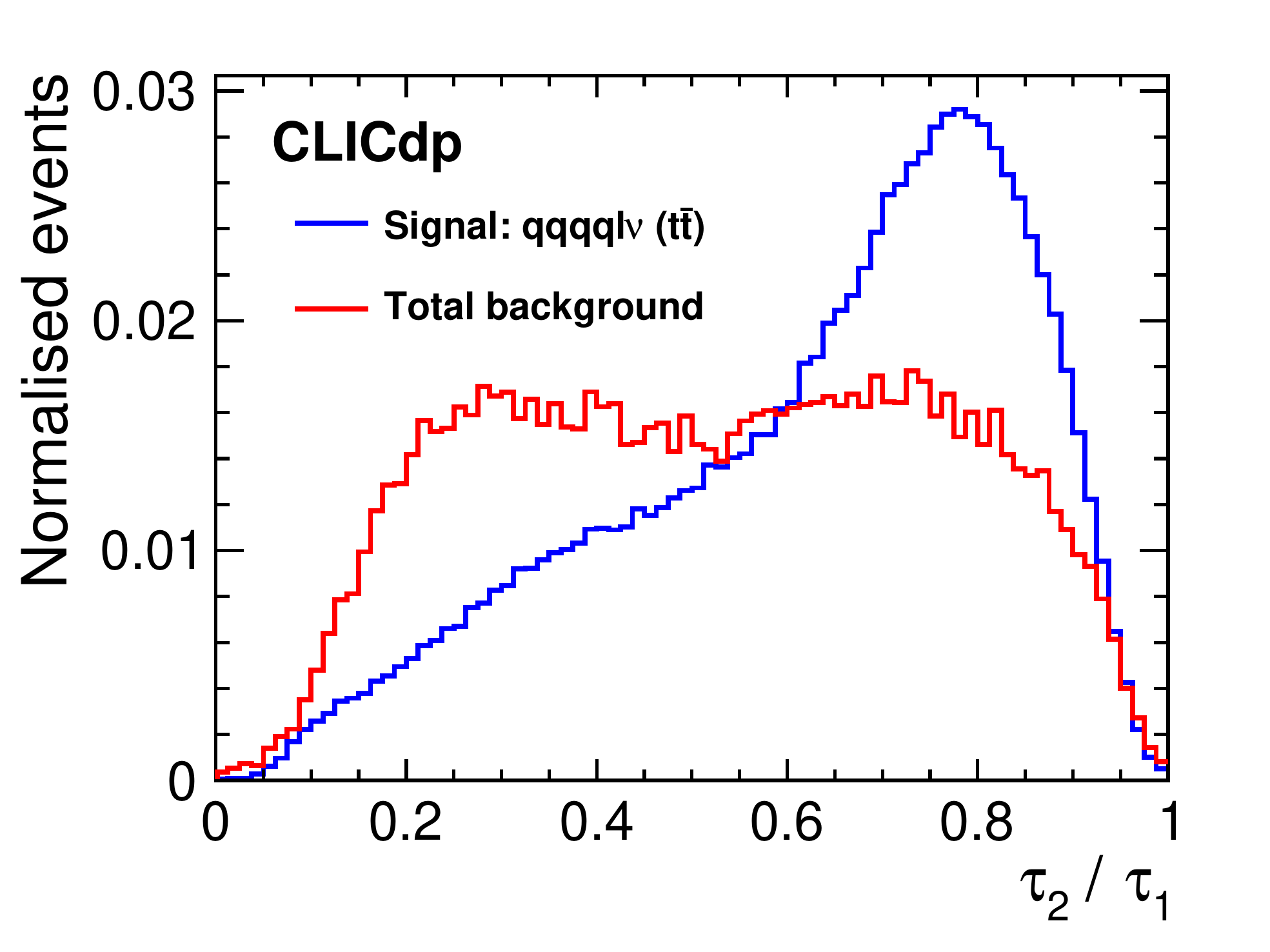}
  ~~~~
  \includegraphics[width=0.45\textwidth]{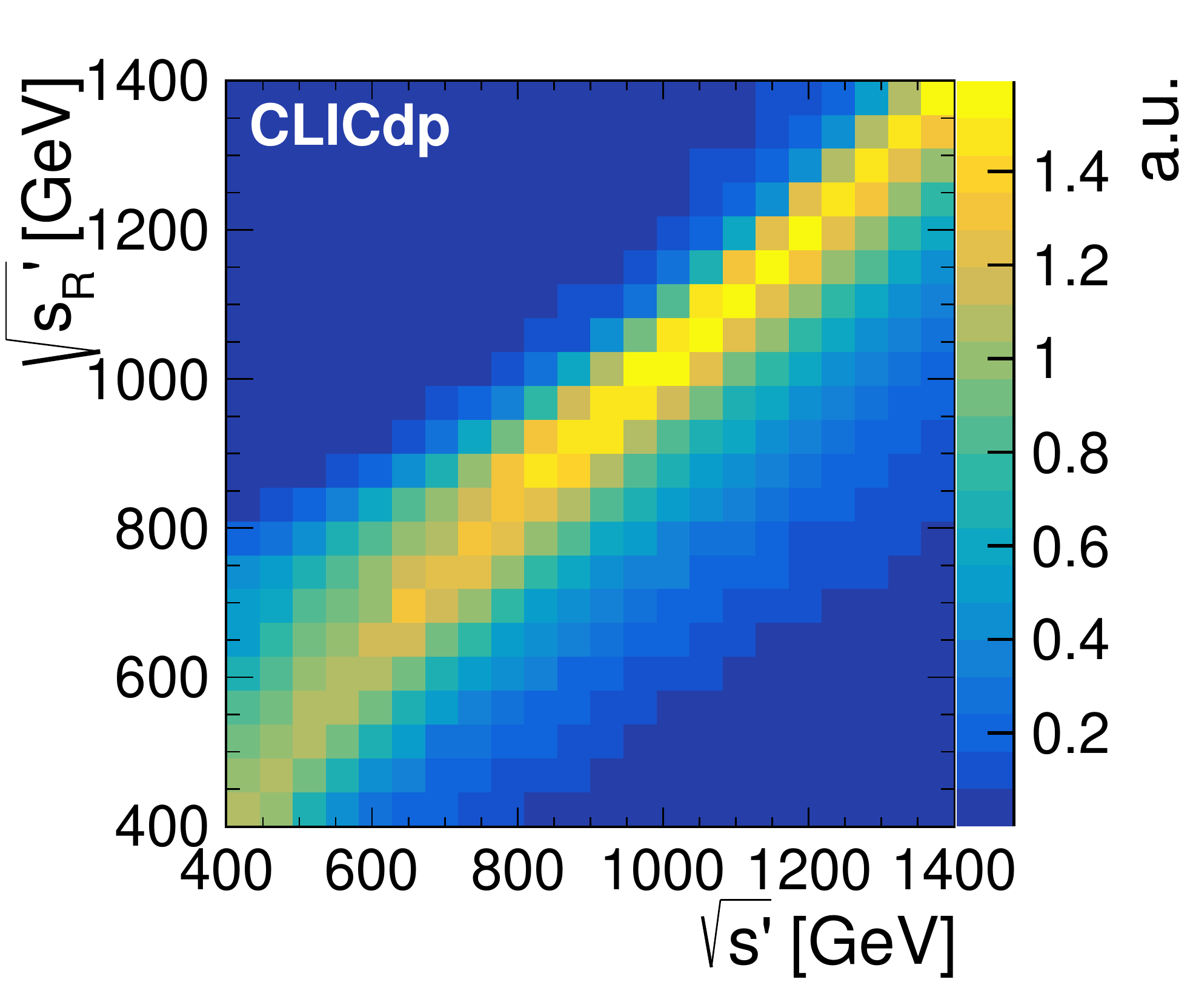} 
  \caption{The N-subjettiness ratio $\tau_2/\tau_1$ for the leptonically decaying large-$R$ jet (left). Reconstructed centre-of-mass energy $\rootsprimereco$ from kinematic fitting vs the generated collision energy, $\rootsprime$, including the effects of the luminosity spectrum and ISR (right). To illustrate the correlation down to lower $\rootsprime$, the $\rootsprime$ distribution is reweighted so that each column contains the same number of entries, leading to a flat distribution in $\rootsprime$. \label{fig:ttbar:radiative:variables}}
\end{figure}

This section describes an analysis using radiative events to extract $\csttbar$ and $\afb$ in regions of $\rootsprime$ below the nominal collision energy of $\roots=1.4\,\TeV$. The $\afb$ is evaluated across three mutually exclusive intervals in $\rootsprime$: 0.40\,--\,0.90\,\tev, 0.90\,--\,1.2\,\tev and $\geq1.2\,\tev$. The analysis is based on the study discussed in detail in \cite{winter:2018}.

This analysis uses jet-shape variables such as N-subjettiness \cite{Thaler:2010tr} as input to a multivariate classifier. These variables are well suited for the identification of boosted objects with multi-body kinematics and discriminate effectively against QCD background jets. 
 
The first stage of the event reconstruction is the identification of isolated charged leptons. In this analysis we apply a jet-based procedure where initially all PFOs are clustered into five jets using the Durham algorithm~\cite{Fastjet}. In the next step all PFOs identified as a charged lepton by the \pandora algorithm are considered as a candidate. The energy ratio between each such candidate and the corresponding jet it was clustered into is evaluated, and the candidate with the highest energy ratio is selected as the isolated lepton. For events without explicit input candidates from \pandora, which amount to $2\%\,(1\%)$ of the events with a final state electron (muon), the PFO with the highest energy ratio is chosen to ensure high efficiency. This method yields a charge tagging efficiency of 93\% (96\%) for electrons (muons).

The remaining PFOs are clustered into two exclusive large-$R$ jets using the VLC algorithm with a jet radius of 1.5 ($\beta,\gamma=1$). The large-$R$ jets are associated with either the hadronically decaying top quark or the $\PQb$-quark from the leptonically decaying top quark. After evaluating the association based on invariant mass, energy, $\PQb$-quark tagging, and separation of jets from the isolated lepton, associating the jet with the highest energy with the hadronically decaying top quark is found to give the best performance for reconstructing the correct top-quark decay angle.

The sub-structure of each large-$R$ jet is characterised by the N-subjettiness~\cite{Thaler:2010tr}, $\tau_{N}$, defined as
\begin{equation*}
\tau_{N} = \frac{1}{d_{0}} \sum_k \ensuremath{p_{\mathrm{T},k}}\xspace \min \{ \Delta R_{1,k}, \Delta R_{2,k}, \ldots, \Delta R_{N,k} \},
\end{equation*}
where $k$ runs over the constituent particles of the jet, each with transverse momentum $\ensuremath{p_{\mathrm{T},k}}$. The distance in the pseudorapidity-azimuth plane, between each candidate subjet $J$ and constituent particle $k$, is denoted $\Delta R_{J,k}^2={\Delta\eta^2+\Delta\phi^2}$ and $d_0 = R_0 \cdot \sum_k \ensuremath{p_{\mathrm{T},k}}$, where $R_0$ is the jet radius used in the large-$R$ jet clustering. $\tau_{N}$ quantifies to what degree a jet can be regarded as composed of $N$ subjets. A large value would indicate that the jets have a large fraction of their energy distributed away from the candidate subjet directions, i.e. that it has at least $N+1$ subjets. A key step for defining N-subjettiness is to make an appropriate choice of the candidate subjets. In this analysis they are produced by reclustering the large-$R$ jets into $J$ exclusive jets (\kT algorithm, $R=0.3$). The analysis studied several ratios $\tau_{N+1}/\tau_N$ for both the hadronically and leptonically decaying top quark. The N-subjettiness ratio $\tau_2/\tau_1$ for the leptonically decaying large-$R$ jet is shown to the left of \autoref{fig:ttbar:radiative:variables}. This variable is used to distinguish the $\PQb$-quark of the leptonic top-quark decay from jets with a multi-pronged sub-structure as present in some of the background processes considered. The distributions are shown after both pre-selection and quality cuts and are normalised to unity.

The jet-multiplicity of each large-$R$ jet is estimated by reclustering the constituent PFOs with a small radius of $R=0.05$ using the \kT algorithm, quoting the number of resulting inclusive micro-jets. The sub-structure of the large-$R$ jets is further characterised by reclustering the large-$R$ jet into three exclusive jets, with a radius $R=0.3$ using the \kT algorithm, and studying the angular separation of the resulting subjets. A small separation is observed for background events due to fake splitting of single-quark jets, while the genuine sub-structure of the top-quark jets yields larger separations. We also define the so-called jet splitting scales, $d_{23}$ and $d_{34}$, representing the jet clustering distance parameters for the two last merging steps in the exclusive large-$R$ jet clustering.

Kinematic fitting, as implemented in MarlinKinFit\,\cite{List:88030}, is used to reconstruct the $\rootsprime$ of each collision, allowing for ISR and beamstrahlung. The fit has four degrees of freedom (the 3-momentum of the neutrino and the $z$ component of the photon momentum) and six constraints (the total 4-momentum of the system, the mass of the leptonically decaying W boson, and the masses of the two top-quark candidates). It is assumed that any unobserved ISR and beamstrahlung contributions have negligible transverse momentum. This method yields a resolution of $\sim$75~GeV on $\rootsprime$. A cut is placed on the resulting reconstructed $\rootsprimereco$, as part of the pre-selection. The cut value corresponds to the definition of the kinematic region for each signal interval considered. The reconstruction performance is illustrated to the right of \autoref{fig:ttbar:radiative:variables}. The figure shows the correlation between the reconstructed and generated values on an event-by-event basis, down to the lowest collision energies. All subsequent references to jet kinematic properties in this section refer to those of the fitted objects.

The full event selection is performed in three stages: a pre-selection to suppress apparent backgrounds, a cut on reconstruction quality, and finally a multivariate classification algorithm. The selection is based on PFOs with \code{tight} timing cuts, as discussed in \autoref{ssec:reco}.\footnote{This choice is motivated by the study of jet substructure variables that have lower contamination from beam jets.}

The pre-selection consists of the following requirements: 
\begin{itemize}[leftmargin=2em, label={\tiny\raisebox{1ex}{\textbullet}}]
\item scalar sum of transverse momenta $>200\,\mathrm{GeV}$, 
\item energy of the hadronically decaying top quark $>100\,\mathrm{GeV}$, 
\item transverse momentum of $\PQb$-quark jet $>20\,\mathrm{GeV}$, 
\item jet splitting scales $-\log_{10}(d_{23})<7$ and $-\log_{10}(d_{34})<9$. 
\end{itemize}
These cuts are followed by a series of quality cuts aimed at removing events in which the polar angle of the hadronically decaying top quark is poorly reconstructed, for example due to proximity to the edge of the detector acceptance. The cuts used are: 
\begin{itemize}[leftmargin=2em, label={\tiny\raisebox{1ex}{\textbullet}}]
\item invariant mass and transverse momentum of the hadronically decaying top quark $>100\,\mathrm{GeV}$, 
\item invariant mass of the leptonically decaying top quark $<100\,\mathrm{GeV}$,
\item angle $\theta_{12}$ between the leading and next-to-leading energy subjets of the hadronically decaying top quark $0.2<\cos\theta_{12}<0.9$,
\item jet splitting scale $-\log_{10}(d_{23})>3$,
\item $z$ component of the total event momentum from the kinematic fitter $<100\,\mathrm{GeV}$ (used as a quality cut for the kinematic fitter routine).
\end{itemize}

The final stage of the event selection was performed by training two BDTs to search for $\ttbar$ events, one with a centre-of-mass energy $\rootsprime$ above $1.2\,\tev$ and one with a centre-of-mass energy $\rootsprime$ below or equal to $1.2\,\tev$. Each BDT uses 21 variables based on the kinematics of the hadronically decaying top quark, lepton and $\PQb$-jet, the substructure of both large-$R$ jets, the number of lepton candidates with energy $>30\,\mathrm{GeV}$, $\PQb$-quark tagging information, and event shapes.

\begin{figure}
\vspace{5mm}
\centering
\includegraphics[width=0.48\columnwidth]{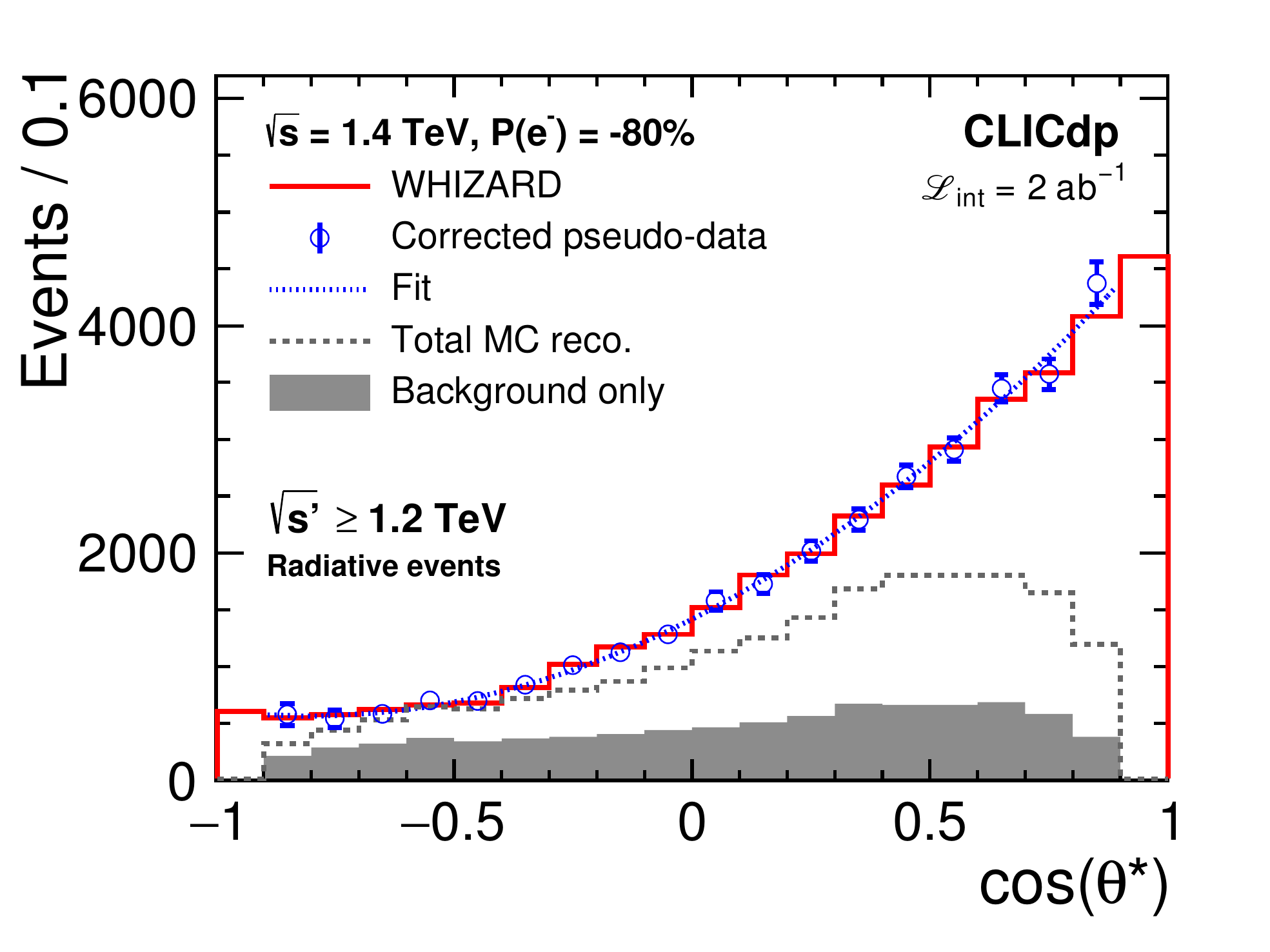}
~~~~
\includegraphics[width=0.48\columnwidth]{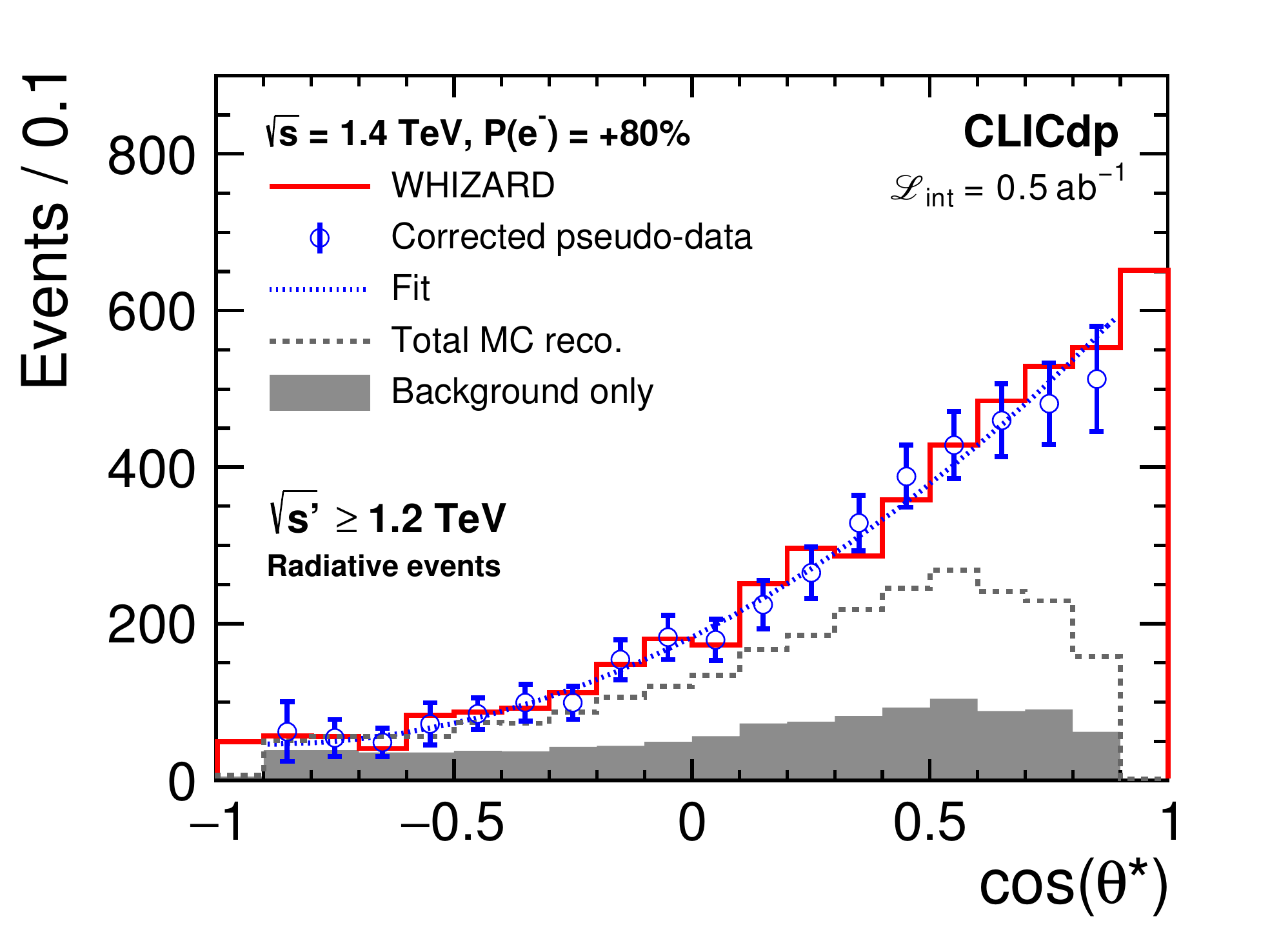}
~~~~
\includegraphics[width=0.48\columnwidth]{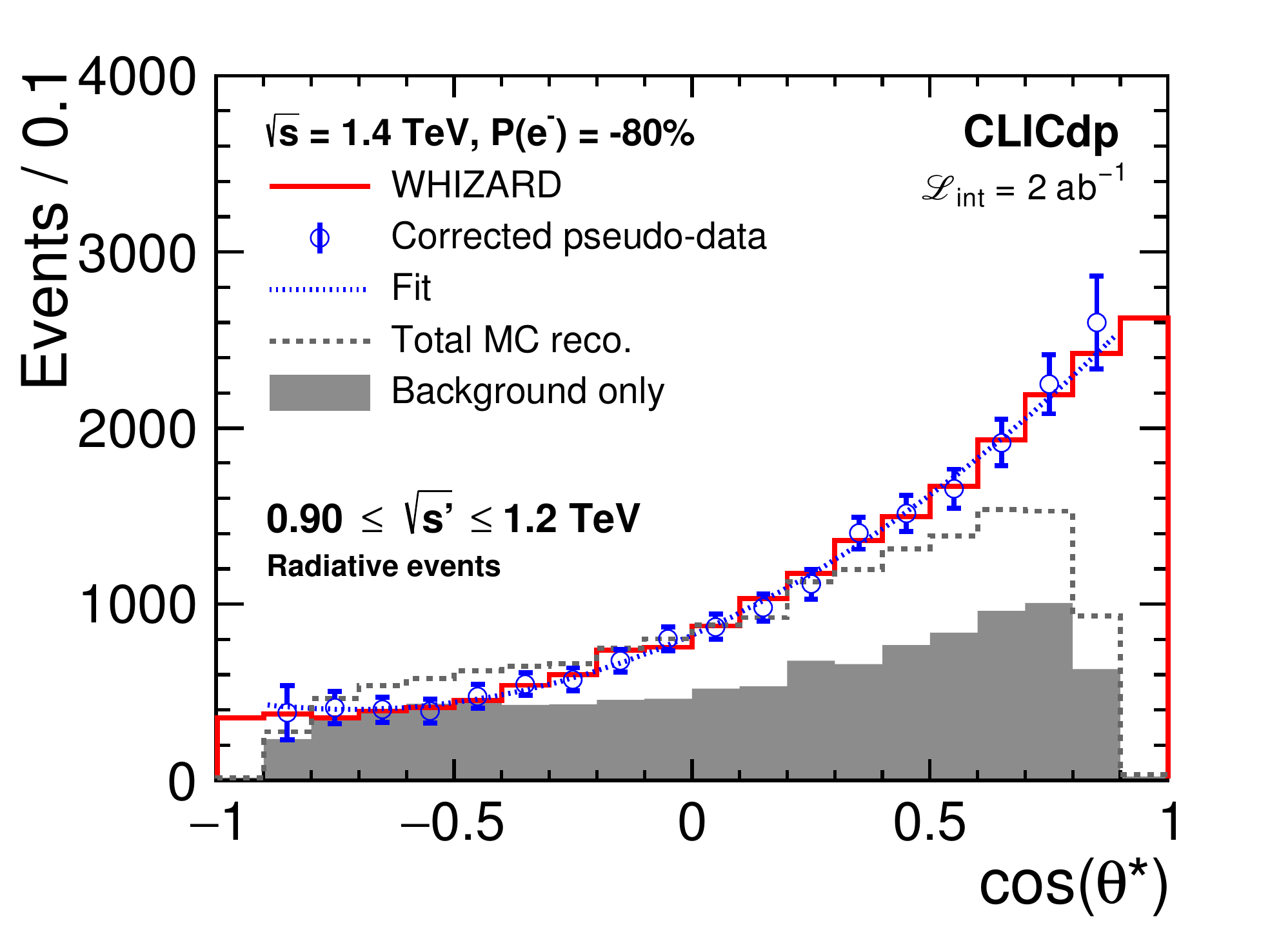}
~~~~
\includegraphics[width=0.48\columnwidth]{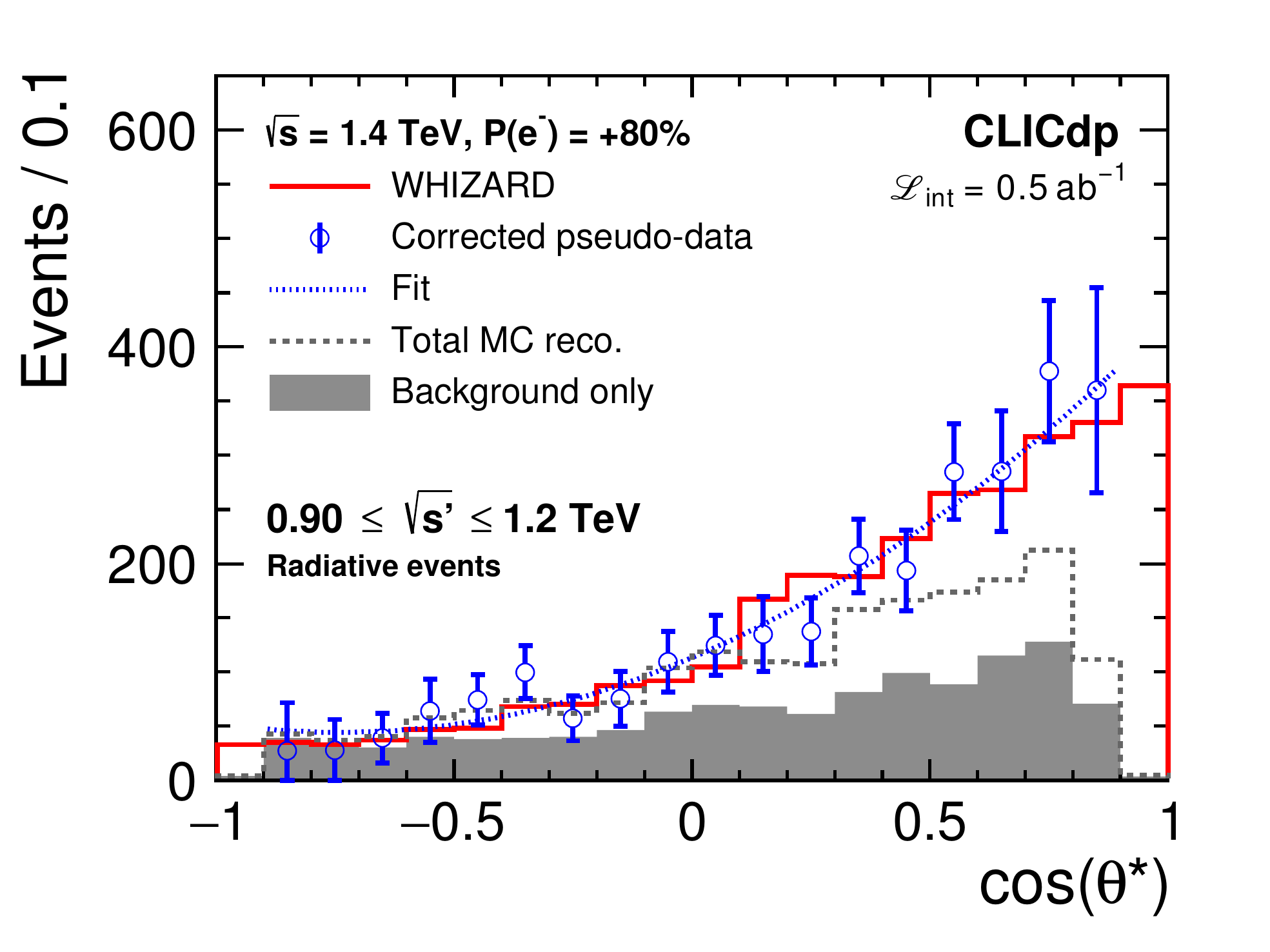}
~~~~
\includegraphics[width=0.48\columnwidth]{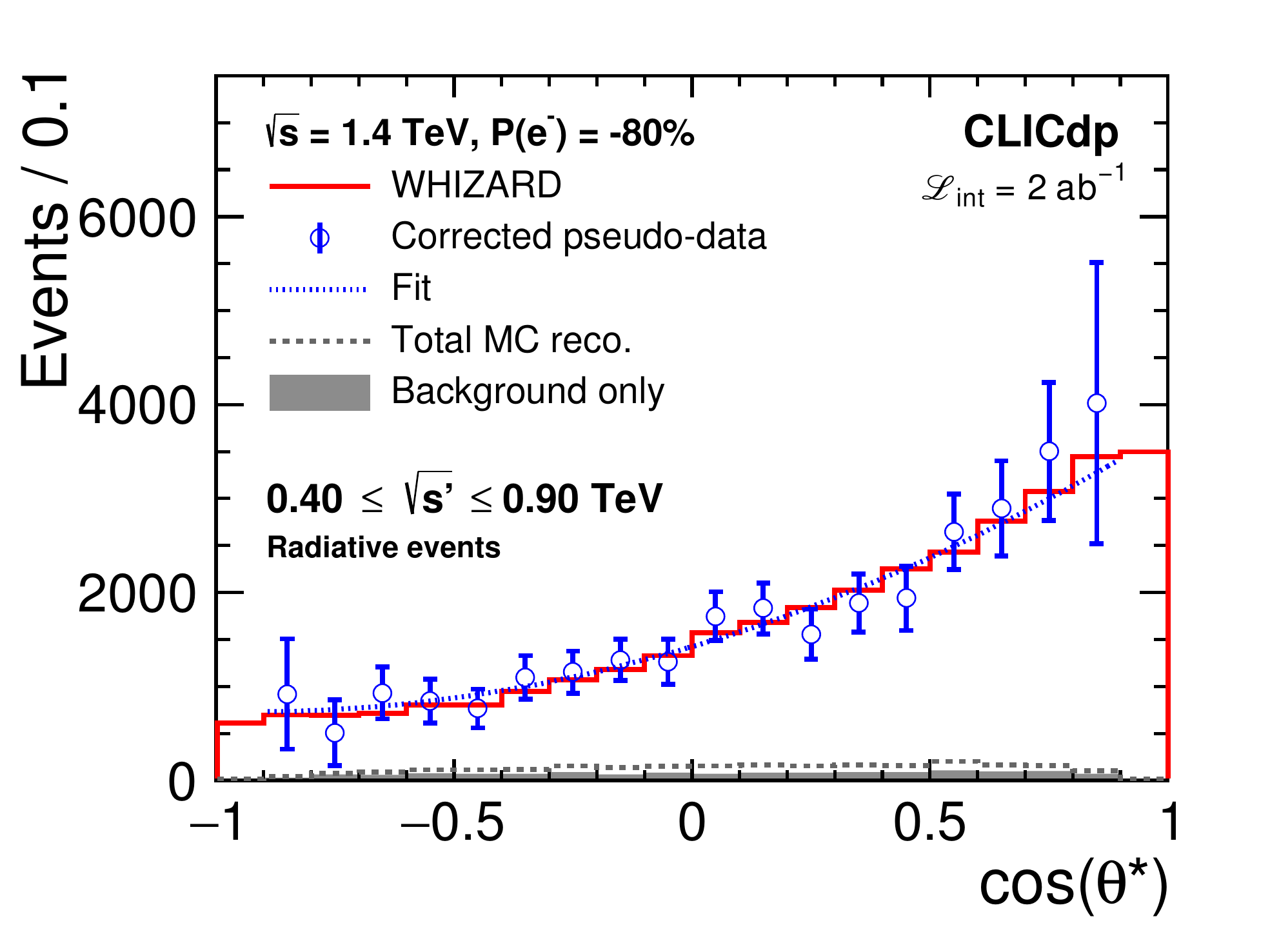}
~~~~
\includegraphics[width=0.48\columnwidth]{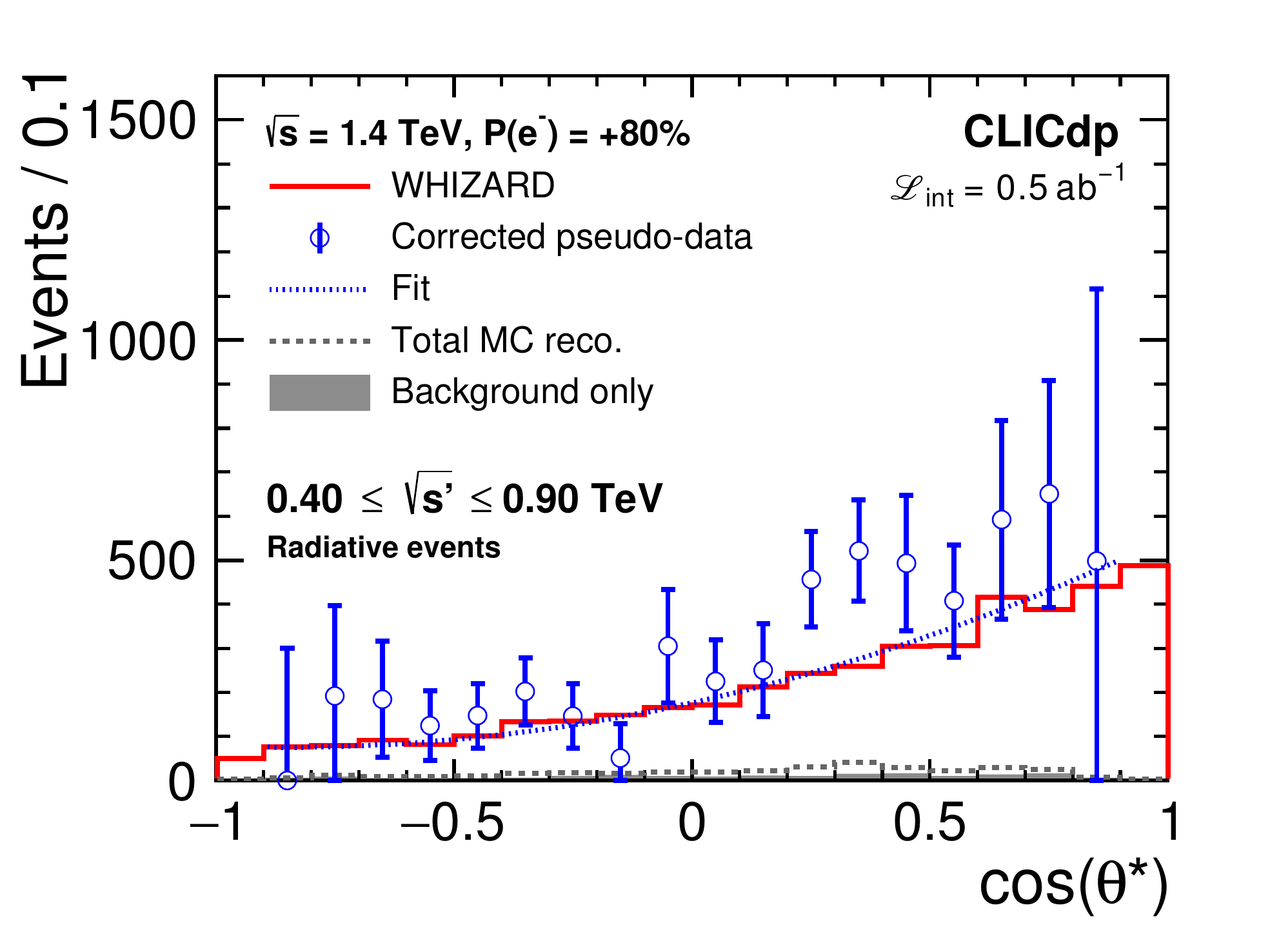}
\caption{The top-quark polar angle distributions for the analysis of radiative semi-leptonic $\ttbar$ events at $\roots=1.4\,\TeV$, for $P(\Pem)=\text{-}80\%$ (left column) and $P(\Pem)=\text{+}80\%$ (right column), and an integrated luminosity of $2.0\,\abinv$ and $0.5\,\abinv$, respectively. Distributions are shown for the three signal regions in $\rootsprime$. The dashed black curve shows the reconstructed total MC distribution, while the grey area indicates the level of background only. The blue data points and dotted line represent one pseudo-experiment after subtraction of background and correction for finite selection efficiencies, and the corresponding fit, respectively. The red solid line represents the simulated parton-level distribution. \label{fig:ttbar:radiative:distributions}}
\end{figure}

The event selection efficiencies for the signal and dominant background processes are shown in \autoref{tab:ttbar:radiative:selection:bin3}, \autoref{tab:ttbar:radiative:selection:bin2}, and \autoref{tab:ttbar:radiative:selection:bin1} in the Appendix, along with the total number of selected events. Contributions from additional backgrounds such as $\epem\to\PQq\PQq\PGn\PGn$ and additional six-fermion processes are found to be negligibly small. Note that most events with a large energy loss due to photon radiation also have a large net boost along the $z$-direction which makes them more difficult to reconstruct. In particular, this results in a significantly lower overall efficiency for the selection in the interval $0.40 \leq \rootsprime \leq 0.90\,\tev$. In the same interval we impose a strict cut on the lepton momentum, $>70\,\GeV$.

The polar-angle distributions of the hadronically decaying top-quark candidates are shown in \autoref{fig:ttbar:radiative:distributions}. 
The dashed black curve shows the reconstructed distribution for the total MC, while the grey area indicates the level of background only. These include the effects of detector modelling, event reconstruction, and candidate selection. The blue data points represent one pseudo-experiment performed for the given luminosity, after subtraction of background and correction for finite selection efficiencies. The blue dotted line shows the fit performed to the pseudo-experiment data and is used to extract $\csttbar$ and $\afb$ as defined in \autoref{eq:totcs} and \autoref{eq:afb}. The red solid line displays the simulated distribution at parton-level (\whizard). The distributions are shown for the fiducial region $-0.9\leq\cos\theta^{*}\leq0.9$.
The selection efficiency in the region $-0.6\leq\cos\theta^{*}\leq0.6$ is generally flat with a central value of 40\% in the interval $\rootsprime\geq1.2\,\tev$, 35\% in the interval $0.90\leq\rootsprime\leq1.2\,\tev$, and 4\% in the interval $0.40\leq\rootsprime\leq0.90\,\tev$. In the forward regions the efficiency drops by a factor of 2 in all three regions of $\rootsprime$, as visible in \autoref{fig:ttbar:radiative:distributions}.

%===================================================================================================
%===================================================================================================

\subsection{Boosted event topologies}
\label{ssec:pairprodboosted}

\begin{figure}
  \centering
  \includegraphics[width=0.75\columnwidth]{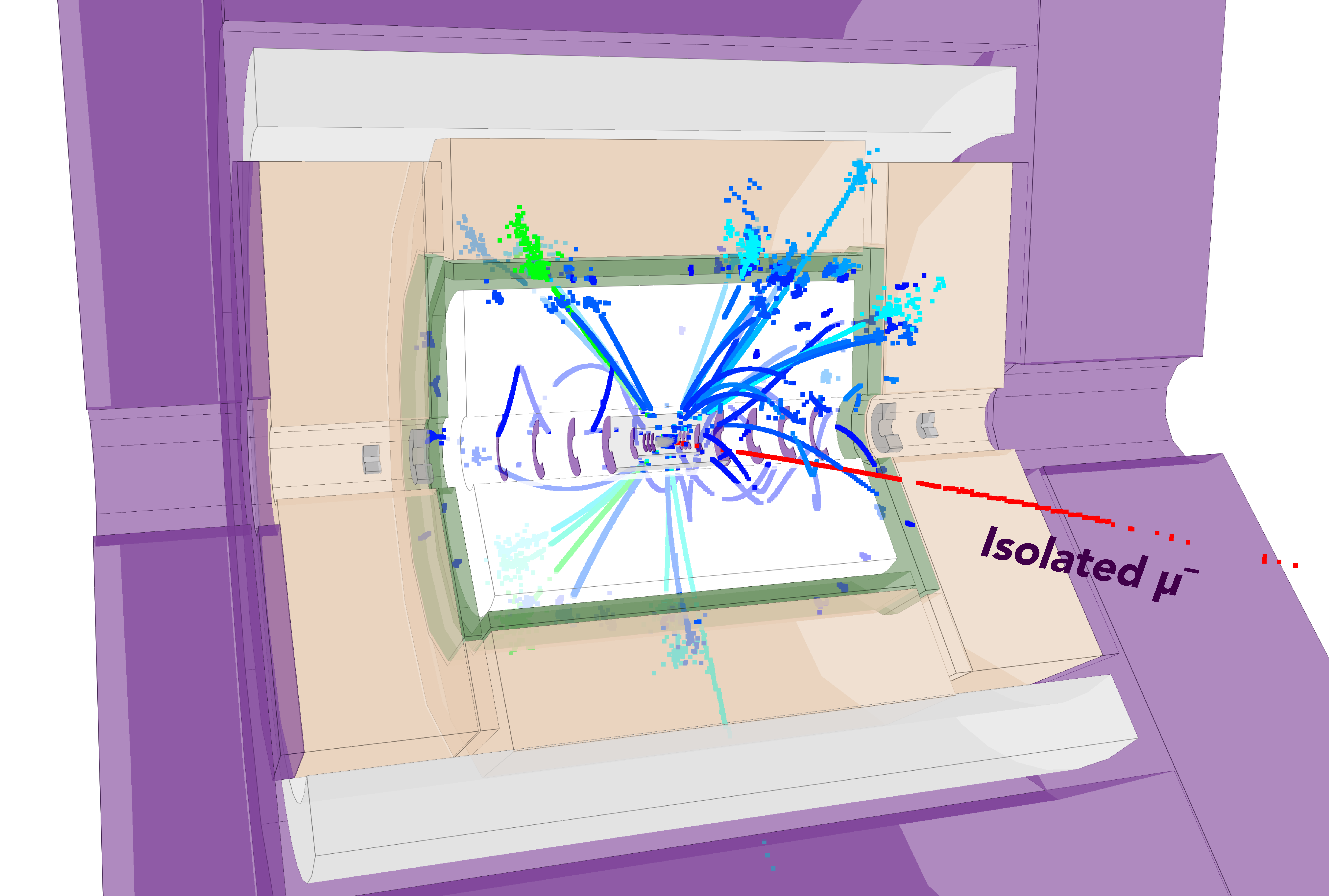}\\
  \vspace{1.2cm}
    \includegraphics[width=0.75\columnwidth]{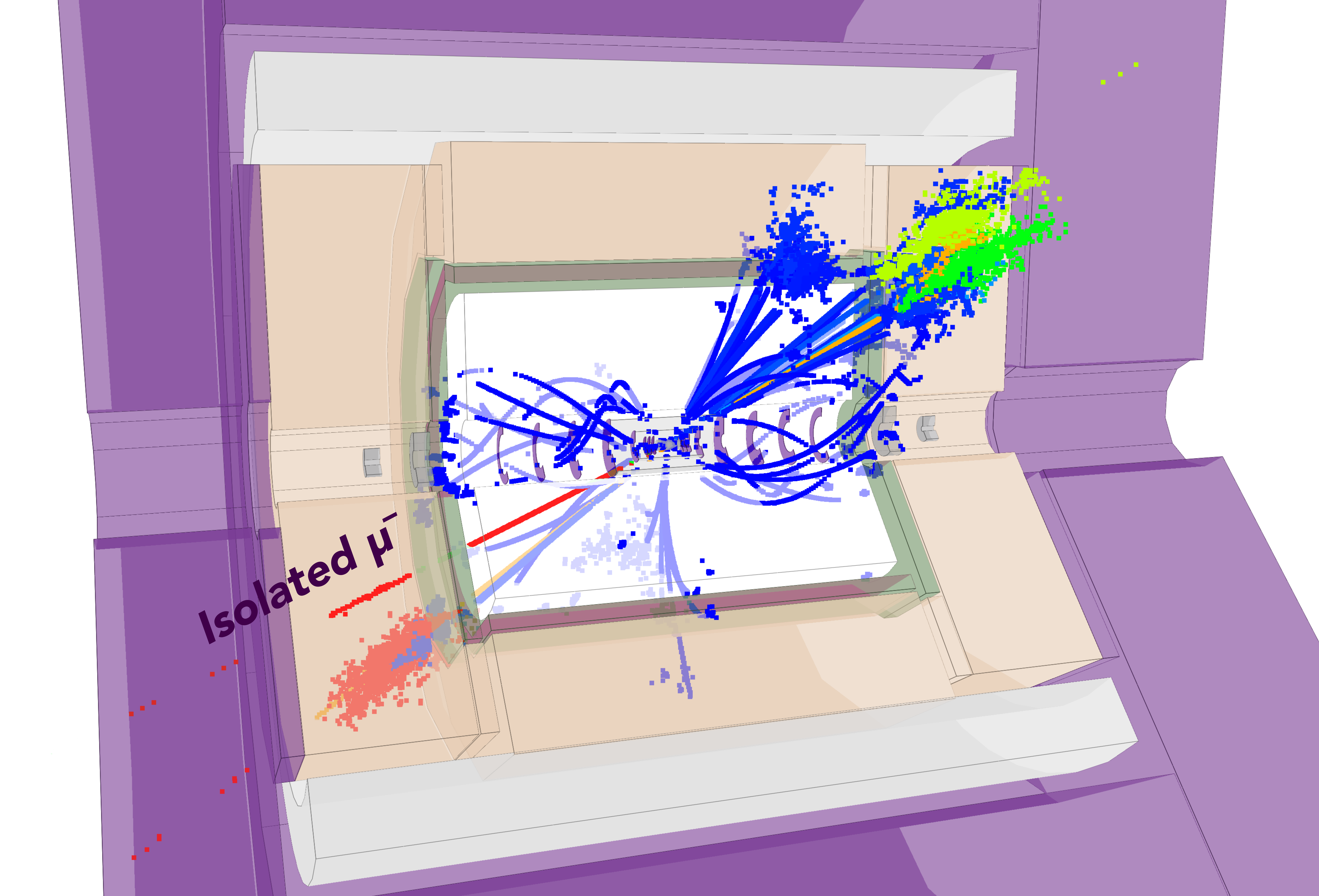}
  \vspace{0.7cm}
  \caption{Example displays of $\ttbar\to\PQq\PQq\PQq\PQq\PGm\PGnGm$ events in \clicild at $\roots=380\,\gev$ (top) and $\roots=3\,\tev$ (bottom). The events include overlay of $\gghadrons$ background as described in \autoref{ssec:detsim}. An isolated lepton is clearly seen along with four separate jets (top) or two larger boosted jets (bottom).\label{fig:ttbar:boosted:eventdisplay}}
\end{figure}

For operation above $\sim1\,\tev$ a large fraction of the top quarks will be produced with significant boosts. In particular, the event topology is very different from that of the analysis described in \autoref{ssec:pairprod380gev}, where the top quarks are produced close to the threshold with a resulting isotropic event topology, as illustrated by an example event in the upper panel of \autoref{fig:ttbar:boosted:eventdisplay}. In contrast, the lower panel of \autoref{fig:ttbar:boosted:eventdisplay} shows a boosted semi-leptonic $\ttbar$ event at 3\,TeV with clear separation between the decay products of the top- and anti-top quark respectively. Owing to the boost, the top-quark candidates are more easily distinguishable from each other and the relative effect of migrations, as discussed in \autoref{ssec:pairprod380gev}, is therefore expected to be smaller. This section describes the event selection and results for an analysis targeting semi-leptonic $\ttbar$ events ($\Pl=\Pe,\PGm$) at the collision energies of $1.4\,\TeV$ and $3\,\TeV$. The signal events are restricted to the kinematic region defined as $\rootsprime\geq1.2\,\tev$ and $\rootsprime\geq2.6\,\tev$, respectively. A corresponding cut is applied to the reconstructed collision energy, $\rootsprimereco$, as part of the pre-selection defined below.

The event selection proceeds through the identification of one isolated charged lepton in association with one boosted top quark, the latter being identified using the dedicated top-quark tagger algorithm whose details and performance are described in \autoref{sec:boosted}. Events with isolated high-energy photons are removed and a cut is placed on the reconstructed centre-of-mass energy, $\rootsprimereco$, corresponding to the kinematic regions of the signal outlined above. The selection is based on PFOs with \code{default} timing cuts at 1.4\tev and \code{tight} timing cuts at 3\tev; see the discussion in \autoref{ssec:reco}. 

The charged final state lepton is identified using the isolated lepton finding procedure described in \autoref{ssec:reco}. In addition we require that the $\pT$ of the isolated lepton candidate is larger than $10\,\mathrm{GeV}$. In cases where several candidates exist, the candidate with the highest $\pT$ is selected. For events where both top quarks fulfils $\cos(\theta)\leq0.8$ at parton-level, the efficiency for identifying the muon (electron) in the final state is about 90\% (80\%), out of which 99\% (98\%) are reconstructed with the correct charge.

The remaining PFOs are clustered in two subsequent steps following the approach outlined in \autoref{sec:boosted}. The resulting two exclusive large-$R$ jets are used as input to the top-quark tagging algorithm that constitutes the basis for identification of the hadronically decaying top quark in the following analysis.

\begin{table}[t!]
\centering
\begin{minipage}{\columnwidth}
\resizebox{1.0\textwidth}{!}{
\begin{tabular}{lcccccccc}
\toprule
\vspace{1.0mm}
{} & \multicolumn{2}{c}{$\sigma\,[\fb]$} & \multicolumn{2}{c}{$\epsilon_{\,\mathrm{Pre}}\,[\%]$} & \multicolumn{2}{c}{$\epsilon_{\,\mathrm{MVA}}\,[\%]$} & \multicolumn{2}{c}{$N$} \rule{0pt}{3ex} \\
\vspace{1.0mm}
P(\Pem) & -80\% & +80\% & -80\% & +80\% & -80\% & +80\% & -80\% & +80\% \\
Process \\
\midrule
$\epem(\to\ttbar)\to\PQq\PQq\PQq\PQq\Pl\PGn\,(\Pl=\Pe,\PGm)$\footnote{Kinematic region defined as $\rootsprime\geq1.2\,\tev$} & 18.4 & 9.83 & 43 & 44 & 85 & 87 & 13,469 & 1,902 \\
\midrule
$\epem(\to\ttbar)\to\PQq\PQq\PQq\PQq\Pl\PGn\,(\Pl=\Pe,\PGm)$\footnote{$\rootsprime<1.2\,\tev$} & 28.5 & 14.9 & 2.5 & 2.7 & 68 & 56 & 952 & 111 \\
$\epem(\to\ttbar)\to\PQq\PQq\PQq\PQq\Pl\PGn\,(\Pl=\PGt)$ &23.2 & 12.3 & 4.7 & 4.8 & 63 & 57 & 1,379 & 167 \\
$\epem(\not\to\ttbar)\to\PQq\PQq\PQq\PQq\Pl\PGn$ & 72.2 & 16.5 & 6.0 & 7.2 & 35 & 59 & 3,032 & 348 \\
$\epem\to\PQq\PQq\PQq\PQq\PQq\PQq$ & 116 & 44.9 & 2.3 & 2.4 & 9.2 & 9.5 & 499 & 51 \\
$\epem\to\PQq\PQq\Pl\PGn\Pl\PGn$ & 44.1 & 15.3 & 1.2 & 1.5 & 27 & 40 & 285 & 45 \\
$\epem\to\PQq\PQq\PQq\PQq$ & 2,300 & 347 & 0.31 & 0.47 & 0.22 & 0.56 & 32 & 5 \\
$\epem\to\PQq\PQq\Pl\PGn$ & 6,980 & 1,640 & 0.02 & 0.01 & 0.00 & 0.00 & - & - \\
$\epem\to\PQq\PQq\Pl\Pl$ & 2,680 & 2,530 & 0.01 & 0.08 & 0.00 & 0.00 & - & - \\
$\epem\to\PQq\PQq$ & 4,840 & 3,170 & 0.21 & 0.16 & 1.3 & 0.00 & 259 & - \\
\bottomrule
\end{tabular}
}
\end{minipage}
\caption{Event selection summary for the analysis of $\ttbar$ events at $\roots=1.4\,\tev$, assuming $2.0\,\abinv$ and $0.5\,\abinv$ for $P(\Pem)=\text{-}80\%$ and $P(\Pem)=\text{+}80\%$, respectively. The cross section quoted for the signal sample in the uppermost row is defined in the kinematic region $\rootsprime\geq1.2\,\tev$. The fractional pre-selection and MVA selection efficiencies are shown in the subsequent columns along with the number of events in the final sample. \label{tab:ttbar:boosted:selection1400}}
\end{table}

\begin{table}[h]
\centering
\begin{minipage}{\columnwidth}
\resizebox{1.0\textwidth}{!}{
\begin{tabular}{lcccccccc}
\toprule
\vspace{1.0mm}
{} & \multicolumn{2}{c}{$\sigma\,[\fb]$} & \multicolumn{2}{c}{$\epsilon_{\,\mathrm{Pre}}\,[\%]$} & \multicolumn{2}{c}{$\epsilon_{\,\mathrm{MVA}}\,[\%]$} & \multicolumn{2}{c}{$N$} \rule{0pt}{3ex} \\
\vspace{1.0mm}
P(\Pem) & -80\% & +80\% & -80\% & +80\% & -80\% & +80\% & -80\% & +80\% \\
Process \\
\midrule
$\epem(\to\ttbar)\to\PQq\PQq\PQq\PQq\Pl\PGn\,(\Pl=\Pe,\PGm)$\footnote{Kinematic region defined as $\rootsprime\geq2.6\,\tev$} & 3.48 & 1.89 & 41 & 43 & 80 & 85 & 4,563 & 692 \\
\midrule
$\epem(\to\ttbar)\to\PQq\PQq\PQq\PQq\Pl\PGn\,(\Pl=\Pe,\PGm)$\footnote{$\rootsprime<2.6\,\tev$} & 13.7 & 7.26 & 0.98 & 0.86 & 65 & 76 & 352 & 48 \\
$\epem(\to\ttbar)\to\PQq\PQq\PQq\PQq\Pl\PGn\,(\Pl=\PGt)$ & 8.45 & 4.51 & 3.6 & 3.8 & 58 & 47 & 699 & 81 \\
$\epem(\not\to\ttbar)\to\PQq\PQq\PQq\PQq\Pl\PGn$ & 99.6 & 22.6 & 1.4 & 1.4 & 23 & 51 & 1,344 & 155 \\
$\epem\to\PQq\PQq\PQq\PQq\PQq\PQq$ & 54.0 & 18.0 & 3.4 & 3.8 & 4.7 & 6.1 & 344 & 41 \\
$\epem\to\PQq\PQq\Pl\PGn\Pl\PGn$ & 59.7 & 14.9 & 0.28 & 0.37 & 23 & 40 & 155 & 22 \\
$\epem\to\PQq\PQq\PQq\PQq$ & 963 & 130 & 0.36 & 0.38 & 0.21 & 0.39 & 29 & 2 \\
$\epem\to\PQq\PQq\Pl\PGn$ & 8,810 & 2,310 & 0.01 & 0.01 & 0.00 & 0.00 & - & - \\
$\epem\to\PQq\PQq\Pl\Pl$ & 3,230 & 3,060 & 0.02 & 0.02 & 0.44 & 0.00 & 13 & - \\
$\epem\to\PQq\PQq$ & 3,510 & 2,390 & 0.15 & 0.11 & 0.29 & 0.00 & 61 & - \\
\bottomrule
\end{tabular}
}
\end{minipage}
\caption{Event selection summary for the analysis of $\ttbar$ events at $\roots=3\,\tev$, assuming $4.0\,\abinv$ and $1.0\,\abinv$ for $P(\Pem)=\text{-}80\%$ and $P(\Pem)=\text{+}80\%$, respectively. The cross section quoted for the signal sample in the uppermost row is defined in the kinematic region $\rootsprime\geq2.6\,\tev$. The fractional pre-selection and MVA selection efficiencies are shown in the subsequent columns along with the number of events in the final sample. \label{tab:ttbar:boosted:selection3000}}
\end{table}

Isolated high-energy photons are identified as photons from the particle flow reconstruction with a $\pT$ in excess of $75\,\mathrm{GeV}$, a polar angle in the range $10^\circ\leq\theta\leq170^\circ$, and low activity in a cone around the candidate PFO. 

To reconstruct the effective centre-of-mass $\rootsprime$ we first assume that the missing transverse momentum, estimated by adding up the 4-vectors of the two large-$R$ jets and the isolated charged lepton, can be used as an estimator for the neutrino transverse momentum components. Here we neglect the effect from unidentified ISR and beamstrahlung photons. The $z$-component of the neutrino momentum, $p_{\PGn,z}$, is retrieved by solving 
\begin{equation}\label{eq:boosted:sprimereco}
M_{\PW}^2 = m_{\Pl}^2 + 2(E_{\Pl}E_{\PGn} - \vec{p}_{\Pl}\cdot\vec{p}_{\PGn}),
\end{equation}
given a constraint on $M_{\PW}$, the mass of the leptonically decaying $\PW$ boson. Here, the indices $\Pl$ and $\PGn$ denote the lepton and neutrino candidate quantities, respectively. \autoref{eq:boosted:sprimereco} is quadratic in $p_{\PGn,z}$ and has no solution if the observed missing transverse energy fluctuates such that the invariant mass of the combined neutrino-lepton system is above $M_{\PW}$. In such cases the missing transverse energy is scaled to provide a real solution. The resulting neutrino-lepton system solutions are combined with each of the large-$R$ jets and the final candidate is chosen as the one that yields a mass closest to the generated top-quark mass. This method yields an RMS on $\rootsprime$ of $\sim$140\,\GeV.

\begin{figure}[t!]
  \centering
  \includegraphics[width=0.48\columnwidth]{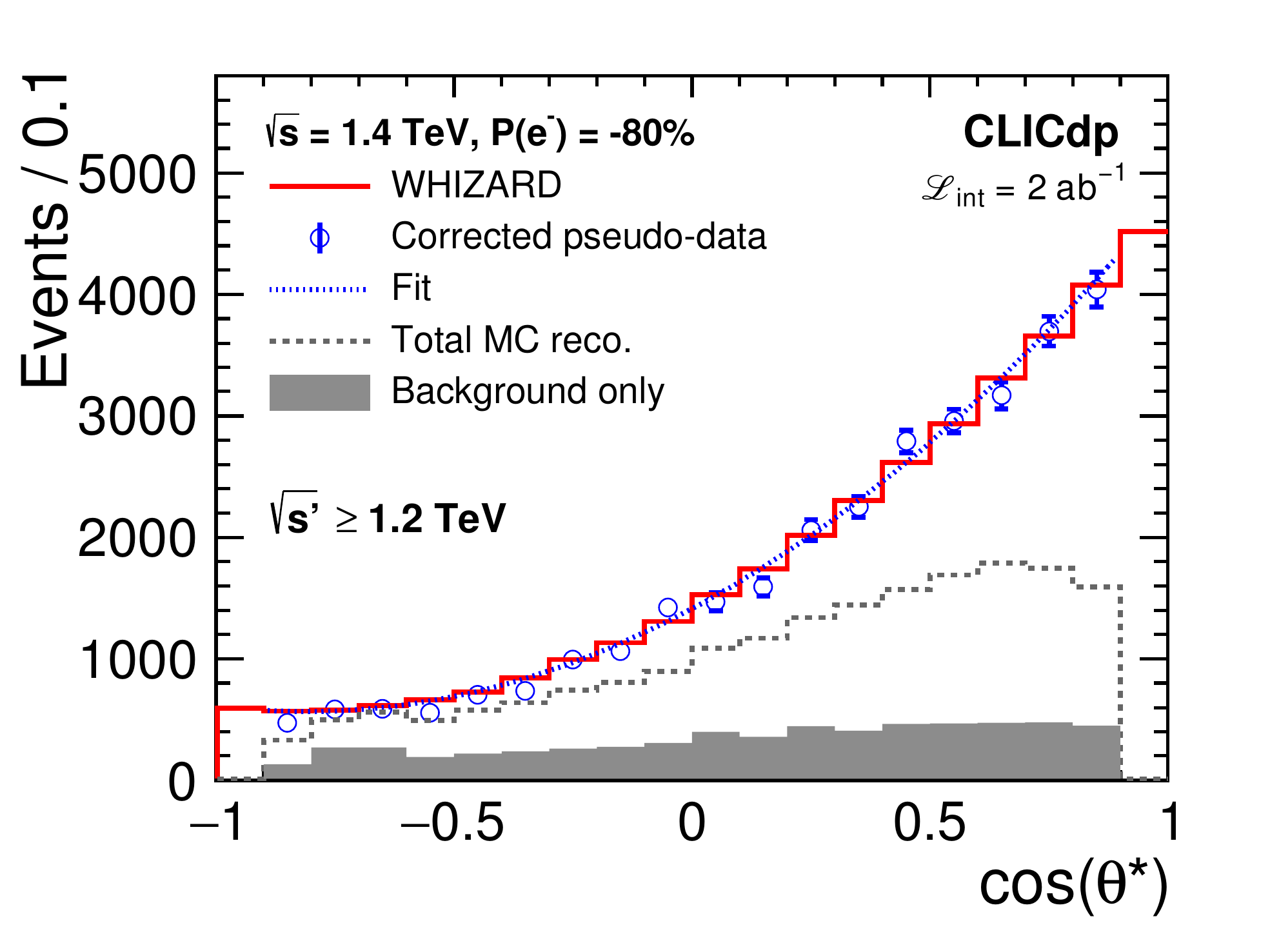}
  ~~~~
   \includegraphics[width=0.48\columnwidth]{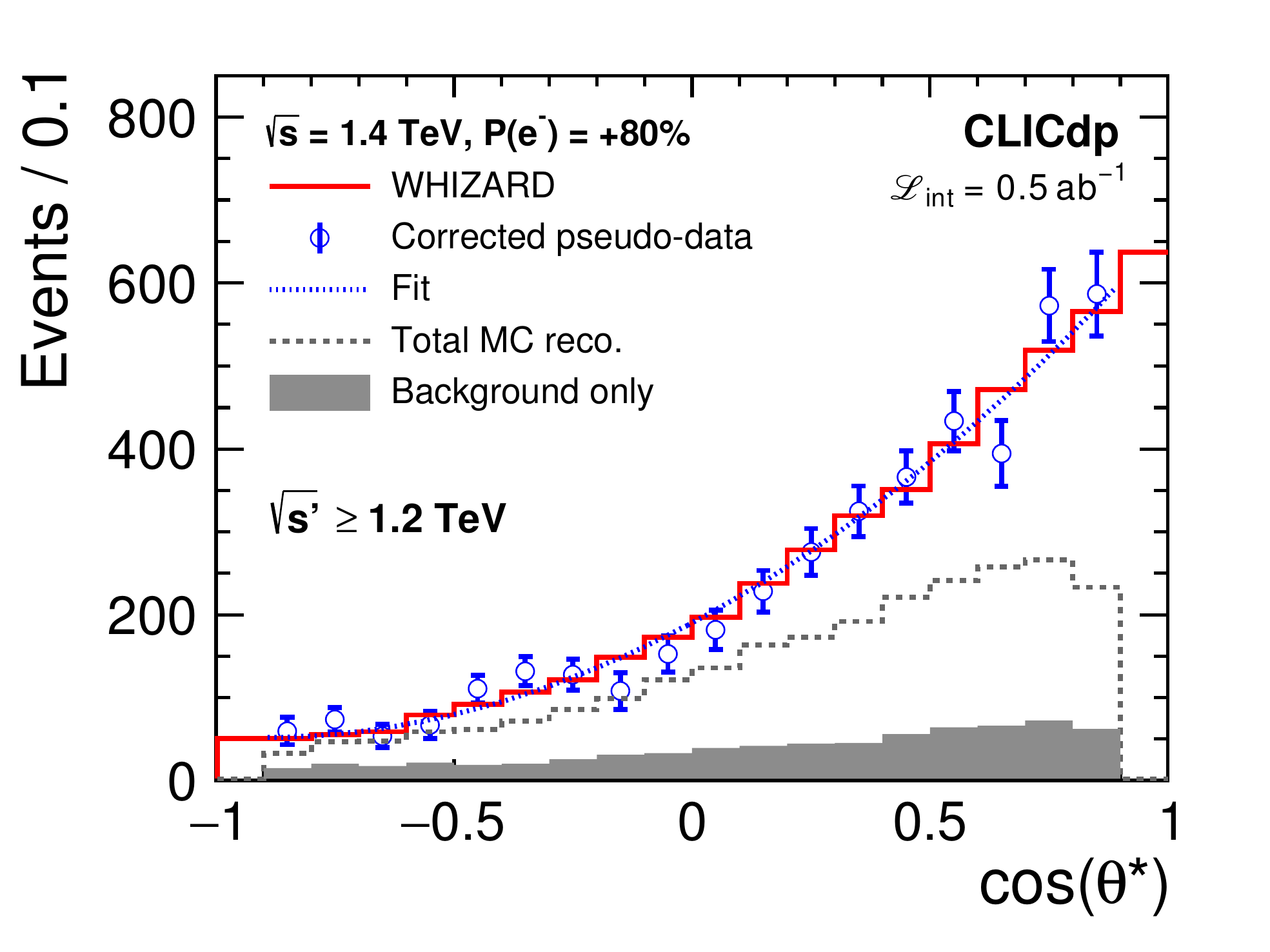}
  \caption{Top-quark polar angle distributions from the analysis of boosted semi-leptonic $\ttbar$ events, at a nominal collision energy of 1.4\,\tev for $P(\Pem)=\text{-}80\%$ (left) and $P(\Pem)=\text{+}80\%$ (right), and an integrated luminosity of $2.0\,\abinv$ and $0.5\,\abinv$, respectively. The dashed black curve shows the reconstructed total MC distribution, while the grey area indicates the level of background only. The blue data points and dotted line represent one pseudo-experiment after subtraction of background and correction for finite selection efficiencies, and the corresponding fit, respectively. The red solid line represents the simulated parton-level distribution.\label{fig:ttbar:boosted:costheta}}
\end{figure}

\begin{figure}[t!]
  \centering
    \includegraphics[width=0.48\columnwidth]{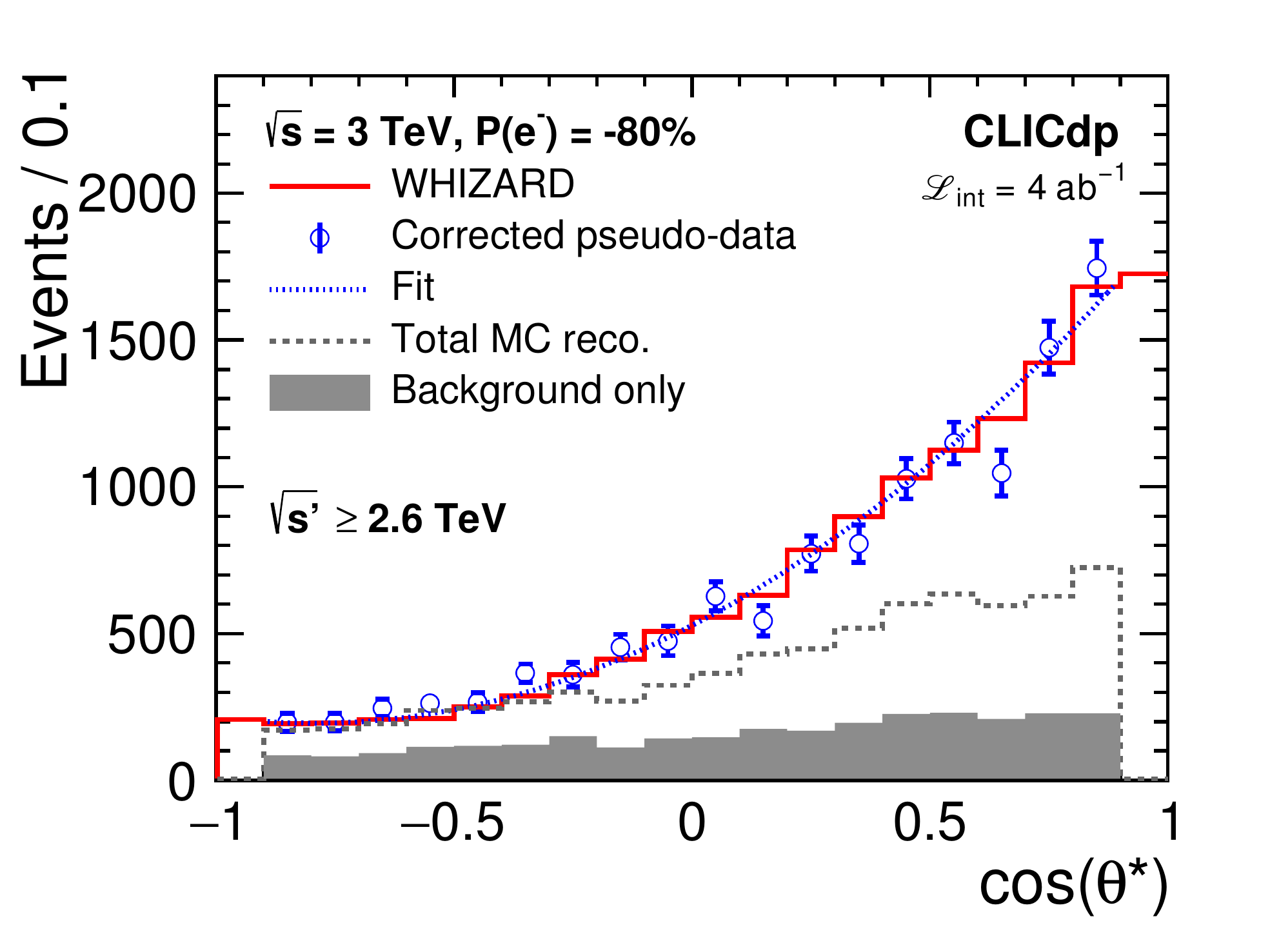}
   ~~~~
  \includegraphics[width=0.48\columnwidth]{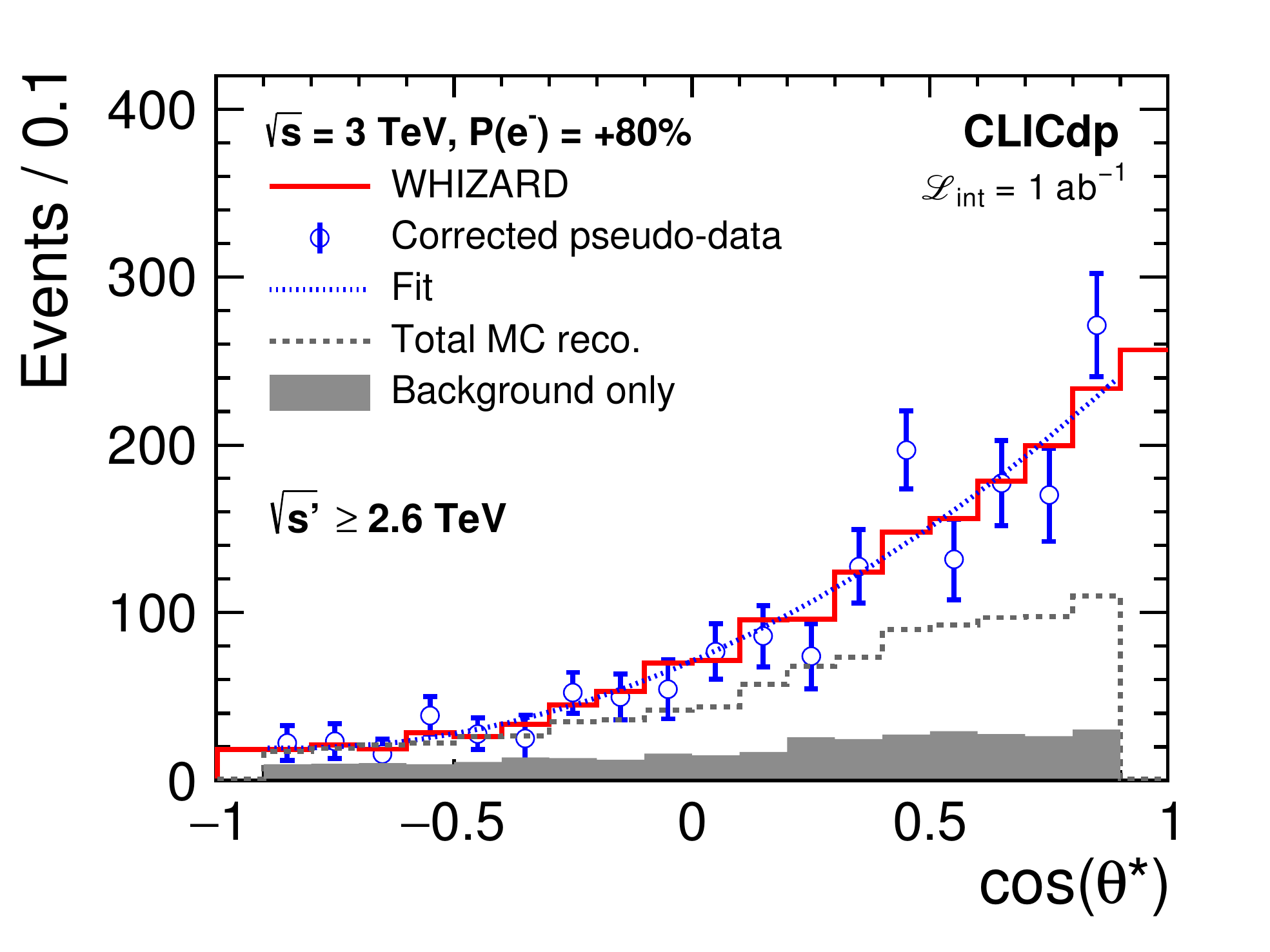}
  \caption{Top-quark polar angle distributions from the analysis of boosted semi-leptonic $\ttbar$ events, at a nominal collision energy of 3\,\tev for $P(\Pem)=\text{-}80\%$ (left) and $P(\Pem)=\text{+}80\%$ (right), and an integrated luminosity of $4.0\,\abinv$ and $1.0\,\abinv$, respectively. See further details in \autoref{fig:ttbar:boosted:costheta}. \label{fig:ttbar:boosted:costheta:polpos}}
\end{figure}

The remaining events are analysed using multivariate classification algorithms based on BDTs. In light of the large variety of the different backgrounds considered, two initial MVAs are trained focussing on slightly different topologies. The first MVA is trained using backgrounds with two quarks and either 0, 1, or 2 leptons, while the second MVA focuses on fully-hadronic four-quark and six-quark jet topologies. The final MVA considers all relevant backgrounds and includes the score from the two initial MVAs. Each MVA is trained on the 20 most important variables and the parameters of the algorithm are tuned to reduce overtraining. In addition to the scores from the initial MVAs, the most important variables include those derived from the kinematics of both the hadronically and leptonically decaying top quark (including the detailed output from the top-tagger), event missing $\pT$, visible energy and event shape, lepton kinematics, flavour tagging information, jet splitting scales, and substructure variables such as N-subjettiness discussed in \autoref{ssec:pairprodradiative}. Separate BDTs are applied for the 1.4\,\tev and 3\,\tev samples and for the two different polarisations considered. The cut applied on the classification score is chosen to minimise the statistical uncertainty on the two extracted observables $\afb$ and $\csttbar$, which are defined in \autoref{eq:totcs} and \autoref{eq:afb}.

The fractional event selection efficiencies for the signal and dominant background processes along with the total number of events selected are shown in \autoref{tab:ttbar:boosted:selection1400} and \autoref{tab:ttbar:boosted:selection3000} for the samples at nominal collision energies of 1.4 and 3\,TeV. Note that the cross sections quoted include the effect of ISR and the CLIC luminosity spectrum. The signal samples are further defined in the kinematic regions $\rootsprime\geq1.2\,\tev$ and $\rootsprime\geq2.6\,\tev$, respectively. Contributions from other backgrounds such as $\epem\to\PQq\PQq\PGn\PGn$ and additional six-fermion processes are found to be negligibly small.

The polar-angle distributions of the hadronically decaying top-quark candidates are shown in \autoref{fig:ttbar:boosted:costheta} and \autoref{fig:ttbar:boosted:costheta:polpos}. See \autoref{ssec:pairprodradiative} for a full description of the different distributions shown.
The selection efficiency in the region $-0.7\leq\cos\theta^{*}\leq0.7$ is generally flat with a central value of about 50\% for both analyses, at $\roots=1.4\,\tev$ and $\roots=3.0\,\tev$. In the forward regions the efficiency drops to 30\%.

%===================================================================================================
%===================================================================================================
\subsection{Cross section and asymmetry measurements}
\label{ssec:pairprodresults}

The total production cross section $\csttbar$\footnote{Note that the extracted cross sections represent a convolution of $\csttbar$ with the luminosity spectrum in a range of effective collision energies.} and forward-backward asymmetry $\afb$ are extracted from the polar-angle distribution in each analysis. \autoref{eq:AFBFit} is assumed to correctly describe the shape of the distributions in the full range, $-1.0\leq\cos(\theta^{*})\leq1.0$, and is fitted in the fiducial region $-0.9\leq\cos(\theta^{*})\leq0.9$, motivated by the limited acceptance in the very forward region. The resulting parameters $\sigma_{1,2,3}$ are used to extract the observables, $\csttbar$ and $\afb$, in the full range, through \autoref{eq:totcs} and \autoref{eq:afb}.

The polar-angle distributions are fitted after background subtraction and correction for finite selection efficiencies. For the analysis at $\roots=380\,\gev$, the overall shape of the polar-angle distribution is restored by the quality ($D^2$) cut, and an overall constant factor is used to correct for the selection efficiency. For the analyses at higher centre-of-mass energy the limited acceptance of the event selection in the forward region significantly distorts the reconstructed polar-angle distributions as seen for example in \autoref{fig:ttbar:boosted:costheta}. To compensate for the selection efficiencies, an efficiency correction estimated bin-by-bin is applied. To avoid a bias from statistical fluctuations it is estimated using half of the available sample and applied to the other half, and vice versa. This procedure assumes that the MC correctly describes the selection efficiency in the polar-angle distribution. Note further that the extracted observables, by construction, reproduce the corresponding generator level results, for the full range $-1.0\leq\cos(\theta^{*})\leq1.0$, up to statistical fluctuations introduced by the procedure described above.

\begin{table}[b!]
\begin{minipage}{\columnwidth}
\begin{center}
\begin{tabular}{l|cccccc}
\toprule
$\roots$ & \multicolumn{2}{c}{380\,\gev\footnote{Results from \autoref{ssec:pairprod380gev}}} & \multicolumn{2}{c}{1.4\,\tev\footnote{Results from \autoref{ssec:pairprodboosted}\label{footnote:ttbarresultstab}}} & \multicolumn{2}{c}{3\,\tev~\footref{footnote:ttbarresultstab}} \rule{0pt}{3ex} \\
P(\Pem) & -80\% & +80\% & -80\% & +80\% & -80\% & +80\% \rule{0pt}{3ex} \\
\midrule
$\csttbar$\footnote{Convolution of $\csttbar$ with the CLIC luminosity spectrum in the kinematic region studied.
}
~[fb] & 161.00 & 75.97 & 18.44 & 9.84 & 3.52 & 1.91 \rule{0pt}{3ex} \\
stat. unc.~[fb] & 0.77 & 0.52 & 0.21 & 0.29 & 0.07 & 0.09 \rule{0pt}{3ex} \\
\midrule
\afb & 0.1761 & 0.2065 & 0.567 & 0.620 & 0.596 & 0.645 \rule{0pt}{3ex}\\
stat. unc. & 0.0067 & 0.0059 & 0.008 & 0.020 & 0.014 & 0.034 \rule{0pt}{3ex}\\
\bottomrule
\end{tabular}
\end{center}
\caption{Results from the analysis of semi-leptonically decaying top quarks at the three stages of CLIC. The values are obtained from full simulation studies using the \clicild detector concept. Note that the cross section, $\csttbar$, and \afb are defined in the kinematic region of $\rootsprime\geq1.2\,(2.6)\,\tev$ for operation at $\roots=1.4\,\tev\,(3\,\tev)$. For operation at $\roots=380\,\gev$ the $\afb$ for $P(\Pem)$ = -80\% is extracted using the event sample defined by $D^2<1$; the other results at $\roots=380\,\gev$ are obtained using the sample with a looser selection cut, $D^2<15$. \label{tab:ttbar:results}}
\end{minipage}
\end{table}

The results from the analyses discussed in \autoref{sec:pairprod} are summarised in \autoref{tab:ttbar:results} and \autoref{tab:ttbar:radiative:results}. The tables show the reconstructed quantities $\csttbar$ and \afb. Results at 1.4\tev, for the region $\rootsprime\geq1.2\,\tev$, were computed for each of the two analyses presented in \autoref{ssec:pairprodradiative} and \autoref{ssec:pairprodboosted}, respectively. These show good agreement taking into account the different event selection efficiencies, and serve as a useful cross-check.

The results presented in this section are the first studies of $\ttbar$ production in full simulation for a multi-TeV $\epem$ collider. Further improvements can be made for example by including fully-hadronic final states, or semi-leptonic tau events where the tau decays leptonically. However, for the former, the jet charge reconstruction, needed for the reconstruction of observables such as the \afb, is challenging and needs to be studied in more detail. The BSM reach of these results is illustrated in \autoref{sec:phenom_interp} where the sensitivity of top-philic operators is presented.

\begin{table}
\begin{minipage}{\columnwidth}
\begin{center}
\begin{tabular}{l|cccccc}
\toprule
{} & \multicolumn{2}{c}{$\rootsprime\in[400,900)\,\tev$} & \multicolumn{2}{c}{$\rootsprime\in[900,1200)\,\tev$} & \multicolumn{2}{c}{$\rootsprime\geq1200\,\mathrm{GeV}$} \rule{0pt}{3ex} \\
P(\Pem) & -80\% & +80\% & -80\% & +80\% \rule{0pt}{3ex} & -80\% & +80\% \rule{0pt}{3ex} \\
\midrule
$\csttbar$\footnote{Convolution of $\csttbar$ with the CLIC luminosity spectrum in the kinematic region studied.}~[fb] & 16.56 & 8.63 & 11.01 & 5.87 \rule{0pt}{3ex} & 18.41 & 9.84 \rule{0pt}{3ex} \\
stat. unc.~[fb] & 0.73 & 0.88 & 0.23 & 0.36 \rule{0pt}{3ex} & 0.22 & 0.33 \rule{0pt}{3ex} \\
\midrule
\afb & 0.458 & 0.514 & 0.546 & 0.588 \rule{0pt}{3ex} & 0.562 & 0.621 \rule{0pt}{3ex}\\
stat. unc. & 0.050 & 0.127 & 0.021 & 0.055 \rule{0pt}{3ex} & 0.011 & 0.029 \rule{0pt}{3ex}\\
\bottomrule
\end{tabular}
\end{center}
\caption{Results for radiative events at $\roots=1.4\,\tev$ with an integrated luminosity of $2.0\,\abinv$ and $0.5\,\abinv$ for $P(\Pem)=\text{-}80\%$ and $P(\Pem)=\text{+}80\%$, respectively. Values are shown from three intervals of $\rootsprime$ below the nominal collision energy. \label{tab:ttbar:radiative:results}}
\end{minipage}
\end{table}

%===================================================================================================
\subsection{Systematic uncertainties}
\label{ssec:pairprodresults:sys}

The expected uncertainties given in \autoref{tab:ttbar:results} and \autoref{tab:ttbar:radiative:results} are purely statistical and do not include potential sources of systematic uncertainty. The results presented illustrate the level of precision desirable for the control of systematic effects. Although a full investigation of systematic uncertainties is beyond the scope of this paper, the impacts of some ad-hoc variations are discussed for the analysis of radiative events at $\roots=1.4\tev$, presented in \autoref{ssec:pairprodradiative}. This analysis is used as an example because the statistical uncertainties are generally lower at $\roots=1.4\tev$ than at $\roots=3\tev$. 
Further, events with signifiant energy loss due to ISR and beamstrahlung, as studied in this analysis, are potentially more likely to be sensitive to systematic uncertainties as background rejection is more challenging. In this sense, systematic effects estimated for the radiative event analysis can be considered as conservative estimates for the other analyses.
In each case studied, the effects on $\csttbar$ and $\afb$ are considered.

\begin{itemize}
\item The normalisation of the background is varied by $\pm$5\%. We consider only the dominant background processes: $\PQq\PQq\PQq\PQq\Pl\PGn$ (non-$\PQt\PQt$), $\PQq\PQq\Pl\PGn\Pl\PGn$, and $\PQq\PQq$. This leads to an effect on the cross section of around 1--3\%, and around 0.4--1.2\% for $\afb$;
\item The background shape modelling is studied by applying a linear gradient of $\pm$2\% to the shape of the total background in $\cos(\theta^*)$. This leads to an uncertainty of about 0.2--0.8\% on $\csttbar$ and 0.9--2.9\% on $\afb$;
\item To check for a possible bias in the event selection towards the generated $\afb$, the MC datasets are reweighted to different values of $\afb$ according to \autoref{eq:AFBFit} and \autoref{eq:afb}.  The relation between the reconstructed and generated values of $\afb$ is found to be linear and hence such an effect could be corrected for.
\end{itemize}

The analysis is found to be insensitive to the choice of fit range or MVA score cut value.
In addition, it is expected that the integrated luminosity will be known with an accuracy of a few per mille using the luminometer envisaged for CLIC \cite{Lukic:2013fw, Bozovic-Jelisavcic:2013aca}, and therefore does not represent a significant systematic uncertainty for this analysis.
In summary, the estimates presented above indicate that this analysis is not limited by systematic effects.

\section{Associated \texorpdfstring{\ttbar}{ttbar} production processes at high energy}

At the higher CLIC energy stages, top-quark pairs can be produced in additional processes beyond $\epem\to\PQt\PAQt$. The top Yukawa coupling can be directly obtained from the $\epem\to\ttH$ cross section. This process also allows a study of the CP properties of the Higgs boson in the $\PQt\PQt\PH$ coupling. The second CLIC stage at $\roots=1.5\,\TeV$ (or the previous baseline of 1.4\,\TeV as used here) is well suited for making these measurements. At $\roots=3\,\TeV$ the production of top-quark pairs in the VBF process $\epem\to\PQt\PAQt\PGne\PAGne$ can also be studied.

\subsection{Study of \texorpdfstring{\ttH}{ttH} production}
\label{sec:ttH}

Results from a first study of $\PQt\PAQt\PH$ production at CLIC and projections for the precision on the top Yukawa coupling were presented in~\cite{LCD:tth_1400, LCD:tth_backgrounds_1400, Abramowicz:2016zbo}. In the following, a refined version of this analysis is described. An improved version of the flavour tagging is used. The estimated precision on the top Yukawa coupling is based on NLO QCD calculations~\cite{Nejad:2016bci}, whereas the previous analysis used LO predictions. The sensitivity to CP mixing in the $\PQt\PQt\PH$ coupling is presented.

\paragraph{Cross section measurement and top Yukawa coupling}

The $\epem\to\PQt\PAQt\PH$ process has been studied for the \clicsid detector concept using $\PH\to\PQb\PAQb$ decays at $\roots=1.4\,\TeV$. The analysis focuses on fully-hadronic and semi-leptonic top-quark pair decays, which lead to final states with eight or six jets, respectively, including four $\PQb$-quark jets. The study assumes unpolarised beams and an integrated luminosity of 1.5\,\abinv. At the end of this section, we also give the expected uncertainty on the top Yukawa coupling for the baseline scenario described in \autoref{ssec:staging}.

The two channels are differentiated by the presence of an isolated electron, muon or tau lepton. If zero leptons are found, the event is classified as fully-hadronic. If one isolated lepton is found, the event is classified as semi-leptonic. Events where more than one isolated lepton is found are not considered further.
The longitudinally-invariant \kT algorithm with $R=1.0$ is used to cluster the particles of each event into a specific number of jets. Events classified as fully-hadronic are clustered into eight jets. In semi-leptonic events, the lepton candidate is removed and the remaining particles are clustered into six jets. The jets are then combined to form the $\PW$ boson, top-quark, and Higgs boson candidates. For example, in the case of the semi-leptonic channel, the jet assignment with the minimum value of
\begin{equation} \label{eq:ttH:chi2}
    \chi^{2} = \frac{(m_{ij}-m_{\PW})^{2}}{\sigma_{\PW}^{2}} + \frac{(m_{ijk} - m_{\PQt})^{2}}{\sigma_{\PQt}^{2}} + \frac{(m_{lm} - m_{\PH})^{2}}{\sigma_{\PH}^{2}} \,,
\end{equation}
gives the \PW boson, top-quark, and Higgs boson candidates, where $m_{ij}$ is the invariant mass of the jet pair 
used to reconstruct the $\PW$ candidate, $m_{ijk}$ is the invariant mass of the three jets used to 
reconstruct the hadronically decaying top-quark candidate, and $m_{lm}$ is the invariant mass of the jet pair used to reconstruct the 
Higgs boson candidate. The expected invariant mass resolutions $\sigma_{\PW, \PQt, \PH}$ have been estimated from
combinations of two or three reconstructed jets matched to the $\PW$ boson, top-quark, and Higgs boson decay products at hadron level. A similar construction is used for the fully-hadronic channel with additional terms corresponding to the second hadronically decaying top quark.

Multivariate BDT classifiers are used in the final step of the analysis to separate signal and background events. These are constructed  individually for the semi-leptonic and fully-hadronic event candidates. The classifiers were trained using variables related to flavour tagging and event kinematics, as well as variables derived following the pairing in \autoref{eq:ttH:chi2}: the reconstructed Higgs mass, the $\chi^2$, and angular separations between the event constituents. For the semi-leptonic channel we also include lepton variables, while the fully-hadronic channel considers additional jet variables. Cuts on the BDT classifier outputs are chosen to maximise the signal significances, estimated as $S/\sqrt{S+B}$, where $S\,(B)$ represent the signal (background) sample. 

The expected numbers of selected events for $1.5\,\abinv$ at $\roots=1.4\,\TeV$ are listed in \autoref{tab:tth:sel_eff}. The $\ttH$ cross section can be measured with a precision of $11.1\,\%$ in the semi-leptonic channel and $9.6\,\%$ in the hadronic channel. The combined precision of the two channels is $7.3\,\%$. Note that all $\ttH$ processes are considered as signal in the final calculations. 

The benchmark analyses described here use LO Monte Carlo samples. The K-factor defined as the ratio of the NLO to the LO cross section is 0.938~\cite{ReuterRothe} including the effects of ISR and beamstrahlung. Scaling the projected precision to the NLO cross section leads to an uncertainty of $7.5\,\%$.

\begin{table}[htbp]
\centering
\begin{tabular}{lccc}
\toprule
Process & $N$ & \multicolumn{2}{c}{Selected as} \\
              &  & fully-hadronic & semi-leptonic \\
\midrule
$\epem\to\PQt \PAQt \PH$, 6 jet, $\PH \to \PQb \PAQb$ & 647 & 367 & 38  \\
$\epem\to\PQt \PAQt \PH$, 4 jet, $\PH \to \PQb \PAQb$ & 623 & 1 & 270  \\
\midrule
$\epem\to\PQt \PAQt \PH$, 2 jet, $\PH \to \PQb \PAQb$ & 150 & 2 & 22  \\
$\epem\to\PQt \PAQt \PH$, 6 jet, $\PH \not\to \PQb \PAQb$ & 473 & 54 & 11  \\
$\epem\to\PQt \PAQt \PH$, 4 jet, $\PH \not\to \PQb \PAQb$ & 455 & 8 & 22  \\
$\epem\to\PQt \PAQt \PH$, 2 jet, $\PH \not\to \PQb \PAQb$ & 110 & 0 & 1  \\
$\epem\to\PQt \PAQt \PQb \PAQb$, 6 jet & 824 & 326 & 26  \\
$\epem\to\PQt \PAQt \PQb \PAQb$, 4 jet & 794 & 57 & 226  \\
$\epem\to\PQt \PAQt \PQb \PAQb$, 2 jet & 191 & 2 & 18  \\
$\epem\to\PQt \PAQt \PZ$, 6 jet & 2,843 & 345 & 34  \\
$\epem\to\PQt \PAQt \PZ$, 4 jet & 2,738 & 59 & 217  \\
$\epem\to\PQt \PAQt \PZ$, 2 jet & 659 & 1 & 16  \\
$\epem\to\PQt \PAQt$ & 203,700 & 498 & 742  \\
\bottomrule
\end{tabular}
\caption{Expected numbers of signal and background events in the fully-hadronic and semi-leptonic channels 
for $1.5\,\abinv$ at $\roots=1.4\,\TeV$. The columns show the total numbers of events before selection and the numbers of events 
passing the fully-hadronic and semi-leptonic BDT selections. No preselection is applied in the analysis. \label{tab:tth:sel_eff}}
\end{table}

When extracting the top Yukawa coupling value from the $\ttH$ cross section, a small contribution from the Higgsstrahlung diagram, where the Higgs boson is radiated off the intermediate $\PZ$ boson, has to be taken into account~\cite{Djouadi:1992gp}.
The factor to translate the uncertainty of the $\PQt\PAQt\PH$ production cross section into an uncertainty on the top Yukawa coupling was calculated including NLO QCD corrections, ISR and beamstrahlung~\cite{ReuterRothe}:

\begin{equation*}
\frac{\Delta y_{\PQt}}{y_{\PQt}} = 0.503~\frac{\Delta \sigma}{\sigma} \, . \nonumber
\end{equation*}

Thus, the expected precision on the top Yukawa coupling is $\Delta y_{\PQt}/y_{\PQt} = 3.8\%$, for $1.5\,\abinv$ of data at $\roots=1.4\,\TeV$ without beam polarisation. The corresponding precision for 2\,\abinv of data with -80\% electron beam polarisation plus 0.5\,\abinv of data with +80\% electron beam polarisation is:

\begin{equation*}
\frac{\Delta y_{\PQt}}{y_{\PQt}} = 2.7\%. \nonumber
\end{equation*}

It was recently demonstrated that an even better statistical precision on $y_{\PQt}$ can be achieved indirectly using loop contributions to decays such as $\PH\to\Pg\Pg$~\cite{Boselli:2018zxr}. However, this approach implies additional model dependence compared to the direct extraction described here.

\paragraph{CP mixing in the \texorpdfstring{\ttH}{ttH} coupling}

\begin{figure}
\begin{center}
\includegraphics[width=0.48\columnwidth]{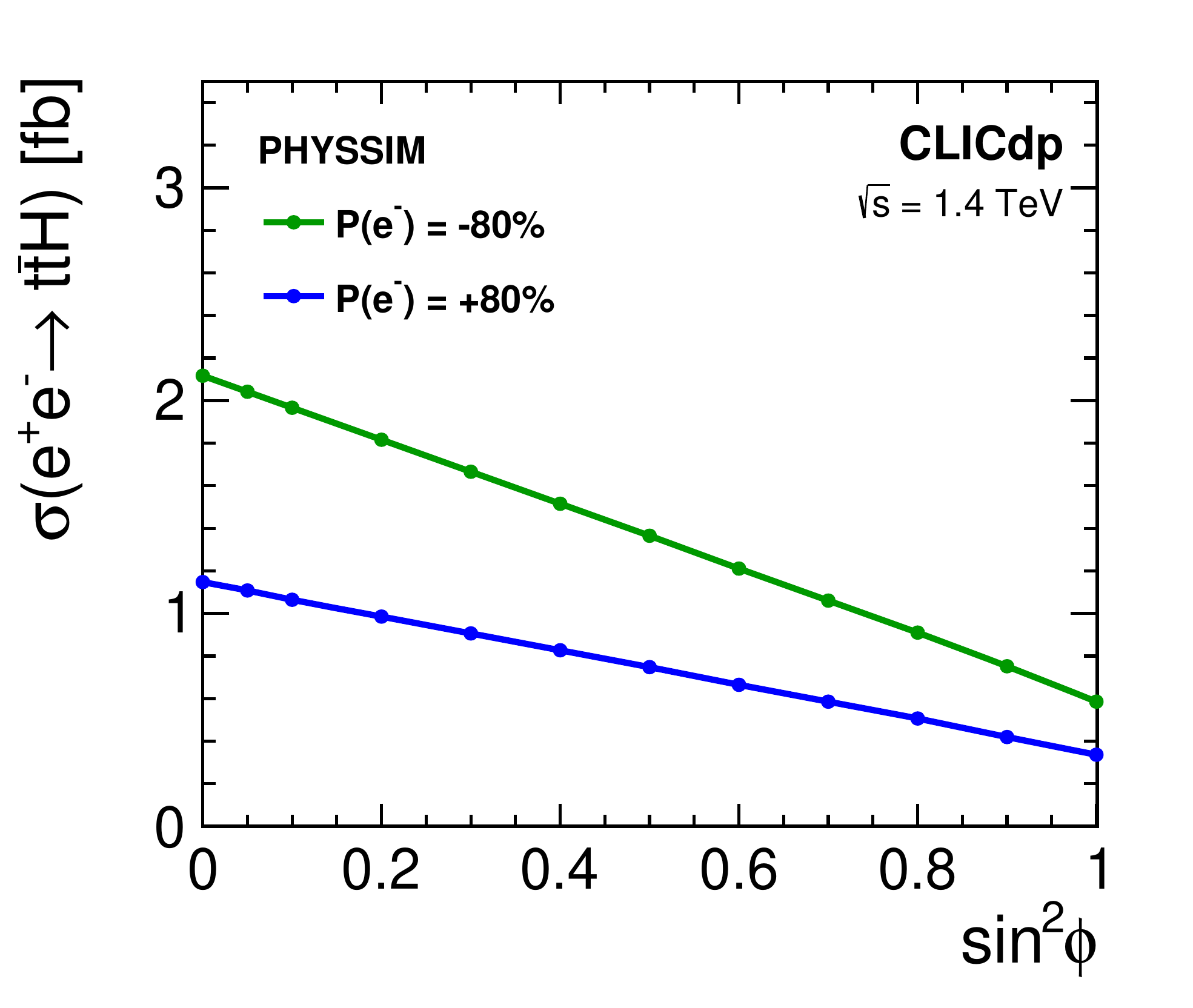}
~~~~
\includegraphics[width=0.48\columnwidth]{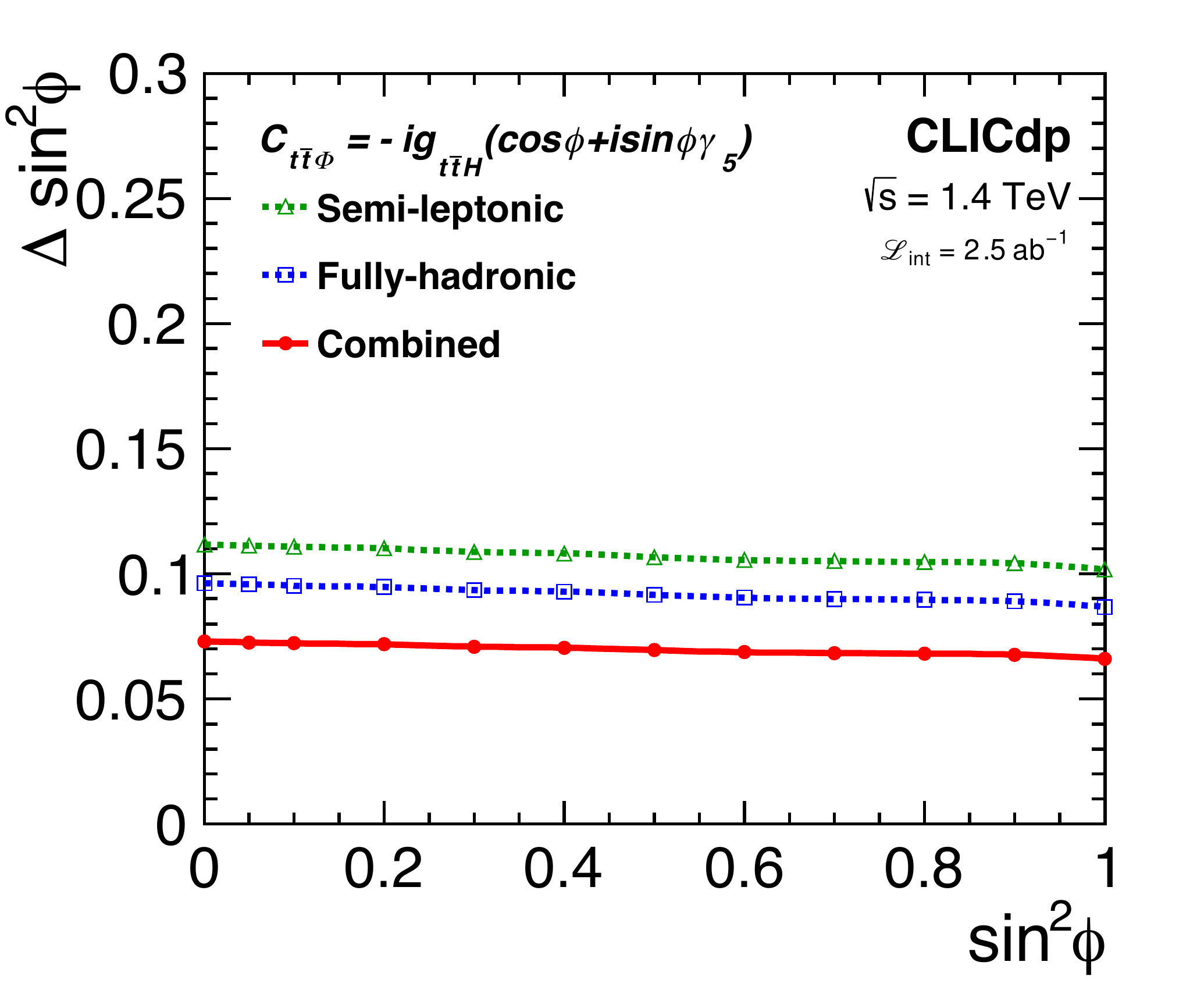}
\caption{Cross section for the process $\epem\to\PQt\PAQt\PH$ at $\roots = 1.4\,\tev$ including the effects of ISR and beamstrahlung as a function of $\sin^{2}{\phi}$ (left). Sensitivity to the CP mixing angle $\sin^{2}{\phi}$ as a function of $\sin^{2}{\phi}$ from the $\PQt\PAQt\PH$ cross section at $\roots=1.4\,\TeV$, assuming an integrated luminosity of 2\,\abinv with -80\% electron beam polarisation plus 0.5\,\abinv with +80\% electron beam polarisation (right). The sensitivities for both considered final states are shown separately in addition to the combined projection.\label{fig:tth:sin2phi}}
\end{center}
\end{figure}

The measurement of the $\PQt\PAQt\PH$ cross section can be used to search for a CP-odd contribution to the $\PQt\PQt\PH$ coupling. Here, CP mixing can be parameterised as
\begin{equation*}
-i g_{\PQt\PQt\PH} (\cos{\phi} + i\,\sin{\phi}\,\PGg_{5}),
\end{equation*}
where $\phi$ denotes the mixing angle ($\phi=0$ for the SM case). Note that $\cos^{2}{\phi} + \sin^{2}{\phi} = 1$ is assumed. The SM is given by $\sin^{2}{\phi} = 0$ while $\sin^{2}{\phi} = 1$ corresponds to a pure CP-odd coupling. The dependence of the $\PQt\PAQt\PH$ production cross section as a function of $\sin^{2}{\phi}$ is shown in \autoref{fig:tth:sin2phi} \cite{gen:physsim}. The cross section decreases linearly with increasing $\sin^{2}{\phi}$. Similarly to the approach used for the extraction of the top Yukawa coupling discussed above, the uncertainty on the cross section can be translated into an uncertainty on $\sin^{2}{\phi}$.

Signal event samples were generated assuming different values of $\sin^{2}{\phi}$ for the semi-leptonic and fully-hadronic final states. The semi-leptonic analysis described for the cross section measurement above was repeated for each $\sin^{2}{\phi}$ value, while for the fully-hadronic analysis the cross section uncertainty for the SM assumption was extrapolated.

The expected precision as a function of $\sin^{2}{\phi}$ is shown in \autoref{fig:tth:sin2phi}. Although the cross section decreases by about a factor 3.5 for a pure CP-odd coupling compared with the SM, the  expected uncertainty on $\sin^{2}{\phi}$ of about 0.07, for 2\,\abinv of data with -80\% electron beam polarisation plus 0.5\,\abinv of data with +80\% electron beam polarisation at $\roots=1.4\,\TeV$, is almost independent of $\sin^{2}{\phi}$. This precision can be improved further using additional information provided by differential distributions~\cite{Godbole:2011hw}.

\subsection{Vector boson fusion production}
\label{sec:WWtt}

The high-energy stages of CLIC allow the study of top-quark pair production initiated by low-virtuality (nearly on-shell) and highly energetic vector bosons, in the so-called vector boson fusion topology; see for example \autoref{fig:production_diagram:vbf}. This production mode is particularly interesting because it gives direct access to on-shell $\PWp\PWm\ttbar$ production, which might reveal large BSM effects. A particularly important role in this context is played by VBF production initiated by longitudinally polarised vector bosons, which are effectively equivalent to the Higgs field at high energy, owing to the Equivalence Theorem. In several new physics scenarios aimed at addressing the Naturalness problem, the Higgs boson and the top-quark interactions are largely modified, hence it is natural to study processes that are directly sensitive to such interactions. A specific example is the enhancement of ``Higgs current'' type operators (namely $Q_{\varphi{\PQt}}$, $Q_{\varphi{\PQq}}^{(1)}$, $Q_{\varphi{\PQq}}^{(3)}$, in \autoref{tab:operators}) in the top-quark compositeness scenario \cite{Pomarol:2008bh} presented further in \autoref{sec:phenom_interp}.

In this section we present a parton-level study of the CLIC sensitivity to EFT operators in VBF top-quark pair production focusing on the $\PWp\PWm\rightarrow \ttbar$ process and the four operators $Q_{\varphi{\PQt}}$, $Q_{\varphi{\PQq}}^{(1)}$, $Q_{\varphi{\PQq}}^{(3)}$ and $Q_{\PQt\PW}$. The reason for this choice is that these operators are the only ones, out of the nine in \autoref{tab:operators}, that give contributions to $\PWp\PWm\rightarrow\ttbar$ with an effect that grows quadratically with energy. Consequently these are the ones that are best probed by this channel. 

The process $\Pep\Pem\rightarrow\ttbar\PGn\PAGn$ was simulated at tree-level using {\sc{MadGraph}} \cite{Alwall:2014hca} and the EFT UFO implemenation described in~\cite{Alloul:2013naa}. \whizard \cite{Kilian:2007gr} was used for the $\Pep\Pem\rightarrow\ttbar$ process in order to include the effects from beamsstrahlung and ISR. The study was performed assuming unpolarised beams. For the baseline scenario described in \autoref{ssec:staging}, a somewhat better result is expected as the signal cross section is more enhanced for the -80\% electron polarisation configuration compared to the main background from top-quark pair production.

\begin{figure}
\centering
\includegraphics[width=0.48\columnwidth]{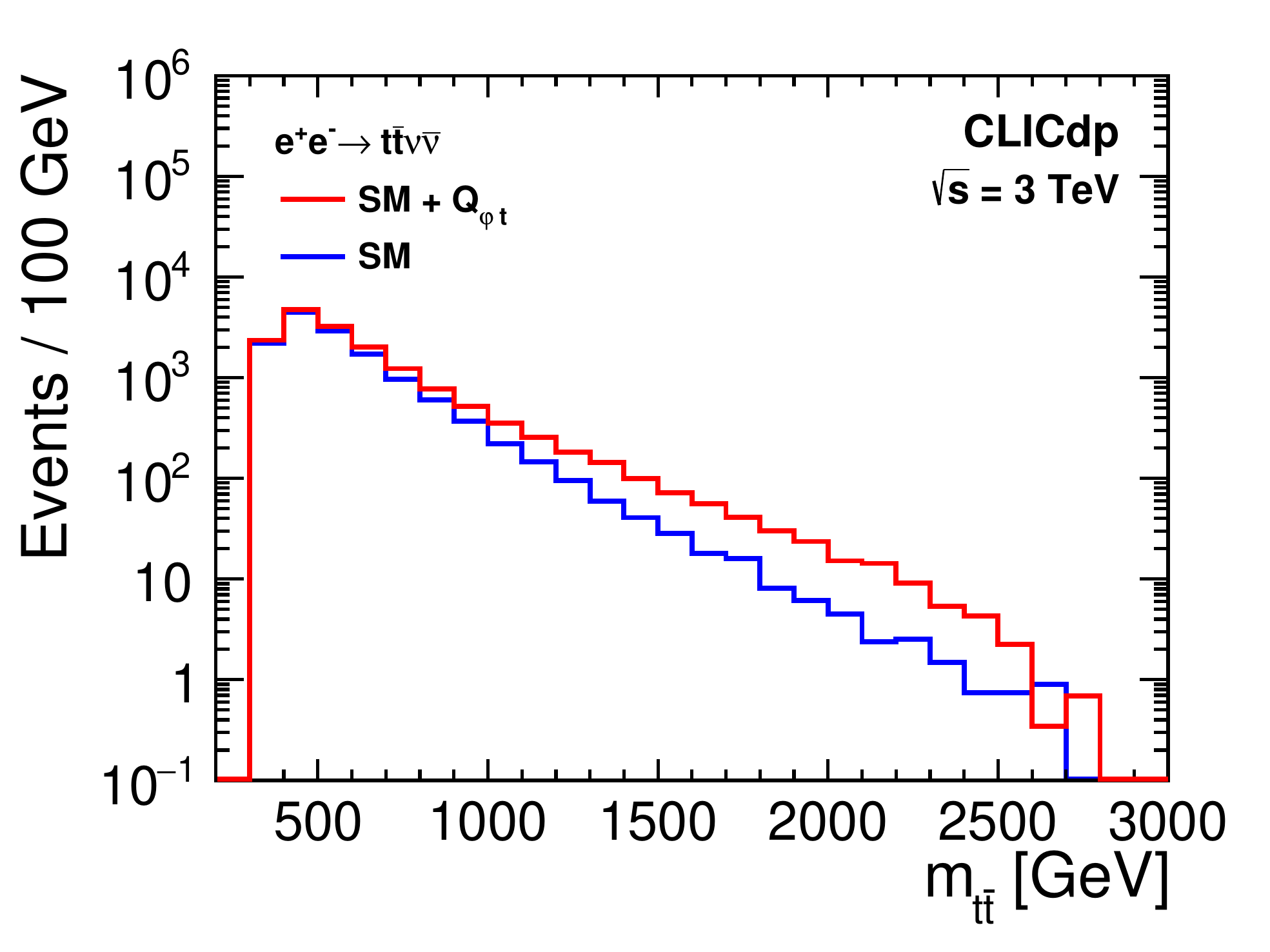}
~~~~
\includegraphics[width=0.48\columnwidth]{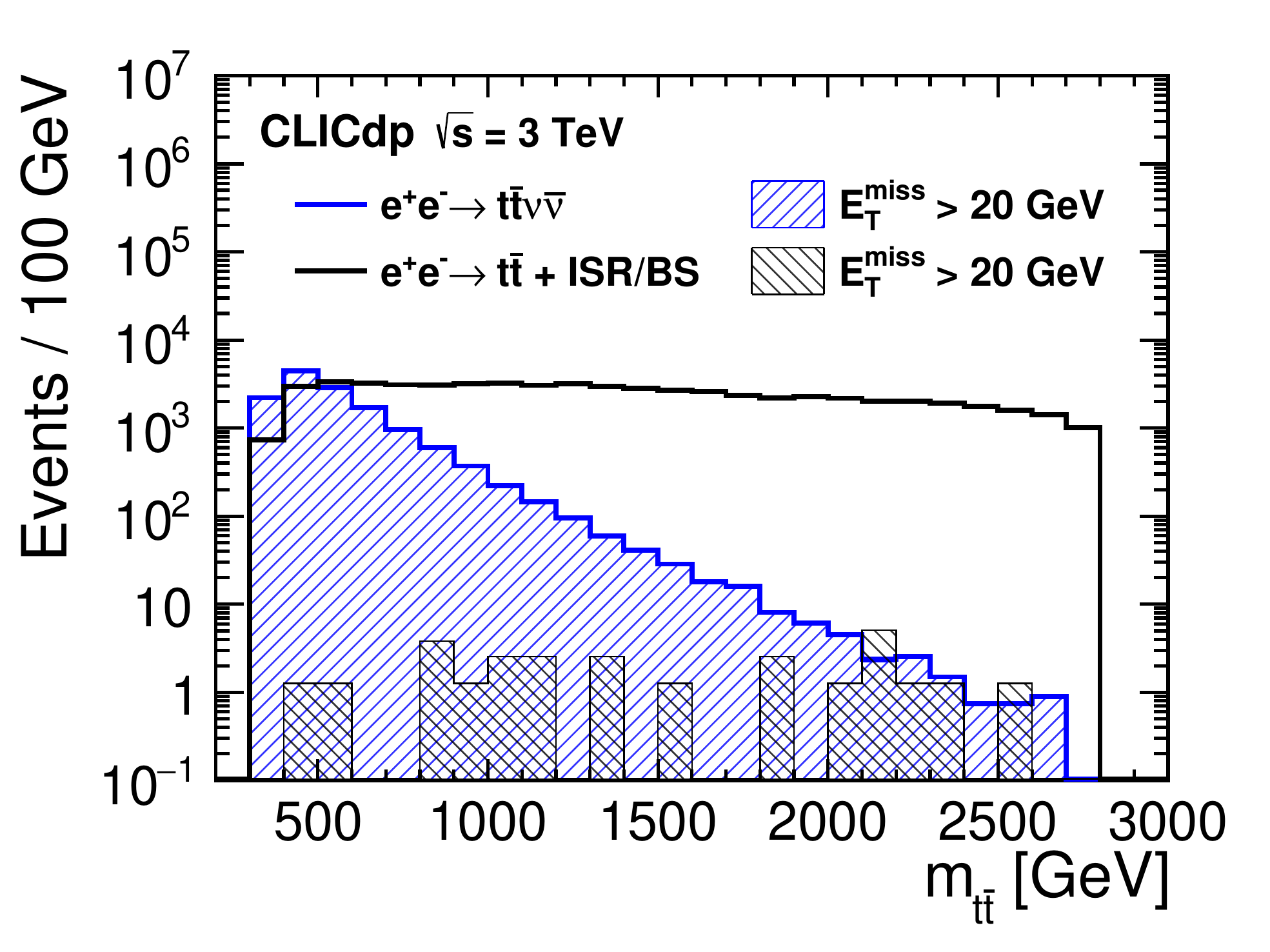}
\caption{Top-quark pair invariant mass distributions for CLIC at $\roots=3\,\tev$ with $5\,\invab$. $\Pep\Pem\rightarrow\ttbar\PAGn\PGn$ (VBF) after a cut on the invariant mass of the two neutrinos, illustrating the effect of including one example EFT operator, $Q_{\varphi{\PQt}}$, that grows with energy (left). VBF signal distribution and the $\epem\rightarrow\ttbar$ SM background process before and after a cut on the total missing transverse energy, $E_{\mathrm{T}}^{\mathrm{miss}}>20\,\gev$ (right). Note that $\epem\rightarrow\ttbar$ includes both ISR and beamstrahlung (BS) effects. \label{fig:VBF:ttdist}}
\end{figure}

The starting point for the analysis is to isolate the VBF topology in the complete $2\rightarrow 4$ process $\Pep\Pem\rightarrow\ttbar\PGn\PAGn$. This is achieved by reconstructing the invariant mass of the two neutrinos from the final state top quarks and the initial state momenta, and requiring this to be above $200\,~\gev$. The main role of this cut is to suppress $\ttbar\PZ$ production with the $\PZ$ decaying to neutrinos, which constitutes a significant fraction of the total $\ttbar\PGn\PAGn$ cross section. \autoref{fig:VBF:ttdist} shows the \ttbar invariant mass distribution at $\roots=3\,\TeV$ for an integrated luminosity of $5\,\invab$. The left plot shows the distribution after the cut on the invariant mass of the two neutrinos, for the SM case and for a BSM scenario including one example EFT operator, $Q_{\varphi{\PQt}}$ ($C_{\varphi{\PQt}}=-0.41\,\TeV^{-2}$), that grows with energy. In addition, acceptance cuts $p_T>20$~\,GeV and $|\cos\theta|\leq0.9$ are applied to the top-quark candidates. While a perfect top-quark reconstruction efficiency is assumed in this fiducial region for the figure, a $70\%$ efficiency, compatible with the results of \autoref{sec:boosted}, is included in the fit. A potentially significant background originates from $\epem\rightarrow\ttbar$, if the nominal 3\,TeV collision energy is strongly reduced by ISR or by beamstrahlung effects. However, a loose requirement on the total missing transverse energy, $E_T^{\rm{miss}}>20$\,GeV, is sufficient to reduce this background to a negligible level, as shown to the right of \autoref{fig:VBF:ttdist}. At the same time, the cut has a negligible effect on the VBF signal.

\begin{figure}[t!]
\begin{center}
\includegraphics[scale=1.0]{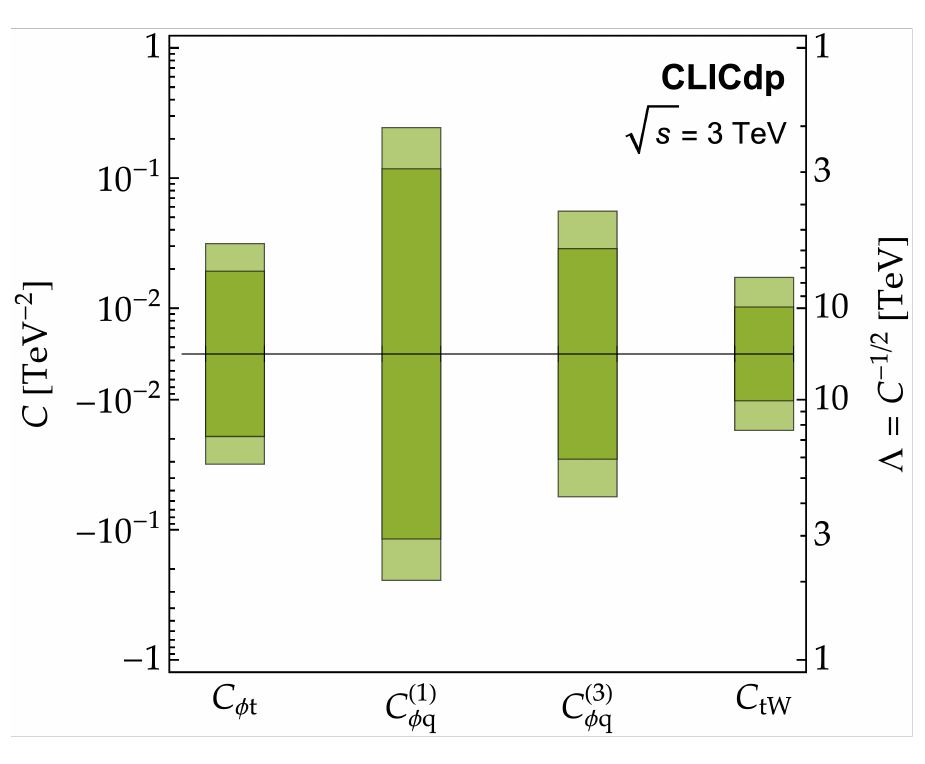}
\caption{Single-operator sensitivity (68\%) from vector boson fusion top-quark pair production at the 3\,TeV CLIC. Both upper and lower limits are reported simultaneously on each side of the horizontal line. The stronger limits (dark green) assume perfect top-quark reconstruction and no systematic uncertainties; the weaker ones (light green) are derived with $50\%$ $\ttbar$ reconstruction efficiency and $3\%$ systematics. \label{fig:wwttreach}}
\end{center}
\end{figure}

The sensitivity to the EFT operators is estimated by performing a doubly-differential binned likelihood fit to the $\ttbar$ invariant mass and to the cosine of the $\ttbar$ scattering angle in the centre-of-mass frame. The latter variable improves the sensitivity because the contribution of the EFT operators has a different angular dependence than that of the SM. Uncorrelated $3\%$ systematic relative uncertainties are assumed, and summed in quadrature with the statistical error. The $1\sigma$ sensitivity to the four EFT operator coefficients is displayed in \autoref{fig:wwttreach}. Note that the results reported in the figure are single-operator sensitivities, obtained by including only one operator at a time. The result should thus be interpreted with care, bearing in mind that cancellations are possible for certain combinations of operators, resulting in insensitive flat directions of the $4$-dimensional operator space. 

In conclusion, the result of this simplified analysis is that VBF top-quark pair production is a promising EFT probe that merits further study. A more detailed investigation, based on more realistic simulations and aimed at a more in-depth assessment of the impact of this channel in the CLIC EFT fit, is currently being performed and will be presented elsewhere.

\section{Flavour-changing neutral current top-quark decays}
\label{sec:fcnc}

The experimental sensitivity to rare top-quark decays is determined by
the expected number of \ttbar pairs produced, the efficiency of the rare decay
reconstruction, and the effectiveness of the background suppression.
As the cross section for top-quark pair production at higher-energy
stages drops significantly (see \autoref{fig:cross_sections_ttbar}),
we focus on the measurement of the FCNC top-quark decays at $\roots=380\,\gev$.
We also assume that data samples collected with -80\% and +80\%  electron beam
polarisations, corresponding to the integrated luminosity of 500\,fb$^{-1}$ each,
are combined for the analysis. 

We study FCNC couplings involving the charm quark, as many BSM models
enhance these channels~\cite{Agashe:2013hma}. 
Channels involving the charm quark can be well reconstructed at CLIC thanks
to good c-tagging capabilities, while the expected limits at the
HL-LHC, when based on the searches for single top production, are expected
to be significantly weaker than for the corresponsing channels involving the up quark.
For the \tcg decay, the measurement of the cross section for single top
production in association with a photon at the HL-LHC can be translated to:
\begin{eqnarray*}
\text{BR}(\tcg) & < & 7.4 \times 10^{-5},
\end{eqnarray*}
at 95\% C.L. assuming $3\,\abinv$ collected at 14\,TeV~\cite{Collaboration:2293646}.
If the c-quark dominates $\tqh$ decays, the following 95\% C.L. limit is expected
using top-quark pair production events at the HL-LHC~\cite{ATL-PHYS-PUB-2016-019}:
\begin{eqnarray*}
\text{BR}(\tch) & < & 2 \times 10^{-4}.
\end{eqnarray*}

\subsection{\texorpdfstring{\PQt$\rightarrow\PQc\PGg$}{t$\rightarrow$c$\gamma$}}

\begin{figure}
  \centering
  \includegraphics[width=0.65\columnwidth]{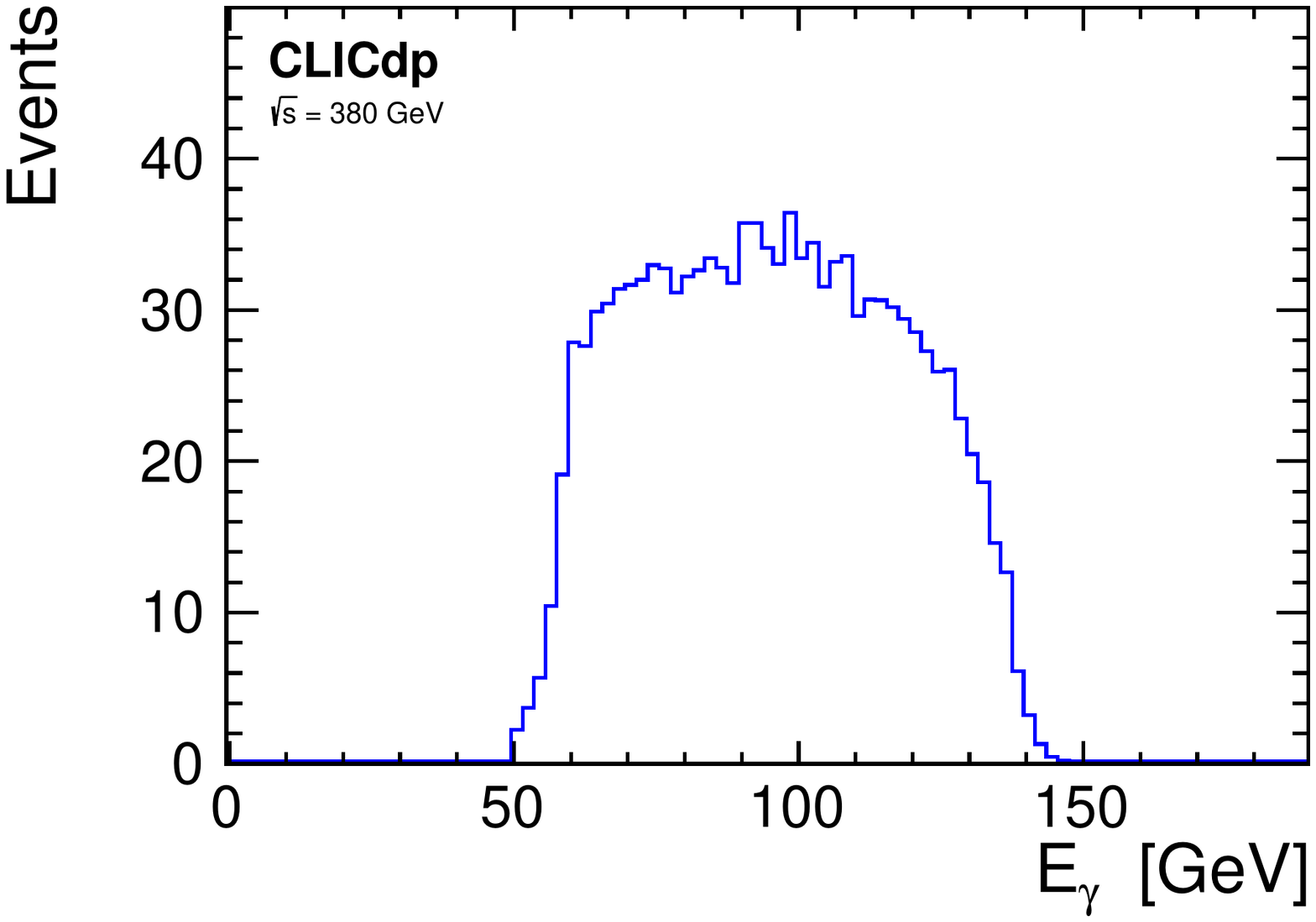}
  \caption{Distribution of the reconstructed photon energy for events with 
    FCNC top-quark decay  \tcg  at  380~GeV
    CLIC. The signal sample is normalised to 1.0\,ab$^{-1}$ and
    BR(\tcg) $=10^{-3}$. 
\label{fig:fcnc_cgamma_egamma}}
\end{figure}

Top-quark pair production events for the signal sample were generated
with \whizard 2.2.8~\cite{Kilian:2007gr,Moretti:2001zz} using a model
with anomalous top-quark couplings (\verb|SM_top_anom|).
The vector coefficient of the tensor $\PQt\PQc\PGg$ coupling was tuned to
obtain BR(\tcg) $=10^{-3}$. This ensures that the contribution from
the FCNC decay to the total top-quark width as well as the possibility
of having two FCNC decays in the same event are negligible.
The same BR value is also used when comparing signal and
background event distributions. This choice is arbitrary and has no
influence on the expected exclusion limits.
Either the top or the anti-top quark (referred to as the ``signal top
quark'' in the following) decays via the FCNC channel, 
and the other (denoted the ``spectator top quark'') via the standard
hadronic or leptonic decay. 
The FCNC decay channel $\PQt\rightarrow\PQc\PGg$ is 
characterised by the presence of a high-energy photon, with an energy 
between about 50\,GeV and 140\,GeV, resulting from the decay kinematics
at $\roots=380\,\gev$ (see \autoref{fig:fcnc_cgamma_egamma}). 
This gives a very clear signature, allowing for efficient separation
of signal events from possible backgrounds.
The analysis only considers the fully-hadronic decay channel where the
spectator top quark decays into a $\PQb$-quark and a $\PW$ boson, the
latter decaying hadronically.
With the signal top quark decaying to a $\PQc$-quark and a photon,
the target events should contain a high-energy photon and a $\PQc$-quark jet
as well as one $\PQb$-quark jet and two jets from the
$\PW$ decay of the spectator top quark.
The background sample considered consists of \epem events
compatible with top-quark pair production (6-fermion sample),
four-fermion production events (dominated by $\PWp\PWm$ contribution)
and pair production of quarks other than the top quark.

\begin{figure}
  \centering
  \includegraphics[width=0.65\columnwidth]{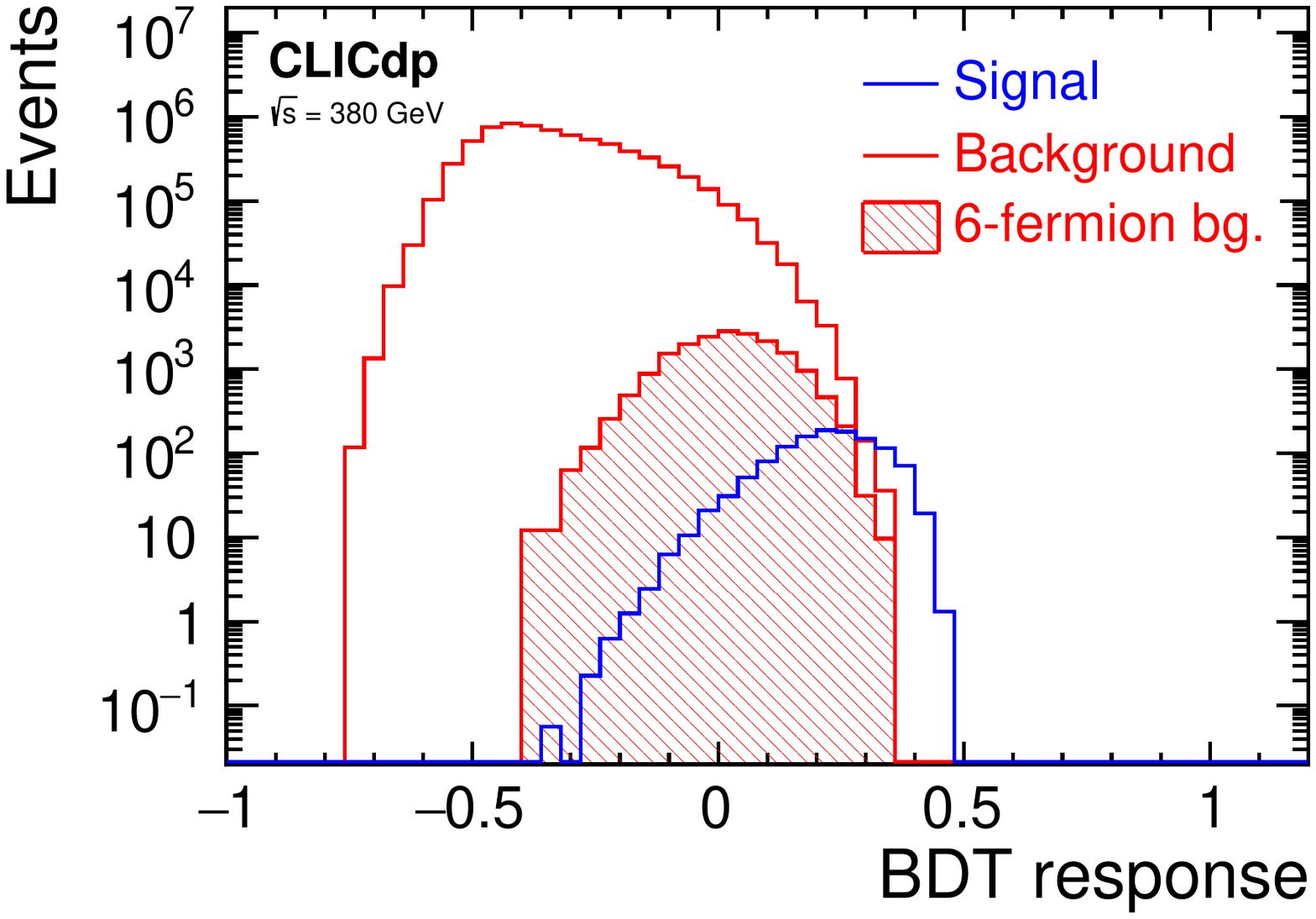}
  \caption{Distribution of the BDT classifier response for events with 
    FCNC top-quark decay  \tcg (signal, blue histogram)
    and SM events (background, red histogram), for FCNC selection at  380~GeV
    CLIC. The background sample is normalised to 1.0\,ab$^{-1}$ while the
    signal (events with FCNC decay) is normalised to
    BR(\tcg) $=10^{-3}$. The hatched histogram indicates the
    contribution from the six-fermion background.
\label{fig:fcnc_cgamma_bdt}}
\end{figure}

\begin{figure}
  \centering
  \includegraphics[width=0.65\columnwidth]{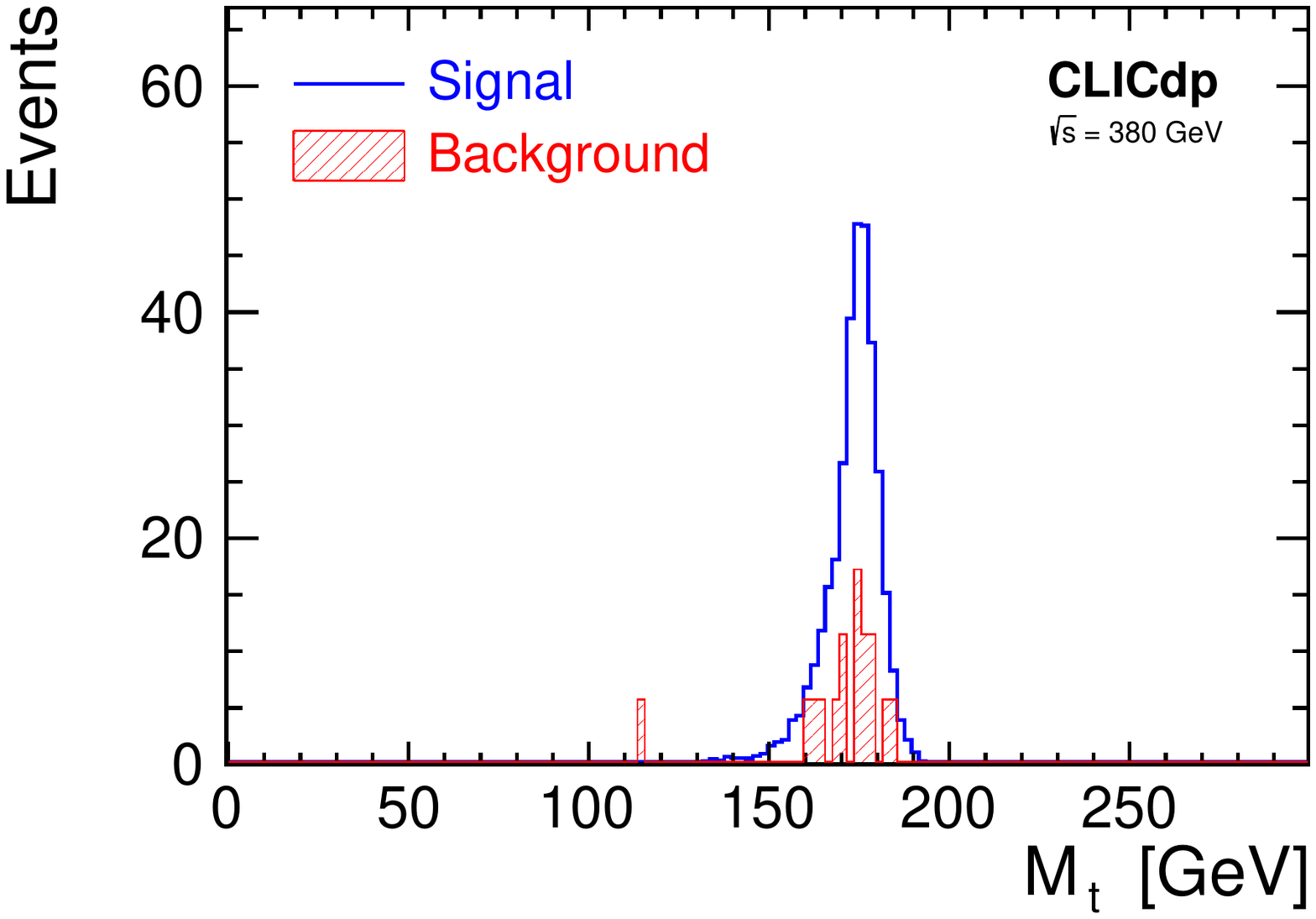}
  \caption{Invariant mass distribution of the top quark from the
    FCNC decay \tcg reconstructed at 380~GeV CLIC
    after selection based on the BDT response.
    The distribution is normalised to 1.0\,ab$^{-1}$ and
    BR(\tcg) $=10^{-3}$ for the signal events.
    The hatched histogram  indicates the SM background contribution. 
\label{fig:fcnc_cgamma_mtop}}
\end{figure}

The analysis uses a relatively loose event pre-selection
based on the requirement of an isolated photon with at least 50\,GeV of
energy.
This requirement reduces the background from standard
\ttbar decays by a factor of 20 while keeping 92\% of the signal
events.
Contributions from four-fermion and quark-pair background events are
reduced by factors of about 6 and 4, respectively.
For the selected events, a reconstruction of the event kinematics is
performed for a signal ($\PGg$ + 4 jets) and a background (6 jet)
hypothesis.
Jets are reconstructed using the VLC algorithm in exclusive mode with
a radius of 1.6 and $\beta=\gamma=0.8$. For each hypothesis, all the
possible jet combinations are considered, and the configuration that
minimises a $\chi^2$ value for the event is chosen. 
The $\chi^2$ formula includes constraints on the invariant masses of
the two reconstructed top-quark candidates and one (for signal
hypothesis) or two (for background hypothesis) reconstructed $\PW$
bosons.  
The discrimination of background events from signal events is based
on a multivariate classifier analysis using a BDT approach with 42 input
variables.
The variables giving the largest impact on the classifier response
include the photon properties, reconstructed invariant mass of the
signal top quark, reconstructed $\PQb$ and $\PQc$ jet energies and the
total energy of the event, flavour tagging results and the ratio of
$\chi^2$ values for the signal and background hypotheses. 
The resulting distributions of the BDT classifier response for the 
signal and background (SM top-quark decays plus other SM processes) samples are
shown in \autoref{fig:fcnc_cgamma_bdt}.
Although there is a significant overlap of the two distributions,
an almost clean sample of signal events can be selected by imposing a tight cut
on the BDT response.
This is illustrated in  \autoref{fig:fcnc_cgamma_mtop},
where the reconstructed invariant mass distribution for the signal top-quark decays
is shown after imposing a BDT response cut, BDT score $>0.29$
(corresponding to the highest signal significance for the test
scenario with BR(\tcg) $=10^{-3}$).
With this cut, 28\% of the signal events are selected while the
background contributions are reduced by three to five orders of magnitude.
The total selection efficiency for signal events is 26\%.
Details on the selection efficiency for the signal and considered
background processes are presented in \autoref{tab:fcnc_cg_effi}. 
The expected limit on the branching ratio of the FCNC top-quark decay
\tcg is extracted from a comparison of the measured BDT response
distribution with the distributions expected for the background 
and signal+background hypotheses.
The expected 95\% C.L. limit calculated using the \CLs approach \cite{Read:2002hq} is
\begin{eqnarray*}
  \text{BR}(\tcg)  & < & 2.6 \times 10^{-5}  
  \end{eqnarray*}
for 1.0\,ab$^{-1}$ collected at 380\,GeV.
The limit was calculated with \roostats \cite{Moneta:2010pm} using the 
frequentist limit calculator method, assuming no signal contribution.

\begin{table}[t]
\centering
\begin{tabular}{lcccc}
\toprule
Sample   & $\sigma$  & $\epsilon_{\mathrm{Pre}}$ (\%) & $\epsilon_{\mathrm{BDT}>0.29}$ (\%)
                     &  $N_{\mathrm{BDT}>0.29}$   \\
\midrule
FCNC \tcg  & 1.32 fb & 92 & 28  & 340 \\
\midrule
6-fermion   & 691 fb  & 2.7 & 0.14     & 26     \\
4-fermion   &  13 pb  & 16  & 0.003    & 65     \\
$\PQq\PAQq$ &  21 pb  & 24  & $<$0.001 & -  \\
\bottomrule
\end{tabular}
\caption{
  Cross section values, selection efficiencies and numbers of events
  expected for signal and background processes 
  in the analysis searching for FCNC decay \tcg at
  CLIC at 380\,GeV.
  Numbers of events correspond to a luminosity of
  1.0\,ab$^{-1}$, assuming unpolarised beams, and BR(\tcg) $=10^{-3}$ for signal events. 
  \label{tab:fcnc_cg_effi}
}
\end{table}

FCNC top-quark couplings to $\PGg\PQc$ and $\PZ\PQc$ can also be constrained from
the limit on the single top-quark production
$\epem \rightarrow \PQt \PAQc$.
Although these measurements have not been studied in detail for CLIC,
estimates based on a fast simulation approach presented for the FCC-ee
running at 350\,GeV \cite{Khanpour:2014xla} indicate that the limit
expected from a search for single-top production at 380\,GeV 
CLIC would be weaker than the one resulting from the direct search for
the decay \tcg, presented above.

%%%%%%%%%%%%%%%%%%%%%%%%%%%%%%%%%%%%%%%%%%%%%%%%%%%%%%%%%%%%%%%%%%%%%%%%%%%%%%%

\subsection{\texorpdfstring{\PQt$\rightarrow\PQc\PH$}{t$\rightarrow$cH}}
\label{sec:fcnc_tch}

For the top-quark FCNC decay \tch we consider only the final state
with the Higgs boson decaying to two $\PQb$ quarks, \hbb, which has the
dominant contribution of about 58\% in the SM.
This FCNC decay channel is challenging as the expected final state is
the same as for the SM top-quark pair decays (six jets for the fully-hadronic channel or four
jets, an isolated lepton and missing energy for the semi-leptonic channel).
Signal-background discrimination can only be based on the kinematic
event properties and the flavour tagging. 

Signal samples were generated with \whizard~2.2.8~\cite{Kilian:2007gr,Moretti:2001zz}
using the 2HDM(III) model~\cite{Atwood:1996vj} implemented in {\sc SARAH}~\cite{Staub:2015kfa}.
The background sample considered in the analysis includes a full set of
6-fermion event samples produced for the study of top-quark pair production at
$\sqrt{s} = 380$\,GeV described in \autoref{sec:pairprod}.
Backgrounds from four-fermion final state events (dominated by
\PWp\PWm production) and from pair production of quarks other than the 
top quark are also included.
Signal event selection was studied assuming 500\,fb$^{-1}$ of data collected with
electron beam polarisation of -80\% and the results were scaled to the integrated
luminosity of 1.0\,ab$^{-1}$ collected with equal sharing of -80\% and +80\%
polarisations.

Searches for FCNC decays of the top quark are made in both the fully-hadronic 
and semi-leptonic event samples.
The analysis is divided into two steps: pre-selection and
classification of \ttbar candidate events, and final discrimination between FCNC
and SM top-quark decays optimised for limit setting.
For the first step, the pre-selection cut and BDT algorithm
developed for top-quark pair event classification are used, as described in
\autoref{sec:Mass:cont}.
As expected, almost all FCNC signal events are classified as hadronic
or semi-leptonic top-quark pair events, with both the  pre-selection 
efficiency ($\epsilon_{\mathrm{Pre}}$) and classification efficiency ($\epsilon_{\ttbar}$) 
of about 99\%, see \autoref{tab:fcnc_ch_effi}. 

When reconstructing the decay kinematics we look for the jet combination that minimises the
$\chi^2$ value for the corresponding hypothesis.
However, to reduce the number of possible jet configurations, an additional,
tighter cut on the flavour tagging results is applied first.
Only events with at least three jets with $\PQb$-tag > 0.4 and one jet
with $\PQc$-tag > 0.4 are considered as FCNC signal candidates.
Two of these $\PQb$-jets and a $\PQc$-jet are then considered as candidates for
the FCNC top-quark decay products (with the Higgs boson decaying to two $\PQb$-jets).
While the efficiency of these cuts ($\epsilon_{\mathrm{FCNC}}$) for
signal events is about 45\%, all backgrounds from SM processes are
significantly suppressed.
Moreover, the number of jet configurations fitting the signal or background
hypothesis is reduced.
The valid configurations are then compared based on the
$\chi^2$ value for the event. 
For the SM top-quark pair decay hypothesis, the $\chi^2$ formula from 
the top-quark mass reconstruction is used, as described in \autoref{sec:Mass:cont}.
A similar $\chi^2$ formula is used for the FCNC top-quark pair decay hypothesis,
but with one of the $\PW$ boson masses replaced by the mass of the Higgs boson.

\begin{figure}
  \centering
  \includegraphics[width=0.48\columnwidth]{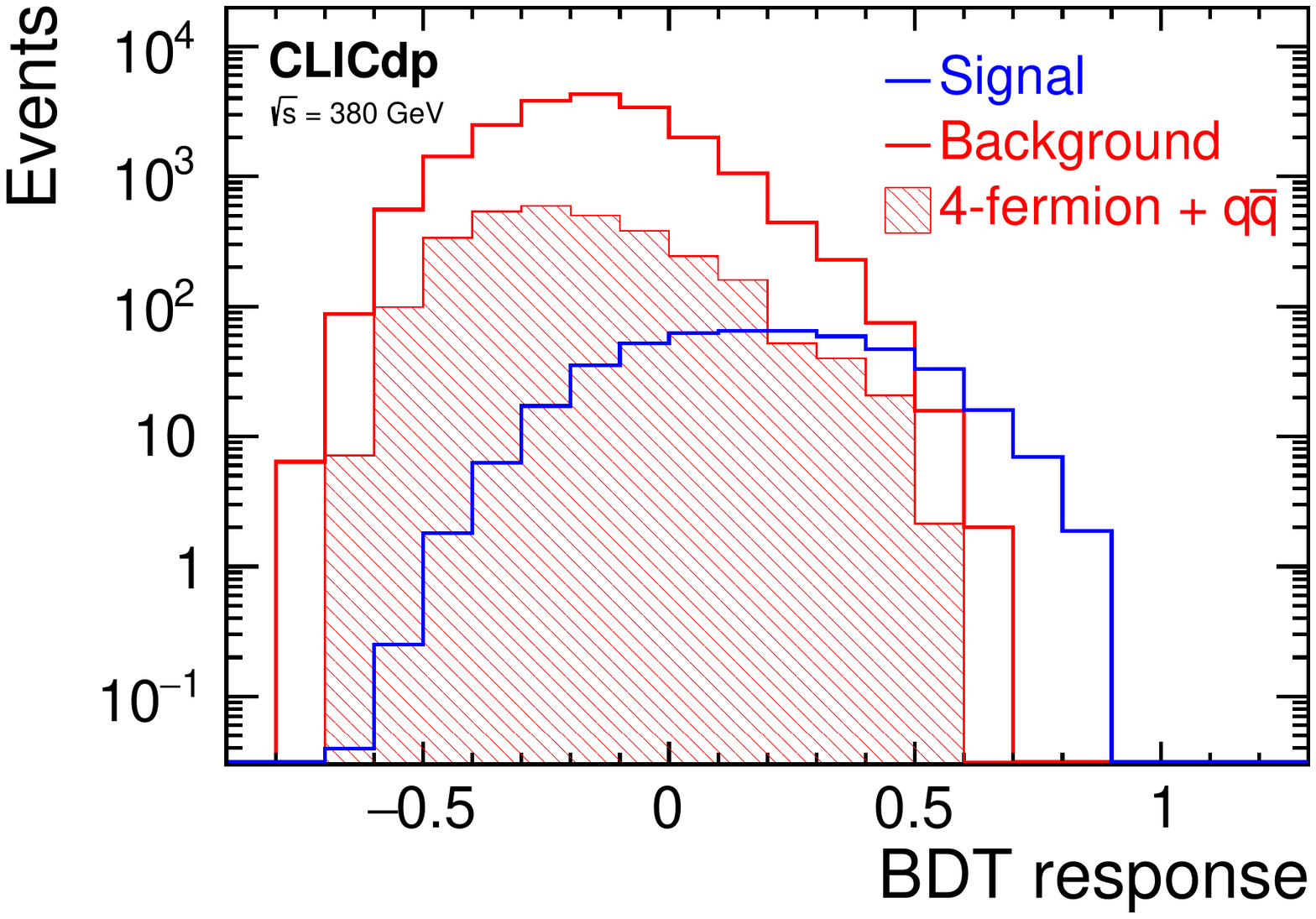}
  \includegraphics[width=0.48\columnwidth]{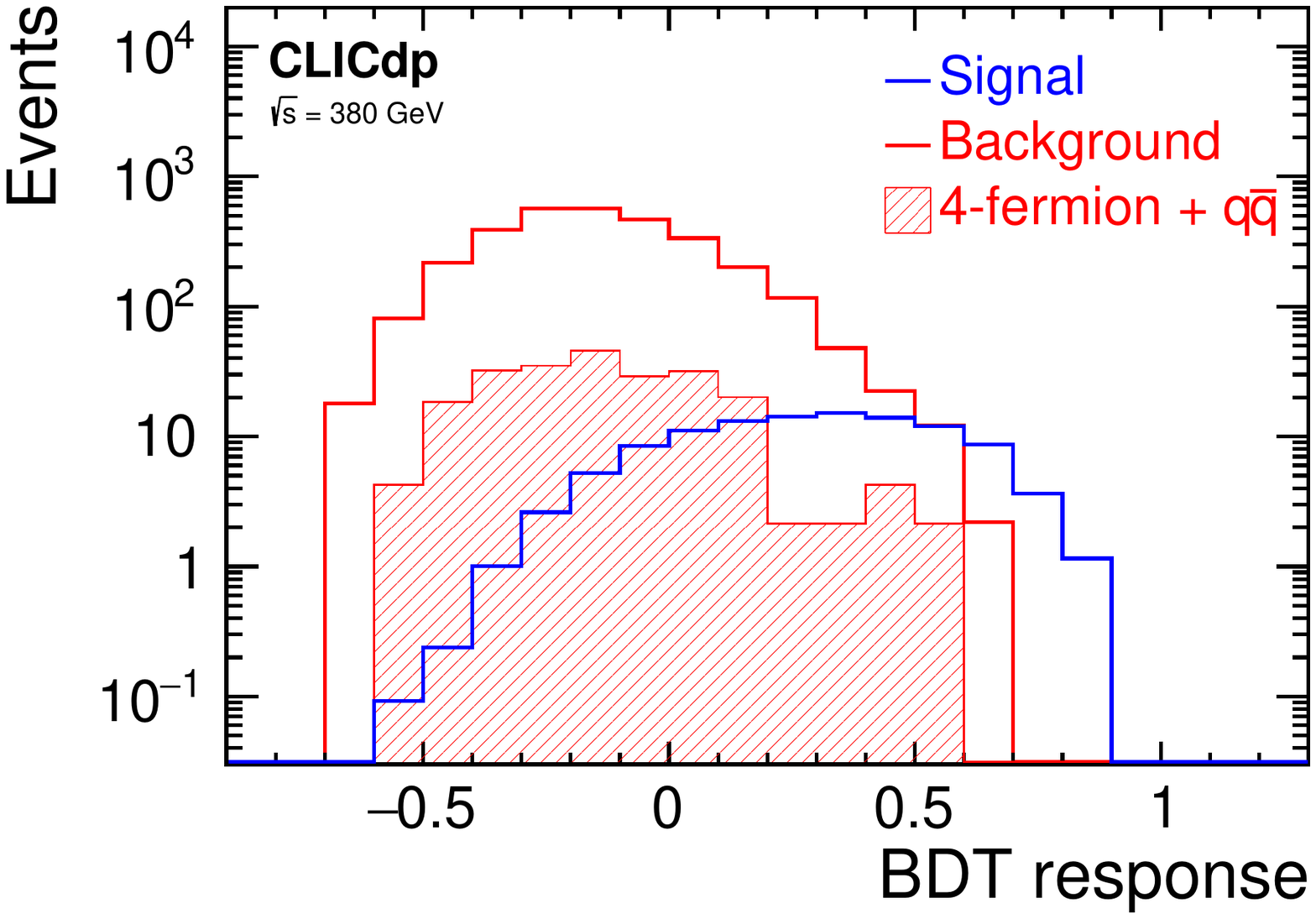}
  \caption{Response distribution of the BDT classifier used for the final \tch event selection at $\roots=380\,\gev$, for hadronic (left) and semi-leptonic
    (right) event samples. The total background (red histogram) is
    normalised to 1.0\,ab$^{-1}$ while the signal (events with a FCNC
    top quark decay; blue histogram) is normalised to BR(\tch)$\times$BR(\hbb)
    $=10^{-3}$. The hatched histogram indicates the contribution from four
    fermion and quark-pair backgrounds.               
\label{fig:fcnc_ch_bdt}}
\end{figure}

The reconstructed final state kinematics and the flavour tagging results are
used as an input for the final BDT selection optimised to discriminate
between signal and background events.
The following variables are used for BDT training:
  $\chi^2$ values for signal and background hypotheses,
  reconstructed Higgs boson mass and 
  $\PW$ boson mass from the spectator top-quark decay,
  the smaller of the two $\PQb$-tag values for the jets from Higgs boson decay,
  the $\PQc$-tag and $\PQb$-tag value for the $\PQc$-quark from FCNC decay,
  the $\PQb$-tag value for the $\PQb$-jet from the spectator top-quark decay,
  the smaller of the two $\PQb$-tag values for jets from top-quark decays (for
  the background hypothesis) and the responses of the BDT classifier used
  at the event classification stage (for hadronic and semi-leptonic
  event selection, see \autoref{sec:Mass:cont}).
The resulting response distributions from the BDT classifier are
presented in \autoref{fig:fcnc_ch_bdt}, separately for the hadronic and
the semi-leptonic samples.
The background sample is normalised to 1.0\,ab$^{-1}$ while the signal
(events with a FCNC top quark decay) are normalised to BR(\tch)$\times$BR(\hbb) $=10^{-3}$.
To select a signal-dominated sample a relatively tight selection cut on
the BDT response is required.
Details on the selection efficiency for the different event samples
considered in the analysis are presented in \autoref{tab:fcnc_ch_effi}. 
For the cut on the response of the final BDT score, $> 0.4$, 
the total selection efficiency for FCNC events is 11\% while
the suppression of the SM \ttbar background is at the level of
$1.4\cdot10^{-4}$ and the non-\ttbar backgrounds are suppressed at the level
of $10^{-6}$. 

The expected limit on the branching ratio of the FCNC top-quark decay
\tch is extracted from a comparison of the measured BDT response
distribution with the distributions for the background 
and signal+background hypotheses.
The 95\% C.L. limit calculated with the \CLs approach \cite{Read:2002hq} is
\begin{eqnarray*}
  \text{BR}(\tch) \times \text{BR}(\hbb) & < & 8.8 \times 10^{-5},
\end{eqnarray*}   
for 1.0\,ab$^{-1}$ collected at 380\,GeV.

\begin{table}[t]
\centering
\begin{tabular}{lcccccc}
\toprule
Sample   & $\sigma$  & $\epsilon_{\mathrm{Pre}}$ (\%) & $\epsilon_{\ttbar}$ (\%) &
                       $\epsilon_{\mathrm{FCNC}}$ (\%) & $\epsilon_{\mathrm{BDT}>0.4}$ (\%) &
                       $N_{\mathrm{BDT}>0.4}$   \\ 
\midrule
FCNC  \tch  & 1.32 fb & 99 & 99 &   45 & 25   & 145 \\
\midrule
6-fermion   & 691 fb  &  88  & 90   &  3.6 & 0.51 &   100 \\
4-fermion   &  13 pb  & 8.5  & 5.1  &  2.8 & 0.97 &   15 \\
$\PQq\PAQq$ &  21 pb  & 20 & 1.1  &  3.3 & 0.94 &    14 \\
\bottomrule
\end{tabular}
\caption{
  Cross section values, selection efficiencies, and numbers of events
  expected for signal and background processes 
  in the analysis searching for FCNC decay \tch at
  CLIC at 380\,GeV.
  Selection efficiencies are quoted for  pre-selection
  ($\epsilon_{\mathrm{Pre}}$) and classification ($\epsilon_{\ttbar}$) of
  \ttbar events, as well as for selection of FCNC candidate events
  ($\epsilon_{\mathrm{FCNC}}$) and selection of signal dominated sample with
  cut on BDT response ($\epsilon_{\mathrm{BDT}>0.4}$).
  Numbers of events correspond to a luminosity of
  1.0\,ab$^{-1}$ with equal amounts of -80\% and +80\% polarisation running,
  and BR(\tch)$\times$BR(\hbb) $=10^{-3}$ for signal events. 
  \label{tab:fcnc_ch_effi}
}
\end{table}

%%%%%%%%%%%%%%%%%%%%%%%%%%%%%%%%%%%%%%%%%%%%%%%%%%%%%%%%%%%%%%%%%%%%%%%%%%%%%%%%%%%%5

\subsection{\texorpdfstring{\PQt$\rightarrow\PQc$E$\!\!\!\!\!\slash$}{t$\rightarrow$c+missing energy}}

We search for events where the top quark decays into a charm quark
and a heavy stable particle, which escapes from the detector, giving a characteristic `missing energy` signature.
As the mass of the produced heavy state has to be reconstructed from
energy-momentum conservation, only the hadronic decay channel is
considered for the spectator top quark.
For the expected final state consisting of four jets, the dominant
background contribution is from processes with four fermions in the
final state, primarily from \PWp\PWm decays.
To model the decay to an invisible scalar particle, dedicated samples of
events with \tch decay were generated, as described in
\autoref{sec:fcnc_tch}, with the Higgs boson defined as a stable particle
in \pythia (and thus invisible in the detector). 
Signal samples for scalar masses from 25 to 150\,GeV were generated
assuming an electron beam polarisation of -80\%  and the results were scaled to the
number of top-quark pairs expected for the integrated luminosity of 1.0\,ab$^{-1}$ collected
with equal sharing of -80\% and +80\% polarisations.

To reduce large backgrounds coming from four-fermion and quark-pair
production processes a set of pre-selection cuts is applied.
We require the total invariant mass of the hadronic final state
to be above 140~GeV, the total transverse momentum above 20\,GeV, and
the absolute value of the longitudinal momentum below 100\,GeV.
After clustering the hadronic final state into four jets, using the VLC
algorithm with a radius of 1.6 and $\beta = \gamma = 0.8$, 
we require one of the resulting jets to have a b-tag value of at least 0.6
(for the $\PQb$-jet candidate from the decay of the spectator top quark) and
all other jets to have a b-tag value below 0.4. 
We also reject all events with a reconstructed isolated lepton.

\begin{figure}
  \centering
  \includegraphics[width=0.65\columnwidth]{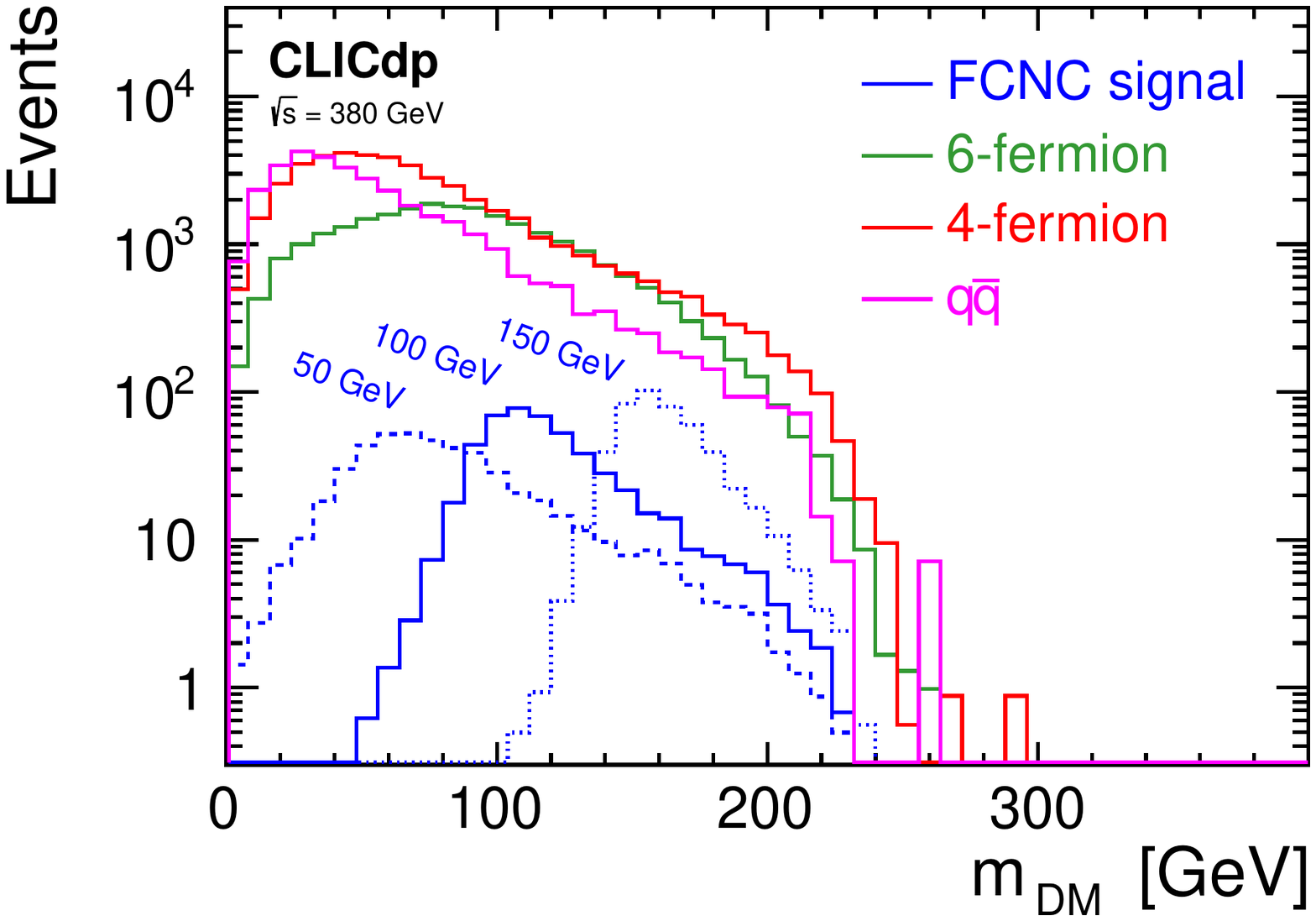}
  \caption{Distribution of the reconstructed invariant mass of the
    invisible decay product for the FCNC decay \tcx, reconstructed at
    380~GeV CLIC  after preselection cuts. The background
    contributions coming from 6-fermion (green histogram),
    four-fermion (red) and quark pair (magenta) production processes
    are normalised to 1.0\,ab$^{-1}$ while the signal samples, for the
    assumed mass of the invisible scalar of 50, 100 and 150 GeV (blue
    histograms) are normalised to BR(\tcx) $=10^{-3}$. 
\label{fig:fcnc_cdm_minv}}
\end{figure}

To reconstruct the kinematics of the event
we assume that the jet with the highest c-tag value comes from the
FCNC top-quark decay while the two remaining jets result from the \PW
boson decay. 
This choice of jet configuration matches the true decay kinematics in
about 70\% to 75\% of signal events (depending on the scalar mass).
The invariant mass of the signal top quark as well as the invariant
mass of the invisible decay product can then be reconstructed from the
energy-momentum conservation.
\autoref{fig:fcnc_cdm_minv} shows the distribution of the
reconstructed invariant mass of the invisible decay product for signal
events (for the scalar masses of 50, 100 and 150\,GeV) and for different
background samples.
For the signal samples the efficiency of the pre-selection cuts
described above varies between 35\% and 42\% (depending on the scalar mass),
while for the four-fermion and quark-pair background samples it is
about 0.35\% and 0.16\%, respectively.

\begin{figure}
  \centering
  \includegraphics[width=0.48\columnwidth]{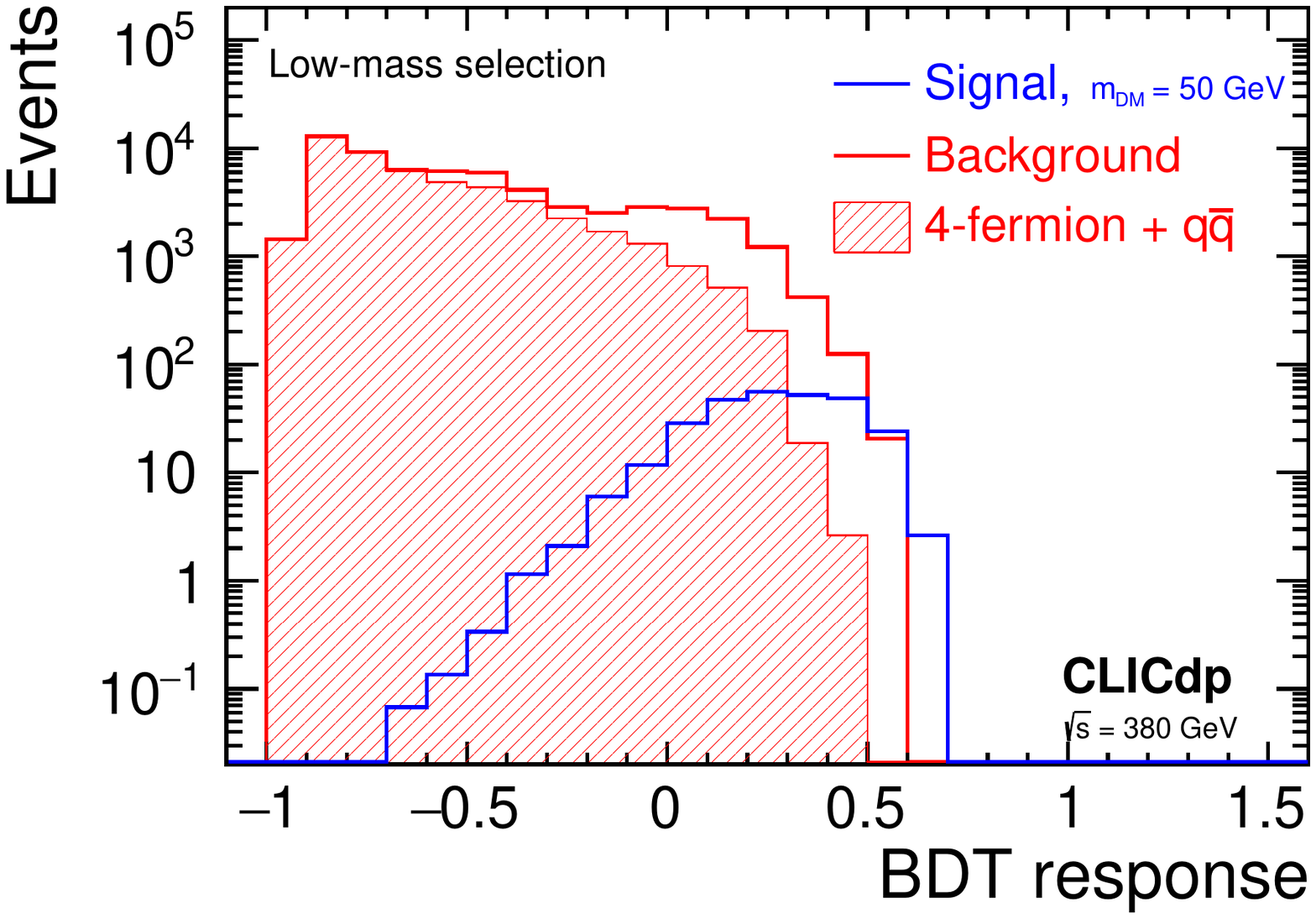}
  \includegraphics[width=0.48\columnwidth]{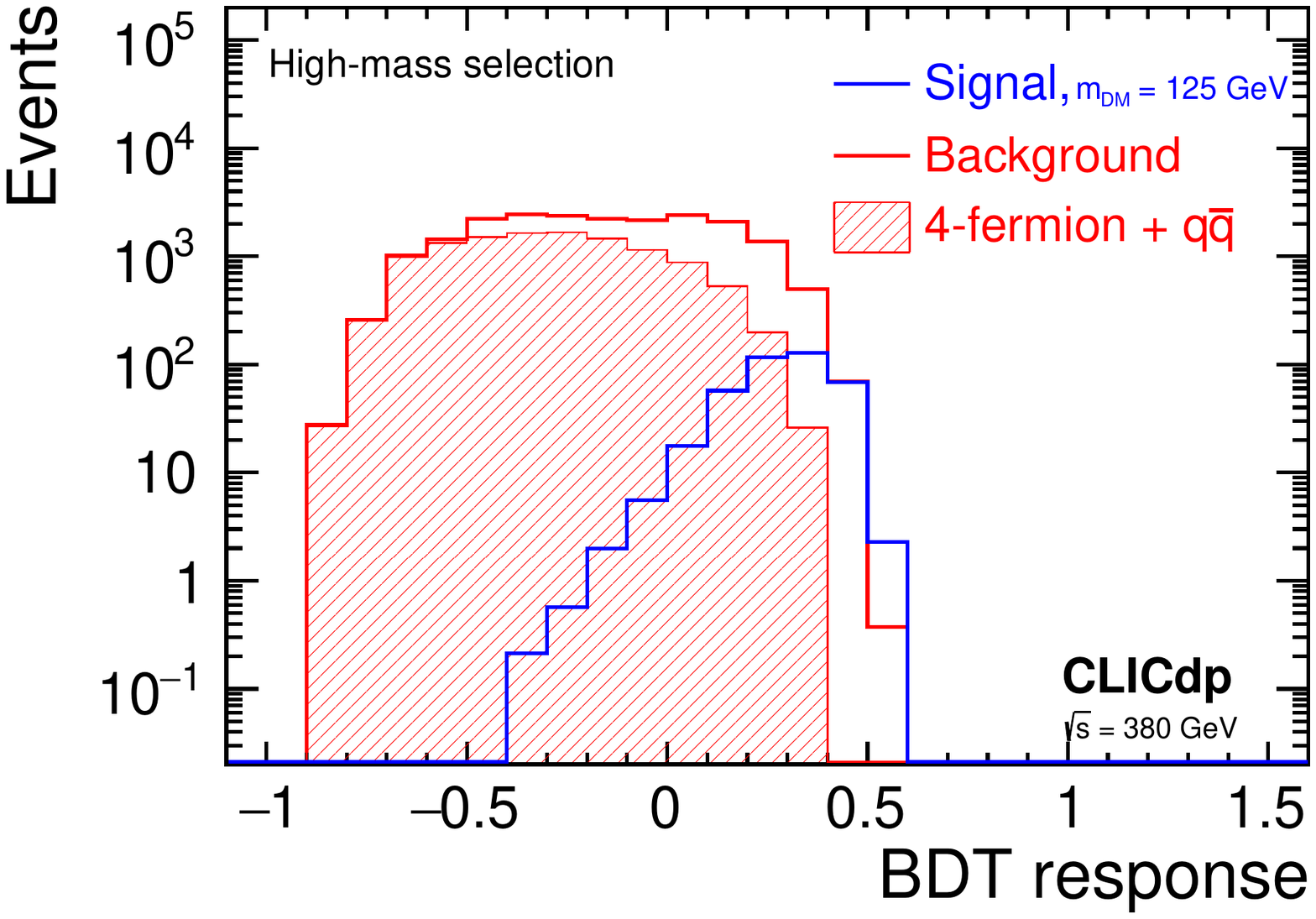}
  \caption{Response distributions of the BDT classifiers used for the
    final selection of \tcx events at $\roots=380\,\GeV$, for an assumed mass of the
    invisible decay product of 50\,GeV (low mass selection; left) and 125\,GeV
    (high mass selection; right). The total background (red histogram) is
    normalised to 1.0\,ab$^{-1}$ while the signal (events with a FCNC
    top quark decay; blue histogram) is normalised to BR(\tcx)
    $=10^{-3}$. The hatched histogram indicates the contribution from four
    fermion and quark-pair backgrounds.               
\label{fig:fcnc_cx_bdt}}
\end{figure}

For the final discrimination between the
FCNC decay events and the SM background processes, a multivariate analysis is used.
The BDT classification algorithm is trained, separately
for low scalar masses (signal scalar masses of 25~GeV,
50~GeV and 75~GeV) and high scalar masses (signal scalar masses of 100~GeV,
125~GeV and 150~GeV). 
The set of variables includes:
  total energy of the event $E_{\mathrm{tot}}$,
  total transverse momentum $\pT$,
  total invariant mass $M_{\mathrm{inv}}$,
  missing mass $M_{\mathrm{miss}}$,
  sphericity and acoplanarity of the event,
  minimum ($y_{\mathrm{min}}$) and maximum ($y_{\mathrm{max}}$)
  distance cuts for four-jet reconstruction with the VLC algorithm,
  b-jet energy and invariant mass,
  reconstructed masses of the two top quarks,
  reconstructed mass and energy of the invisible scalar, 
  and the $\chi^2$ value calculated from the reconstructed masses of
  the $\PW$ boson and two top quarks.
  For each considered value of the invisible scalar particle mass the
  BDT response distribution was plotted for events in the $\pm$30~GeV
  window in the reconstructed particle mass.
Examples of the response distributions for the low mass and high mass
BDT classifiers, for the selected masses of 50~GeV and 125~GeV, are
shown in \autoref{fig:fcnc_cx_bdt}.
Details on the selection efficiency for the two selected values of the
invisible scalar mass are summarised in \autoref{tab:fcnc_cx_effi}. 
For high values of BDT response, BDT$>$0.25, the background is
dominated by the six-fermion sample, while the suppression factor for
four-fermion and quark-pair background is at the level of $10^{-5}$
and $10^{-6}$, respectively.

\begin{figure}
  \centering
  \includegraphics[width=0.65\columnwidth]{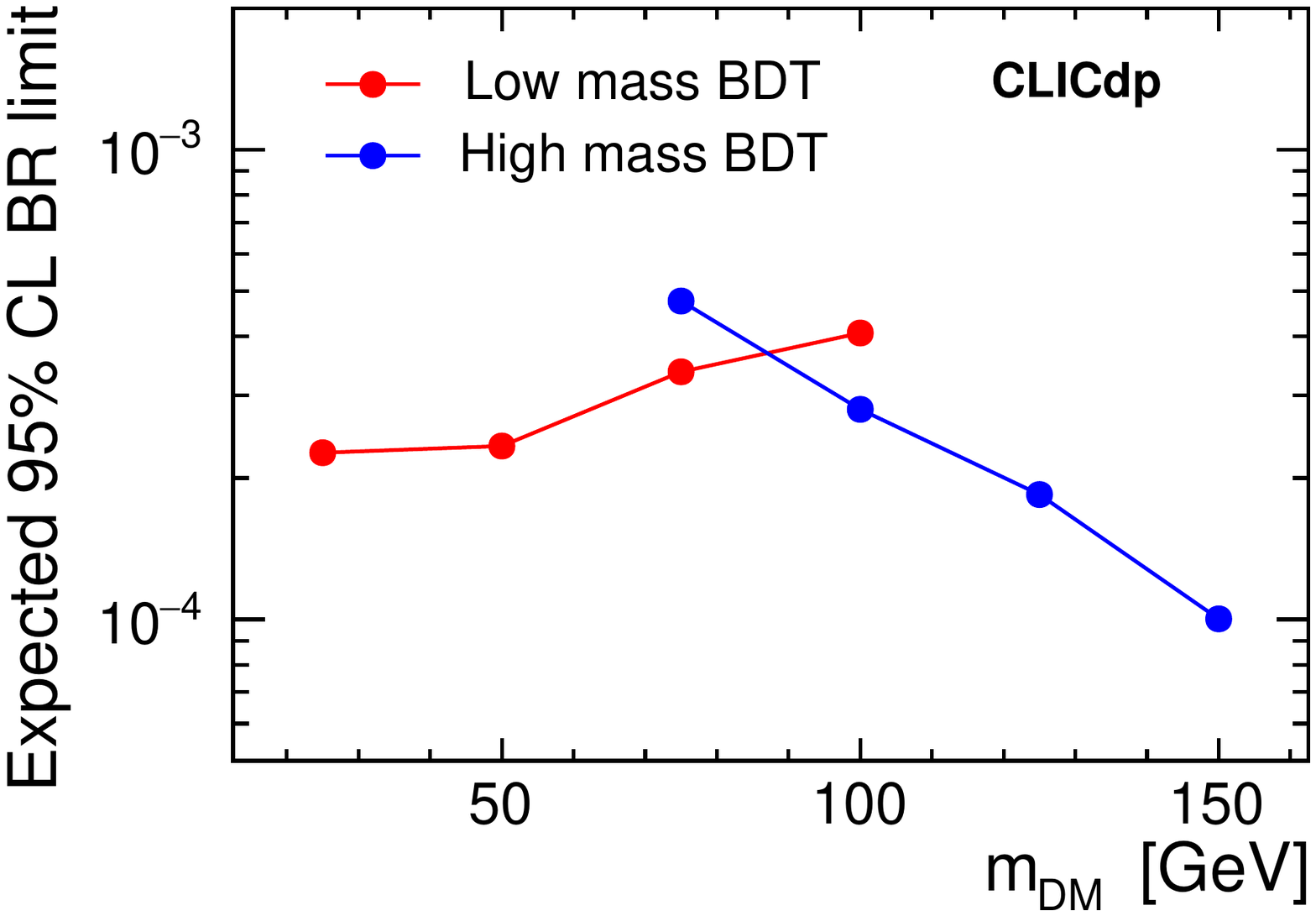}
  \caption{Limits at 95\% C.L. on the top quark FCNC decay \tcx\,
    expected for 1.0\,ab$^{-1}$ collected at 380~GeV CLIC equal amounts of -80\% and +80\% 
    electron beam polarisation, as a function of the assumed mass of the invisible decay
    product, $m_\text{DM}$. Limits are calculated from the BDT response distribution
    in the $\pm$30~GeV window in the reconstructed particle mass.
    Two BDT algorithms are trained separately for low and high mass range.
\label{fig:fcnc_cdm_limits}}
\end{figure}

Expected limits on the branching ratio of the FCNC top-quark decay
\tcx were calculated as a function of the scalar particle mass
from a comparison of the measured BDT response distribution (in the
$\pm$30~GeV reconstructed mass window) with the distributions expected
for the background only hypothesis and the one including signal contribution.
The limits calculated using the \CLs approach \cite{Read:2002hq} 
for 1.0\,ab$^{-1}$ collected at 380\,GeV are summarised in
\autoref{fig:fcnc_cdm_limits}. 

\begin{table}
\centering
\begin{tabular}{lcccc}
\toprule
Sample   & $\sigma$  & $\epsilon_{\mathrm{Pre}}$ (\%) & $\epsilon_{\mathrm{BDT}>0.25}$ (\%)
                     &  $N_{BDT>0.25}$   \\

\midrule
\multicolumn{5}{l}{ {{\it Low mass selection}, $m_\text{DM}=50$}\,GeV} \\
\midrule  
FCNC \tcx  & 1.32 fb & 41 & 29 & 155 \\
\midrule
6-fermion   & 691 fb  & 4.0  & 3.3 & 935 \\
4-fermion   &  13 pb  & 0.35  & 0.17  &   77 \\
$\PQq\PAQq$ &  21 pb  & 0.16 & 0.11 &    36 \\
\midrule
\multicolumn{5}{l}{ {{\it High mass selection}, $m_\text{DM}=125$}\,GeV  } \\
\midrule  
FCNC \tcx   & 1.32 fb & 40 & 51 &  266 \\
\midrule
6-fermion   & 691 fb  & 4.0  & 4.0 &   1080 \\
4-fermion   &  13 pb  & 0.35  & 0.20    & 92 \\
$\PQq\PAQq$ &  21 pb  & 0.16 & 0.042   &   14 \\
\bottomrule
\end{tabular}
\caption{
  Cross section values, selection efficiencies and numbers of events
  expected for signal and background processes 
  in the analysis searching for FCNC decay \tcx at
  CLIC at 380\,GeV.
  Results are presented for a mass of the invisible scalar particle,  $m_\text{DM}$, of
  50\,GeV (upper part of the table) and 125\,GeV (lower part).
  Selection efficiencies are quoted for  pre-selection
  ($\epsilon_{\mathrm{Pre}}$) and the final selection of the signal enhanced sample with
  the BDT response cut, BDT$>$0.25.
  Numbers of events correspond to a luminosity of
  1.0\,ab$^{-1}$ with equal amounts of -80\% and +80\% polarisation running,
  and BR(\tcx) $=10^{-3}$ for signal events. 
  \label{tab:fcnc_cx_effi}
}
\end{table}

\section{Phenomenological interpretations}
\label{sec:phenom_interp}

The capabilities of CLIC to improve our knowledge of top-quark physics can be directly illustrated by studying the sensitivity to new physics in the top-quark sector. This section describes the BSM
implications of such top-quark studies by introducing a general EFT interpretation, followed by a more specific analysis in the context of top-quark compositeness scenarios.

The results presented are based on full simulation studies of top-quark pair production at all three stages of CLIC. Note however that the results, to some extent, are partial in the sense that they do not include all the top-quark related measurements possible at CLIC. Furthermore, the results should be studied in the broader context of the global sensitivity of CLIC to the SM EFT, discussed in~\cite{deBlas:2018mhx} and references therein.

\subsection{General top-philic interpretation}
\label{ssec:topphilicinterp}
Under the simple hypothesis of top-philic BSM, motivated and outlined in \autoref{ssec:interpretations_theory}, it is possible to restrict the global EFT interpretation of top-quark physics to the nine operators given in \autoref{tab:operators}. For simplicity, we further restrict the analysis presented below to CP-conserving BSM physics, which implies real operator coefficients, leading to a total of nine real parameters to be determined.\footnote{CP-breaking new physics could also be studied, but this would require taking into account potentially strong constraints from electric dipole moments and other low-energy probes. This goes beyond the scope of the present studies, see however \cite{PerelloVosZhang}.}
The top-quark pair production process is sensitive to seven of these operators. The two missing ones are the combination $Q_{\varphi{\PQq}}^{(1)}+Q_{\varphi{\PQq}}^{(3)}$, which does not affect the electroweak interactions of the top-quark, and $Q_{\PQt\varphi}$. The latter operator modifies the top Yukawa coupling by an amount $\Delta{y_{\PQt}}=v^2 C_{\PQt\varphi}$, where $v=246\,\gev$ and $C_{\PQt\varphi}$ is the Wilson coefficient of the corresponding operator $Q_{\PQt\varphi}$. Hence it is probed at $1\sigma$ level by the $\ttH$ analysis reported in \autoref{sec:ttH}:\footnote{This corresponds to the single-operator sensitivity to $C_{\PQt\varphi}$, i.e. the sensitivity when $Q_{\PQt\varphi}$ is the only operator with a BSM effect. A global EFT interpretation of the $\ttH$ analysis, duly combined with $\ttbar$ production and with the other probes of top-quark physics, would be needed for a more robust assessment of the sensitivity.}
\begin{equation}\label{deltaytop}
[C_{\PQt\varphi}]_{1\sigma}=\frac1{v^2}[\Delta y_{\PQt}]_{1\sigma}\simeq0.4\,{\tev}^{-2}\,.
\end{equation}
While it corresponds to a rather low ``operator scale'' $\Lambda_{\PQt\varphi}\equiv1/\sqrt{C_{\PQt\varphi}}\simeq1.5\,\tev$, we will see in the next section that this sensitivity allows to probe a much higher new physics scale in top-quark compositeness scenarios, where $C_{\PQt\varphi}$ can be enhanced by a strong coupling. While the operator combination $Q_{\varphi{\PQq}}^{(1)}+Q_{\varphi{\PQq}}^{(3)}$ does not affect the electroweak interactions of the top-quark, it does modify the left-handed $\PQb\PAQb\PZ$ coupling as $\delta{g_{\PQb,L}}/g_{\PQb,L}=(C_{\varphi{\PQq}}^{(1)}+C_{\varphi{\PQq}}^{(3)})v^2/(\cW^2+\sW^2/3)$. The sensitivity of LEP to the bottom couplings was at the per mille level, corresponding to
\begin{equation}\label{zbb}
[C_{\varphi{\PQq}}^{(1)}+C_{\varphi{\PQq}}^{(3)}]_{1\sigma}\simeq1.4\times10^{-2}\,{\tev}^{-2}\simeq[8\,{\rm{TeV}}]^{-2}\,.
\end{equation}
It is thus legitimate to restrict the EFT analysis of the top-quark pair production to the orthogonal combination $Q_{\varphi{\PQq}}^{(1)}-Q_{\varphi{\PQq}}^{(3)}$, i.e. to set $C_{\varphi{\PQq}}^{(1)}=-C_{\varphi{\PQq}}^{(3)}$ and report results for $C_{\varphi{\PQq}}^{-}=C_{\varphi{\PQq}}^{(1)}-C_{\varphi{\PQq}}^{(3)}$. Note that $Q_{\varphi{\PQq}}^{(1)}+Q_{\varphi{\PQq}}^{(3)}$ is the only combination of the top-philic operators that modifies the bottom quark interactions with the gauge fields, hence no further LEP constraint needs to be taken into account.

\begin{table}
\begin{minipage}{\columnwidth}
\begin{center}
\begin{tabular}{r|cccccc}
\toprule
$\roots$ & \multicolumn{2}{c}{380\,\gev\footnote{Results from \autoref{ssec:pairprod380gev}}} & \multicolumn{2}{c}{1.4\,\tev\footnote{Results from \autoref{ssec:pairprodboosted}\label{footnote:ttbarresultstab}}} & \multicolumn{2}{c}{3\,\tev~\footref{footnote:ttbarresultstab}} \rule{0pt}{3ex} \\
P(\Pem) & -80\% & +80\% & -80\% & +80\% & -80\% & +80\% \rule{0pt}{3ex} \\
\midrule
$\csttbar$ & 12.9\% & 12.1\% & 6.0\% & 5.8\% & 4.6\% & 4.7\% \rule{0pt}{3ex}\\
$\afb$ & 4.7\% & 12.1\% & 6.0\% & 5.8\% & 4.6\% & 4.7\% \rule{0pt}{3ex}\\
Statistically opt. obs. & 7.8\% & 12.1\% & 6.0\% & 5.8\% & 4.6\% & 4.7\% \rule{0pt}{3ex}\\
\bottomrule
\end{tabular}
\end{center}
\caption{Equivalent fractions of the theoretical $\epem \to \ttbar$ rate at the nominal centre-of-mass energies that are assumed in the EFT fit. When multiplied by the inclusive $\Pep\Pem\rightarrow\ttbar$ cross section, excluding the energy loss from ISR or beamstrahlung, and by the corresponding integrated luminosity, these fractions yield the numbers of events at final level of the analyses of semi-leptonic final states, as studied in full simulation at the three stages of CLIC and presented in \autoref{ssec:pairprod380gev} and \autoref{ssec:pairprodboosted}. Different variants of the final event selection are used for  $\afb$, $\csttbar$ and the set of statistically optimal observables, as explained in the text. \label{tab:eft:efficiencies}}
\end{minipage}
\end{table}
 
For each EFT operator ${\mathcal{O}}_i$, the corresponding observable is defined~\cite{Atwood:1991ka}, as the sum over the data sample of the derivative of the logarithm of the differential cross section with respect to its Wilson coefficient $C_i$, evaluated for vanishing $C$'s. The fully differential cross section for the $\epem\to\ttbar\to\PQb\PWp\PAQb\PWm$ process is employed in order to maximise the sensitivity. It is computed analytically at LO~\cite{PerelloVosZhang}, for a massless $\PQb$-quark, in the narrow-width approximation for the top quark, ignoring ISR and beamstrahlung. The results are verified with numerical results obtained with the \texttt{TEFT\_EW} model~\cite{teft_EW} in the \texttt{MG5\_aMC@NLO} package~\cite{Alwall:2014hca}, which is also used to obtain an estimate of the NLO correction. The expected sensitivities and the covariance matrix in the EFT parameter space are obtained, for simplicity, by asymptotic formulas that rely on the same LO expression. Note that this approach requires truncating the EFT prediction at the linear level in the Wilson coefficients. This approximation is found to be accurate for this set of observables and a full linear collider physics programme.

The seven operators that contribute to top-quark pair production are probed either using the observables $\csttbar$ and $\afb$ defined in \autoref{sec:pairprod}, or using statistically optimal observables. Realistic event selection efficiencies, derived from the results of \autoref{ssec:pairprod380gev} and \autoref{ssec:pairprodboosted}, and the energy loss from ISR and beamstrahlung are included in the EFT fits using the equivalent event fractions given in \autoref{tab:eft:efficiencies}. Hence our results show the potential of semi-leptonic final states with an electron or muon~\footnote{Including final-states with a tau lepton, or the fully-hadronic channel, could ultimately strengthen the constraints presented.}. For operation at $\roots=380\,\gev$, the $\afb$ for $P(\Pem)$ = -80\% is extracted using the event sample defined by $D^2<1$, while the results used for $\csttbar$ are obtained using the looser selection cut, defined by $D^2<15$. Further, at $\roots=380\,\gev$ the results used for the statistically optimal observables use a somewhat different selection as introduced at the end of \autoref{ssec:pairprod380gev}. As demonstrated in the individual analysis sections of \autoref{sec:pairprod}, the background levels are small compared to the signals; for technical reasons we have neglected both systematics and contributions to the statistical uncertainty from backgrounds in the following EFT studies.

\begin{figure}[t!]
\begin{center}
\includegraphics[scale=.92]{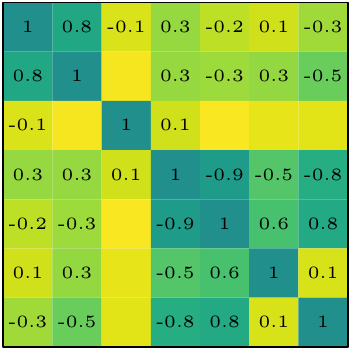}\hspace{20pt}
\raisebox{-4.5mm}{\includegraphics[scale=.88]{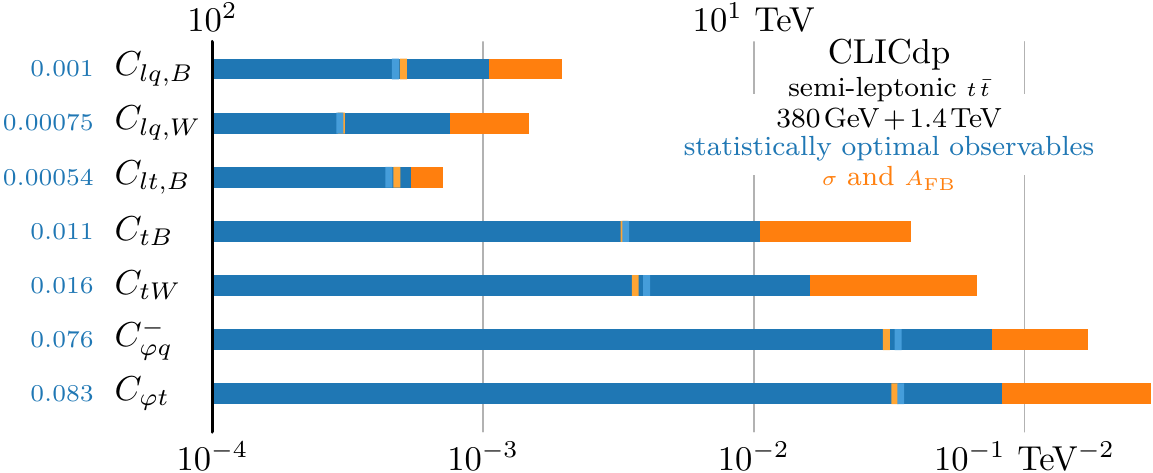}}
\caption{Global EFT analysis results for top-quark pair production at the $380\,\gev$ and $1.4\,\tev$ CLIC, using respectively $1.0\,\abinv$ and $2.5\,\abinv$, equally shared between $\pm$80\% electron beam polarisation configurations at the initial stage and split in the ratio 80:20 at the higher-energy stage. $1\sigma$ sensitivities to the seven Wilson coefficients for the combination of $\afb$ and $\csttbar$ are shown in orange bars, and for the statistically optimal observables in blue bars. The corresponding single-operator sensitivities are shown as ticks of the same colour for the two set of operators. The correlation matrix among the Wilson coefficients for the fits using the statistically optimal observables (left). Empty matrix elements indicate values closer to 0 than to 0.1. \label{fig:limits_no_high}}
\end{center}
\end{figure}

\begin{figure}[t!]
\begin{center}
\includegraphics[scale=.92]{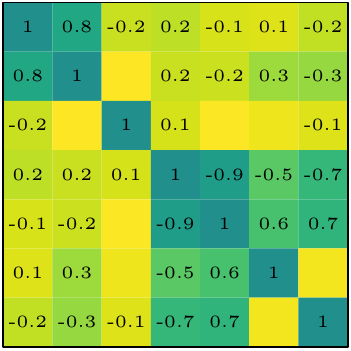}\hspace{20pt}
\raisebox{-4.5mm}{\includegraphics[scale=.88]{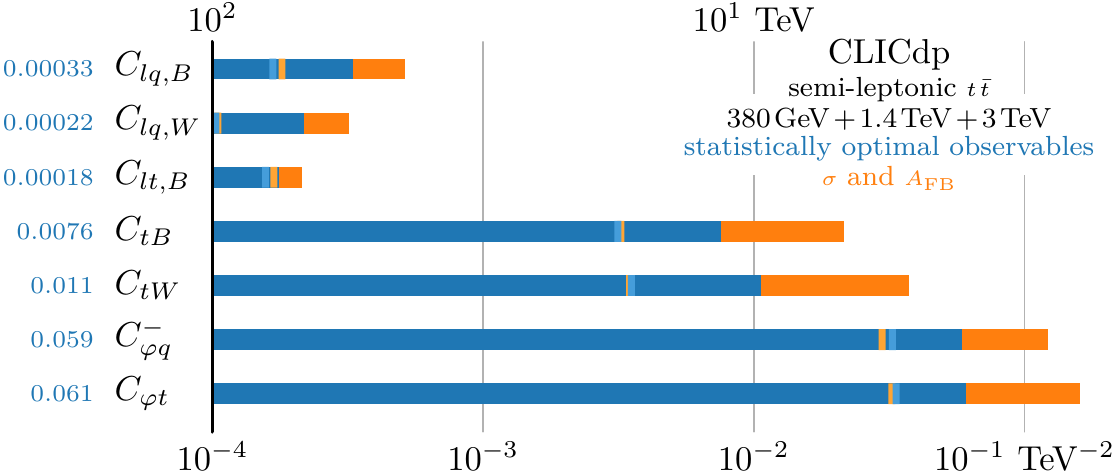}}
\caption{Same as \autoref{fig:limits_no_high}, but including operation at $3\,\tev$ CLIC using $5.0\,\abinv$, split in the ratio 80:20 for operation with -80\% and +80\% electron beam polarisation, respectively. \label{fig:limits}}
\end{center}
\end{figure}

The resulting global and individual $1\sigma$ constraints on the seven Wilson coefficients obtained for CLIC operation at $\sqrt{s}=380\,\gev$ and $\sqrt{s}=1.4\,\tev$ are presented in orange in \autoref{fig:limits_no_high}. Note that the reach of the initial CLIC stage alone does not allow for the simultaneous determination of the seven Wilson coefficients. While the global reach is illustrated in bars, the corresponding single-operator sensitivities are shown as ticks of the same colour. The sensitivities are measured in ${\rm{TeV}}^{-2}$ on the lower horizontal axis (and reported in numbers of the same unit), while the upper horizontal axis reports the sensitivity to the operator scale $\Lambda\equiv 1/\sqrt{C}$. \autoref{fig:limits} shows the sensitivity of CLIC when also including the results at the highest energy stage with collisions at $\sqrt{s}=3\,\tev$.

These observables alone provide an excellent reach, but are insufficient to simultaneously constrain all the directions in the seven-dimensional EFT parameter space and leave a number of flat (poorly constrained) directions in the likelihood. The analysis of radiative events at $\sqrt{s}=1.4\,\tev$, presented in \autoref{ssec:pairprodradiative}, could have been included to improve the situation, however as illustrated below, the use of statistically optimal observables in the $\ttbar$ final state, as described in more detail in \cite{PerelloVosZhang} and in \cite{Grzadkowski:2000nx, Janot:2015yza, Khiem:2015ofa}, efficiently eradicates all such flat directions, leading to a nearly-diagonal correlation matrix as seen in \autoref{fig:limits_no_high} and \autoref{fig:limits}. Detailed studies of the reconstruction and systematics of the optimal observables are left for the future.

The resulting constraints are presented in blue in \autoref{fig:limits_no_high} and \autoref{fig:limits}, and constitute a significant improvement compared to the combination of $\csttbar$ and $\afb$ alone for all coefficients considered. Correlation matrices are displayed for the fits using the statistically optimal observables.

\begin{figure}[t!]
\begin{center}
\includegraphics[scale=1.08]{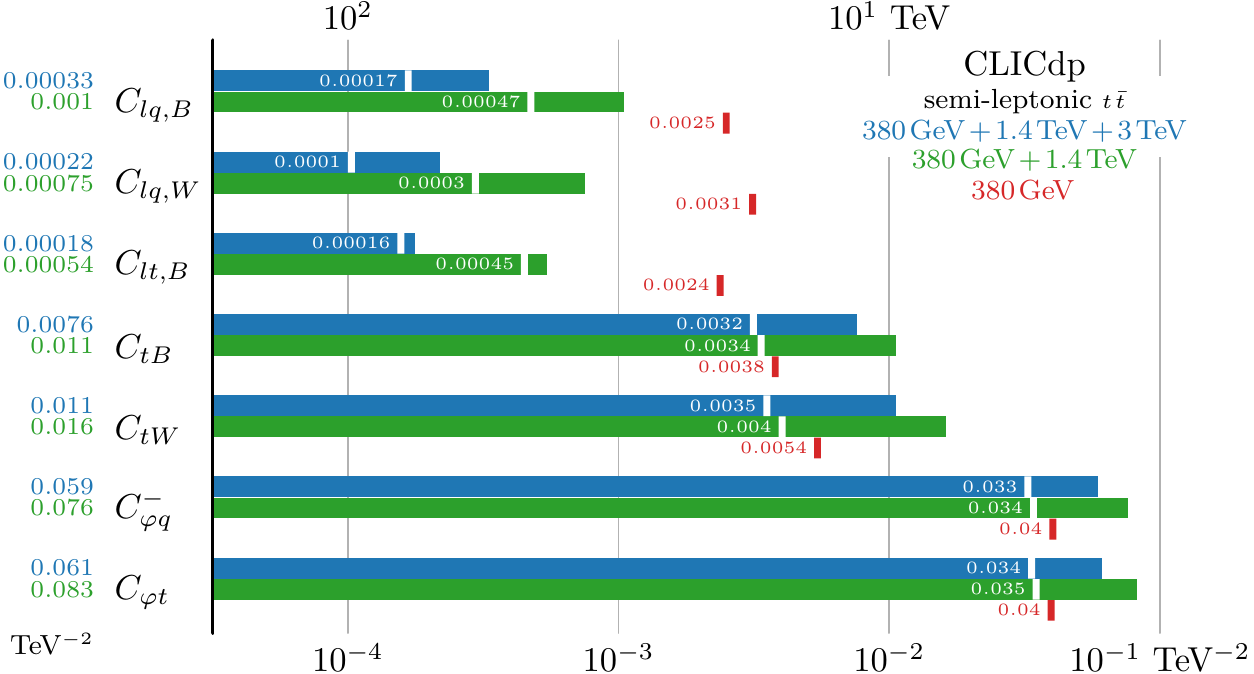}
\caption{Summary of the global EFT analysis results using statistically optimal observables for the three CLIC energy stages. The colour bars indicate the $1\sigma$ constraints on each of the seven Wilson coefficients. The blue bars illustrate the results for the full CLIC programme of consecutive operation at $380\,\gev$, $1.4\,\tev$, and $3\,\tev$, using respectively $1.0\,\abinv$, $2.5\,\abinv$, and $5.0\,\abinv$, equally shared between $\pm$80\% electron beam polarisation configurations at the initial stage and split in the ratio 80:20 at the higher-energy stages. Similarly, the green bars give the results of operation at $380\,\gev$ and $1.4\,\tev$, excluding the $3\,\tev$ stage. The corresponding individual operator sensitivities are shown as ticks. Note that a global fit is not possible for operation at $380\,\gev$ alone, for which the individual operator constraints are shown in red. \label{fig:limits_summary}}
\end{center}
\end{figure}

\autoref{fig:limits_summary} displays the global and individual $1\sigma$ constraints on the seven Wilson coefficients obtained for operation at $\sqrt{s}=380\,$GeV only, in combination with the $1.4\,$TeV stage, and including also the $3\,$TeV stage. In conclusion, we find that operation at high energy dramatically improves the sensitivity to the four-fermion operators $Q_{\Pl\PQq,\mathrm{B}}$, $Q_{\Pl\PQq,\PW}$, and $Q_{\Pl\PQt,\mathrm{B}}$, and to a lesser extent, to $Q_{\PQt\mathrm{B}}$ and $Q_{\PQt\PW}$. This is due to the fact that the four-fermion operators give a contribution to the amplitude that grows quadratically with the centre-of-mass energy as explained in \autoref{ssec:interpretations_theory}. The contributions to the amplitude of the operators $Q_{\PQt\mathrm{B}}$ and $Q_{\PQt\PW}$ also grow with energy, but only linearly, and so the improvement is less significant. The sensitivity to the four-fermion operators improves by more than one order of magnitude with the high-energy stages.

By combining the sensitivities to the seven Wilson coefficients and the correlations among them, one can reconstruct the likelihood and use it for a rigorous reinterpretation of the result in explicit BSM scenarios where theoretical correlations are present among the operator coefficients. This is done for two top-quark compositeness scenarios as presented in \autoref{ssec:topcompositeness}.

\subsection{Top-quark compositeness}
\label{ssec:topcompositeness}

Scenarios where supposedly elementary SM particles are actually composite bound states originating from more fundamental dynamics are particularly well-motivated for the Higgs boson since Higgs compositeness at the TeV scale would arguably be the simplest solution to the Electroweak Naturalness Problem. The same is true also for the top quark, whose dynamics are strongly tied to that of the Higgs boson. Here we focus on the ``canonical'' scenario (see \cite{Panico:2015jxa} for a recent review) where the Higgs boson is a pseudo-Nambu--Goldstone boson of an underlying strongly-interacting composite sector, broadly characterised by a mass scale $m_*$ and by a coupling strength parameter $g_*$ \cite{Giudice:2007fh}. Top-quark compositeness emerges naturally in this framework \cite{Pomarol:2008bh}, and it is actually believed to be the only viable option to generate the top Yukawa coupling \cite{Rattazzi:2008pe}.

In this model, top-quark compositeness is characterised by two couplings, $y_L$ and $y_R$, that control the strength of the mixing of the $\PQq_L$ doublet and of the $\PQt_R$ singlet to the composite sector dynamics, respectively. In order for $y_{L,R}$ to produce a realistic top Yukawa coupling, they have to fulfil parametrically (i.e., up to dimensionless factors) the relation
\begin{equation}\label{ytop_par}
\displaystyle
y_t\simeq\frac{y_Ly_R}{g_*}.
\end{equation}
There are infinitely many ways to satisfy \autoref{ytop_par}; the only constraint is $y_{L,R}\leq g_*$ (out of which one derives $g_*>y_{\PQt}$) for internal consistency of the construction. It is however sufficient to focus on two benchmark scenarios:
\begin{eqnarray}
\displaystyle
&{\textrm{Partial compositeness:}}\;\; & y_L=y_R=\sqrt{y_{\PQt} g_*}\,, \label{part}\\
&{\textrm{Total $t_R$ compositeness:}}\;\; & y_L=y_{\PQt},\;y_R=g_*\,. \label{tot}
\end{eqnarray}
The names reflect the fact that the presence of the $y_{L,R}$ mixings make the physical $\PQq_L$ and $\PQt_R$ particles an admixture of elementary and composite degrees of freedom, with a fraction of compositeness $y_L/g_*$ and $y_R/g_*$, respectively. For the partial compositeness scenarios in \autoref{part}, the two chiralities possess the same compositeness fraction, while for total $\PQt_R$ compositeness scenarios in \autoref{tot}, the right-handed component is a fully composite state and the left-handed one is mostly elementary. The third logical possibility would be fully composite $\PQq_L=\{\PQt_L,\PQb_L\}$; however, this is disfavoured by the constraints from bottom quark physics.

The couplings $y_L$, $y_R$ and $g_*$ control, together with the mass scale $m_*$ of the BSM particles, the expected magnitude of the $d=6$ EFT operators that are obtained in models of this sort by integrating out the composite sector dynamics. The power-counting formula reads
\begin{equation}\label{power_counting}
{\mathcal{L}}^{\,d=6}=\frac1{m_*^2}\frac1{g_*^2}\,{\widehat{\hspace{.001em}\mathcal{L}}}\,[y_L q_L,\,y_R t_R,\,g_*\varphi,\,g_V V_\mu]\,,
\end{equation}
where ${\widehat{\hspace{.001em}\mathcal{L}}}$ is a generic (gauge-invariant) $d=6$ local functional of the fields and their derivatives with dimensionless coefficients. The Higgs field $\varphi$ is fully composite, hence it appears with coupling $g_*$. The SM gauge fields, which are collectively denoted as $V$, are elementary and couple through gauge interactions, i.e. with the ordinary SM gauge couplings $g_V$. The composite sector coupling $g_*$ is necessarily larger than $y_t$, as previously mentioned, but apart from this it can a priori assume any value up to the maximal coupling strength of $4\pi$. Large $g_*$ values are perfectly plausible or even expected because of the underlying strongly interacting nature of the composite sector. It is thus important to explore the whole range $g_*\in [y_t,\,4\pi]$, including values well above $1$, for which \autoref{power_counting} predicts large departures from the naive $1/m_*^2$ estimate of the Wilson coefficients. 

Based on \autoref{power_counting}, the coefficients of the top-philic operators in \autoref{tab:operators} can be estimated up to dimensionless coefficients, denoted as ``$\PGg$''. The coefficients are grouped into four categories:

\noindent{\bf{Higgs current operators:}}
\begin{equation}
C_{\varphi{t}}=y_R^2\PGg_{\varphi{t}}\frac1{m_*^2}\,,\;\;C_{\varphi{q}}^{(1)}=y_L^2\PGg_{\varphi{q}}^{(1)}\frac1{m_*^2}\,,\;\;C_{\varphi{q}}^{(3)}=y_L^2\PGg_{\varphi{q}}^{(3)}\frac1{m_*^2}\,
\end{equation}
Most models enjoy an accidental symmetry, dubbed $P_{LR}$ custodial \cite{Agashe:2006at}, that enforces $C_{\varphi{q}}^{(1)}+C_{\varphi{q}}^{(3)}=0$ up to small loop corrections. Hence in what follows we will set $\PGg_{\varphi{q}}^{(1)}=-\PGg_{\varphi{q}}^{(3)}$ and consequently we ignore the $\PQb\PAQb\PZ$ constraint in \autoref{zbb}. Ordinary custodial symmetry can also suppress $\PGg_{\varphi{t}}$, which are thus also set conservatively to zero. At large $g_*$ these operators (or at least the remaining combination) are significantly enhanced, particularly so in the total $\PQt_R$ compositeness scenario in \autoref{tot}.

\noindent{\bf{Modified top Yukawa:}} 
\begin{equation}
C_{t\varphi}=y_Ly_R g_* \PGg_{t\varphi}\frac1{m_*^2}=y_t g_*^2 \PGg_{t\varphi}\frac1{m_*^2} \,
\end{equation}
where we used \autoref{ytop_par}. Note the quadratic enhancement with $g_*$, which boosts the sensitivity to $m_*$ of the $y_t$ determination in \autoref{deltaytop}. 

\noindent{\bf{Four-fermion operators:}} 
\begin{equation}
C_{lt,B}={g^\prime}^2 \frac{y_R^2}{g_*^2} \PGg_{lt,B}\frac1{m_*^2}\,\;\;\;\;\;
C_{lq,B}={g^\prime}^2 \frac{y_L^2}{g_*^2} \PGg_{lq,B}\frac1{m_*^2}\,\;\;\;\;\;
C_{lq,W}=g^2 \frac{y_L^2}{g_*^2} \PGg_{lq,W}\frac1{m_*^2}\,
\end{equation}
These operators emerge from couplings with the gauge fields, as indicated in \autoref{tab:operators}, hence their coefficient is proportional to the corresponding gauge coupling. They are subsequently converted into four-fermion operators through equations of motion, and this brings a second power of the gauge coupling. Their coefficients are all suppressed at large $g_*$ in the partial compositeness scenario, while for total $t_R$ compositeness one of them, $C_{lt,B}$, is unsuppressed. 

\noindent{\bf{Dipole operators:}} 
\begin{equation}
C_{tB}=y_t {g^\prime} \PGg_{tB}\frac1{m_*^2}\left(\frac{g_*^2}{16\pi^2}\right)\,\;\;\;\;\;
C_{tW}=y_t {g} \PGg_{tW}\frac1{m_*^2}\left(\frac{g_*^2}{16\pi^2}\right)\,
\end{equation}
The loop suppression factor clearly does not emerge from the power-counting formula in \autoref{power_counting}. It is included because in all known perturbative (i.e., $g_*<4\pi$) realisations of the composite Higgs scenario these operators cannot be generated at tree-level. While we cannot exclude the existence of perturbative models that do not experience such a suppression, we include the loop factor in order to obtain a conservative estimate of the sensitivity.

\begin{figure}
\begin{center}
\includegraphics[width=.48\columnwidth]{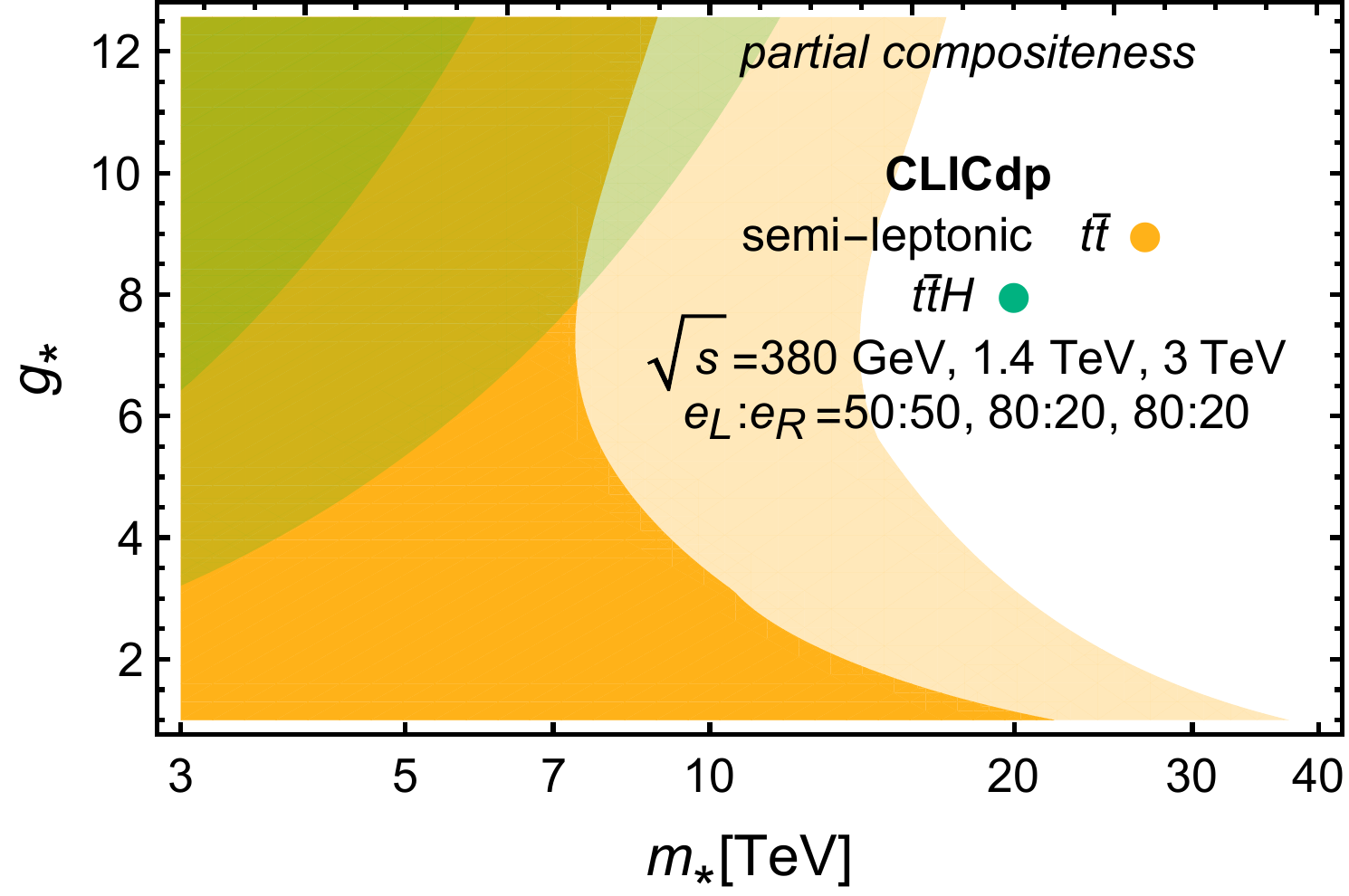}
~~~~
\includegraphics[width=.48\columnwidth]{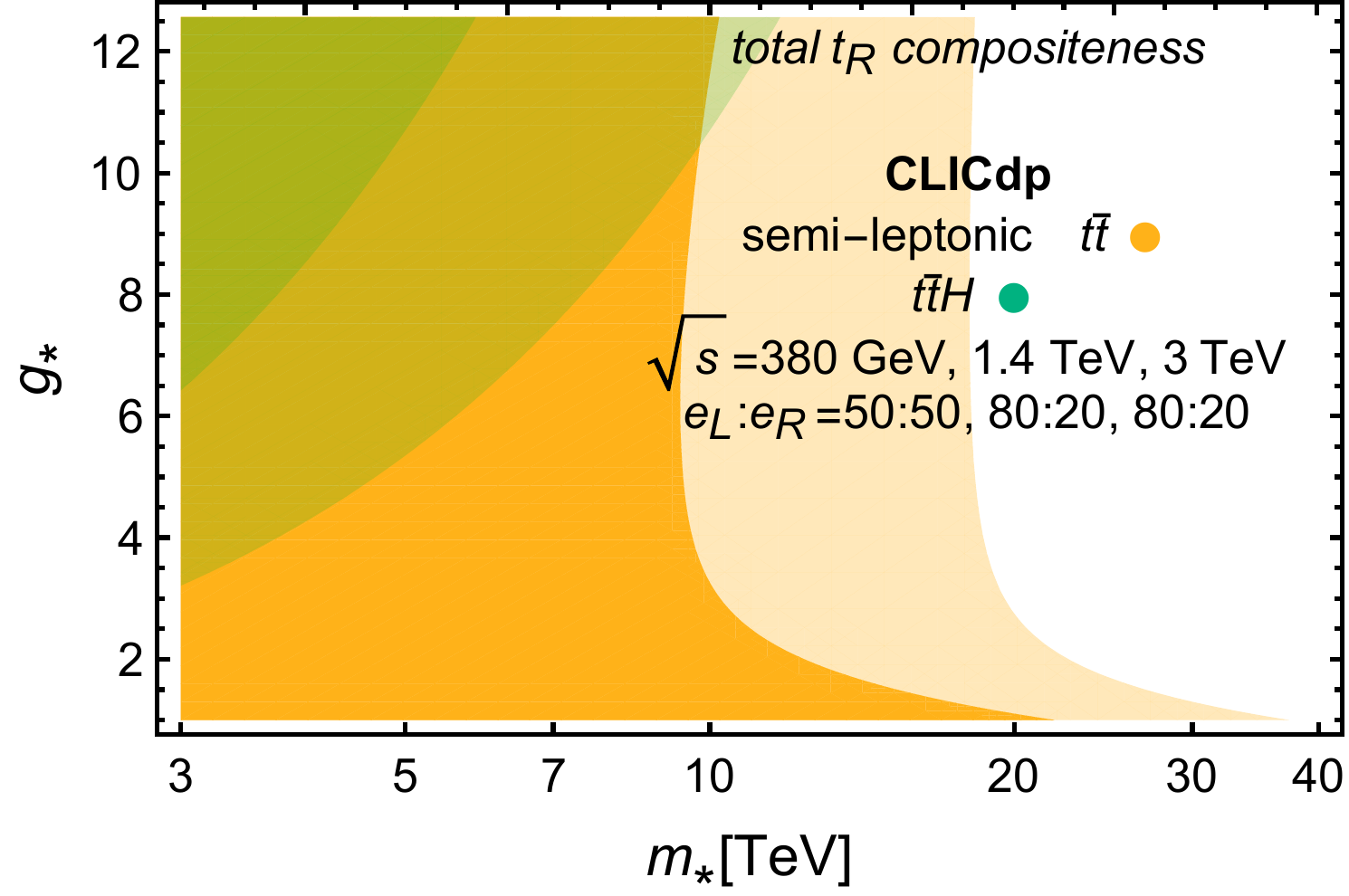}
\caption{``Optimistic'' (light color) and ``pessimistic'' (dark color) $5\sigma$ discovery regions for the partial compositeness (left) and the total $t_R$ compositeness (right) scenarios. The orange contours are derived from the $\ttbar$ global fit in \autoref{ssec:topphilicinterp}, while the green contour is derived from the top Yukawa analysis in \autoref{deltaytop}. \label{fig:limits_2}}
\end{center}
\end{figure}

The results from the global EFT fit reported in \autoref{ssec:topphilicinterp} were employed for an assessment of the CLIC exclusion and discovery reach of the top-quark compositeness scenarios at 3\,TeV. The results, focussing on $5\sigma$ discovery for brevity, are reported in \autoref{fig:limits_2}. The figure is obtained by varying the dimensionless parameters $\PGg$ in the range $[1/2,2]$ independently, and taking the values that minimise or maximise the likelihood at each point of the $(m_*,g_*)$ plane. This produces two discovery contours, that each correspond to optimistic and pessimistic assumptions on the values of $\PGg$ that will be obtained in the underlying UV theory. The impact of the top-quark pair production study (from \autoref{fig:limits_summary}) is shown separately from the top Yukawa determination in \autoref{deltaytop}.

The four-fermions operators, due to the very strong sensitivity from operation at $\roots=3\,\tev$ play a dominant role in the reach. The conclusion is that top-quark compositeness can be discovered at CLIC up to $8$~TeV, and more than $20$~TeV can be reached in favourable configurations. In particular, top-quark compositeness emerging in connection to the Naturalness Problem can be conclusively probed at CLIC.

\section{Summary and conclusions}
\label{sec:summary}

A detailed study of the top-quark physics reach of CLIC has been presented in this paper, in the context of CLIC operating at three energy stages: 1.0\,\abinv at \roots=380\,\gev, 2.5\,\abinv at 1.4\,\tev and 5.0\,\abinv at 3\,\tev.

The initial stage of operation includes an energy scan in the top-quark pair production threshold region, which allows the top-quark mass to be extracted in a theoretically well-defined manner with a precision of around 50\,\mev assuming an integrated luminosity of 100\,\fbinv. Additional mass measurements at 380\,\gev, based on the reconstruction of hadronic top-quark decays or the energy spectrum of ISR photons with complementary systematic uncertainties, might improve the overall understanding of the top-quark mass further. The projections presented are limited by theoretical uncertainties and CLIC would benefit from future theoretical work focusing on the top-quark pair production threshold region.

The large number of top quarks produced at 380\,\gev in combination with the relatively low background levels and $\PQc$-tagging capabilities of the CLIC detector concepts allows competitive searches for FCNC decays with charm quarks in the final state, such as $\PQt\to\PQc\PGg$ and $\PQt\to\PQc\PH$.

Pair production of the top quark in electron-positron collisions gives indirect sensitivity to new physics contributions. At 380\,\gev the jets from top-quark decays are well separated and can be reconstructed individually. Boosted top tagging techniques based on jet substructure information are needed to reconstruct top-quark pair production events with sufficient precision at the higher-energy CLIC stages. Tagging efficiencies for hadronically decaying top quarks in the boosted regime of 70\% are achieved thanks to the low background levels, and the high granularity and excellent jet energy resolution of the detector concepts optimised for PFA.

A global interpretation of top-quark pair production using seven Wilson coefficients requires at least two energy stages. New physics scales of the order of tens of \tev can be reached. The results of the EFT fit have been used to assess the CLIC sensitivity for top-quark compositeness, where the reach extends to compositeness scales of up to about 10\,\tev.

The higher-energy stages also allow the study of top-quark pair production in association with other particles. At 1.4\,\tev the top Yukawa coupling can be measured with a precision of 2.7\% using \ttH events. In addition, the \ttH process allows searches to be made for a CP-odd contribution to the $\ttH$ coupling. Further, operation at 3\,\tev gives access to top-quark pair production in vector boson fusion. 

Final states with top-quark pairs are very complex and the performance is often limited by reconstruction issues like the confusion in the jet clustering. Here, CLIC would profit from further improvements of the jet reconstruction algorithms and boosted top-tagging strategies.

Overall, the assumed energy staging scenario is well suited for exploiting the CLIC potential in the area of top physics. The energy of the first stage is high enough to avoid large theoretical uncertainties from threshold effects in the interpretation of top-quark pair production. While the maximum of the \ttH cross section at about 800\,\gev is below the centre-of-mass energy of the second CLIC stage, the decreasing cross section is compensated by fact that the luminosity of a linear collider rises with energy. A similar precision of the top Yukawa coupling measurement is possible at 1.4\,\tev compared to operation near the maximum of the \ttH cross section. The sensitivity of top-quark pair production, including vector boson fusion, to new physics contributions benefits from the largest possible centre-of-mass energy. The unique capability of CLIC to reach an energy of 3\,\tev substantially enhances the reach of these measurements.

\acknowledgments
The CLICdp collaboration gratefully acknowledges CERN for its continued support;
This work benefited from services provided by the ILC Virtual Organisation, supported by the national resource providers of the EGI Federation. This research was done using resources provided by the Open Science Grid, which is supported by the National Science Foundation and the U.S. Department of Energy's Office of Science.
This work was supported by
the European Union's Horizon 2020 Research and Innovation programme under Grant Agreement No.\,654168;
the European Commission through the Marie Curie Career Integration Grant 631962;
the Helmholtz Association, Germany;
the DFG Collaborative Research Centre ``Particles, Strings and the Early Universe'', Germany;
the DFG cluster of excellence ``Origin and Structure of the Universe'', Germany;
the German -- Israel Foundation (GIF); the Israel Science Foundation (ISF);
the I-CORE Program, Israel;
the Israel Academy of Sciences;
the Research Council of Norway;
the Institute of High Energy Physics, China under Contract No.\,Y7515540U1;
the National Science Centre, Poland, HARMONIA project under contract UMO-2015/18/M/ST2/00518 and OPUS project under contract UMO-2017/25/B/ST2/00496;
the Spanish Ministry of Economy, Industry and Competitiveness under projects MINEICO/FEDER-UE, FPA2015-65652-C4-3-R, FPA2015-71292-C2-1-P and FPA2015-71956-REDT;
the IFT Centro de Excelencia Severo Ochoa program, under grants SEV-2012-0249 and SEV-2014-0398, Spain;
the Spanish MINECO Ram\'on y Cajal program (RYC-2014-16022) and the MECD grant FPA2016-78645-P, Spain;
the Ministry of Education, Science and Technological Development of the Republic of Serbia under contract No.\,OI171012;
the UK Science and Technology Facilities Council (STFC), United Kingdom;
Gonville and Caius College, United Kingdom;
and the U.S. Department of Energy, Office of Science under contract DE-AC02-06CH11357.
% The bibliography will probably be heavily edited during typesetting.
% We'll parse it and, using the arxiv number or the journal data, will
% query inspire, trying to verify the data (this will probalby spot
% eventual typos) and retrive the document DOI and eventual errata.
% We however suggest to always provide author, title and journal data:
% in short all the informations that clearly identify a document.

\bibliographystyle{JHEP}
\bibliography{./bibliography/bibliography.bib}

\providecommand{\href}[2]{#2}\begingroup\raggedright\begin{thebibliography}{100}

\bibitem{staging_baseline_yellow_report}
P.~Burrows et~al., eds., \emph{{Updated baseline for a staged Compact Linear
  Collider, \href{http://dx.doi.org/10.5170/CERN-2016-004}{CERN-2016-004}}}.
  {CERN}, 2016, [\href{https://arxiv.org/abs/1608.07537}{{\ttfamily
  1608.07537}}].

\bibitem{Corsini:CTF3}
R.~Corsini, \emph{{Final Results from the CLIC Test Facility (CTF3)}},  in
  \emph{{\href{https://dx.doi.org/10.18429/JACoW-IPAC2017-TUZB1}{Proc. 8th Int.
  Particle Accelerator Conf.}, Copenhagen, Denmark}}, 2017.

\bibitem{CLICCDR_vol1}
M.~Aicheler et~al., eds., \emph{{A Multi-TeV Linear Collider based on CLIC
  Technology: CLIC Conceptual Design Report,
  \href{http://dx.doi.org/10.5170/CERN-2012-007}{CERN-2012-007}}}. {CERN},
  2012.

\bibitem{CLIC_PhysDet_CDR}
L.~Linssen et~al., eds., \emph{{Physics and Detectors at CLIC: CLIC Conceptual
  Design Report,
  \href{http://dx.doi.org/10.5170/CERN-2012-003}{CERN-2012-003}}}. {CERN},
  2012.

\bibitem{Robson:2018zje}
A.~Robson and P.~Roloff, \emph{{Updated CLIC luminosity staging baseline and
  Higgs coupling prospects}}, {\emph{CLICdp-Note-2018-002} (2018) }
  [\href{https://arxiv.org/abs/1812.01644}{{\ttfamily 1812.01644}}].

\bibitem{Abramowicz:2016zbo}
H.~Abramowicz et~al., \emph{{Higgs physics at the CLIC electron-positron linear
  collider}}, \href{https://doi.org/10.1140/epjc/s10052-017-4968-5}{\emph{Eur.
  Phys. J.} {\bfseries C77} (2017) 475}
  [\href{https://arxiv.org/abs/1608.07538}{{\ttfamily 1608.07538}}].

\bibitem{ildloi:2009}
{\scshape Linear Collider ILD Concept Group -} collaboration, \emph{{The
  International Large Detector: Letter of Intent}},
  \href{https://doi.org/10.2172/975166}{\emph{FERMILAB-LOI-2010-03,
  FERMILAB-PUB-09-682-E, DESY-09-87, KEK-REPORT-2009-6} (2010) }
  [\href{https://arxiv.org/abs/1006.3396}{{\ttfamily 1006.3396}}].

\bibitem{ilctdrvol4:2013}
H.~Abramowicz et~al., \emph{{The International Linear Collider Technical Design
  Report - Volume 4: Detectors}}, {\emph{ILC-REPORT-2013-040} (2013) }
  [\href{https://arxiv.org/abs/1306.6329}{{\ttfamily 1306.6329}}].

\bibitem{Aihara:2009ad}
H.~Aihara, P.~Burrows, M.~Oreglia, E.~L. Berger, V.~Guarino, J.~Repond et~al.,
  \emph{{SiD Letter of Intent}}, {\emph{SLAC-R-989, FERMILAB-LOI-2009-01,
  FERMILAB-PUB-09-681-E} (2009) }
  [\href{https://arxiv.org/abs/0911.0006}{{\ttfamily 0911.0006}}].

\bibitem{Poss:2013oea}
S.~Poss and A.~Sailer, \emph{{Luminosity Spectrum Reconstruction at Linear
  Colliders}}, \href{https://doi.org/10.1140/epjc/s10052-014-2833-3}{\emph{Eur.
  Phys. J.} {\bfseries C74} (2014) 2833}
  [\href{https://arxiv.org/abs/1309.0372}{{\ttfamily 1309.0372}}].

\bibitem{Bach:2017ggt}
F.~Bach, B.~C. Nejad, A.~H. Hoang, W.~Kilian, J.~Reuter, M.~Stahlhofen et~al.,
  \emph{Fully-differential top-pair production at a lepton collider: from
  threshold to continuum},
  \href{https://doi.org/10.1007/JHEP03(2018)184}{\emph{JHEP} {\bfseries 03}
  (2018) 184} [\href{https://arxiv.org/abs/1712.02220}{{\ttfamily
  1712.02220}}].

\bibitem{Fuster2015}
J.~Fuster, I.~Garc{\'\i}a, P.~Gomis, M.~Perell{\'o}, E.~Ros and M.~Vos,
  \emph{{Study of single top production at high energy electron positron
  colliders}}, \href{https://doi.org/10.1140/epjc/s10052-015-3453-2}{\emph{Eur.
  Phys. J.} {\bfseries C75} (2015) 223}
  [\href{https://arxiv.org/abs/1411.2355}{{\ttfamily 1411.2355}}].

\bibitem{Butenschoen:2016lpz}
M.~Butenschoen, B.~Dehnadi, A.~H. Hoang, V.~Mateu, M.~Preisser and I.~W.
  Stewart, \emph{{Top Quark Mass Calibration for Monte Carlo Event
  Generators}},
  \href{https://doi.org/10.1103/PhysRevLett.117.232001}{\emph{Phys. Rev. Lett.}
  {\bfseries 117} (2016) 232001}
  [\href{https://arxiv.org/abs/1608.01318}{{\ttfamily 1608.01318}}].

\bibitem{Hoang:1999zc}
A.~H. Hoang and T.~Teubner, \emph{{Top-quark pair production close to
  threshold: Top-quark mass, width and momentum distribution}},
  \href{https://doi.org/10.1103/PhysRevD.60.114027}{\emph{Phys. Rev.}
  {\bfseries D60} (1999) 114027}
  [\href{https://arxiv.org/abs/hep-ph/9904468}{{\ttfamily hep-ph/9904468}}].

\bibitem{Beneke:1998rk}
M.~Beneke, \emph{{A quark mass definition adequate for threshold problems}},
  \href{https://doi.org/10.1016/S0370-2693(98)00741-2}{\emph{Phys. Lett.}
  {\bfseries B434} (1998) 115}
  [\href{https://arxiv.org/abs/hep-ph/9804241}{{\ttfamily hep-ph/9804241}}].

\bibitem{Marquard:2015qpa}
P.~Marquard, A.~V. Smirnov, V.~A. Smirnov and M.~Steinhauser, \emph{{Quark Mass
  Relations to Four-Loop Order in Perturbative QCD}},
  \href{https://doi.org/10.1103/PhysRevLett.114.142002}{\emph{Phys. Rev. Lett.}
  {\bfseries 114} (2015) 142002}
  [\href{https://arxiv.org/abs/1502.01030}{{\ttfamily 1502.01030}}].

\bibitem{Aad:2015waa}
{\scshape ATLAS} collaboration, \emph{{Determination of the top-quark pole mass
  using $ t\overline{t} $ + 1-jet events collected with the ATLAS experiment in
  7 TeV pp collisions}},
  \href{https://doi.org/10.1007/JHEP10(2015)121}{\emph{JHEP} {\bfseries 10}
  (2015) 121} [\href{https://arxiv.org/abs/1507.01769}{{\ttfamily
  1507.01769}}].

\bibitem{Khachatryan:2016mqs}
{\scshape CMS} collaboration, \emph{{Measurement of the $ t\overline{t} $
  production cross section in the e-$\mu$ channel in proton-proton collisions
  at $\sqrt{s}$ = 7 and 8 TeV}},
  \href{https://doi.org/10.1007/JHEP08(2016)029}{\emph{JHEP} {\bfseries 08}
  (2016) 029} [\href{https://arxiv.org/abs/1603.02303}{{\ttfamily
  1603.02303}}].

\bibitem{Fadin:1987wz}
V.~S. Fadin and V.~A. Khoze, \emph{{Threshold Behavior of Heavy Top Production
  in e+ e- Collisions}}, {\emph{JETP Lett.} {\bfseries 46} (1987) 525}.

\bibitem{Fadin:1988fn}
V.~S. Fadin and V.~A. Khoze, \emph{{Production of a pair of heavy quarks in e+
  e- annihilation in the threshold region}}, {\emph{Sov. J. Nucl. Phys.}
  {\bfseries 48} (1988) 309}.

\bibitem{Strassler:1990nw}
M.~J. Strassler and M.~E. Peskin, \emph{{The Heavy top quark threshold: QCD and
  the Higgs}}, \href{https://doi.org/10.1103/PhysRevD.43.1500}{\emph{Phys.
  Rev.} {\bfseries D43} (1991) 1500}.

\bibitem{Beneke:2016kkb}
M.~Beneke, Y.~Kiyo, A.~Maier and J.~Piclum, \emph{{Near-threshold production of
  heavy quarks with QQbar\_threshold}},
  \href{https://doi.org/10.1016/j.cpc.2016.07.026}{\emph{Comput. Phys. Commun.}
  {\bfseries 209} (2016) 96}
  [\href{https://arxiv.org/abs/1605.03010}{{\ttfamily 1605.03010}}].

\bibitem{Hoang:2013uda}
A.~H. Hoang and M.~Stahlhofen, \emph{{The Top-Antitop Threshold at the ILC:
  NNLL QCD Uncertainties}},
  \href{https://doi.org/10.1007/JHEP05(2014)121}{\emph{JHEP} {\bfseries 05}
  (2014) 121} [\href{https://arxiv.org/abs/1309.6323}{{\ttfamily 1309.6323}}].

\bibitem{Gusken:1985nf}
S.~Gusken, J.~H. Kuhn and P.~M. Zerwas, \emph{{Threshold behavior of top
  production in $e^+ e^-$ annihilation}},
  \href{https://doi.org/10.1016/0370-2693(85)90983-9}{\emph{Phys. Lett.}
  {\bfseries B155} (1985) 185}.

\bibitem{Bigi:1986jk}
I.~I.~Y. Bigi, Y.~L. Dokshitzer, V.~A. Khoze, J.~H. Kuhn and P.~M. Zerwas,
  \emph{{Production and decay properties of ultraheavy quarks}},
  \href{https://doi.org/10.1016/0370-2693(86)91275-X}{\emph{Phys. Lett.}
  {\bfseries B181} (1986) 157}.

\bibitem{Martinez:2002st}
M.~Martinez and R.~Miquel, \emph{{Multiparameter fits to the $t\bar{t}$
  threshold observables at a future $\epem$ linear collider}},
  \href{https://doi.org/https://doi.org/10.1140/epjc/s2002-01094-1}{\emph{Eur.
  Phys. J.} {\bfseries C27} (2003) 49}
  [\href{https://arxiv.org/abs/hep-ph/0207315}{{\ttfamily hep-ph/0207315}}].

\bibitem{Seidel:2013sqa}
K.~Seidel, F.~Simon, M.~Tesar and S.~Poss, \emph{{Top quark mass measurements
  at and above threshold at CLIC}},
  \href{https://doi.org/10.1140/epjc/s10052-013-2530-7}{\emph{Eur. Phys. J.}
  {\bfseries C73} (2013) 2530}
  [\href{https://arxiv.org/abs/1303.3758}{{\ttfamily 1303.3758}}].

\bibitem{Horiguchi:2013wra}
T.~Horiguchi, A.~Ishikawa, T.~Suehara, K.~Fujii, Y.~Sumino, Y.~Kiyo et~al.,
  \emph{{Study of top quark pair production near threshold at the ILC}},
  {\emph{{}} (2013) } [\href{https://arxiv.org/abs/1310.0563}{{\ttfamily
  1310.0563}}].

\bibitem{Gao:2014eea}
J.~Gao and H.~X. Zhu, \emph{{Top Quark Forward-Backward Asymmetry in $e^+e^-$
  Annihilation at Next-to-Next-to-Leading Order in QCD}},
  \href{https://doi.org/10.1103/PhysRevLett.113.262001}{\emph{Phys. Rev. Lett.}
  {\bfseries 113} (2014) 262001}
  [\href{https://arxiv.org/abs/1410.3165}{{\ttfamily 1410.3165}}].

\bibitem{Gao:2014nva}
J.~Gao and H.~X. Zhu, \emph{{Electroweak prodution of top-quark pairs in
  $e^+e^-$ annihilation at NNLO in QCD: the vector current contributions}},
  \href{https://doi.org/10.1103/PhysRevD.90.114022}{\emph{Phys. Rev.}
  {\bfseries D90} (2014) 114022}
  [\href{https://arxiv.org/abs/1408.5150}{{\ttfamily 1408.5150}}].

\bibitem{Chen:2016zbz}
L.~Chen, O.~Dekkers, D.~Heisler, W.~Bernreuther and Z.-G. Si, \emph{{Top-quark
  pair production at next-to-next-to-leading order QCD in electron positron
  collisions}}, \href{https://doi.org/10.1007/JHEP12(2016)098}{\emph{JHEP}
  {\bfseries 12} (2016) 098}
  [\href{https://arxiv.org/abs/1610.07897}{{\ttfamily 1610.07897}}].

\bibitem{Lei:2008ii}
L.~Guo, W.-G. Ma, R.-Y. Zhang and S.-M. Wang, \emph{{One-loop QCD corrections
  to the $e^+ e^- \to W^+ W^- b \overline{b}$ process at the ILC}},
  \href{https://doi.org/10.1016/j.physletb.2008.02.058}{\emph{Phys. Lett.}
  {\bfseries B662} (2008) 150}
  [\href{https://arxiv.org/abs/0802.4124}{{\ttfamily 0802.4124}}].

\bibitem{Liebler:2015ipp}
S.~Liebler, G.~Moortgat-Pick and A.~S. Papanastasiou, \emph{{Probing the
  top-quark width through ratios of resonance contributions of
  $e^+e^-\rightarrow W^+W^-b\bar{b}$}},
  \href{https://doi.org/10.1007/JHEP03(2016)099}{\emph{JHEP} {\bfseries 03}
  (2016) 099} [\href{https://arxiv.org/abs/1511.02350}{{\ttfamily
  1511.02350}}].

\bibitem{Nejad:2016bci}
B.~Chokouf{\'e}~Nejad, W.~Kilian, J.~M. Lindert, S.~Pozzorini, J.~Reuter and
  C.~Weiss, \emph{{NLO QCD predictions for off-shell $ t\overline{t} $ and $
  t\overline{t}H $ production and decay at a linear collider}},
  \href{https://doi.org/10.1007/JHEP12(2016)075}{\emph{JHEP} {\bfseries 12}
  (2016) 075} [\href{https://arxiv.org/abs/1609.03390}{{\ttfamily
  1609.03390}}].

\bibitem{Weiss:2015npa}
C.~Weiss, B.~Chokoufe~Nejad, W.~Kilian and J.~Reuter, \emph{{Automated NLO QCD
  Corrections with WHIZARD}},
  \href{https://doi.org/10.22323/1.234.0466}{\emph{PoS EPS-HEP2015} (2015) 466}
  [\href{https://arxiv.org/abs/1510.02666}{{\ttfamily 1510.02666}}].

\bibitem{Alwall:2014hca}
J.~Alwall, R.~Frederix, S.~Frixione, V.~Hirschi, F.~Maltoni, O.~Mattelaer
  et~al., \emph{{The automated computation of tree-level and next-to-leading
  order differential cross sections, and their matching to parton shower
  simulations}}, \href{https://doi.org/10.1007/JHEP07(2014)079}{\emph{JHEP}
  {\bfseries 07} (2014) 079} [\href{https://arxiv.org/abs/1405.0301}{{\ttfamily
  1405.0301}}].

\bibitem{Beenakker:1991ca}
W.~Beenakker, S.~C. van~der Marck and W.~Hollik, \emph{{$e^+ e^-$ annihilation
  into heavy fermion pairs at high-energy colliders}},
  \href{https://doi.org/10.1016/0550-3213(91)90606-X}{\emph{Nucl. Phys.}
  {\bfseries B365} (1991) 24}.

\bibitem{Fleischer:2003kk}
J.~Fleischer, A.~Leike, T.~Riemann and A.~Werthenbach, \emph{{Electroweak one
  loop corrections for $e^+ e^-$ annihilation into $t \overline{t}$ including
  hard bremsstrahlung}},
  \href{https://doi.org/10.1140/epjc/s2003-01263-8}{\emph{Eur. Phys. J.}
  {\bfseries C31} (2003) 37}
  [\href{https://arxiv.org/abs/hep-ph/0302259}{{\ttfamily hep-ph/0302259}}].

\bibitem{Hahn:2003ab}
T.~Hahn, W.~Hollik, A.~Lorca, T.~Riemann and A.~Werthenbach, \emph{{O($\alpha$)
  electroweak corrections to the processes $e^+ e^- \to \tau^-
  \tau^+,\,\,c\overline{c},\,\,b\overline{b},\,\,t\overline{t}$: a
  comparison}},  in \emph{{\href{http://cds.cern.ch/record/628077}{Proc. 4th
  ECFA / DESY Workshop on Physics and Detectors for a 90-GeV to 800-GeV Linear
  e$^+$ e$^-$ Collider}, Amsterdam, Netherlands}}, 2003.

\bibitem{Khiem:2012bp}
P.~H. Khiem, J.~Fujimoto, T.~Ishikawa, T.~Kaneko, K.~Kato, Y.~Kurihara et~al.,
  \emph{{Full $\mathcal{O}(\alpha)$ electroweak radiative corrections to
  $e^+e^- \rightarrow t \bar{t} \gamma$ with GRACE-Loop}},
  \href{https://doi.org/10.1140/epjc/s10052-013-2400-3}{\emph{Eur. Phys. J.}
  {\bfseries C73} (2013) 2400}
  [\href{https://arxiv.org/abs/1211.1112}{{\ttfamily 1211.1112}}].

\bibitem{Pecjak:2016nee}
B.~D. Pecjak, D.~J. Scott, X.~Wang and L.~L. Yang, \emph{{Resummed differential
  cross sections for top-quark pairs at the LHC}},
  \href{https://doi.org/10.1103/PhysRevLett.116.202001}{\emph{Phys. Rev. Lett.}
  {\bfseries 116} (2016) 202001}
  [\href{https://arxiv.org/abs/1601.07020}{{\ttfamily 1601.07020}}].

\bibitem{Fleming:2007qr}
S.~Fleming, A.~H. Hoang, S.~Mantry and I.~W. Stewart, \emph{{Jets from massive
  unstable particles: Top-mass determination}},
  \href{https://doi.org/10.1103/PhysRevD.77.074010}{\emph{Phys. Rev.}
  {\bfseries D77} (2008) 074010}
  [\href{https://arxiv.org/abs/hep-ph/0703207}{{\ttfamily hep-ph/0703207}}].

\bibitem{Fleming:2007tv}
S.~Fleming, A.~H. Hoang, S.~Mantry and I.~W. Stewart, \emph{{Factorization
  approach for top mass reconstruction at high energies}}, {\emph{eConf}
  {\bfseries C0705302} (2007) LOOP06}
  [\href{https://arxiv.org/abs/0710.4205}{{\ttfamily 0710.4205}}].

\bibitem{Grzadkowski:2010es}
B.~Grzadkowski, M.~Iskrzynski, M.~Misiak and J.~Rosiek, \emph{{Dimension-six
  terms in the Standard Model Lagrangian}},
  \href{https://doi.org/10.1007/JHEP10(2010)085}{\emph{JHEP} {\bfseries 10}
  (2010) 085} [\href{https://arxiv.org/abs/1008.4884}{{\ttfamily 1008.4884}}].

\bibitem{Barbieri:2004qk}
R.~Barbieri, A.~Pomarol, R.~Rattazzi and A.~Strumia, \emph{{Electroweak
  symmetry breaking after LEP1 and LEP2}},
  \href{https://doi.org/10.1016/j.nuclphysb.2004.10.014}{\emph{Nucl. Phys.}
  {\bfseries B703} (2004) 127}
  [\href{https://arxiv.org/abs/hep-ph/0405040}{{\ttfamily hep-ph/0405040}}].

\bibitem{Wells:2015uba}
J.~D. Wells and Z.~Zhang, \emph{{Effective theories of universal theories}},
  \href{https://doi.org/10.1007/JHEP01(2016)123}{\emph{JHEP} {\bfseries 01}
  (2016) 123} [\href{https://arxiv.org/abs/1510.08462}{{\ttfamily
  1510.08462}}].

\bibitem{Wells:2015cre}
J.~D. Wells and Z.~Zhang, \emph{{Renormalization group evolution of the
  universal theories EFT}},
  \href{https://doi.org/10.1007/JHEP06(2016)122}{\emph{JHEP} {\bfseries 06}
  (2016) 122} [\href{https://arxiv.org/abs/1512.03056}{{\ttfamily
  1512.03056}}].

\bibitem{AguilarSaavedra:2018nen}
D.~Barducci et~al., \emph{{Interpreting top-quark LHC measurements in the
  standard-model effective field theory}},
  {\emph{{\href{http://cds.cern.ch/record/2305783}{CERN-LPCC-2018-01}}} (2018)
  } [\href{https://arxiv.org/abs/1802.07237}{{\ttfamily 1802.07237}}].

\bibitem{AguilarSaavedra:2010zi}
J.~A. Aguilar-Saavedra, \emph{{Effective four-fermion operators in top physics:
  A roadmap}},
  \href{https://doi.org/10.1016/j.nuclphysb.2010.10.015}{\emph{Nucl. Phys.}
  {\bfseries B843} (2011) 638}
  [\href{https://arxiv.org/abs/1008.3562}{{\ttfamily 1008.3562}}].

\bibitem{Schmidt:1995mr}
C.~R. Schmidt, \emph{{Top quark production and decay at next-to-leading order
  in $e^+ e^-$ annihilation}},
  \href{https://doi.org/10.1103/PhysRevD.54.3250}{\emph{Phys. Rev.} {\bfseries
  D54} (1996) 3250} [\href{https://arxiv.org/abs/hep-ph/9504434}{{\ttfamily
  hep-ph/9504434}}].

\bibitem{Glashow:1970gm}
S.~L. Glashow, J.~Iliopoulos and L.~Maiani, \emph{{Weak Interactions with
  Lepton-Hadron Symmetry}},
  \href{https://doi.org/10.1103/PhysRevD.2.1285}{\emph{Phys. Rev.} {\bfseries
  D2} (1970) 1285}.

\bibitem{Cabibbo:1963yz}
N.~Cabibbo, \emph{{Unitary Symmetry and Leptonic Decays}},
  \href{https://doi.org/10.1103/PhysRevLett.10.531}{\emph{Phys. Rev. Lett.}
  {\bfseries 10} (1963) 531}.

\bibitem{Kobayashi:1973fv}
M.~Kobayashi and T.~Maskawa, \emph{{CP Violation in the Renormalizable Theory
  of Weak Interaction}}, \href{https://doi.org/10.1143/PTP.49.652}{\emph{Prog.
  Theor. Phys.} {\bfseries 49} (1973) 652}.

\bibitem{Agashe:2013hma}
K.~Agashe et~al., \emph{{Snowmass 2013 Top quark working group report}},  in
  \emph{{\href{https://arxiv.org/abs/1311.2028}{Proc. Community Summer Study
  2013: Snowmass on the Mississippi (CSS2013)}, Minneapolis, MN, USA}}, 2013.

\bibitem{deBlas:2018mhx}
J.~de~Blas et~al., eds., \emph{{The CLIC Potential for New Physics}}. {CERN},
  2018, [\href{https://arxiv.org/abs/1812.02093}{{\ttfamily 1812.02093}}].

\bibitem{Bejar:2001sj}
S.~Bejar, J.~Guasch and J.~Sola, \emph{{FCNC top quark decays beyond the
  standard model}},  in
  \emph{{\href{http://www.slac.stanford.edu/econf/C000911/}{Proc. 5th Int.
  Symp. on Radiative Corrections (RADCOR 2000)} Carmel, CA, USA}}, 2000.

\bibitem{DiazCruz:2006qy}
L.~Diaz-Cruz and C.~Pagliarone, \emph{{Perspectives of detecting CKM-suppressed
  top quark decays at ILC}},  in
  \emph{{\href{https://inspirehep.net/record/977829}{Proc. Int. Conf.: New
  Trends in High-Energy Physics}, Yalta, Crimea, Ukraine}}, 2006.

\bibitem{Bardhan:2016txk}
D.~Bardhan, G.~Bhattacharyya, D.~Ghosh, M.~Patra and S.~Raychaudhuri,
  \emph{{Detailed analysis of flavor-changing decays of top quarks as a probe
  of new physics at the LHC}},
  \href{https://doi.org/10.1103/PhysRevD.94.015026}{\emph{Phys. Rev.}
  {\bfseries D94} (2016) 015026}
  [\href{https://arxiv.org/abs/1601.04165}{{\ttfamily 1601.04165}}].

\bibitem{Mele:1999zx}
B.~Mele, \emph{{Top quark rare decays in the standard model and beyond}},  in
  \emph{{\href{https://inspirehep.net/record/537965}{Proc. 14th Int. Workshop,
  High energy physics and quantum field theory (QFTHEP'99)}, Moscow, Russia}},
  1999.

\bibitem{Kalinowski:2018ylg}
J.~Kalinowski, W.~Kotlarski, T.~Robens, D.~Sokolowska and A.~F. Zarnecki,
  \emph{{Benchmarking the Inert Doublet Model for $e^+ e^-$ colliders}},
  \href{https://doi.org/10.1007/JHEP12(2018)081}{\emph{JHEP} {\bfseries 12}
  (2018) 081} [\href{https://arxiv.org/abs/1809.07712}{{\ttfamily
  1809.07712}}].

\bibitem{Agostinelli2003}
{\scshape GEANT4} collaboration, \emph{{GEANT4: A simulation toolkit}},
  \href{https://doi.org/10.1016/S0168-9002(03)01368-8}{\emph{Nucl.\ Instrum.\
  Meth.} {\bfseries A506} (2003) 250}.

\bibitem{Allison2006}
J.~Allison et~al., \emph{{GEANT4 developments and applications}},
  \href{https://doi.org/10.1109/TNS.2006.869826}{\emph{IEEE Trans.\ Nucl.\
  Sci.} {\bfseries 53} (2006) 270}.

\bibitem{Grefe:2014sca}
{\scshape {{CLICdp}}} collaboration, \emph{{ILCDIRAC, a DIRAC extension for the
  Linear Collider community}},
  \href{https://doi.org/10.1088/1742-6596/513/3/032077}{\emph{J. Phys. Conf.
  Ser.} {\bfseries 513} (2014) 032077}.

\bibitem{Tsaregorodtsev:2008zz}
A.~Tsaregorodtsev et~al., \emph{{DIRAC: a community grid solution}},
  \href{https://doi.org/10.1088/1742-6596/119/6/062048}{\emph{J. Phys. Conf.
  Ser.} {\bfseries 119} (2008) 062048}.

\bibitem{Kilian:2007gr}
W.~Kilian, T.~Ohl and J.~Reuter, \emph{{WHIZARD: simulating multi-particle
  processes at LHC and ILC}},
  \href{https://doi.org/10.1140/epjc/s10052-011-1742-y}{\emph{Eur.\ Phys.\ J.}
  {\bfseries C71} (2011) 1742}
  [\href{https://arxiv.org/abs/0708.4233}{{\ttfamily 0708.4233}}].

\bibitem{Skrzypek:1990qs}
M.~Skrzypek and S.~Jadach, \emph{{Exact and approximate solutions for the
  electron nonsinglet structure function in QED}},
  \href{https://doi.org/10.1007/BF01483573}{\emph{Z. Phys.} {\bfseries C49}
  (1991) 577}.

\bibitem{Sjostrand2006}
T.~Sj{\"o}strand, S.~Mrenna and P.~Z. Skands, \emph{{PYTHIA 6.4 physics and
  manual}}, \href{https://doi.org/10.1088/1126-6708/2006/05/026}{\emph{JHEP}
  {\bfseries 0605} (2006) 026}
  [\href{https://arxiv.org/abs/hep-ph/0603175}{{\ttfamily hep-ph/0603175}}].

\bibitem{Alexander:1995bk}
{\scshape OPAL} collaboration, \emph{{A comparison of b and uds quark jets to
  gluon jets}}, \href{https://doi.org/10.1007/s002880050059}{\emph{Z. Phys.}
  {\bfseries C69} (1996) 543}.

\bibitem{Chekanov:2289960}
S.~Chekanov, M.~Demarteau, A.~Fischer and J.~Zhang, \emph{{Effect of PYTHIA8
  tunes on event shapes and top-quark reconstruction in e$^+$e$^-$ annihilation
  at CLIC, \href{http://cds.cern.ch/record/2289960}{CLICdp-Note-2017-005}}}.
  {CERN}, 2017, [\href{https://arxiv.org/abs/1710.07713}{{\ttfamily
  1710.07713}}].

\bibitem{tauola}
Z.~Was, \emph{{TAUOLA the library for tau lepton decay, and KKMC / KORALB /
  KORALZ /... status report}},
  \href{https://doi.org/10.1016/S0920-5632(01)01200-2}{\emph{Nucl.\ Phys.\
  Proc.\ Suppl.} {\bfseries 98} (2001) 96}
  [\href{https://arxiv.org/abs/hep-ph/0011305}{{\ttfamily hep-ph/0011305}}].

\bibitem{gen:physsim}
K.~Fujii, \emph{{Physics Study Libraries
  \url{http://www-jlc.kek.jp/subg/offl/physsim/}}}. {Accessed 25 Aug}, {2016}.

\bibitem{Dittmaier:2012vm}
S.~Dittmaier et~al., \emph{{Handbook of LHC Higgs Cross Sections: 2.
  Differential Distributions}},
  \href{https://doi.org/10.5170/CERN-2012-002}{\emph{CERN-2012-002} (2012) }
  [\href{https://arxiv.org/abs/1201.3084}{{\ttfamily 1201.3084}}].

\bibitem{Mokka}
P.~Mora~de Freitas and H.~Videau, \emph{{Detector simulation with MOKKA /
  GEANT4: Present and future
  \href{http://flc.desy.de/lcnotes/notes/LC-TOOL-2003-010.ps.gz}{LC-TOOL-2003-010}}},
  in \emph{Proc. Int. Workshop on Linear Colliders (LCWS 2002): JeJu Island,
  Korea}, 2002.

\bibitem{Graf:2006ei}
N.~Graf and J.~McCormick, \emph{{Simulator for the linear collider (SLIC): A
  tool for ILC detector simulations}},
  \href{https://doi.org/10.1063/1.2396991}{\emph{AIP Conf.\ Proc.} {\bfseries
  867} (2006) 503}.

\bibitem{MarlinLCCD}
F.~Gaede, \emph{{Marlin and LCCD: Software tools for the ILC}},
  \href{https://doi.org/10.1016/j.nima.2005.11.138}{\emph{Nucl.\ Instrum.\
  Meth.} {\bfseries A559} (2006) 177}.

\bibitem{Graf:2011zzc}
N.~A. Graf, \emph{{org.lcsim: Event reconstruction in Java}},
  \href{https://doi.org/10.1088/1742-6596/331/3/032012}{\emph{J.\ Phys.\ Conf.\
  Ser.} {\bfseries 331} (2011) 032012}.

\bibitem{guineapig}
D.~Schulte, \emph{{Study of Electromagnetic and Hadronic Background in the
  Interaction Region of the TESLA Collider,
  \href{DESY-TESLA-97-08}{DESY-TESLA-97-08}}}, Ph.D. thesis, Universit\"{a}t
  Hamburg, 1997.

\bibitem{thomson:pandora}
M.~Thomson, \emph{{Particle Flow Calorimetry and the PandoraPFA Algorithm}},
  \href{https://doi.org/10.1016/j.nima.2009.09.009}{\emph{Nucl.\ Instrum.\
  Meth.} {\bfseries A611} (2009) 25}
  [\href{https://arxiv.org/abs/0907.3577}{{\ttfamily 0907.3577}}].

\bibitem{Marshall2013153}
J.~Marshall, A.~M{\"u}nnich and M.~Thomson, \emph{{Performance of Particle Flow
  Calorimetry at CLIC}},
  \href{https://doi.org/10.1016/j.nima.2012.10.038}{\emph{Nucl.\ Instrum.\
  Meth.} {\bfseries A700} (2013) 153}
  [\href{https://arxiv.org/abs/1209.4039}{{\ttfamily 1209.4039}}].

\bibitem{Marshall:2015rfa}
J.~S. Marshall and M.~A. Thomson, \emph{{The Pandora Software Development Kit
  for Pattern Recognition}},
  \href{https://doi.org/10.1140/epjc/s10052-015-3659-3}{\emph{Eur. Phys. J.}
  {\bfseries C75} (2015) 439}
  [\href{https://arxiv.org/abs/1506.05348}{{\ttfamily 1506.05348}}].

\bibitem{LCDnote_TauFinder}
A.~M\"unnich, \emph{{Taufinder: A Reconstruction Algorithm for $\tau$ Leptons
  at Linear Colliders,
  \href{http://cds.cern.ch/record/1443551/}{LCD-Note-2010-009}}},  2010.

\bibitem{Fastjet}
M.~Cacciari, G.~P. Salam and G.~Soyez, \emph{{FastJet User Manual}},
  \href{https://doi.org/10.1140/epjc/s10052-012-1896-2}{\emph{Eur. Phys. J.}
  {\bfseries C72} (2012) 1896}
  [\href{https://arxiv.org/abs/1111.6097}{{\ttfamily 1111.6097}}].

\bibitem{Catani:1993hr}
S.~Catani, Y.~Dokshitzer, M.~Seymour and B.~Webber,
  \emph{{Longitudinally-invariant $k_\perp$-clustering algorithms for
  hadron-hadron collisions}},
  \href{https://doi.org/10.1016/0550-3213(93)90166-M}{\emph{Nucl. Phys. B}
  {\bfseries 406} (1993) 187 }.

\bibitem{Ellis:1993tq}
S.~D. Ellis and D.~E. Soper, \emph{{Successive combination jet algorithm for
  hadron collisions}},
  \href{https://doi.org/10.1103/PhysRevD.48.3160}{\emph{Phys. Rev.} {\bfseries
  D48} (1993) 3160} [\href{https://arxiv.org/abs/hep-ph/9305266}{{\ttfamily
  hep-ph/9305266}}].

\bibitem{Boronat:2016tgd}
M.~Boronat, J.~Fuster, I.~Garcia, P.~Roloff, R.~Simoniello and M.~Vos,
  \emph{{Jet reconstruction at high-energy lepton colliders}},
  \href{https://doi.org/10.1140/epjc/s10052-018-5594-6}{\emph{Eur.\ Phys.\ J.}
  {\bfseries C78} (2016) 144}
  [\href{https://arxiv.org/abs/1607.05039}{{\ttfamily 1607.05039}}].

\bibitem{Simon2015}
F.~Simon and L.~Weuste, \emph{{Light-flavor squark reconstruction at CLIC}},
  \href{https://doi.org/10.1140/epjc/s10052-015-3607-2}{\emph{Eur. Phys. J.}
  {\bfseries C75} (2015) 379}
  [\href{https://arxiv.org/abs/1505.01129}{{\ttfamily 1505.01129}}].

\bibitem{Suehara:2015ura}
T.~Suehara and T.~Tanabe, \emph{{LCFIPlus: A framework for jet analysis in
  linear collider studies}},
  \href{https://doi.org/10.1016/j.nima.2015.11.054}{\emph{Nucl. Instrum. Meth.}
  {\bfseries A808} (2016) 109}
  [\href{https://arxiv.org/abs/1506.08371}{{\ttfamily 1506.08371}}].

\bibitem{Nachman:2014kla}
B.~Nachman, P.~Nef, A.~Schwartzman, M.~Swiatlowski and C.~Wanotayaroj,
  \emph{{Jets from Jets: Re-clustering as a tool for large radius jet
  reconstruction and grooming at the LHC}},
  \href{https://doi.org/10.1007/JHEP02(2015)075}{\emph{JHEP} {\bfseries 02}
  (2015) 075} [\href{https://arxiv.org/abs/1407.2922}{{\ttfamily 1407.2922}}].

\bibitem{Kaplan:2008ie}
D.~E. Kaplan, K.~Rehermann, M.~D. Schwartz and B.~Tweedie, \emph{{Top Tagging:
  A Method for Identifying Boosted Hadronically Decaying Top Quarks}},
  \href{https://doi.org/10.1103/PhysRevLett.101.142001}{\emph{Phys. Rev. Lett.}
  {\bfseries 101} (2008) 142001}
  [\href{https://arxiv.org/abs/0806.0848}{{\ttfamily 0806.0848}}].

\bibitem{Fastjet:2006}
M.~Cacciari and G.~P. Salam, \emph{{Dispelling the $N^{3}$ myth for the $k_t$
  jet-finder}},
  \href{https://doi.org/10.1016/j.physletb.2006.08.037}{\emph{Phys. Lett.}
  {\bfseries B641} (2006) 57}.

\bibitem{CLIC_beam_web}
D.~Arominski, \emph{{CLIC Beam-Beam Interactions
  (\url{http://clic-beam-beam.web.cern.ch/clic-beam-beam/)}}}. {CERN}, 2018.

\bibitem{Simon:2014hna}
F.~Simon, \emph{{Perspectives for Top Quark Physics at the (I)LC}},  in
  \emph{{\href{http://www.slac.stanford.edu/econf/C1409281/}{Proc. 7th Int.
  Workshop on Top Quark Physics (TOP2014)}, Cannes, France}}, 2014.

\bibitem{Hoang:2004tg}
A.~H. Hoang and C.~J. Reisser, \emph{{Electroweak absorptive parts in NRQCD
  matching conditions}},
  \href{https://doi.org/10.1103/PhysRevD.71.074022}{\emph{Phys. Rev.}
  {\bfseries D71} (2005) 074022}
  [\href{https://arxiv.org/abs/hep-ph/0412258}{{\ttfamily hep-ph/0412258}}].

\bibitem{Hoang:2008ud}
A.~H. Hoang, C.~J. Reisser and P.~Ruiz-Femenia, \emph{{Implementing invariant
  mass cuts and finite lifetime effects in top-antitop production at
  threshold}},
  \href{https://doi.org/10.1016/j.nuclphysbps.2008.12.091}{\emph{Nucl. Phys.
  Proc. Suppl.} {\bfseries 186} (2009) 403}
  [\href{https://arxiv.org/abs/0810.2934}{{\ttfamily 0810.2934}}].

\bibitem{Hoang:2010gu}
A.~H. Hoang, C.~J. Reisser and P.~Ruiz-Femenia, \emph{{Phase Space Matching and
  Finite Lifetime Effects for Top-Pair Production Close to Threshold}},
  \href{https://doi.org/10.1103/PhysRevD.82.014005}{\emph{Phys. Rev.}
  {\bfseries D82} (2010) 014005}
  [\href{https://arxiv.org/abs/1002.3223}{{\ttfamily 1002.3223}}].

\bibitem{Beneke:2010mp}
M.~Beneke, B.~Jantzen and P.~Ruiz-Femenia, \emph{{Electroweak non-resonant NLO
  corrections to $e^+ e^- \to W^+ W^- b\bar{b}$ in the $t\bar{t}$ resonance
  region}}, \href{https://doi.org/10.1016/j.nuclphysb.2010.07.006}{\emph{Nucl.
  Phys.} {\bfseries B840} (2010) 186}
  [\href{https://arxiv.org/abs/1004.2188}{{\ttfamily 1004.2188}}].

\bibitem{Beneke:2017rdn}
M.~Beneke, A.~Maier, T.~Rauh and P.~Ruiz-Femenia, \emph{{Non-resonant and
  electroweak NNLO correction to the $e^+ e^-$ top anti-top threshold}},
  \href{https://doi.org/10.1007/JHEP02(2018)125}{\emph{JHEP} {\bfseries 02}
  (2018) 125} [\href{https://arxiv.org/abs/1711.10429}{{\ttfamily
  1711.10429}}].

\bibitem{Simon:2016htt}
F.~Simon, \emph{{A First Look at the Impact of NNNLO Theory Uncertainties on
  Top Mass Measurements at the ILC}},  in
  \emph{{\href{http://www.slac.stanford.edu/econf/C151102.1/}{Proc. Int.
  Workshop on Future Linear Colliders (LCWS15)}, Whistler, B.C., Canada}},
  2016.

\bibitem{Simon:2016pwp}
F.~Simon, \emph{{Impact of Theory Uncertainties on the Precision of the Top
  Quark Mass in a Threshold Scan at Future e+e- Colliders}}, {\emph{PoS
  ICHEP2016} (2017) 872} [\href{https://arxiv.org/abs/1611.03399}{{\ttfamily
  1611.03399}}].

\bibitem{Lukic:2013fw}
S.~Luki\'c, I.~Bo\v{z}ovi\'{c}~Jelisav\v{c}i\'{c}, M.~Pandurovi\'c and
  I.~Smiljani\'c, \emph{{Correction of beam-beam effects in luminosity
  measurement in the forward region at CLIC}},
  \href{https://doi.org/10.1088/1748-0221/8/05/P05008}{\emph{JINST} {\bfseries
  8} (2013) P05008} [\href{https://arxiv.org/abs/1301.1449}{{\ttfamily
  1301.1449}}].

\bibitem{Bozovic-Jelisavcic:2013aca}
I.~Bo\v{z}ovi\'{c}~Jelisav\v{c}i\'{c}, S.~Luki\'c,
  G.~Milutinovi\'c~Dumbelovi\'c, M.~Pandurovi\'c and I.~Smiljani\'c,
  \emph{{Luminosity measurement at ILC}},
  \href{https://doi.org/10.1088/1748-0221/8/08/P08012}{\emph{JINST} {\bfseries
  8} (2013) P08012} [\href{https://arxiv.org/abs/1304.4082}{{\ttfamily
  1304.4082}}].

\bibitem{Hoang:2008qy}
A.~H. Hoang, V.~Mateu and S.~Mohammad~Zebarjad, \emph{{{Heavy Quark Vacuum
  Polarization Function at O($\alpha^2(s)$) and O($\alpha^3(s)$)}}},
  \href{https://doi.org/10.1016/j.nuclphysb.2008.12.005}{\emph{Nucl. Phys.}
  {\bfseries B813} (2009) 349}
  [\href{https://arxiv.org/abs/0807.4173}{{\ttfamily 0807.4173}}].

\bibitem{Maier:2017ypu}
A.~Maier and P.~Marquard, \emph{{{Validity of Pad\'e approximations in vacuum
  polarization at three- and four-loop order}}},
  \href{https://doi.org/10.1103/PhysRevD.97.056016}{\emph{Phys. Rev.}
  {\bfseries D97} (2018) 056016}
  [\href{https://arxiv.org/abs/1710.03724}{{\ttfamily 1710.03724}}].

\bibitem{Widl:2018}
A.~Widl, A.~H. Hoang and V.~Matheu, \emph{{Private communication}},  2018.

\bibitem{Hoang:1998ng}
A.~H. Hoang, Z.~Ligeti and A.~Manohar, \emph{{B Decay and the $\Upsilon$
  Mass}}, \href{https://doi.org/10.1103/PhysRevLett.82.277}{\emph{Phys. Rev.
  Lett.} {\bfseries 82} (1999) 277}.

\bibitem{Hoang:1998hm}
A.~H. Hoang, Z.~Ligeti and A.~Manohar, \emph{B decays in the upsilon
  expansion}, \href{https://doi.org/10.1103/PhysRevD.59.074017}{\emph{Phys.
  Rev. D} {\bfseries 59} (1999) 074017}.

\bibitem{Hoang:1999ye}
A.~H. Hoang, \emph{{1S and $\overline{\textrm{MS}}$ bottom quark mass from
  $\Upsilon$ sum rules}},
  \href{https://doi.org/10.1103/PhysRevD.61.034005}{\emph{Phys. Rev. D}
  {\bfseries 61} (1999) 034005}.

\bibitem{Hoang:2008yj}
A.~H. Hoang, A.~Jain, I.~Scimemi and I.~W. Stewart, \emph{{Infrared
  Renormalization Group Flow for Heavy Quark Masses}},
  \href{https://doi.org/10.1103/PhysRevLett.101.151602}{\emph{Phys. Rev. Lett.}
  {\bfseries 101} (2008) 151602}
  [\href{https://arxiv.org/abs/0803.4214}{{\ttfamily 0803.4214}}].

\bibitem{Hoang:2017suc}
A.~H. Hoang, A.~Jain, C.~Lepenik, V.~Mateu, M.~Preisser, I.~Scimemi et~al.,
  \emph{{The MSR mass and the $
  \mathcal{O}\left({\Lambda}_{\mathrm{QCD}}\right) $ renormalon sum rule}},
  \href{https://doi.org/10.1007/JHEP04(2018)003}{\emph{JHEP} {\bfseries 04}
  (2018) 003} [\href{https://arxiv.org/abs/1704.01580}{{\ttfamily
  1704.01580}}].

\bibitem{Hoang:2017btd}
A.~H. Hoang, C.~Lepenik and M.~Preisser, \emph{{On the light massive flavor
  dependence of the large order asymptotic behavior and the ambiguity of the
  pole mass}}, \href{https://doi.org/10.1007/JHEP09(2017)099}{\emph{JHEP}
  {\bfseries 09} (2017) 099}
  [\href{https://arxiv.org/abs/1706.08526}{{\ttfamily 1706.08526}}].

\bibitem{Adloff:2008aa}
{\scshape CALICE} collaboration, \emph{{Response of the CALICE Si-W
  electromagnetic calorimeter physics prototype to electrons}},
  \href{https://doi.org/10.1016/j.nima.2009.07.026}{\emph{Nucl. Instrum. Meth.}
  {\bfseries A608} (2009) 372}
  [\href{https://arxiv.org/abs/0811.2354}{{\ttfamily 0811.2354}}].

\bibitem{Hoang:2012us}
A.~H. Hoang, P.~Ruiz-Femen{\'\i}a and M.~Stahlhofen, \emph{{Renormalization
  group improved bottom mass from $\Upsilon$ sum rules at NNLL order}},
  \href{https://doi.org/10.1007/JHEP10(2012)188}{\emph{JHEP} {\bfseries 2012}
  (2012) } [\href{https://arxiv.org/abs/1209.0450}{{\ttfamily 1209.0450}}].

\bibitem{Randall:1999ee}
L.~Randall and R.~Sundrum, \emph{{A Large mass hierarchy from a small extra
  dimension}}, \href{https://doi.org/10.1103/PhysRevLett.83.3370}{\emph{Phys.
  Rev. Lett.} {\bfseries 83} (1999) 3370}
  [\href{https://arxiv.org/abs/hep-ph/9905221}{{\ttfamily hep-ph/9905221}}].

\bibitem{Pomarol:2008bh}
A.~Pomarol and J.~Serra, \emph{{Top quark compositeness: Feasibility and
  implications}}, \href{https://doi.org/10.1103/PhysRevD.78.074026}{\emph{Phys.
  Rev.} {\bfseries D78} (2008) 074026}
  [\href{https://arxiv.org/abs/0806.3247}{{\ttfamily 0806.3247}}].

\bibitem{ALEPH:2005ab}
{\scshape LEP Electroweak Working Group, SLD Electroweak and Heavy Flavour
  Groups; ALEPH, DELPHI, L3, OPAL, and SLD} collaboration, \emph{{Precision
  electroweak measurements on the $Z$ resonance}},
  \href{https://doi.org/10.1016/j.physrep.2005.12.006}{\emph{Phys. Rept.}
  {\bfseries 427} (2006) 257}
  [\href{https://arxiv.org/abs/hep-ex/0509008}{{\ttfamily hep-ex/0509008}}].

\bibitem{Amjad:2015mma}
M.~S. Amjad et~al., \emph{{A precise characterisation of the top quark
  electro-weak vertices at the ILC}},
  \href{https://doi.org/10.1140/epjc/s10052-015-3746-5}{\emph{Eur. Phys. J.}
  {\bfseries C75} (2015) 512}
  [\href{https://arxiv.org/abs/1505.06020}{{\ttfamily 1505.06020}}].

\bibitem{AguilarSaavedra:2012vh}
J.~A. Aguilar-Saavedra, M.~C.~N. Fiolhais and A.~Onofre, \emph{{Top Effective
  Operators at the ILC}},
  \href{https://doi.org/10.1007/JHEP07(2012)180}{\emph{JHEP} {\bfseries 07}
  (2012) 180} [\href{https://arxiv.org/abs/1206.1033}{{\ttfamily 1206.1033}}].

\bibitem{PerelloVosZhang}
G.~Durieux, M.~Perell{\'o}, M.~Vos and C.~Zhang, \emph{{Global and optimal
  probes for the top-quark effective field theory at future lepton colliders}},
  \href{https://doi.org/10.1007/JHEP10(2018)168}{\emph{JHEP} {\bfseries 10}
  (2018) 168} [\href{https://arxiv.org/abs/1807.02121}{{\ttfamily
  1807.02121}}].

\bibitem{Bernreuther:2017cyi}
W.~Bernreuther, L.~Chen, I.~Garc{\'\i}a, M.~Perell{\'o}, R.~Poeschl, F.~Richard
  et~al., \emph{{CP-violating top quark couplings at future linear $e^+e^-$
  colliders}}, \href{https://doi.org/10.1140/epjc/s10052-018-5625-3}{\emph{Eur.
  Phys. J.} {\bfseries C78} (2018) 155}
  [\href{https://arxiv.org/abs/1710.06737}{{\ttfamily 1710.06737}}].

\bibitem{Atwood:1991ka}
D.~Atwood and A.~Soni, \emph{{Analysis for magnetic moment and electric dipole
  moment form-factors of the top quark via $e^+ e^- \to t\overline{t}$}},
  \href{https://doi.org/10.1103/PhysRevD.45.2405}{\emph{Phys. Rev.} {\bfseries
  D45} (1992) 2405}.

\bibitem{Davier:1992nw}
M.~Davier, L.~Duflot, F.~Le~Diberder and A.~Rouge, \emph{{The optimal method
  for the measurement of tau polarization}},
  \href{https://doi.org/10.1016/0370-2693(93)90101-M}{\emph{Phys. Lett.}
  {\bfseries B306} (1993) 411}.

\bibitem{Diehl:1993br}
M.~Diehl and O.~Nachtmann, \emph{{Optimal observables for the measurement of
  three gauge boson couplings in $e^+ e^- \to W^+ W^-$}},
  \href{https://doi.org/10.1007/BF01555899}{\emph{Z. Phys.} {\bfseries C62}
  (1994) 397}.

\bibitem{Grzadkowski:2000nx}
B.~Grzadkowski and Z.~Hioki, \emph{{Optimal observable analysis of the angular
  and energy distributions for top quark decay products at polarized linear
  colliders}}, \href{https://doi.org/10.1016/S0550-3213(00)00385-0}{\emph{Nucl.
  Phys.} {\bfseries B585} (2000) 3}
  [\href{https://arxiv.org/abs/hep-ph/0004223}{{\ttfamily hep-ph/0004223}}].

\bibitem{Janot:2015yza}
P.~Janot, \emph{{Top-quark electroweak couplings at the FCC-ee}},
  \href{https://doi.org/10.1007/JHEP04(2015)182}{\emph{JHEP} {\bfseries 04}
  (2015) 182} [\href{https://arxiv.org/abs/1503.01325}{{\ttfamily
  1503.01325}}].

\bibitem{Khiem:2015ofa}
P.~H. Khiem, E.~Kou, Y.~Kurihara and F.~Le~Diberder, \emph{{Probing New Physics
  using top quark polarization in the $e^+e^- \to t\overline{t}$ process at
  future Linear Colliders}},  in
  \emph{{\href{https://arxiv.org/abs/1503.04247}{Proc. TYL-FJPPL workshops on
  "Top Physics at ILC"}}}, 2015.

\bibitem{garcia:2016}
I.~Garc{\'\i}a, \emph{Future Linear Colliders: Detector R\&D, Jet
  Reconstruction and Top Physics potential}, Ph.D. thesis, Universidad de
  Valencia, Valencia,
  \href{http://cds.cern.ch/record/2239794}{CERN-THESIS-2016-214}, 2016.

\bibitem{winter:2018}
A.~Winter, \emph{{Prospects for Higgs boson \& top quark measurements and
  applications of digital calorimetry at future linear colliders}}, Ph.D.
  thesis, {University of Birmingham, Birmingham},
  \href{http://etheses.bham.ac.uk/id/eprint/8458}{URI}, 2018.

\bibitem{Thaler:2010tr}
J.~Thaler and K.~Van~Tilburg, \emph{{Identifying Boosted Objects with
  N-subjettiness}}, \href{https://doi.org/10.1007/JHEP03(2011)015}{\emph{JHEP}
  {\bfseries 03} (2011) 15} [\href{https://arxiv.org/abs/1011.2268}{{\ttfamily
  1011.2268}}].

\bibitem{List:88030}
B.~List and J.~List, \emph{Marlin{K}infit: An object--oriented kinematic
  fitting package,
  \href{http://bib-pubdb1.desy.de/record/88030}{LC-TOOL-2009-001}},  2009.

\bibitem{LCD:tth_1400}
S.~Redford, P.~Roloff and M.~Vogel, \emph{{Physics potential of the top Yukawa
  coupling measurement at a 1.4\,\TeV Compact Linear Collider using the
  \clicsid detector,
  \href{http://cds.cern.ch/record/1690648}{CLICdp-Note-2014-001}}},  2014.

\bibitem{LCD:tth_backgrounds_1400}
S.~Redford, P.~Roloff and M.~Vogel, \emph{{Study of the effect of additional
  background channels on the top Yukawa coupling measurement at a 1.4 TeV CLIC
  \href{http://cds.cern.ch/record/1982243/}{CLICdp-Note-2015-001}}},  2015.

\bibitem{ReuterRothe}
J.~Reuter and V.~Rothe, \emph{{Private communication}},  2018.

\bibitem{Djouadi:1992gp}
A.~Djouadi, J.~Kalinowski and P.~M. Zerwas, \emph{{Measuring the $\PH\PQt\PAQt$
  coupling in $\epem$ collisions}},
  \href{https://doi.org/10.1142/S0217732392001464}{\emph{Mod. Phys. Lett.}
  {\bfseries A7} (1992) 1765}.

\bibitem{Boselli:2018zxr}
S.~Boselli, R.~Hunter and A.~Mitov, \emph{{Prospects for the determination of
  the top-quark Yukawa coupling at future $e^+e^-$ colliders}},
  {\emph{\rm{Submitted to JHEP}} (2018) }
  [\href{https://arxiv.org/abs/1805.12027}{{\ttfamily 1805.12027}}].

\bibitem{Godbole:2011hw}
R.~M. Godbole, C.~Hangst, M.~Muhlleitner, S.~D. Rindani and P.~Sharma,
  \emph{{Model-independent analysis of Higgs spin and CP properties in the
  process $e^+ e^- \to t \bar t \Phi$}},
  \href{https://doi.org/10.1140/epjc/s10052-011-1681-7}{\emph{Eur. Phys. J.}
  {\bfseries C71} (2011) 1681}
  [\href{https://arxiv.org/abs/1103.5404}{{\ttfamily 1103.5404}}].

\bibitem{Alloul:2013naa}
A.~Alloul, B.~Fuks and V.~Sanz, \emph{{Phenomenology of the Higgs effective
  Lagrangian via \sc{FeynRules}}},
  \href{https://doi.org/10.1007/JHEP04(2014)110}{\emph{JHEP} {\bfseries 04}
  (2014) 110} [\href{https://arxiv.org/abs/1310.5150}{{\ttfamily 1310.5150}}].

\bibitem{Collaboration:2293646}
C.~Collaboration, \emph{{The Phase-2 Upgrade of the CMS Endcap Calorimeter,
  \href{https://cds.cern.ch/record/2293646}{CERN-LHCC-2017-023}}}. {CERN},
  Geneva, Nov, 2017.

\bibitem{ATL-PHYS-PUB-2016-019}
{ATLAS Collaboration}, \emph{{Expected sensitivity of ATLAS to FCNC top quark
  decays $t \rightarrow Zu$ and $t \rightarrow Hq$ at the High Luminosity LHC,
  \href{https://cds.cern.ch/record/2209126}{ATL-PHYS-PUB-2016-019}}},  2016.

\bibitem{Moretti:2001zz}
M.~Moretti, T.~Ohl and J.~Reuter, \emph{{O'Mega: An Optimizing matrix element
  generator}},  in
  \emph{{\href{https://inspirehep.net/record/1409349}{Proceedings, Physics and
  Experimentation at a Linear Electron-Positron Collider, 2nd ECFA/DESY Study},
  Lund, Sweden}}, pp.~1981--2009, 1998-2001.

\bibitem{Read:2002hq}
A.~L. Read, \emph{{Presentation of search results: The CL(s) technique}},
  \href{https://doi.org/10.1088/0954-3899/28/10/313}{\emph{J. Phys.} {\bfseries
  G28} (2002) 2693}.

\bibitem{Moneta:2010pm}
L.~Moneta, K.~Belasco, K.~S. Cranmer, S.~Kreiss, A.~Lazzaro, D.~Piparo et~al.,
  \emph{{The RooStats Project}}, {\emph{PoS ACAT2010} (2010) 057}
  [\href{https://arxiv.org/abs/1009.1003}{{\ttfamily 1009.1003}}].

\bibitem{Khanpour:2014xla}
H.~Khanpour, S.~Khatibi, M.~Khatiri~Yanehsari and M.~Mohammadi~Najafabadi,
  \emph{{Single top quark production as a probe of anomalous $tq\gamma$ and
  $tqZ$ couplings at the FCC-ee}},
  \href{https://doi.org/10.1016/j.physletb.2017.10.047}{\emph{Phys. Lett.}
  {\bfseries B775} (2017) 25}
  [\href{https://arxiv.org/abs/1408.2090}{{\ttfamily 1408.2090}}].

\bibitem{Atwood:1996vj}
D.~Atwood, L.~Reina and A.~Soni, \emph{{Phenomenology of two Higgs doublet
  models with flavor changing neutral currents}},
  \href{https://doi.org/10.1103/PhysRevD.55.3156}{\emph{Phys. Rev.} {\bfseries
  D55} (1997) 3156} [\href{https://arxiv.org/abs/hep-ph/9609279}{{\ttfamily
  hep-ph/9609279}}].

\bibitem{Staub:2015kfa}
F.~Staub, \emph{{Exploring new models in all detail with SARAH}},
  \href{https://doi.org/10.1155/2015/840780}{\emph{Adv. High Energy Phys.}
  {\bfseries 2015} (2015) 840780}
  [\href{https://arxiv.org/abs/1503.04200}{{\ttfamily 1503.04200}}].

\bibitem{teft_EW}
O.~Bessidskaia~Bylund, F.~Maltoni, I.~Tsinikos, E.~Vryonidou and C.~Zhang,
  \emph{{Probing top quark neutral couplings in the Standard Model Effective
  Field Theory at NLO in QCD}},
  \href{https://doi.org/10.1007/JHEP05(2016)052}{\emph{JHEP} {\bfseries 05}
  (2016) 052} [\href{https://arxiv.org/abs/1601.08193}{{\ttfamily
  1601.08193}}].

\bibitem{Panico:2015jxa}
G.~Panico and A.~Wulzer, \emph{{The Composite Nambu-Goldstone Higgs}},
  \href{https://doi.org/10.1007/978-3-319-22617-0}{\emph{Lect. Notes Phys.}
  {\bfseries 913} (2016) } [\href{https://arxiv.org/abs/1506.01961}{{\ttfamily
  1506.01961}}].

\bibitem{Giudice:2007fh}
G.~F. Giudice, C.~Grojean, A.~Pomarol and R.~Rattazzi, \emph{{The
  strongly-interacting light Higgs}},
  \href{https://doi.org/10.1088/1126-6708/2007/06/045}{\emph{JHEP} {\bfseries
  06} (2007) 045} [\href{https://arxiv.org/abs/hep-ph/0703164}{{\ttfamily
  hep-ph/0703164}}].

\bibitem{Rattazzi:2008pe}
R.~Rattazzi, V.~S. Rychkov, E.~Tonni and A.~Vichi, \emph{{Bounding scalar
  operator dimensions in 4D CFT}},
  \href{https://doi.org/10.1088/1126-6708/2008/12/031}{\emph{JHEP} {\bfseries
  12} (2008) 031} [\href{https://arxiv.org/abs/0807.0004}{{\ttfamily
  0807.0004}}].

\bibitem{Agashe:2006at}
K.~Agashe, R.~Contino, L.~Da~Rold and A.~Pomarol, \emph{{A Custodial symmetry
  for $Zb \bar b$}},
  \href{https://doi.org/10.1016/j.physletb.2006.08.005}{\emph{Phys. Lett.}
  {\bfseries B641} (2006) 62}
  [\href{https://arxiv.org/abs/hep-ph/0605341}{{\ttfamily hep-ph/0605341}}].

\end{thebibliography}\endgroup

\clearpage
\appendix
\section{Helicity amplitudes in top-quark pair production} \label{sec:hamp}

We list below the helicity amplitudes $\widehat{\mathcal{M}}$ that appear in \autoref{eq:diffcsttbar}:

\begin{adjustbox}{width=0.85\textwidth, left}
%\resizebox{0.9\textwidth}{!}{
\begin{minipage}{\columnwidth}

\begin{eqnarray}\label{eq:amp}
\left|{\mathcal{\widehat{M}}}\left(-\frac12,+\frac12,-\frac12,+\frac12\right)\right| &=& \left| 2e^2\left({\mathcal{F}}^L_{1V}-\beta{\mathcal{F}}^L_{1A}+{\mathcal{F}}^L_{2V}\right) + (1+\beta)C_{LL}s+(1-\beta) C_{LR}s\right|,\hspace{2cm}\nonumber\\
\left|{\mathcal{\widehat{M}}}\left(-\frac12,+\frac12,+\frac12,-\frac12\right)\right| &=& \left| 2e^2\left({\mathcal{F}}^L_{1V}+\beta{\mathcal{F}}^L_{1A}+{\mathcal{F}}^L_{2V}\right) + (1-\beta)C_{LL}s+(1+\beta) C_{LR}s\right|,\nonumber\\
\left|{\mathcal{\widehat{M}}}\left(+\frac12,-\frac12,+\frac12,-\frac12\right)\right| &=& \left| 2e^2\left({\mathcal{F}}^R_{1V}+\beta{\mathcal{F}}^R_{1A}+{\mathcal{F}}^R_{2V}\right) + (1-\beta)C_{RL}s+(1+\beta) C_{RR}s\right|,\nonumber\\
\left|{\mathcal{\widehat{M}}}\left(+\frac12,-\frac12,-\frac12,+\frac12\right)\right| &=& \left| 2e^2\left({\mathcal{F}}^R_{1V}-\beta{\mathcal{F}}^R_{1A}+{\mathcal{F}}^R_{2V}\right) + (1+\beta)C_{RL}s+(1-\beta) C_{RR}s\right|,\nonumber\\
\left|{\mathcal{\widehat{M}}}\left(-\frac12,+\frac12,-\frac12,-\frac12\right)\right| &=& \left| \frac{2\sqrt{2}m_t}{\sqrt{s}}e^2\left[{\mathcal{F}}^L_{1V}+\frac{s}{4m_t^2}\left(F_{2V}^L+\beta{\mathcal{F}}^L_{2A}\right)\right] + \sqrt{2}m_t C_{LL}\sqrt{s}+\sqrt{2}m_t C_{LR}\sqrt{s} \right|,\nonumber\\
\left|{\mathcal{\widehat{M}}}\left(-\frac12,+\frac12,+\frac12,+\frac12\right)\right| &=& \left| \frac{2\sqrt{2}m_t}{\sqrt{s}}e^2\left[{\mathcal{F}}^L_{1V}+\frac{s}{4m_t^2}\left(F_{2V}^L-\beta{\mathcal{F}}^L_{2A}\right)\right] + \sqrt{2}m_t C_{LL}\sqrt{s}+\sqrt{2}m_t C_{LR}\sqrt{s} \right|,\nonumber\\
\left|{\mathcal{\widehat{M}}}\left(+\frac12,-\frac12,-\frac12,-\frac12\right)\right| &=& \left| \frac{2\sqrt{2}m_t}{\sqrt{s}}e^2\left[{\mathcal{F}}^R_{1V}+\frac{s}{4m_t^2}\left(F_{2V}^R+\beta{\mathcal{F}}^R_{2A}\right)\right] + \sqrt{2}m_t C_{RL}\sqrt{s}+\sqrt{2}m_t C_{RR}\sqrt{s} \right|,\nonumber\\
\left|{\mathcal{\widehat{M}}}\left(+\frac12,-\frac12,+\frac12,+\frac12\right)\right| &=& \left| \frac{2\sqrt{2}m_t}{\sqrt{s}}e^2\left[{\mathcal{F}}^R_{1V}+\frac{s}{4m_t^2}\left(F_{2V}^R-\beta{\mathcal{F}}^R_{2A}\right)\right] + \sqrt{2}m_t C_{RL}\sqrt{s}+\sqrt{2}m_t C_{RR}\sqrt{s} \right|.\hspace{2cm}
\end{eqnarray}

\end{minipage}
%}
\end{adjustbox}

\vspace{5mm}
where the following definitions are used (same as in \cite{Schmidt:1995mr}):
\begin{align}
{\mathcal{F}}^L_{ij} &= -F^{\PGg}_{ij}+\left(\frac{-\frac12+\sW^2}{\sW\cW}\right)\left(\frac{s}{s-m_{\PZ}^2}\right)F_{ij}^{\PZ}, \nonumber\\[1ex]
{\mathcal{F}}^R_{ij} &= -F^{\PGg}_{ij}+\left(\frac{\sW^2}{\sW\cW}\right)\left(\frac{s}{s-m_{\PZ}^2}\right)F_{ij}^{\PZ}.\nonumber
\end{align}

\clearpage

\section{Additional event selection summary tables}

\begin{table}[h]
\centering
\begin{minipage}{\columnwidth}
\resizebox{1.0\textwidth}{!}{
\begin{tabular}{lcccccccc}
\toprule
\vspace{1.0mm}
{} & \multicolumn{2}{c}{$\sigma\,[\fb]$} & \multicolumn{2}{c}{$\epsilon_{\,\mathrm{Pre}}\,[\%]$} & \multicolumn{2}{c}{$\epsilon_{\,\mathrm{MVA}}\,[\%]$} & \multicolumn{2}{c}{$N$} \rule{0pt}{3ex} \\
\vspace{1.0mm}
P(\Pem) & -80\% & +80\% & -80\% & +80\% & -80\% & +80\% & -80\% & +80\% \\
Process \\
\midrule
$\epem(\to\ttbar)\to\PQq\PQq\PQq\PQq\Pl\PGn\,(\Pl=\Pe,\PGm)$\footnote{Kinematic region defined as $\rootsprime\in[400,900)\,\tev$} & 16.6 & 8.7 & 4.0 & 5.0 & 90 & 92 & 1,199 & 200 \\
\midrule
$\epem(\to\ttbar)\to\PQq\PQq\PQq\PQq\Pl\PGn\,(\Pl=\Pe,\PGm)$\footnote{$\rootsprime\not\in[400,900)\,\tev$} & 30.2 & 16.0 & 0.49 & 0.51 & 79 & 82 & 235 & 33 \\
$\epem(\to\ttbar)\to\PQq\PQq\PQq\PQq\Pl\PGn\,(\Pl=\PGt)$ & 23.2 & 12.3 & 0.65 & 0.65 & 45 & 50 & 137 & 20 \\
$\epem(\not\to\ttbar)\to\PQq\PQq\PQq\PQq\Pl\PGn$ & 72.3 & 16.5 & 0.42 & 0.63 & 62 & 73 & 371 & 38 \\
$\epem\to\PQq\PQq\PQq\PQq\PQq\PQq$ & 116 & 44.9 & 0.26 & 0.29 & 17 & 17 & 105 & 11 \\
$\epem\to\PQq\PQq\Pl\PGn\Pl\PGn$ & 44.1 & 15.3 & 0.68 & 1.3 & 73 & 64 & 432 & 66 \\
$\epem\to\PQq\PQq\PQq\PQq$ & 2,300 & 347 & 0.017 & 0.033 & 4.6 & 14 & 35 & 8 \\
$\epem\to\PQq\PQq\Pl\PGn$ & 6,980 & 1,640 & 0.0051 & 0.0044 & 3.4 & 18 & 24 & 7 \\
$\epem\to\PQq\PQq\Pl\Pl$ & 2,680 & 2,530 & 0.0041 & 0.0022 & 13 & 25 & 27 & 7 \\
$\epem\to\PQq\PQq$ & 4,840 & 3,170 & 0.015 & 0.0074 & 10 & 7.9 & 147 & 9 \\
\bottomrule
\end{tabular}
}
\end{minipage}
\caption{Pre-selection and final event selection efficiencies and expected number of events for the analysis of $\ttbar$ production with radiative events in the interval $400\,\gev\leq\rootsprime<900\,\gev$, assuming an integrated luminosity of $2.0\,\abinv$ and $0.5\,\abinv$ for $P(\Pem)=\text{-}80\%$ and $P(\Pem)=\text{+}80\%$, respectively. \label{tab:ttbar:radiative:selection:bin3}}
\end{table}

\begin{table}[h]
\centering
\begin{minipage}{\columnwidth}
\resizebox{1.0\textwidth}{!}{
\begin{tabular}{lcccccccc}
\toprule
\vspace{1.0mm}
{} & \multicolumn{2}{c}{$\sigma\,[\fb]$} & \multicolumn{2}{c}{$\epsilon_{\,\mathrm{Pre}}\,[\%]$} & \multicolumn{2}{c}{$\epsilon_{\,\mathrm{MVA}}\,[\%]$} & \multicolumn{2}{c}{$N$} \rule{0pt}{3ex} \\
\vspace{1.0mm}
P(\Pem) & -80\% & +80\% & -80\% & +80\% & -80\% & +80\% & -80\% & +80\% \\
Process \\
\midrule
$\epem(\to\ttbar)\to\PQq\PQq\PQq\PQq\Pl\PGn\,(\Pl=\Pe,\PGm)$\footnote{Kinematic region defined as $\rootsprime\in[900,1200)\,\tev$} & 11.0 & 5.79 & 33 & 30 & 86 & 85 & 6,271 & 744 \\
\midrule
$\epem(\to\ttbar)\to\PQq\PQq\PQq\PQq\Pl\PGn\,(\Pl=\Pe,\PGm)$\footnote{$\rootsprime\not\in[900,1200)\,\tev$} & 35.8 & 18.9 & 6.3 & 5.5 & 80 & 81 & 3,598 & 420 \\
$\epem(\to\ttbar)\to\PQq\PQq\PQq\PQq\Pl\PGn\,(\Pl=\PGt)$ & 23.2 & 12.3 & 10 & 9.5 & 34 & 27 & 1,647 & 161 \\
$\epem(\not\to\ttbar)\to\PQq\PQq\PQq\PQq\Pl\PGn$ & 72.3 & 16.5 & 3.7 & 9.5 & 49 & 34 & 2,652 & 270 \\
$\epem\to\PQq\PQq\PQq\PQq\PQq\PQq$ & 116 & 44.9 & 2.5 & 2.8 & 7.9 & 7.3 & 454 & 46 \\
$\epem\to\PQq\PQq\Pl\PGn\Pl\PGn$ & 44.1 & 15.3 & 3.0 & 4.8 & 62 & 55.9 & 1,652 & 207 \\
$\epem\to\PQq\PQq\PQq\PQq$ & 2,300 & 347 & 0.24 & 0.39 & 2.2 & 3.2 & 241 & 22 \\
$\epem\to\PQq\PQq\Pl\PGn$ & 6,980 & 1,640 & 0.042 & 0.021 & 3.2 & 8.6 & 185 & 15 \\
$\epem\to\PQq\PQq\Pl\Pl$ & 2,680 & 2,530 & 0.024 & 0.016 & 6.4 & 6.8 & 82 & 14 \\
$\epem\to\PQq\PQq$ & 4,840 & 3,170 & 0.18 & 0.14 & 4.4 & 6.4 & 777 & 145 \\
\bottomrule
\end{tabular}
}
\end{minipage}
\caption{Pre-selection and final event selection efficiencies and expected number of events for the analysis of $\ttbar$ production with radiative events in the interval $900\,\gev\leq\rootsprime<1200\,\gev$, assuming an integrated luminosity of $2.0\,\abinv$ and $0.5\,\abinv$ for $P(\Pem)=\text{-}80\%$ and $P(\Pem)=\text{+}80\%$, respectively. \label{tab:ttbar:radiative:selection:bin2}}
\end{table}

\begin{table}[h]
\centering
\begin{minipage}{\columnwidth}
\resizebox{1.0\textwidth}{!}{
\begin{tabular}{lcccccccc}
\toprule
\vspace{1.0mm}
{} & \multicolumn{2}{c}{$\sigma\,[\fb]$} & \multicolumn{2}{c}{$\epsilon_{\,\mathrm{Pre}}\,[\%]$} & \multicolumn{2}{c}{$\epsilon_{\,\mathrm{MVA}}\,[\%]$} & \multicolumn{2}{c}{$N$} \rule{0pt}{3ex} \\
\vspace{1.0mm}
P(\Pem) & -80\% & +80\% & -80\% & +80\% & -80\% & +80\% & -80\% & +80\% \\
Process \\
\midrule
$\epem(\to\ttbar)\to\PQq\PQq\PQq\PQq\Pl\PGn\,(\Pl=\Pe,\PGm)$\footnote{Kinematic region defined as $\rootsprime\geq1.2\,\tev$} & 18.4 & 9.83 & 37 & 34 & 86 & 88 & 11,598 & 1,496 \\
\midrule
$\epem(\to\ttbar)\to\PQq\PQq\PQq\PQq\Pl\PGn\,(\Pl=\Pe,\PGm)$\footnote{$\rootsprime<1.2\,\tev$} & 28.5 & 14.9 & 3.1 & 3.3 & 83 & 86 & 1,468 & 209 \\
$\epem(\to\ttbar)\to\PQq\PQq\PQq\PQq\Pl\PGn\,(\Pl=\PGt)$ & 23.2 & 12.3 & 12 & 13 & 31 & 27 & 1,705 & 209 \\
$\epem(\not\to\ttbar)\to\PQq\PQq\PQq\PQq\Pl\PGn$ & 72.2 & 16.5 & 4.8 & 7.1 & 47 & 63 & 3,283 & 369 \\
$\epem\to\PQq\PQq\PQq\PQq\PQq\PQq$ & 116 & 44.9 & 2.2 & 2.2 & 5.2 & 5.9 & 272 & 29 \\
$\epem\to\PQq\PQq\Pl\PGn\Pl\PGn$ & 44.1 & 15.3 & 1.5 & 2.4 & 56 & 65 & 740 & 122 \\
$\epem\to\PQq\PQq\PQq\PQq$ & 2,300 & 347 & 0.45 & 0.71 & 1.1 & 2.0 & 217 & 25 \\
$\epem\to\PQq\PQq\Pl\PGn$ & 6,980 & 1,640 & 0.048 & 0.026 & 2.2 & 6.2 & 145 & 13 \\
$\epem\to\PQq\PQq\Pl\Pl$ & 2,680 & 2,530 & 0.031 & 0.021 & 3.8 & 6.5 & 64 & 18 \\
$\epem\to\PQq\PQq$ & 4,840 & 3,170 & 0.30 & 0.21 & 1.6 & 2.7 & 466 & 86 \\
\bottomrule
\end{tabular}
}
\end{minipage}
\caption{Pre-selection and final event selection efficiencies and expected number of events for the analysis of $\ttbar$ production with radiative events in the interval $\rootsprime\geq1200\,\gev$, assuming an integrated luminosity of $2.0\,\abinv$ and $0.5\,\abinv$ for $P(\Pem)=\text{-}80\%$ and $P(\Pem)=\text{+}80\%$, respectively. 
\label{tab:ttbar:radiative:selection:bin1}}
\end{table}

\end{document}